\documentclass[aps,prb,longbibliography,twocolumn]{revtex4-2}

\usepackage{array}
\usepackage{amsmath}
\usepackage{comment}
\usepackage{amssymb} 
\usepackage{amsthm} 
\usepackage{array}
\usepackage{textcomp}
\usepackage{upgreek}
\usepackage{units}
\usepackage[pdftex]{graphicx}
\usepackage{microtype}
\usepackage{enumitem}
\usepackage{xcolor}
\usepackage{gnuplottex}    
\usepackage{multirow} 
\usepackage{sansmath}
\usepackage{txfonts}
\usepackage{float}
\usepackage{adjustbox}

\usepackage{tikz}
\usepackage{pgfplots}
\pgfplotsset{compat=1.18} 
\usetikzlibrary{ decorations.markings, patterns}
\usetikzlibrary{decorations.markings}
\usetikzlibrary{arrows, positioning, shapes.geometric}
\usepackage[bookmarksnumbered,pdfpagelabels=true,plainpages=false,colorlinks=true,linkcolor=blue,citecolor=red,urlcolor=blue]{hyperref}
\usetikzlibrary { decorations.pathmorphing, decorations.pathreplacing, decorations.shapes} 
\usetikzlibrary{decorations.pathreplacing,calligraphy}
\usetikzlibrary{intersections,calc}
\usetikzlibrary{angles,quotes, decorations.pathreplacing}
\usetikzlibrary{fit}
\usetikzlibrary{arrows}
\usepackage{hyperref}
\hypersetup{
    colorlinks=true,
    linkcolor=blue,
    filecolor=magenta,      
    urlcolor=cyan
    }

\renewcommand{\v}[1]{\ensuremath{\mathbf{#1}}} 
 
\newcommand{\ket}[1]{\left| #1 \right>} 
\newcommand{\bra}[1]{\left< #1 \right|} 
\let\baraccent=\= 
\renewcommand{\=}[1]{\stackrel{#1}{=}} 

\theoremstyle{definition}

\theoremstyle{remark}

\renewcommand{\vec}[1]{\ensuremath{\boldsymbol{#1}}}

\usepackage{pifont}
\begin{document}

\title{
Thermal Hall Effect of Magnons from Many-Body Skew Scattering
}

\author{Dimos Chatzichrysafis}
\affiliation{Institute of Physics, Johannes Gutenberg University Mainz, 55128 Mainz, Germany}

\author{Alexander Mook}
\affiliation{Institute of Physics, Johannes Gutenberg University Mainz, 55128 Mainz, Germany}

\begin{abstract}
    We present a theory of magnonic thermal Hall transport driven by many-body skew scattering. In field-polarized chiral magnets, the Dzyaloshinskii-Moriya interaction induces three-magnon interactions that violate time-reversal symmetry and interfere with the four-magnon scattering, thereby breaking microscopic detailed balance and causing thermal Hall currents. This mechanism operates without disorder or Berry curvature and is intrinsic to interacting spin systems. We argue that it can be as significant as the band-geometric anomalous velocity, highlighting the importance of magnon interactions in thermal Hall physics.
\end{abstract}

\maketitle 

\section{Introduction}
Spin and heat transport in quantum magnets offer insights into a material's quantum many-body ground state. In this context, one intriguing phenomenon is the thermal Hall effect (THE) \cite{Guo2022review, Zhang2024-reviewTHE}, where a transverse heat current emerges in response to a temperature gradient. This effect occurs spontaneously in materials with broken time-reversal symmetry, such as those with magnetic order. Their collective magnetic excitations known as magnons play a crucial role in contributing to the thermal Hall conductivity, denoted as $\kappa_\text{H}$. Recent research has focused on the \emph{intrinsic} thermal Hall conductivity, $\kappa_\text{H}^\text{int}$, which provides a window into the quantum geometric properties of magnon bands, specifically their Berry curvature \cite{Katsura2010, Matsumoto2011PRL, Matsumoto2011}, as obtained in the limit of non-interacting quasiparticles. 
Thus, the THE serves as a tool for probing the geometry or topology of magnon bands, aiding the detection of topological magnon insulators \cite{Hoogdalem2013, Shindou13, Shindou13b, Shindou14, zhang2013, Mook2014, mook2014edge, Kim2016honey, mcclarty2022ar, owerre2016first, Mook2016Weyl, Mook2019Coplanar, Neumann2022, Hoyer2024}, e.g., in pyrochlore ferromagnets \cite{Onose2010, Ideue2012}, Cu(1,3-benzenedicarboxylate) \cite{chisnell2015topological,Hirschberger2015}, the Kitaev material $\alpha$-RuCl$_3$ \cite{Czajka2022}, and skyrmion crystals \cite{Takeda2024}.

However, discrepancies between experimental  $\kappa_\text{H}$ and predictions based on the non-interacting theory, e.g., in ${\mathrm{Cr}}_{2}{\mathrm{Ge}}_{2}{\mathrm{Te}}_{6}$ \cite{Choi2023damping} and SrCu$_2$(BO$_3$)$_2$ \cite{Romhanyi2015, McClarty2017,Suetsugu2022}, suggest that many-body interactions play a critical role in the spectral and topological properties of magnons \cite{zhitomirskychernyshev2013, Chernyshev2016dampedtopo, McClarty2017, McClarty2018, Pershoguba2018, McClarty2019, Park2020inplaneDMI, mook2021, Lu2021Diracmagnon, Nikitin2022, Mook2023multipolar, Koyama2023edgestates, Gohlke2023PRL, Chen2023dampedAFM, Sun2023interactingDirac, Li2023CherninAFM, Zhu2023-ju, Zhu2024interactions, Habel2024breakdown, Koyama2024, Sourounis2024}. Indeed, Refs.~\onlinecite{Park2020inplaneDMI, mook2021, Lu2021Diracmagnon, Li2023CherninAFM, Mook2023multipolar, Zhu2023-ju, Koyama2024} have shown that within an interacting theory, $\kappa_{\text{H}}^{\text{int}}$ is significantly renormalized. Nonetheless, a comprehensive understanding of interacting magnon thermal Hall physics remains an aspirational goal. Notably, previous studies assumed that the quasiparticles' anomalous velocity, induced by the Berry curvature, remains the sole source of the transverse current even in the presence of interactions.

Herein, we study the magnonic THE beyond band geometry, and argue that many-body interactions of magnons provide a separate source of thermal Hall currents. We develop a theory of magnon-magnon skew-scattering (MMSS)---a dynamical chirality in magnon collisions. Using semi-classical Boltzmann transport theory, we show how magnon-magnon scattering channels interfere to break time-reversal symmetry and microscopic detailed balance. We illustrate our findings with a two-dimensional chiral ferromagnet on a square lattice, where a single magnon band with zero Berry curvature makes MMSS the main contributor to the THE. MMSS is an intrinsic property of the interacting spin system and thus does not require disorder. We propose the experimental observation of MMSS by thermal Hall measurements in ferromagnetic Janus van der Waals materials \cite{Hou2022, JanusMaterialsReview} whose magnetic ions form a Bravais lattice.

\section{Model}
We consider a two-dimensional chiral magnet described by a spin Hamiltonian 
\begin{align}
	H = \sum_i^N \left [ \sum_{\mu = \pm x, \pm y} \left( -\frac{J}{2} \vec{S}_i \cdot \vec{S}_{i+\mu} + \frac{D}{2} \hat{\vec{\mu}} \cdot \vec{S}_i \times \vec{S}_{i+\mu} \right) - \Delta S^z_i \right],
    \label{eq:ham-chiral-magnet}
\end{align}
where $i$ labels sites at position $\vec{r}_i$ on a square lattice, and $i \pm \mu$ is shorthand for the neighbor of site $i$ along $\pm \mu$ direction. $N$ is the total number of sites.
A positive exchange $J$ stabilizes ferromagnetism, which is counteracted by the relativistic Dzyaloshinskii-Moriya interaction (DMI) $D$ pointing along the bond directions $\hat{\vec{\mu}}$ \footnote{Our results also apply to the case of interfacial DMI. See also the Appendix, App.~\ref{sec:NSWT} for more information.}. The magnet becomes field polarized for a Zeeman field $\Delta > \Delta_\text{c} \approx 0.84 S D^2/J$ \cite{Ezawa2011, Park2020inplaneDMI}. 

We use the Holstein-Primakoff transformation \cite{holsteinprimakoff1940}, 
$
    S_i^+ = S_i^x + \mathrm{i} S_i^y = (2S-a^\dagger_i a_i)^{1/2} a_i
$
and
$
S_i^z = S - a^\dagger_i a_i
$ ($\hbar = 1$), to expand in fluctuations about the field-polarized state. Up to an irrelevant constant, the expansion is given by
\begin{equation}
    H= H_2 + H_3 + H_4 + \ldots,
    \label{eq:HP decomposition}
\end{equation}
where $H_n$ contains $n$ bosonic creation/annihilation operators ($a^\dagger_i$/$a_i$). 
Upon Fourier transformation, $
a_i = \frac{1}{\sqrt{N}} \sum_{\vec{k}} \mathrm{e}^{\mathrm{i} \vec{k} \cdot \vec{r}_i} a_{\vec{k}}
$, the bilinear $H_2 = \sum_{\vec{k}} \varepsilon_{\vec{k}} a^\dagger_{\vec{k}} a_{\vec{k}}$ yields the free magnon dispersion
\begin{align}
	\varepsilon_{\vec{k}} &= 2JS (2 - \cos k_x - \cos k_y) + \Delta,
    \label{eq:dispersion}
\end{align}
where the lattice constant has been set to one. The terms in Eq.~\eqref{eq:HP decomposition} beyond bilinear order contain magnon-magnon interactions. 
The cubic term,
\begin{equation}
   H_3 = \frac{1}{2\sqrt{N}} \sum_{\vec{k}_1, \vec{k}_2, \vec{k}_3}\delta_{\vec{k}_1 + \vec{k}_2, \vec{k}_3} \left[ V_{\vec{k}_3;\vec{k}_1, \vec{k}_2}a^{\dagger}_{\vec{k}_1}a^{\dagger}_{\vec{k}_2}a_{\vec{k}_3} + \text{h.c.}\right], 
   \label{eq:threemagnon}
\end{equation}
encompasses three-magnon interactions, with the complex three-magnon vertex 
\begin{align}	  
    V_{\vec{k}_3;\vec{k}_1,\vec{k}_2} = \sqrt{2S} D \left[ 
	\sin k_1^x + \sin k_2^x + \mathrm{i} \left( \sin k_1^y + \sin k_2^y \right)	
	\right].
    \label{eq:Vvertex}
\end{align}
It arises from the DMI, which breaks spin and magnon number conservation. In contrast, the four-magnon interaction, 
\begin{align}
	H_4 
	&= \frac{1}{4 N} \sum_{\vec{k}_1, \vec{k}_2, \vec{k}_3, \vec{k}_4}
    \delta_{\vec{k}_1 + \vec{k}_2, \vec{k}_3 + \vec{k}_4}
	W_{\vec{k}_1, \vec{k}_2; \vec{k}_3, \vec{k}_4} a^\dagger_{\vec{k}_1} a^\dagger_{\vec{k}_2} a_{\vec{k}_3} a_{\vec{k}_4},
	\label{eq:fourboson}
\end{align}
derives from the exchange interaction and conserves the magnon number \cite{Oguchi}. Its interaction vertex is real,
\begin{align}
	W_{\vec{k}_1, \vec{k}_2; \vec{k}_3, \vec{k}_4} =
	J \sum_{\mu=x,y} \left[ \sum_{a=1}^4 \cos k_a^\mu - \sum_{b=1}^2 \sum_{c=3}^4 \cos (k_b^\mu-k_c^\mu) \right]
	\label{eq:Wvertex}.
\end{align}
In Eqs.~\eqref{eq:threemagnon} and \eqref{eq:fourboson}, the momentum conservation is implicitly understood modulo a reciprocal lattice vector. For a derivation of Eqs.~\eqref{eq:dispersion}-\eqref{eq:Wvertex}, see the Appendix (App.) \ref{sec:NSWT}.

\section{Breaking of Time-Reversal Symmetry}
A THE arises when the system is \emph{not} time-reversal (TR) symmetric. 
The action of the TR operator $\varTheta$ on an annihilation operator is 
$
	\varTheta a_{\vec{k}} \varTheta^{-1}  = \mathrm{e}^{\mathrm{i}\varphi} a_{-\vec{k}}
$
($
	\varTheta a^\dagger_{\vec{k}} \varTheta^{-1}  = a^\dagger_{-\vec{k}} \mathrm{e}^{-\mathrm{i}\varphi}
$), for some real but fixed $\varphi$, originating from the unitary part of the TR operator. Being antiunitary, $\varTheta$ contains complex conjugation.
A Hamiltonian $H$ is TR invariant, if
$
[H,\varTheta] = 0
$
and thus
$
	H = \varTheta H \varTheta^{-1}
$.
Given a decomposition of $H = \sum_{n=2}^\infty H_n$ as in Eq.~\eqref{eq:HP decomposition}, TR invariance requires
$
	H_n = \varTheta H_n \varTheta^{-1}
$
for each $n$ separately, implying
\begin{subequations}
\begin{align}
    H_2: && \varepsilon_{\vec{k}} &= \varepsilon_{-\vec{k}}, \label{eq:H2condition} \\
    H_3: &&  V^\ast_{\vec{k}_3;\vec{k}_1, \vec{k}_2} \mathrm{e}^{\mathrm{i} \varphi} &= V_{-\vec{k}_3 ;-\vec{k}_1, -\vec{k}_2}, \label{eq:H3condition}\\
    H_4: && W^\ast_{ \vec{k}_1, \vec{k}_2; \vec{k}_3, \vec{k}_4} &= W_{ -\vec{k}_1, -\vec{k}_2; -\vec{k}_3, - \vec{k}_4}.
    \label{eq:H4condition}
\end{align}
\end{subequations}
For the case with multiple magnon bands, see App.~\ref{sec: TR}.

According to Eqs.~\eqref{eq:dispersion} and \eqref{eq:Wvertex} the TR conditions for $H_2$ [Eq.~\eqref{eq:H2condition}] and $H_4$ [Eq.~\eqref{eq:H4condition}] are met. On the contrary, for the three-magnon vertex in Eq.~\eqref{eq:Vvertex}, we find no $\varphi$
for which the TR condition of $H_3$ in Eq.~\eqref{eq:H3condition} is met. Thus, even though the harmonic magnon gas is TR symmetric, the chiral anharmonic interactions in Eq.~\eqref{eq:Vvertex} are not \footnote{This is an instance of spurious symmetries in linear spin-wave theory \cite{Gohlke2023PRL}.}, and we expect magnon-magnon scattering to contribute to the THE.

\section{Transport Theory}
To calculate the interaction-induced magnon THE we use the semi-classical Boltzmann transport equation. This choice is inspired by Refs.~\onlinecite{Mangeolle2022PRB, Mangeolle2022PRX}, which studied the THE of phonons scattering off collective magnetic excitations. With details provided in App.~\ref{section:Transport theory}, we present the main steps to derive the thermal conductivity tensor $\kappa$.

The constitutive equation for $\kappa$ is Fourier's law, $\vec{q} = \kappa (- \vec{\nabla} T)$, relating a heat current density $\vec{q}$ to a temperature gradient $\vec{\nabla} T$. Microscopically, the leading harmonic part of $\vec{q}$ reads as
\begin{align}
    \vec{q} = \frac{1}{V} \sum_{\vec{k}} \varepsilon_{\vec{k}} \vec{v}_{\vec{k}}
    N_{\vec{k}}
    \label{eq:current}
\end{align}
for a sample of volume $V$,
where $\vec{v}_{\vec{k}} = \left( 1 / \hbar\right)\partial \varepsilon_{\vec{k}} / \partial \vec{k}$ is the group velocity of a magnon with momentum $\vec{k}$. Its non-equilibrium population $N_{\vec{k}} = \overline{N}_{\vec{k}}+\delta N_{\vec{k}}$ is the sum of the equilibrium population $\overline{N}_{\vec{k}}$ and a small deviation $\delta N_{\vec{k}}$ responsible for transport. To compute $\delta N_{\vec{k}}$ we invoke the Boltzmann transport equation, 
$
    \dot{\vec{r}} \cdot \partial N_{\vec{k}}/ \partial \vec{r}
    =
    I^\text{coll}_{\vec{k}},
$
where $I^\text{coll}_{\vec{k}}$ is the collision integral. After linearization, it reads as
$
    \vec{v}_{\vec{k}} \cdot \vec{\nabla} T \partial \overline{N}_{\vec{k}} / \partial T 
       = \sum_{\vec{k}'}C_{\vec{k} \vec{k}'} \delta N_{\vec{k'}}
$.
For the collision kernel $C_{\vec{k} \vec{k}'}$, we use Hardy's basis \cite{Hardy1970, Cepellotti2016}, i.e., $ \mathcal{C}_{\vec{k} \vec{k}'} = C_{\vec{k} \vec{k}'} G_{\vec{k}'} / G_{\vec{k}} $, where  $G_{\vec{k}}=\sqrt{\overline{N}_{\vec{k}}(\overline{N}_{\vec{k}} + 1)}$, such that 
\begin{equation}
    G^{-1}_{\vec{k}} \frac{\partial \overline{N}_{\vec{k}}}{\partial T} 
   \vec{v}_{\vec{k}} \cdot \vec{\nabla} T = \sum_{\vec{k}'} \mathcal{C}_{\vec{k} \vec{k}'} \delta \mathcal{N}_{\vec{k}'}
   ,
   \label{eq: symmetrized collision integral}
\end{equation}
with $\delta \mathcal{N}_{\vec{k}'} = G^{-1}_{\vec{k}'} \delta N_{\vec{k}'}$.
After inverting Eq.~\eqref{eq: symmetrized collision integral} with respect to $\delta \mathcal{N}_{\vec{k}}$, plugging the solution into Eq.~\eqref{eq:current}, and comparing with Fourier's law, we extract the thermal conductivity tensor
\begin{equation}
 \kappa = -\frac{1}{k_\text{B} T^2 V} \sum_{\vec{k},\vec{k}'}  \vec{v}_{\vec{k}} \otimes \vec{v}_{\vec{k}'} 
    \varepsilon_{\vec{k}} \varepsilon_{\vec{k}'} 
    G_{\vec{k}} G_{\vec{k}'}
    \left[ \mathcal{C}^{-1}\right]_{\vec{k}\vec{k}'},
    \label{eq:conductivity2}
\end{equation}
where $\otimes$ denotes the dyadic product. Neglecting many-body corrections to the dispersion, we have adopted Bose-Einstein statistics in equilibrium, 
$
    \overline{N}_{\vec{k}} = (\mathrm{e}^{\varepsilon_{\vec{k}} /(k_\text{B}T)} - 1)^{-1}$ with Boltzmann's constant $k_\text{B}
$, 
so that 
$
    \partial \overline{N}_{\vec{k}} / \partial T = - (\varepsilon_{\vec{k}}/T) (\partial \overline{N}_{\vec{k}} / \partial \varepsilon_{\vec{k}}) = \varepsilon_{\vec{k}} G^2_{\vec{k}} /(k_\text{B} T^2)
$. 
We obtain the Hall conductivity tensor 
$
    \kappa_\text{H} = (\kappa-\kappa^\text{T})/2
$ 
as the antisymmetric part of $\kappa$, which can be arranged as a Hall vector $\vec{\kappa}_\text{H} = (  \kappa^{yz}_\text{H}, \kappa^{zx}_\text{H}, \kappa^{xy}_\text{H})^\text{T}$; explicitly
\begin{equation}
    \vec{\kappa}_\text{H} 
    = 
    -\frac{1}{4k_{B}T^2 V}\sum_{\vec{k},\vec{k}'} \vec{v}_{\vec{k}} \times \vec{v}_{\vec{k}'}\varepsilon_{\vec{k}}\varepsilon_{\vec{k}'}G_{\vec{k}}G_{\vec{k}'}
    \left(
    \left[ \mathcal{C}^{-1}\right]_{\vec{k}\vec{k}'}
    -
    \left[ \mathcal{C}^{-1}\right]_{\vec{k}'\vec{k}}
    \right).
    \label{eq:Hall conductivity general}
\end{equation}
Thus, $\vec{\kappa}_\text{H}$ is finite only if the inverse of the collision kernel has a finite anti-symmetric part: 
$[ \mathcal{C}^{-1}]_{\vec{k}\vec{k}'}
    \ne
    [ \mathcal{C}^{-1}]_{\vec{k}'\vec{k}}$.

For analytical progress, we follow Refs.~\onlinecite{Mangeolle2022PRB, Mangeolle2022PRX} and split the collision kernel 
$
    \mathcal{C}_{\vec{k} \vec{k}'} = -\delta_{\vec{k} \vec{k}'} \tau^{-1}_{\vec{k}} + \mathcal{O}_{\vec{k} \vec{k}'}
$
into a diagonal part, which is the inverse of the relaxation time $\tau_{\vec{k}}$, and an off-diagonal part, $\mathcal{O}_{\vec{k} \vec{k}'}$. Assuming dominant diagonal scattering, the inverse of $\mathcal{C}_{\vec{k} \vec{k}'}$ is
$
    [\mathcal{C}^{-1}]_{\vec{k} \vec{k}'} \approx -\delta_{\vec{k} \vec{k}'} \tau_{\vec{k}} - \tau_{\vec{k}} \tau_{\vec{k}'} \mathcal{O}_{\vec{k} \vec{k}'}
$, which we plug into Eqs.~\eqref{eq:conductivity2} and \eqref{eq:Hall conductivity general} to find
\begin{subequations}
\begin{align}
    \kappa 
    &= 
    \frac{1}{k_\text{B} T^2 V} \sum_{\vec{k},\vec{k}'}  \vec{v}_{\vec{k}} \otimes \vec{v}_{\vec{k}'} 
    \varepsilon_{\vec{k}} \varepsilon_{\vec{k}'} 
    G_{\vec{k}} G_{\vec{k}'}
    \left( \delta_{\vec{k} \vec{k}'} \tau_{\vec{k}} + \tau_{\vec{k}} \tau_{\vec{k}'} \mathcal{O}_{\vec{k} \vec{k}'} \right)
    \label{eq:full conductivity final},
    \\
    \vec{\kappa}_\text{H} 
    &=
     \frac{1}{2k_{B}T^2 V}\sum_{\vec{k}, \vec{k}'} \vec{v}_{\vec{k}} \times \vec{v}_{\vec{k}'}\varepsilon_{\vec{k}} \varepsilon_{\vec{k}'} \tau_{\vec{k}}\tau_{\vec{k}'} G_{\vec{k}} G_{\vec{k}'} \mathcal{A}_{\vec{k}\vec{k}'},
    \label{eq:Hall conductivity final}
\end{align}
\end{subequations}
where 
$
    \mathcal{A}_{\vec{k}\vec{k}'} =
    (\mathcal{O}_{\vec{k} \vec{k}'}-\mathcal{O}_{\vec{k}'\vec{k}})/2
$
is the antisymmetric part of the collision kernel.
Thus, the THE can only occur if $\mathcal{C}$ itself has a finite antisymmetric part. This finding is equivalent to the Hall criterion in Refs.~\onlinecite{Mangeolle2022PRB, Mangeolle2022PRX} given in terms of $C$ rather than $\mathcal{C}$.

To clarify what it takes for $\mathcal{A}_{\vec{k} \vec{k}'}$ to be nonzero, note that 
$
    \mathcal{O}_{\vec{k}\vec{k}'} = \mathcal{O}_{\vec{k}\vec{k}'}^\text{in} - \mathcal{O}_{\vec{k}\vec{k}'}^\text{out}
$
is the difference of scattering rates of processes where mode $\vec{k}$ is created, $\mathcal{O}_{\vec{k}\vec{k}'}^\text{in}$, and destroyed, $\mathcal{O}_{\vec{k}\vec{k}'}^\text{out}$. Furthermore, 
$
    \mathcal{O}_{\vec{k}\vec{k}'}^\text{in} = \mathcal{O}_{\vec{k}\vec{k}'}^{++} + \mathcal{O}_{\vec{k}\vec{k}'}^{+-}
$ 
and 
$
    \mathcal{O}_{\vec{k}\vec{k}'}^\text{out} = \mathcal{O}_{\vec{k}\vec{k}'}^{-+} + \mathcal{O}_{\vec{k}\vec{k}'}^{--} 
$ 
are both sums of processes where mode $\vec{k}'$ is created and destroyed. The ``$\pm$'' superscripts indicate creation ($+$) and destruction ($-$); their order coincides with that of the momenta subscripts. One can use relations between these scattering channels to find (App.~\ref{subsection: DB})
\begin{align}
    \mathcal{A}_{\vec{k}\vec{k}'} 
    &= 
    \frac{\overline{N}_{\vec{k}'}-\overline{N}_{\vec{k}}}{2\overline{N}_{\vec{k}'}} \left(\mathcal{O}_{\vec{k}\vec{k}'}^{++}-\mathrm{e}^{-\beta \varepsilon_{\vec{k}'}}\mathcal{O}_{\vec{k}\vec{k}'}^{--} \right)
    \nonumber 
    \\
    &\quad +
    \frac{\overline{N}_{\vec{k}} + \overline{N}_{\vec{k}'} + 1}{2\overline{N}_{\vec{k}'}} \left(\mathrm{e}^{-\beta \varepsilon_{\vec{k}'}}\mathcal{O}_{\vec{k}\vec{k}'}^{+-}-\mathcal{O}_{\vec{k}\vec{k}'}^{-+} \right).
\end{align}
Thus, a THE can only arise, if the microscopic detailed balance relations,
\begin{align}
    \frac{\mathcal{O}_{\vec{k}\vec{k}'}^{++}}{\mathcal{O}_{\vec{k}\vec{k}'}^{--}} = \mathrm{e}^{-\beta \varepsilon_{\vec{k}'}} 
    \quad
    \text{and}
    \quad
    \frac{\mathcal{O}_{\vec{k}\vec{k}'}^{+-}}{\mathcal{O}_{\vec{k}\vec{k}'}^{-+}} = \mathrm{e}^{\beta \varepsilon_{\vec{k}'}},
    \label{eq: detailed balance}
\end{align}
are violated, a conclusion also reached in Refs.~\onlinecite{Mangeolle2022PRB, Mangeolle2022PRX}, and known from the electronic extrinsic anomalous Hall effect in the context of elastic scattering \cite{Nagaosa2010AHEReview}.

\section{Scattering Theory}
We proceed by deriving a microscopic theory for the collision kernel introduced in Eq.~\eqref{eq: symmetrized collision integral}. Given a Hamiltonian $H$ as in Eq.~\eqref{eq:HP decomposition}, we treat $H_2$ as the unperturbed piece and $H' = H_3 + H_4$ as a perturbation.
In a scattering event due to $H'$, we start from an initial Fock state $|\text{i}\rangle$ (with energy $E_{\text{i}}$ and particle number $N_{\text{i}}$) and enter a final state $ |\text{f} \rangle $ (with energy $E_{\text{f}}$ and particle number $N_{\text{f}}$), in which the number of particles is equal or differs by one, $\Delta N = N_{\text{f}} - N_{\text{i}} \in \{-1,0,+1\}$. The corresponding scattering rate is given within the T-matrix approximation as,
\begin{align}
	\Gamma_{\text{f}\text{i}}\left[ \left\{N_{\vec{k}' } \right \}\right] = \frac{ 2 \pi }{ \hbar } | T_{\text{fi}} |^2 \delta(E_{\text{i}}-E_{\text{f}}),
    \label{eq:FGR}
\end{align}
where $\left\{N_{\vec{k}' } \right \}$ denotes its dependence on the out-of-equilibrium populations $N_{\vec{k}'}$ of magnons with momenta $\vec{k}'$ participating in the studied scattering event. 
Up to second order in $H'$, the $T$-matrix element $T_{\text{fi}} = \langle \text{f} | T | \text{i} \rangle$ contains the operator
\begin{align}
	T = H' 
	  + H' \left( \sum_{v} \frac{| v \rangle  \langle v | }{E_{\text{i}} - E_v + \mathrm{i}\eta } \right) H' , \quad \eta > 0,\label{eq:interaction-matrix-elements}
\end{align}
and $v$ labels intermediate states $| v \rangle$ with energy $E_v$; they involve virtual and real magnons. Pictorially, we write
\begin{align}
    | T_{\text{fi}} |^2
    =
    \begin{cases}
    \left|
    \vcenter{\hbox{
    \begin{tikzpicture}[decoration={markings, 
        mark= at position 0.65 with {\arrow{stealth}},
        mark= at position 2cm with {\arrow{stealth}}}
        ]   
            \begin{scope}[xshift = -0.2cm]
            \draw[postaction={decorate}] (-1,0.2)--(-0.7,0.0);
            \draw[postaction={decorate}] (-1,-0.2)--(-0.7,0.0);
            \draw[postaction={decorate}] (-0.7,0)--(-0.2,0.0);
            \draw[fill=black] (-0.7,0) circle (0.07);
            \end{scope}
            \node at (-0.2,0) {$+$};
            \draw[postaction={decorate}] (0,0.2)--(0.3,0);
            \draw[postaction={decorate}] (0,-0.2)--(0.3,0);
            \draw [postaction={decorate}] (0.3,0) to [out=70,in=110] (0.8,0);
            \draw [postaction={decorate}] (0.3,0) to [out=-70,in=-110] (0.8,0);
            \draw[postaction={decorate}] (0.8,0)--(1.3,0.0);
            \draw[fill=white] (0.3,0) circle (0.07);
            \draw[fill=black] (0.8,0) circle (0.07);
    \end{tikzpicture}
    }}
    \right|^2
    &
    \text{if }\Delta N=-1
    \\
    \left|
    \vcenter{\hbox{
    \begin{tikzpicture}[decoration={markings, 
        mark= at position 0.65 with {\arrow{stealth}},
        mark= at position 2cm with {\arrow{stealth}}}
        ]   
            \begin{scope}[xshift = -0.2cm]
            \draw[postaction={decorate}] (-0.8,0.2)--(-0.5,0);
            \draw[postaction={decorate}] (-0.8,-0.2)--(-0.5,0);
            \draw[postaction={decorate}] (-0.5,0)--(-0.2,-0.2);
            \draw[postaction={decorate}] (-0.5,0)--(-0.2,0.2);
            \draw[fill=white] (-0.5,0) circle (0.07);
            \end{scope}
            \node at (-0.2,0) {$+$};
            \begin{scope}[xshift = -0.2cm]
            \draw[postaction={decorate}] (0.2,0.2)--(0.5,0);
            \draw[postaction={decorate}] (0.2,-0.2)--(0.5,0);
            \draw [postaction={decorate}] (0.5,0)--(1,0);
            \draw[postaction={decorate}] (1,0)--(1.3,-0.2);
            \draw[postaction={decorate}] (1,0)--(1.3,0.2);
            \draw[fill=black] (0.5,0) circle (0.07);
            \draw[fill=black] (1,0) circle (0.07);
            \end{scope}
            \node at (1.3,0) {$+$};
            \begin{scope}[xshift = 1.3cm]
            \draw[postaction={decorate}] (0.2,0.2)--(0.5,0);
            \draw[postaction={decorate}] (0.2,-0.2)--(0.5,0);
            \draw [postaction={decorate}] (0.5,0) to [out=70,in=110] (1,0);
            \draw [postaction={decorate}] (0.5,0) to [out=-70,in=-110] (1,0);
            \draw[postaction={decorate}] (1,0)--(1.3,-0.2);
            \draw[postaction={decorate}] (1,0)--(1.3,0.2);
            \draw[fill=white] (0.5,0) circle (0.07);
            \draw[fill=white] (1,0) circle (0.07);
            \end{scope}
    \end{tikzpicture}
    }}
    \right|^2
    &
    \text{if }\Delta N = 0
    \\
    \left|
    \vcenter{\hbox{
    \begin{tikzpicture}[decoration={markings, 
        mark= at position 0.65 with {\arrow{stealth}},
        mark= at position 2cm with {\arrow{stealth}}}
        ]   
            \begin{scope}[xshift = -0.2cm]
            \draw[postaction={decorate}] (-1,0)--(-0.5,0);
            \draw[postaction={decorate}] (-0.5,0)--(-0.2,-0.2);
            \draw[postaction={decorate}] (-0.5,0)--(-0.2,0.2);
            \draw[fill=black] (-0.5,0) circle (0.07);
            \end{scope}
            \node at (-0.2,0) {$+$};
            \draw[postaction={decorate}] (0,0)--(0.5,0);
            \draw [postaction={decorate}] (0.5,0) to [out=70,in=110] (1,0);
            \draw [postaction={decorate}] (0.5,0) to [out=-70,in=-110] (1,0);
            \draw[postaction={decorate}] (1,0)--(1.3,-0.2);
            \draw[postaction={decorate}] (1,0)--(1.3,0.2);
            \draw[fill=black] (0.5,0) circle (0.07);
            \draw[fill=white] (1,0) circle (0.07);
    \end{tikzpicture}
    }}
    \right|^2
    &
    \text{if }\Delta N = +1
    \end{cases}
    ,
    \label{eq:pictorial_scattering}
\end{align}
where full (empty) circles indicate three-magnon (four-magnon) vertices. In Eq.~\eqref{eq:pictorial_scattering}, the first diagram of each row derives from the linear-in-$\hat{H}'$ term in Eq.~\eqref{eq:interaction-matrix-elements}. Each of the second-order diagrams in Eq.~\eqref{eq:pictorial_scattering} represents an entire class of diagrams associated with the sum over $v$ in Eq.~\eqref{eq:interaction-matrix-elements}. 
When evaluating the absolute squares, we keep terms with up to three circles (as they capture the leading contribution to the THE). For example, for $\Delta N = -1$, we have 
\begin{equation}
\begin{split}
    | T_{\text{fi}} |^2
    \approx
    &
    \left|
    \vcenter{\hbox{
    \begin{tikzpicture}[decoration={markings, 
        mark= at position 0.65 with {\arrow{stealth}},
        mark= at position 2cm with {\arrow{stealth}}}
        ]   
            \begin{scope}[xshift = -0.2cm]
            \draw[postaction={decorate}] (-1,0.2)--(-0.7,0.0);
            \draw[postaction={decorate}] (-1,-0.2)--(-0.7,0.0);
            \draw[postaction={decorate}] (-0.7,0)--(-0.2,0.0);
            \draw[fill=black] (-0.7,0) circle (0.07);
            \end{scope}
    \end{tikzpicture}
    }}
    \right|^2 + 
    2 \mathrm{Re} \left(
    \vcenter{\hbox{
    \begin{tikzpicture}[decoration={markings, 
        mark= at position 0.65 with {\arrow{stealth}},
        mark= at position 2cm with {\arrow{stealth}}}
        ]   
            \begin{scope}[xshift = -0.2cm]
            \draw[postaction={decorate}] (-1,0.2)--(-0.7,0.0);
            \draw[postaction={decorate}] (-1,-0.2)--(-0.7,0.0);
            \draw[postaction={decorate}] (-0.7,0)--(-0.2,0.0);
            \draw[fill=black] (-0.7,0) circle (0.07);
            \end{scope}
    \end{tikzpicture}
    }}
    \right)
    \text{Re}
    \left(
    \vcenter{\hbox{
    \begin{tikzpicture}[decoration={markings, 
        mark= at position 0.65 with {\arrow{stealth}},
        mark= at position 2cm with {\arrow{stealth}}}
        ]   
            \draw[postaction={decorate}] (0,0.2)--(0.3,0);
            \draw[postaction={decorate}] (0,-0.2)--(0.3,0);
            \draw [postaction={decorate}] (0.3,0) to [out=70,in=110] (0.8,0);
            \draw [postaction={decorate}] (0.3,0) to [out=-70,in=-110] (0.8,0);
            \draw[postaction={decorate}] (0.8,0)--(1.3,0.0);
            \draw[fill=white] (0.3,0) circle (0.07);
            \draw[fill=black] (0.8,0) circle (0.07);
    \end{tikzpicture}
    }}
    \right)  
    \\
    &
    \quad
    + 2 \mathrm{Im} \left(
    \vcenter{\hbox{
    \begin{tikzpicture}[decoration={markings, 
        mark= at position 0.65 with {\arrow{stealth}},
        mark= at position 2cm with {\arrow{stealth}}}
        ]   
            \begin{scope}[xshift = -0.2cm]
            \draw[postaction={decorate}] (-1,0.2)--(-0.7,0.0);
            \draw[postaction={decorate}] (-1,-0.2)--(-0.7,0.0);
            \draw[postaction={decorate}] (-0.7,0)--(-0.2,0.0);
            \draw[fill=black] (-0.7,0) circle (0.07);
            \end{scope}
    \end{tikzpicture}
    }}
    \right)
    \text{Im}
    \left(
    \vcenter{\hbox{
    \begin{tikzpicture}[decoration={markings, 
        mark= at position 0.65 with {\arrow{stealth}},
        mark= at position 2cm with {\arrow{stealth}}}
        ]   
            \draw[postaction={decorate}] (0,0.2)--(0.3,0);
            \draw[postaction={decorate}] (0,-0.2)--(0.3,0);
            \draw [postaction={decorate}] (0.3,0) to [out=70,in=110] (0.8,0);
            \draw [postaction={decorate}] (0.3,0) to [out=-70,in=-110] (0.8,0);
            \draw[postaction={decorate}] (0.8,0)--(1.3,0.0);
            \draw[fill=white] (0.3,0) circle (0.07);
            \draw[fill=black] (0.8,0) circle (0.07);
    \end{tikzpicture}
    }}
    \right).
    \label{eq:pictorial T-matrix}
\end{split}
\end{equation}

Each scattering channel comes with an in- and out-version, where the magnon with momentum $\vec{k}$ is created and destroyed, respectively; their $T$-matrix elements are $T_{\text{fi}}^{\text{in}}$ and $T_{\text{fi}}^{\text{out}}$, respectively. The associated scattering rates $\Gamma_{\text{fi}}^{\text{in}/\text{out}}$ are given by replacing $T_{\text{fi}} \to T_{\text{fi}}^{\text{in}/\text{out}}$ in Eq.~\eqref{eq:FGR}. The total rate, given by the difference of the in- and out-rates, enters the collision integral, 
\begin{equation}
    \begin{split}
    I^\text{coll}_{\vec{k}}
    & 
    = \sum_{\text{i},\text{f},\vec{k}'} 
    \left(
    \Gamma^{\text{in}}_{\text{fi}}\left[ \left\{N_{\vec{k}' } \right \}\right] - \Gamma^{\text{out}}_{\text{fi}}\left[ \left\{N_{\vec{k}' } \right \}\right]
    \right)
    \approx 
    G_{\vec{k}}
    \sum_{\vec{k}'} \mathcal{C}_{\vec{k}\vec{k}'} \delta \mathcal{N}_{\vec{k}'}.
    \label{eq: full collision integral}
    \end{split}
\end{equation}
In the last step, we linearized the collision integral with respect to the deviations $\delta N_{\vec{k}'}$, transformed into Hardy's basis, and absorbed the summation over initial and final states into the collision kernel $\mathcal{C}_{\vec{k}\vec{k}'}$, where
$
    \mathcal{C}_{\vec{k}\vec{k}'}=\mathcal{C}_{\vec{k}\vec{k}'}^{\text{in}}-\mathcal{C}_{\vec{k}\vec{k}'}^{\text{out}}
$.
We can thus connect the microscopic scattering theory to the transport theory in Eq.~\eqref{eq: symmetrized collision integral}. By calculating $\mathcal{O}_{\vec{k}\vec{k'}}^{\text{in/out}}$ for the scattering channels shown in Eq.~\eqref{eq:pictorial_scattering}, we observe that processes resulting in a finite $\kappa_\text{H}$ come, in leading order in $1/S$ and $D$, from interference terms that contain one four-magnon vertex (white circle) and two TR-violating three-magnon vertices (black circle), i.e., $\mathcal{A}_{\vec{k}\vec{k}'} \propto (D/J)^2$ \footnote{In contrast, there is no contribution to $\kappa_\text{H}$ from the interference of the $\Delta N = 0$ diagrams in Eq.~\eqref{eq:pictorial_scattering} that contain only white circles, as they do not include any time-reversal-breaking vertices.}. For these terms, the detailed balance relations in Eq.~\eqref{eq: detailed balance} are violated by a minus sign, 
\begin{align}
    \frac{\mathcal{O}_{\vec{k}\vec{k}'}^{++}}{\mathcal{O}_{\vec{k}\vec{k}'}^{--}} = -\mathrm{e}^{-\beta \varepsilon_{\vec{k}'}} 
    \quad
    \text{and}
    \quad
    \frac{\mathcal{O}_{\vec{k}\vec{k}'}^{+-}}{\mathcal{O}_{\vec{k}\vec{k}'}^{-+}} = -\mathrm{e}^{\beta \varepsilon_{\vec{k}'}},
    \label{eq: broken detailed balance}
\end{align}
representing an example of \emph{anti}-detailed balance \cite{Mangeolle2022PRB, Mangeolle2022PRX}.

In total, we found $\sim$ 20 diagrams that make up $I^\text{coll}_{\vec{k}}$. For a complete list and a detailed discussion on deriving the mathematical expressions of their part of $\mathcal{O}_{\vec{k} \vec{k}'}$ that contributes to $\mathcal{A}_{\vec{k}\vec{k}'}$ and the THE, see App.~\ref{subsection: diagrams}. Here, we note two relevant observations for the leading order diagrams in $1/S$ that we considered in the calculation: 

(i) They contain \emph{two} delta-functions, one enforcing global energy conservation [see Eq.~\eqref{eq:FGR}], and the other coming from \emph{resonant} intermediate scattering associated with the imaginary part of $\left( E_{\text{i}}- E_{\text{f}} + \mathrm{i}\eta\right)^{-1} $ in the $T$-matrix [see Eq.~\eqref{eq:interaction-matrix-elements}].

(ii) They vanish in the presence of TR symmetry, leading to $\mathcal{A}_{\vec{k}\vec{k}'} = 0$. Thus, in model \eqref{eq:ham-chiral-magnet}, processes that break the detailed balance also break TR symmetry (see proof in App.~\ref{sub: TR and DB}). This finding is special to model \eqref{eq:ham-chiral-magnet} and does not carry over to the general case, where TR symmetry only enforces $\mathcal{A}_{\vec{k}\vec{k}'} = - \mathcal{A}_{-\vec{k},-\vec{k}'}$ [also leading to $\vec{\kappa}_\text{H} = 0$ in Eq.~\eqref{eq:Hall conductivity final}] \cite{Mangeolle2022PRB, Mangeolle2022PRX}.

\begin{figure*}
    \centering
    \includegraphics[scale=1]{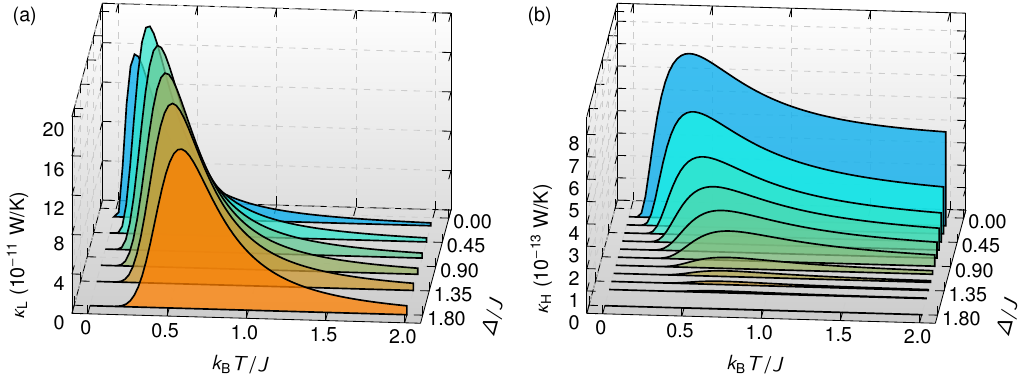}
    \caption{The longitudinal thermal conductivity $\kappa_\text{L}$ (a) and the thermal Hall conductivity $\kappa_\text{H}$ (b) in the field-polarized phase of a chiral magnet on the square-lattice as a function of temperature and magnetic field. The conductivities were obtained from Eqs.~\eqref{eq:full conductivity final} and \eqref{eq:Hall conductivity final}; see the Appendix for details on how the relaxation times and off-diagonal scattering rates were calculated. Note the rescaled axes: $\tilde{T}= k_\text{B}T / J$ and $\tilde{\Delta}= \Delta / J$. Parameters read $J = 1\,\text{meV}$, $D/J=0.1$, $S=1$, and $\alpha_\text{G} = 10^{-3}$.} 
    \label{fig:results}
\end{figure*}

\section{Results}
We have used Eqs.~\eqref{eq:full conductivity final} and \eqref{eq:Hall conductivity final} to compute the longitudinal conductivity $\kappa_\text{L} \equiv \kappa^{xx} (=\kappa^{yy})$ and the Hall conductivity $\kappa_\text{H} \equiv \kappa^{xy}_\text{H}$ in the field-polarized phase of model \eqref{eq:ham-chiral-magnet}. We set $S=1$, $D/J=0.1$, and $J=1\text{ meV}$. For details on the numerical implementation, see App.~\ref{subsection: D's in the calculation}. There, we also explain how to compute the relaxation time $\tau_{\vec{k}}$, for which we have included only the lowest-order scattering events and added to it by hand a phenomenological Gilbert damping-type lifetime $\tau_{\vec{k}}^\text{G} = \hbar/(\alpha_\text{G} \varepsilon_{\vec{k}} )$, as obtained from a linearized Landau-Lifschitz-Gilbert equation \cite{Gurevich1996-pk}, to capture crystal imperfections and other sources of relaxation, but there is no disorder skew scattering. We have set the Gilbert damping to $\alpha_\text{G} = 10^{-3}$ (for other values, see App.~\ref{subsec: values of GD}).
In Fig.~\ref{fig:results}, we present $\kappa_\text{L}$ and $\kappa_\text{H}$ as functions of the rescaled magnetic field $\tilde{\Delta} = \Delta / J$ and temperature $\tilde{T} = k_{\text{B}}T / J$. 

First, we discuss $\kappa_\text{L}$ in Fig.~\ref{fig:results}(a). As a function of $\tilde{T}$, $\kappa_\text{L} \propto \exp(-\tilde{\Delta}/\tilde{T})$ exhibits an activation-like behavior due to the spin-wave gap $\tilde{\Delta}$ and the finite  $\tau_{\vec{k}}^\text{G}$ which dominates scattering at small $\tilde{T}$. After peaking, $\kappa_\text{L}$ decreases as $\kappa_\text{L} \propto 1/\tilde{T}^2$ because of Umklapp scattering (see App.~\ref{sec:Temperaturedependence}), with four-magnon scattering  being the primary contributor to thermal resistivity. The magnetic field shifts the $\kappa_\text{L}$ peak to higher $\tilde{T}$. 

Since $(D/J)^2 \ll 1$, the DMI-induced three-magnon scattering provides only a minor correction to $\kappa_\text{L}$. However, it is crucial for MMSS and the finite $\kappa_\text{H}$ in Fig.~\ref{fig:results}(b). We observe an activation behavior $\kappa_\text{H} \propto \exp(-\gamma \tilde{\Delta}/\tilde{T})$ with a different exponent ($\gamma \ge 2$) because interference of first- and second-order scattering---and, thereby, a larger number of Bose factors---is required for a finite $\mathcal{A}_{\vec{k}\vec{k}'}$.
As $\tilde{T} \to \infty$, $\kappa_\text{H}$ approaches a constant, which is an artifact of the approximations made. Besides not accounting for the paramagnetic phase transition and assuming a constant spectrum, our roughest approximation is having scattering up to different orders included in $\tau_{\vec{k}}$ (only first-order scattering) and $\mathcal{A}_{\vec{k}\vec{k}'}$ (interference scattering). By including interference scattering also in $\tau_{\vec{k}}$, we can show analytically that both $\kappa_\text{L}$ and $\kappa_\text{H}$ vanish with $1/T^4$ as $T \to \infty$ (see App.~\ref{sec:Temperaturedependence}). We refrained from a numerical implementation because the required evaluation of principal parts would lead to a huge computational overhead.

We identify a marked decrease of $\kappa_\text{H}$ with increasing $\tilde{\Delta}$ in Fig.~\ref{fig:results}(b). As noted earlier, only \emph{resonant} intermediate scattering contributes to $\kappa_\text{H}$. All contributions to $\kappa_\text{H}$ involve three-magnon vertices (black circles), requiring three-magnon energy conservation, which occurs when single-particle energies $\varepsilon_{\vec{k}}$ overlap with the two-magnon decay or collision continua, whose density of states reads 
$
    \mathcal{D}_{\vec{k}}(\varepsilon_{\vec{k}}) \propto \sum_{\vec{q}} \delta(\varepsilon_{\vec{k}} - \varepsilon_{\vec{q}} - \varepsilon_{\vec{k}-\vec{q}})
$ 
and 
$
    \mathcal{D}'_{\vec{k}}(\varepsilon_{\vec{k}}) \propto \sum_{\vec{q}} \delta(\varepsilon _{\vec{k}}+ \varepsilon_{\vec{q}} - \varepsilon_{\vec{k}+\vec{q}})
$, 
respectively. As $\Delta$ grows, fewer momenta $\vec{k}$ meet the condition $\mathcal{D}_{\vec{k}}(\varepsilon_{\vec{k}}) \ne 0$ or $\mathcal{D}'_{\vec{k}}(\varepsilon_{\vec{k}}) \ne 0$. The decrease of $\kappa_\text{H}$ observed in Fig.~\ref{fig:results}(b) is thus the result of a field-induced suppression of scattering phase space (see App.~\ref{subsection: frozen phase space}). There is a threshold field $\Delta' = 2 J S$, above which $\mathcal{D}_{\vec{k}}(\varepsilon_{\vec{k}}) = \mathcal{D}'_{\vec{k}}(\varepsilon_{\vec{k}}) = 0$, and $\kappa_\text{H}=0$. Such a threshold can be expected in models where TR violation is exclusively associated with three-magnon terms (or higher odd-number magnon interactions, which are kinematically suppressed at even lower fields \cite{zhitomirskychernyshev2013, Harris1971}), but not in systems where TR violation is due to number-conserving many-body interactions.

\section{Discussion}
In the absence of phonons, $\kappa_\text{L}$ represents the full longitudinal thermal conductivity, yielding a sizable Hall angle $\kappa_\text{H} / \kappa_\text{L}$ of $10^{-3}$ to $10^{-2}$. 
The extracted $\kappa_\text{H} \sim 8 \times 10^{-13}\;\text{W/K}$ corresponds to $\kappa_\text{H}^\text{3D} = \kappa_\text{H} / \ell \sim 10^{-3} \; \text{W/Km}$ in three dimensions (for $\ell = 0.5\,\text{nm}$ layer spacing). According to Eqs.~\eqref{eq:full conductivity final} and \eqref{eq:Hall conductivity final}, scaling $J$ and $D$ by a factor $\lambda$ scales both $\kappa_\text{L}$ and $\kappa_\text{H}$ by the same factor (keeping $\tilde{\Delta}$ and $\tilde{T}$ constant, see App.~\ref{subsub: scaling of conductivities}). Thus, $\kappa_\text{H}$ depends on our choice of $J=1\text{ meV}$ and a large magnonic conductivity that can surpass the intrinsic one is in principle feasible. 
In generally, despite the sizable derived Hall conductivity, our model \eqref{eq:ham-chiral-magnet} was not designed to produce a large THE, but to provide a clear example where MMSS is the leading THE contribution. In this model, the THE is suppressed by the small three-magnon vertex, $\mathcal{A}_{\vec{k}\vec{k}'} \propto (D/J)^2$ with $(D/J)^2 \ll 1$, and the Zeeman field, which kinematically freezes resonant scattering. As this situation changes in systems with strong magnon-magnon interactions, such as spin-orbit dominated (e.g., Kitaev materials, such as $\alpha$-RuCl$_3$) and noncollinear magnets, we conjecture that MMSS contributes to the THE observed in noncollinear antiferromagnets, such as YMnO$_3$ \cite{Kim2019trimerized,Kim2024YMnO3}, in which three-magnon interactions arise from the nonrelativistic exchange interaction. 

For the experimental verification of MMSS, we propose THE experiments on Janus van der Waals materials \cite{ JanusMaterialsReview}. They support interfacial DMI and the magnetic ions can form a Bravais lattice, such as in Cr$YX$ ($Y=\text{S, Se, Te}$; $X=\text{Cl, Br, I}$) \cite{Hou2022}, preventing a magnonic Berry curvature in the harmonic theory. With $J \sim 10\,$meV and $D/J \sim 0.1$ \cite{Hou2022}, 
we expect $\kappa_\text{H}^\text{3D} \sim 10^{-3}-10^{-2}\; \text{W/Km}$. 

The field-polarized phase of Hamiltonian \eqref{eq:ham-chiral-magnet} has been previously studied in the context of the THE in Refs.~\onlinecite{Park2020inplaneDMI, Carnahan2021, Mook2023multipolar}. Classical spin dynamics simulations have revealed a thermal fluctuation induced $\kappa_\text{H}$ peaking with the scalar spin chirality \cite{Carnahan2021} (see also Ref.~\onlinecite{Hou2017thermally}), without clarifying the microscopic origin of the transverse heat current. We propose that the findings of Ref.~\onlinecite{Carnahan2021} represent the classical limit of our quantum theory of MMSS (see App.~\ref{sub: T-B maps} for further details), and note that Ref.~\onlinecite{Carnahan2021} has found thermal Hall conductivities of $\kappa_\text{H} \sim 10^{-12}-10^{-11}\;\text{W/K}$, which is about a factor of ten larger than those found here from MMSS.

Interestingly, as shown in Ref.~\cite{Park2020inplaneDMI}, using a Schwinger boson transformation instead of mapping onto elementary excitations via the Holstein-Primakoff transformation, a THE arises from the Berry curvature of Schwinger boson bands within a self-consistent mean-field theory that captures many-body effects. A thermal Hall conductivity of $\kappa_\text{H} \sim 10^{-15}\;\text{W/K}$ was identified \cite{Park2020inplaneDMI}, which is about a factor of 100 smaller than $\kappa_\text{H}$ from MMSS. 
The finite band geometric thermal Hall conductivity $\kappa_{\text{H}}^{\text{geo}}$ obtained in Ref.~\cite{Park2020inplaneDMI}, inspired us to explore band geometry within the magnon language. In App.~\ref{section: anomalous berry curvature}, we investigate how thermally activated self energy corrections to the bare magnon spectrum can affect its band geometry. We show that anomalous self-energies couple the particle to the hole spectrum within the Bogoliubov-de-Gennes framework and generate a small but finite Berry curvature despite the presence of only one magnon band. For the same parameters as used here for MMSS ($D/J=0.1$ and $S=1$), we find a maximal $\kappa_{\text{H}}^{\text{geo}} \sim 10 ^{-17} \;\text{W/K}$ at $\Delta/J = 0.1$ and $\tilde{T} \sim 1$. For parameters of Ref.~\cite{Park2020inplaneDMI} ($D/J=0.2$ and $S=1/2$), $\kappa_{\text{H}}^{\text{geo}}$ increases up to $10^{-16} \; \text{W/K}$.
We conclude that even though the interaction-renormalized magnon spectrum is capable of generating a Berry curvature contribution to the THE, MMSS is the dominant source of transverse transport in the current model.

The effects of band renormalization on MMSS and additional many-body quantum effects---such as magnon binding  \cite{Mook2023multipolar}---could be carried out systematically 
building on Refs.~\onlinecite{Koyama2024, Saleem2024, Mangeolle2024quantumkinetic}. 
Establishing a correspondence between the semiclassical and quantum theory, similar to the electronic anomalous Hall effect \cite{Nagaosa2010AHEReview}, would significantly enhance our understanding of the THE in magnetic insulators.
\\

\section{Conclusions} 
We showed that many-body interactions between magnons can generate thermal Hall currents independently of band geometry. This proposed mechanism, termed many-body skew scattering, does not rely on Berry curvature or disorder but is intrinsic to interacting spin systems. The magnitude of the THE from many-body skew scattering can match or even surpass that from the band geometric anomalous velocity, with the Bravais lattice case being the most direct example.
Thus, magnon interactions should be considered a potential mechanism when interpreting experimental magnon thermal Hall conductivities. We hope that the theory presented here, which can be applied to other bosonic systems such as magnon-phonon hybrids, will contribute to a comprehensive understanding of thermal Hall transport in magnetic insulators.

\begin{acknowledgments}
We thank L\'{e}o Mangeolle for helpful discussions and insightful comments on the manuscript.
This work was funded by the Deutsche Forschungsgemeinschaft (DFG, German Research Foundation) -- Project No.~504261060 (Emmy Noether Programme). We acknowledge support by the Dynamics and Topology Centre (TopDyn) funded by the State of Rhineland-Palatinate.
\end{acknowledgments} 

\section{Data Availability}
Upon reasonable request, the code and data associated with this article can be made available on Zenodo REF.

\begin{widetext}
    \appendix
\section{{Nonlinear spin-wave theory for a field-polarized chiral magnet}} 
\label{sec:NSWT}
The goal of this section is to provide a detailed derivation of the final Holstein-Primakoff (HP) bosonic Hamiltonian used in the main text. The starting point is the spin Hamiltonian
\begin{equation}
    H=H_{\text{XC}} + H_{\text{DMI}} + H_\text{Z},
    \label{eq:fullhamil}
\end{equation}
where
\begin{equation}
    H_{\text{XC}} = -\frac{J}{2}  \sum_i^N \sum_{\mu = \pm x, \pm y} \vec{S}_i \cdot \vec{S}_{i+\mu} 
\end{equation}
with $\mu = \pm x, \pm y$, accounts for the isotropic Heisenberg exchange between nearest neighbors ($J>0$), and
\begin{equation}
     H_{\text{DMI}} = \frac{D}{2} \sum_i^N \sum_{\mu = \pm x, \pm y} \hat{\vec{\mu}} \cdot \vec{S}_i \times \vec{S}_{i+\mu}
\end{equation}
for the Dzyaloshinskii-Moriya (DM) interaction $D$. The DM vectors are chosen along the bond direction, 
but we emphasize that identical results are obtained for interfacial DM interaction, with DM vectors orthogonal to the bond direction\footnote{ It can be shown that in the more general case where the DM vectors form an angle to the bonds, the $H_{\text{DMI}}$ is modified simply by a phase, $H_{\text{DMI}} \rightarrow \mathrm{e}^{\mathrm{i} \theta} H_{\text{DMI}}$. This phase can be absorbed into the creation/annihilation operators and thus does not affect the physical results.}.

\begin{align}
	   H_\text{Z} =   - \Delta \sum_i^N  S^z_i ,
\end{align}
where $\Delta$ parametrizes the magnetic field. $N$ is the total number of lattice sites. Throughout, we assume that $\Delta$ is large enough to stabilize the field-polarized ground state.

To map the spin Hamiltonian onto a magnonic one expressed in second quantization we use the HP relations \cite{holsteinprimakoff1940}
\begin{equation}
\begin{split}
    & S^{z}_{i} = S - a^{\dagger}_i a_i, \\
    & S^{+}_{i} = \sqrt{2S-a^{\dagger}_{i}a_{i}}a_{i} 
    \approx \sqrt{2S}\left(
    a_{i} - \frac{a^{\dagger}_{i}a_{i}a_{i}}{4S}
    \right), \\ 
    & S^{-}_{i} = a^{\dagger}_{i}\sqrt{2S -a^{\dagger}_{i}a_{i}} \approx \sqrt{2S}
    \left(
    a^{\dagger}_{i} - \frac{a^{\dagger}_{i}a^{\dagger}_{i}a_{i}}{4S}
    \right),
\end{split}
\label{HP}
\end{equation}
with
\begin{equation}
    \begin{split}
        &  S^{x}_{i} = \frac{1}{2} \left(S^{+}_{i} + S^{-}_{i} \right), \\
        &  S^{y}_{i} = \frac{1}{2\mathrm{i}}
        \left(
        S^{+}_{i} - S^{-}_{i}
        \right).
    \end{split}
    \label{plus minus operators}
\end{equation}
In Eq.~\eqref{HP} we have considered small deviations on top of the magnetic ground state and have expanded the square root appearing in $S^{\pm}_{i}$. We proceed by transforming each term in Eq.~\eqref{eq:fullhamil} using the HP relations. 

\subsection{Exchange interaction}
Using Eq.~\eqref{plus minus operators} we write $H_{\text{XC}}$ as
\begin{equation}
\begin{split}
    & H_{\text{XC}} = -\frac{J}{2} \sum_{i, \mu}h_{i, \mu}^{\text{XC}},  \\
    & h_{i, \mu}^{\text{XC}} = S_{i}^{z}S_{i+\mu}^{z} + \frac{1}{2}
    \left( 
    S_{i}^{+}S_{i+\mu}^{-} + S_{i+\mu}^{+}S_{i}^{-}
    \right).
\end{split}
\end{equation}

Implementing the HP transformation of Eq.~\eqref{HP} we get
\begin{equation}
         h_{i , \mu}^{\text{XC}}
        = h_{i, \mu}^{\text{XC},0} + h_{i,\mu}^{\text{XC},2} + h_{i,\mu}^{\text{XC},4},
\end{equation}
where
\begin{equation}
    \begin{split}
        & h_{i, \mu}^{\text{XC},0} = S^2, \\
        & h_{i, \mu}^{\text{XC},2} = S
        \left(
        a^{\dagger}_{i}a_{i + \mu} - a^{\dagger}_{i + \mu}a_{i + \mu} - a^{\dagger}_{i}a_{i} + \text{h.c.}
        \right), \\
        & h_{i, \mu}^{\text{XC},4} = a_{i}^{\dagger}a_{i}a^{\dagger}_{i + \mu}a_{i + \mu} 
        - 
        \left[
        \frac{1}{4} (a_{i+\mu}a^{\dagger}_{i}a^{\dagger}_{i}a_{i} + a^{\dagger}_{i}a^{\dagger}_{i + \mu}a_{i + \mu}a_{i + \mu}) + \text{h.c.} 
        \right].
    \end{split}
\end{equation}
In the following, the constant term $h_{i, \mu}^{\text{XC},0}$ is discarded, and we focus on the free-magnon part $h_{i, \mu}^{\text{XC},2}$ and the four-magnon interaction part $h_{i, \mu}^{\text{XC},4}$.

\subsubsection{Free Hamiltonian from exchange interaction}

To diagonalize the free part of the exchange interaction, $H_{\text{XC}}^{2}= -\frac{JS}{2} \sum_{i, \mu}h_{i, \mu}^{\text{XC},2} $, we utilize the discrete translation invariance of the system and perform a Fourier transformation defined by 
\begin{equation}
    \begin{split}
        & a_{i} = \frac{1}{\sqrt{N}}\sum_{\vec{k}} \mathrm{e}^{\mathrm{i}\vec{k}\cdot \vec{R}_{i}}a_{\vec{k}}, \\ 
        & 
        a^{\dagger}_{i} = \frac{1}{\sqrt{N}}\sum_{\vec{k}} \mathrm{e}^{-\mathrm{i}\vec{k}\cdot \vec{R}_{i}}a^{\dagger}_{\vec{k}},
    \end{split}
\label{FT}
\end{equation}
where $\vec{R}_i$ refers to the position of the $i$-th lattice site. 
In $\vec{k}$-space we get for the free part
\begin{equation}
\begin{split}
    H_{\text{XC}}^{2} 
    &=  \frac{JS}{2} \sum_{i, \mu}(-a^{\dagger}_{i}a_{i + \mu} + a^{\dagger}_{i + \mu}a_{i + \mu} + a^{\dagger}_{i}a_{i} - a^{\dagger}_{i + \mu}a_{i})\\
    & = \frac{JS}{2} \frac{1}{N}\sum_{i, \mu} \sum_{\vec{k}, \vec{k}'}\mathrm{e}^{\mathrm{i} \vec{R}_i \cdot (\vec{k} - \vec{k}')} \left(
    2 - \mathrm{e}^{\mathrm{i}\vec{k} \cdot \vec{\hat{\mu}}} - \mathrm{e}^{-\mathrm{i}\vec{k}' \cdot \vec{\hat{\mu}}}
    \right) 
    a^{\dagger}_{\vec{k}'}a_{\vec{k}} 
    \\ 
    & = 2JS \sum_{\vec{k}}
    \left(
    2 - \cos k_{x} - \cos k_{y}
    \right)
    a^{\dagger}_{\vec{k}}a_{\vec{k}}. 
\end{split}
\end{equation}
In the above, the lattice constant has been assumed to be one. We will keep this convention for the rest of the Appendix unless otherwise stated. 

We combine the result of the exchange interaction together with the Zeeman term 
\begin{equation}
    H_{\text{Z}} = \Delta \sum_{\vec{k}} a^{\dagger}_{\vec{k}}a_{\vec{k}}. 
\end{equation}
Altogether, we find the full harmonic Hamiltonian piece
\begin{align}
    H^2 = H_{\text{XC}}^{2} + H_{\text{Z}} 
    = \sum_{\vec{k}} \varepsilon_{\vec{k}} a^{\dagger}_{\vec{k}}a_{\vec{k}},
\end{align}
where we have defined the free magnon dispersion
 \begin{equation}
     \varepsilon_{\vec{k}} = 2JS \left(
     2 - \cos k_{x} - \cos k_{y}
     \right) 
     + \Delta. 
     \label{magnon spectrum}
 \end{equation}

\subsubsection{Magnon-magnon interactions from the exchange interaction}
The four-magnon interaction arising from the exchange interaction is given by
\begin{equation}
    \begin{split}
         H_{\text{XC}}^4 
         &
         = -\frac{J}{2} \sum_{i, \mu} a^{\dagger}_{i} a_{i} a^{\dagger}_{i + \mu} a_{i + \mu} 
         - 
         \left[
         \frac{1}{4}
         \left(
         a_{i+\mu}a^{\dagger}_{i}a^{\dagger}_{i}a_{i} + a^{\dagger}_{i}a^{\dagger}_{i + \mu}a_{i + \mu}a_{i + \mu}
         \right)
         + \text{h.c.}
         \right] \\
        & = -\frac{J}{2} \sum_{i, \mu} a^{\dagger}_{i} a^{\dagger}_{i + \mu} a_{i}  a_{i + \mu}
         - \frac{1}{4}
        \left(
        a^{\dagger}_{i}a^{\dagger}_{i}a_{i+\mu}a_{i} + a^{\dagger}_{i}a^{\dagger}_{i + \mu}a_{i + \mu}a_{i + \mu} 
        \right.
        \left. + a^{\dagger}_{i}a^{\dagger}_{i+ \mu}a_{i}a_{i} + a^{\dagger}_{i + \mu}a^{\dagger}_{i + \mu}a_{i + \mu}a_{i}
        \right).
    \end{split}
    \label{xc 4}
\end{equation}
The first term in $\vec{k}$-space reads 
\begin{equation}
\begin{split} 
    \sum_{i, \mu} a^{\dagger}_{i} a_{i} a^{\dagger}_{i + \mu} a_{i + \mu} 
    & = \frac{1}{N^2} \sum_{i, \mu}\sum_{\vec{k}_1, \vec{k}_2, \vec{k}_3, \vec{k}_4}  \mathrm{e}^{\mathrm{i}\vec{R}_{i} \cdot 
    \left(
    \vec{k}_3 + \vec{k}_4 - \vec{k}_1 - \vec{k}_2
    \right)
    } \mathrm{e}^{\mathrm{i}
    \left(
    \vec{k}_4 - \vec{k}_{2}
    \right)
    \cdot \hat{\vec{\mu}}} a^{\dagger}_{\vec{k}_{1}}a^{\dagger}_{\vec{k_{2}}} a_{\vec{k}_{3}}  a_{\vec{k}_{4}} \\
    & =
    \frac{N}{N^2} \sum_{\vec{k}_1, \vec{k}_2, \vec{k}_3, \vec{k}_4, \mu}   \delta_{\vec{k}_{1}+\vec{k}_2,\vec{k}_3 +\vec{k}_4} \mathrm{e}^{\mathrm{i}
    \left(
    \vec{k}_4 - \vec{k}_{2}
    \right)
    \cdot \vec{\hat{\mu}}} a^{\dagger}_{\vec{k}_{1}}a^{\dagger}_{\vec{k_{2}}} a_{\vec{k}_{3}}  a_{\vec{k}_{4}} \\
    & =
    \frac{1}{N} \sum_{\vec{k}_1, \vec{k}_2, \vec{k}_3, \vec{k}_4}  \delta_{\vec{k}_{1}+\vec{k}_2,\vec{k}_3 +\vec{k}_4} 2 
    \left(
    \cos 
    \left(
    k_{4}^{x} - k_{2}^{x}
    \right) 
    + \cos
    \left(
    k_{4}^{y} - k_{2}^{y}
    \right)
    \right) 
    a^{\dagger}_{\vec{k}_{1}}a^{\dagger}_{\vec{k}_2} a_{\vec{k}_{3}}  a_{\vec{k}_{4}} \\
    & =
    \frac{1}{N}
    \sum_{\vec{k}_1, \vec{k}_2, \vec{k}_3, \vec{k}_4}  \delta_{\vec{k}_{1}+\vec{k}_2,\vec{k}_3 +\vec{k}_4} 
    \sum_{a=x,y} 
    2 \cos
    \left(
    k_{4}^{a} - k_{2}^{a}
    \right) 
    a^{\dagger}_{\vec{k}_{1}}a^{\dagger}_{\vec{k}_2} a_{\vec{k}_{3}} a_{\vec{k}_{4}}.
\end{split}
\end{equation}
We use that 
$
    \cos
    \left(
    k_{4}^{a}-k_{2}^{a}
    \right) 
    = \cos 
    \left(
    k_{3}^{a}-k_{1}^{a}
    \right)
$ 
due to momentum conservation and that
\begin{align}
    \sum_{\vec{k}_1, \vec{k}_2, \vec{k}_3, \vec{k}_4}  \delta_{\vec{k}_{1}+\vec{k}_2,\vec{k}_3 +\vec{k}_4} \sum_{a=x,y} 2 \cos
    \left(
    k_{4}^{a} - k_{2}^{a}
    \right)
    a^{\dagger}_{\vec{k}_{1}}a^{\dagger}_{\vec{k_{2}}} a_{\vec{k}_{3}} a_{\vec{k}_{4}} \nonumber \\
    = \sum_{\vec{k}_1, \vec{k}_2, \vec{k}_3, \vec{k}_4}  \delta_{\vec{k}_{1}+\vec{k}_2,\vec{k}_3 +\vec{k}_4}\sum_{a=x,y} 2 \cos
    \left(
    k_{4}^{a} - k_{1}^{a}
    \right) 
    a^{\dagger}_{\vec{k}_{2}}a^{\dagger}_{\vec{k}_1} a_{\vec{k}_{3}} a_{\vec{k}_{4}}  \nonumber  \\
    = \sum_{\vec{k}_1, \vec{k}_2, \vec{k}_3, \vec{k}_4}  \delta_{\vec{k}_{1}+\vec{k}_2,\vec{k}_3 +\vec{k}_4} \sum_{a=x,y} 2 \cos
    \left(
    k_{4}^{a} - k_{1}^{a}
    \right)
    a^{\dagger}_{\vec{k}_{1}}a^{\dagger}_{\vec{k}_2} a_{\vec{k}_{3}} a_{\vec{k}_{4}},
\end{align} 
to obtain
\begin{align}
     \sum_{i, \mu} a^{\dagger}_{i} a_{i} a^{\dagger}_{i + \mu} a_{i + \mu} 
     &= 
     \frac{1}{4N} \sum_{\vec{k}_1, \vec{k}_2, \vec{k}_3, \vec{k}_4}   \delta_{\vec{k}_{1}+\vec{k}_2,\vec{k}_3 +\vec{k}_4}
     \sum_{a=x,y} \sum_{b=1}^{2} \sum_{c=3}^{4} 2 
     \cos
     \left(
     k_{b}^{a} - k_{c}^{a}
     \right)
     a^{\dagger}_{\vec{k}_{1}}a^{\dagger}_{\vec{k_{2}}} a_{\vec{k}_{3}} a_{\vec{k}_{4}}.
\end{align}
The other terms in $H^4_{\text{XC}}$ in Eq.~\eqref{xc 4} contain either one term with a different site index or three. The ones with one give rise to a phase factor containing only one momentum in the exponent, while the ones with three, have a phase factor with three momenta. Using momentum conservation the latter can be written with only momentum, too. The final result for the four-magnon interaction Hamiltonian reads in $\vec{k}$-space 
\begin{align}
    H_{\text{XC}}^{4} = \frac{1}{4N}\sum_{\vec{k}_{1}, \vec{k}_2,\vec{k}_3, \vec{k}_4} \delta_{\vec{k}_{1}+\vec{k}_2,\vec{k}_3 +\vec{k}_4}W_{\vec{k}_1, \vec{k}_2; \vec{k}_3, \vec{k}_4} a^{\dagger}_{\vec{k}_{1}}a^{\dagger}_{\vec{k_{2}}} a_{\vec{k}_{3}} a_{\vec{k}_{4}}, 
\end{align}
where the four-magnon vertex is given by
\begin{equation}
     W_{\vec{k}_1, \vec{k}_2; \vec{k}_3, \vec{k}_4} = J\sum_{\mu=x,y}\left(
     \sum_{a=1}^{4} \cos k^{\mu}_{a} - \sum_{b=1}^{2} \sum_{c=3}^{4}  \cos
     \left(
     k_{b}^{a} - k_{c}^{a}
     \right)
     \right).
     \label{4-magnon vertex}
\end{equation}
It comes with a symmetry under the interchange $\vec{k}_1, \vec{k}_2 \leftrightarrow \vec{k}_3, \vec{k}_4$, that is,
\begin{equation}
      W_{\vec{k}_1, \vec{k}_2; \vec{k}_3, \vec{k}_4} =  W_{\vec{k}_3, \vec{k}_4; \vec{k}_1, \vec{k}_2}.
\end{equation}

\subsection{Magnon-magnon interactions from the Dzyaloshinskii-Moriya interaction}
We now perform the HP transformation for the DMI part of the spin Hamiltonian,
\begin{equation}
     H_{\text{DMI}} = \frac{D}{2} \sum_i^N \sum_{\mu = \pm x, \pm y} \hat{\vec{\mu}} \cdot \vec{S}_i \times \vec{S}_{i+\mu}.
\end{equation}
Expanding the cross product we get
\begin{equation}
\begin{split}
    &\sum_{\mu} \hat{\vec{\mu}}\cdot \vec{S}_{i} \times \vec{S}_{i+\mu}  
    = 
    \sum_{\lambda = \pm}
   \lambda\frac{1}{2}
   \left[
   \mathrm{i}
   \left(
   S_{i}^{-}S_{i + \lambda x}^{z}-S_{i + \lambda x}^{-}S_{i}^{z}
   \right)
   + 
   \left(
   S_{i + \lambda y}^{-}S_{i}^{z} - S_{i}^{-}S_{i + \lambda y}^{z}
   \right)
   \right]
   + \text{h.c}.
\end{split}
\end{equation}

The two terms in the parenthesis can be written as
\begin{equation}
\begin{split}
S_{i}^{-}S_{i + \lambda x}^{z}-S_{i + \lambda x}^{-}S_{i}^{z} 
& = \sqrt{2S}
\left[
\left(
a^{\dagger}_{i}-\frac{a^{\dagger}_{i}a^{\dagger}_{i}a_{i}}{4S}
\right)
\left(
S - a^{\dagger}_{i + \lambda x} a_{i + \lambda x}
\right)
- 
\left(
a^{\dagger}_{i + \lambda x}-\frac{a^{\dagger}_{i + \lambda x}a^{\dagger}_{i + \lambda x}a_{i + \lambda x}}{4S}
\right)
\left(
S - a^{\dagger}_{i}a_{i}
\right)
\right] \\
& \approx \sqrt{2S}
\left[
Sa^{\dagger}_{i}- \frac{a_{i}^{\dagger}a_{i}^{\dagger}a_{i}}{4}- a^{\dagger}_{i}a_{i + \lambda x}^{\dagger} a_{i + \lambda x} - Sa^{\dagger}_{i + \lambda x} + \frac{a_{i +\lambda x}^{\dagger}a_{i +\lambda x}^{\dagger}a_{i + \lambda x}}{4} + a^{\dagger}_{i + \lambda x}a^{\dagger}_{i}a_{i}\right],
\end{split}
\end{equation}
and analogously 
\begin{equation}
    \begin{split}
        & S_{i + \lambda y}^{-}S_{i}^{z}-S_{i}^{-}S_{i + \lambda y}^{z} \approx \sqrt{2S}
        \left[
        Sa^{\dagger}_{i + \lambda y} - \frac{a_{i + \lambda y}^{\dagger}a_{i + \lambda y}^{\dagger}a_{i + \lambda y}}{4} - a^{\dagger}_{i +\lambda y} a^{\dagger}_{i}a_{i} - Sa^{\dagger}_{i} + \frac{a^{\dagger}_{i}a_{i}^{\dagger}a_{i}}{4} + a^{\dagger}_{i}a^{\dagger}_{i + \lambda y}a_{i + \lambda y}
        \right].
    \end{split}
    \label{eq: derivation of H3}
\end{equation}
In the above we have discarded terms containing the product of five operators. 
Upon summing over the lattice indices the terms containing three operators that are diagonal in the site index cancel each other out. The same applies for the linear terms. We thus get 
\begin{equation}
\begin{split}
    &  H_{\text{DMI}} = \frac{D\sqrt{2S}}{4} \sum_{i, \lambda = \pm} \lambda 
    \left[
    \mathrm{i} 
    \left(
    a^{\dagger}_{i + \lambda x}a^{\dagger}_{i}a_{i} - a^{\dagger}_{i + \lambda x}a^{\dagger}_{i}a_{i + \lambda x}
    \right)
    +  
    \left(
    a^{\dagger}_{i + \lambda y}a^{\dagger}_{i}a_{i + \lambda y} - a^{\dagger}_{i + \lambda y}a^{\dagger}_{i}a_{i}
    \right)
    + \text{h.c.}
    \right].
\end{split}
\end{equation}
For each term individually we get using Eq.~\eqref{FT}
\begin{equation}
\begin{split}
    \sum_{i} a^{\dagger}_{i + \lambda x}a^{\dagger}_{i}a_{i} 
    & = \frac{1}{N^{3/2}} \sum_{i}\sum_{\vec{k}_{1}, \vec{k}_{2}, \vec{k}_{3}}
    \mathrm{e}^{-\mathrm{i} \vec{k}_{1} \cdot \vec{R}_{i + \lambda x}} 
    \mathrm{e}^{-\mathrm{i}\vec{k}_{2} \cdot \vec{R}_{i}}  
    \mathrm{e}^{\mathrm{i}\vec{k}_{3} \cdot \vec{R}_{i}}
    a^{\dagger}_{\vec{k}_{1}} a^{\dagger}_{\vec{k}_2} a_{\vec{k}_{3}} \\ 
    & = 
    \frac{1}{N^{3/2}} \sum_{i} \mathrm{e}^{\mathrm{i}
    \left(
    \vec{k}_{3}-\vec{k}_{1}-\vec{k}_{2}
    \right) 
    \cdot \vec{R}_{i}} \sum_{\vec{k}_{1}, \vec{k}_{2}, \vec{k}_{3}} \mathrm{e}^{-\lambda \mathrm{i}\vec{k}_{1} \cdot \hat{\vec{x}}} 
    a^{\dagger}_{\vec{k}_{1}} a^{\dagger}_{\vec{k}_2} a_{\vec{k}_{3}} \\
    & = 
    \frac{1}{\sqrt{N}} \sum_{\vec{k}_{1}, \vec{k}_{2}, \vec{k}_{3}} \delta_{\vec{k}_{1} + \vec{k}_{2}, \vec{k}_{3}} \mathrm{e}^{-\mathrm{i}\lambda \vec{k}_{1} \cdot \hat{\vec{x}}}  a^{\dagger}_{\vec{k}_{1}} a^{\dagger}_{\vec{k}_2} a_{\vec{k}_{3}}, 
\end{split}
\end{equation}
and in a similar fashion 
\begin{equation}
\begin{split}
      \sum_{i} a^{\dagger}_{i + \lambda x}a^{\dagger}_{i}a_{i + \lambda x} 
      & = 
      \frac{1}{\sqrt{N}} \sum_{\vec{k}_{1}, \vec{k}_{2}, \vec{k}_{3}} \mathrm{e}^{\mathrm{i}\lambda
      \left(
      \vec{k}_{3}-\vec{k}_{1}
      \right) 
      \cdot \hat{\vec{x}}} \delta_{\vec{k}_{1} + \vec{k}_{2}, \vec{k}_{3}} a^{\dagger}_{\vec{k}_{1}} a^{\dagger}_{\vec{k}_2} a_{\vec{k}_{3}} 
      \\
      & = 
      \frac{1}{\sqrt{N}} \sum_{\vec{k}_{1}, \vec{k}_{2}, \vec{k}_{3}} \mathrm{e}^{\mathrm{i}\lambda\vec{k}_{2} \cdot \hat{\vec{x}}} \delta_{\vec{k}_{1} + \vec{k}_{2}, \vec{k}_{3}} a^{\dagger}_{\vec{k}_{1}} a^{\dagger}_{\vec{k}_2} a_{\vec{k}_{3}}.
\end{split}
\end{equation}
The procedure for the terms with $i$ and $i +\lambda y$ indices is the same. Altogether we have
\begin{equation}
\begin{split}
    & \frac{D\sqrt{2S}}{4\sqrt{N}} \sum_{\vec{k}_{1}, \vec{k}_{2},\vec{k}_{3}} \sum_{\lambda = \pm}  \delta_{\vec{k}_{1} + \vec{k}_{2}, \vec{k}_{3}}
    \lambda
    \left[
    \mathrm{i} 
    \left(
    \mathrm{e}^{-\lambda\mathrm{i}\vec{k}_{1} \cdot \hat{\vec{x}}} - \mathrm{e}^{\mathrm{i} \lambda \vec{k}_{2} \cdot \hat{\vec{x}}}
    \right)
    + 
    \mathrm{e}^{\mathrm{i} \lambda \vec{k}_{2} \cdot \hat{\vec{y}}} - \mathrm{e}^{-\mathrm{i} \lambda \vec{k}_{1} \cdot \hat{\vec{y}}}
    \right] 
    a^{\dagger}_{\vec{k}_{1}} a^{\dagger}_{\vec{k}_2} a_{\vec{k}_{3}} \\
    & = 
    \frac{D\sqrt{2S}}{4\sqrt{N}} \sum_{\vec{k}_{1}, \vec{k}_{2},\vec{k}_{3}} \delta_{\vec{k}_{1} + \vec{k}_{2}, \vec{k}_{3}}
    \left(
    \mathrm{i} \mathrm{e}^{-\mathrm{i}k_{1}^{x}} - \mathrm{i} \mathrm{e}^{\mathrm{i}k_{1}^{x}} + \mathrm{i} \mathrm{e}^{-\mathrm{i}k_{2}^{x}} - \mathrm{i} \mathrm{e}^{\mathrm{i}k_{2}^{x}} + \mathrm{e}^{\mathrm{i}k_{1}^{y}} - \mathrm{e}^{-\mathrm{i}k_{1}^{y}} + \mathrm{e}^{\mathrm{i}k_{2}^{y}} - \mathrm{e}^{-\mathrm{i}k_{2}^{y}}
    \right)
    a^{\dagger}_{\vec{k}_{1}} a^{\dagger}_{\vec{k}_2} a_{\vec{k}_{3}}\\
    & =
    \frac{D\sqrt{2S}}{2\sqrt{N}} \sum_{\vec{k}_{1}, \vec{k}_{2},\vec{k}_{3}} \delta_{\vec{k}_{1} + \vec{k}_{2}, \vec{k}_{3}}
    \left[
    \sin k_{1}^{x} +\sin k_{2}^{x} + \mathrm{i} 
    \left(
    \sin k_{1}^{y} +\sin k_{2}^{y}
    \right)
    \right] 
    a^{\dagger}_{\vec{k}_{1}} a^{\dagger}_{\vec{k}_2} a_{\vec{k}_{3}} \\
    & =
    \frac{1}{2\sqrt{N}} \sum_{\vec{k}_{1}, \vec{k}_{2},\vec{k}_{3}} \delta_{\vec{k}_{1} + \vec{k}_{2}, \vec{k}_{3}} V_{\vec{k}_{3}; \vec{k}_{1},\vec{k}_{2}}  a^{\dagger}_{\vec{k}_{1}} a^{\dagger}_{\vec{k}_2} a_{\vec{k}_{3}},
\end{split}
\end{equation}
where 
\begin{equation}
   V_{\vec{k}_{3}; \vec{k}_{1},\vec{k}_{2}}  =D \sqrt{2S}
   \left[
   \sin k_{1}^{x} +\sin k_{2}^{x} + \mathrm{i} 
   \left(
   \sin k_{1}^{y} +\sin k_{2}^{y}
   \right)
   \right], 
   \label{eq:threemagnonvertex}
\end{equation}
is the three-magnon vertex.

In total, we have for the DMI 
\begin{equation}
    H_{\text{DMI}}= \frac{1}{2\sqrt{N}} \sum_{\vec{k}_{1}, \vec{k}_{2},\vec{k}_{3}} \delta_{\vec{k}_{1} + \vec{k}_{2}, \vec{k}_{3}}
    \left( V_{\vec{k}_{3}; 
    \vec{k}_{1},\vec{k}_{2}}  a^{\dagger}_{\vec{k}_{1}} a^{\dagger}_{\vec{k}_2}a_{\vec{k}_{3}} +  V_{
    \vec{k}_{3};\vec{k}_{1},\vec{k}_{2}}^{*}  a_{\vec{k}_{3}}^{\dagger}a_{\vec{k}_{1}} a_{\vec{k}_2}
    \right)
    . 
    \label{3-magnon potential}
\end{equation}

\subsection{Summary}
We summarize here the results from of this section and introduce the notation for the Hamiltonian to be used in the rest of the Appendix. 
In total we have derived an interacting magnon Hamiltonian that reads
\begin{equation}
    H = H_2 + H_3 + H_4, 
    \label{eq: Full H with indices}
\end{equation}
with the individual pieces given in $\vec{k}$-space, 
\begin{equation}
    \begin{split}
        & H_2 = \sum_{\vec{k}}\varepsilon_{\vec{k}} a_{\vec{k}}^{\dagger}a_{\vec{k}}, \\
        & H_3 = \frac{1}{2\sqrt{N}} \sum_{\vec{k}_{1}, \vec{k}_{2},\vec{k}_{3}} \delta_{\vec{k}_{1} + \vec{k}_{2}, \vec{k}_{3}} 
    \left( V_{\vec{k}_{3}; 
    \vec{k}_{1},\vec{k}_{2}}  a^{\dagger}_{\vec{k}_{1}} a^{\dagger}_{\vec{k}_2}a_{\vec{k}_{3}} +  V_{
    \vec{k}_{3};\vec{k}_{1},\vec{k}_{2}}^{*}  a_{\vec{k}_{3}}^{\dagger}a_{\vec{k}_{1}} a_{\vec{k}_2}
    \right),
    \\
   & 
    H_4 = \frac{1}{4N}\sum_{\vec{k}_{1}, \vec{k}_2,\vec{k}_3, \vec{k}_4} \delta_{\vec{k}_{1}+\vec{k}_2,\vec{k}_3 +\vec{k}_4}W_{\vec{k}_1, \vec{k}_2; \vec{k}_3, \vec{k}_4} a^{\dagger}_{\vec{k}_{1}}a^{\dagger}_{\vec{k_{2}}} a_{\vec{k}_{3}} a_{\vec{k}_{4}}, 
    \end{split}
\end{equation}
and 
\begin{equation}
    \begin{split}
        & \varepsilon_{\vec{k}} = 2JS\left(2 - \cos k_x - \cos k_y \right) + \Delta,\\
         & 
     V_{\vec{k}_{3}; \vec{k}_{1},\vec{k}_{2}}  =D \sqrt{2S}
   \left[
   \sin k_{1}^{x} +\sin k_{2}^{x} + \mathrm{i} 
   \left(
   \sin k_{1}^{y} +\sin k_{2}^{y}
   \right)
   \right],\\
   & 
    W_{\vec{k}_1, \vec{k}_2; \vec{k}_3, \vec{k}_4} = J\sum_{\mu=x,y}\left(
     \sum_{a=1}^{4} \cos k^{\mu}_{a} - \sum_{b=1}^{2} \sum_{c=3}^{4}  \cos
     \left(
     k_{b}^{a} - k_{c}^{a}
     \right)
     \right). 
    \end{split}
\end{equation}
In the notation $H_{n}$ with $n=2,3,4$ introduced in Eq.~\eqref{eq: Full H with indices}, the subscript $n$ refers to the number of magnon operators that each part of the Hamiltonian contains. More specifically, $H_2$ is the non-interacting piece of the Hamiltonian with the energy spectrum given by $\varepsilon_{\vec{k}}$, and $H_3$ and $H_4$ capture three- and four-magnon interactions quantified by the potentials $V_{\vec{k}_3; \vec{k}_1, \vec{k}_2}$ and $W_{\vec{k}_1, \vec{k}_2; \vec{k}_3, \vec{k}}$, respectively. The free magnon spectrum is shown in Fig.~\ref{fig:free spectrum}.

\begin{figure}
    \centering
    \includegraphics[scale=1]{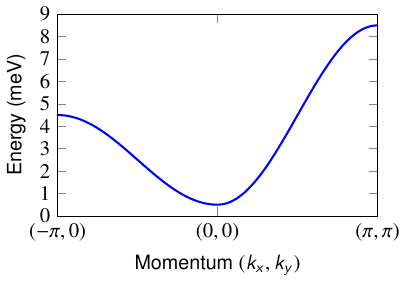}
    \caption{Free magnon dispersion $\varepsilon_{\vec{k}}$ as obtained from the harmonic Hamiltonian piece $H_2$. The dispersion is gapped due to presence of the magnetic field, which  gives rise to the Zeeman energy $\Delta$. Parameters read $S=1$, $J=1\,\text{meV}$, and $\Delta = 0.5\,\text{meV}$.} 
    \label{fig:free spectrum}
\end{figure}

We proceed with commenting on the spin dependence of the terms that we have included in $H$, by taking the non-interacting spectrum $\varepsilon_{\vec{k}}$ as a reference point and comparing the interacting terms $H_3$ and $H_4$ to it. Since the bare spectrum is of the order $O\left(S\right)$, we have that $H_3$ is $O \left( \sqrt{S}^{-1}\right)$ and $H_{4}$ is $O \left(S^{-1}\right)$ relative to $H_2$. Higher order terms that we have neglected in the derivation of $H$ would have an even more subleading dependence on the spin. For example, it can be seen from Eq.~\eqref{eq: derivation of H3} that $H_5$ terms are of the order $O\left(S^{-3/2}\right) $ relative to $H_2$. Thus, $1/\sqrt{S}$ is the perturbative control parameter and we will use it later also in the context of scattering to determine the leading contributions to the thermal Hall effect. 
We also note that in addition to the spin, there is another small parameter in the system: the ratio of the strength of the DM interaction to that of the exchange interaction, denoted as $D/J$. This parameter is small, $D/J \ll 1$, and will play a significant role in the context of transport phenomena, particularly in determining the leading contributions to the thermal Hall effect.

\section{Time reversal symmetry: multi-band case}
\label{sec: TR}
Starting from a magnon expanded Hamiltonian of the form
\begin{equation}
    H = \sum_{n=2}^{\infty}H_{n} = H_2 + H_3 + H_4 + \hdots  ,
    \label{HP expansion}
\end{equation}
we consider the effects of time reversal (TR), assuming  a total of $M$ bands. The different Hamiltonian pieces can be written as 
\begin{equation}
\begin{split}
    &   H_2 = \sum_{\vec{k}} \vec{a}^{\dagger}_{\vec{k}} H_{\vec{k}}\vec{a}_{\vec{k}}, \\
    &   H_3 = \sum_{\vec{k}_1,\vec{k}_2, \vec{k}_3} \delta_{\vec{k}_1 + \vec{k}_2, \vec{k}_3}\sum_{m,n,l} V_{\vec{k}_3;\vec{k}_1, \vec{k}_2}^{l;m,n}a^{\dagger}_{m, \vec{k}_1}a^{\dagger}_{n, \vec{k}_2}a_{l, \vec{k}} + \text{h.c.}, \\
    &   H_4 = \sum_{\vec{k}_1,\vec{k}_2, \vec{k}_3, \vec{k}_4} \delta_{\vec{k}_1 + \vec{k}_2, \vec{k}_3 + \vec{k}_4}\sum_{m,n,l,s} W_{\vec{k}_1, \vec{k}_2; \vec{k}_3, \vec{k}_4}^{l,s;m,n}a^{\dagger}_{m, \vec{k}_1}a^{\dagger}_{n, \vec{k}_2}a_{l, \vec{ k}_3}a_{s, \vec{k}_4}
\end{split}
\label{eq: HP general expansion}
\end{equation}
where $\vec{a}_{\vec{k}}^{\dagger}=(a_{1,\vec{k}}^{\dagger}, a_{2,\vec{k}}^{\dagger}, \hdots, a_{l,\vec{k}} ^{\dagger},\hdots, a_{M,\vec{k}} ^{\dagger})$ is a vector of magnon creation operators, with $a_{l,\vec{k}} ^{\dagger}$ corresponding to the creation operator of the magnon with basis site index $l$, and $H_{\vec{k}}$ is a Hermitian kernel. 
We define the TR operator as an operator $\Theta$ whose action on $\vec{a}_{\vec{k}}^{\dagger}$ and $\vec{a}_{\vec{k}}$ results in
\begin{equation}
         \Theta \vec{a}_{\vec{k}}^{\dagger} \Theta^{-1} = \vec{a}_{-\vec{k}}^{\dagger}U^{\dagger}, \qquad
         \Theta \vec{a}_{\vec{k}} \Theta^{-1} = U \vec{a}_{-\vec{k}},
\end{equation} 
where $U$ is an $M \times M$ unitary matrix. Additionally, since $\Theta$ contains complex conjugation its action on a scalar is given by 
\begin{equation}
    \Theta c \Theta ^{-1}= c^{*}.
\end{equation}
A Hamiltonian as the one given in Eq.~\eqref{HP expansion} is TR-symmetric only if $\Theta H \Theta ^{-1}= H$. For the case where $H$ contains terms up to $H_4$, the conditions for a TR-respecting Hamiltonian are the following: 
\begin{equation}
\begin{split}
       & H_2: \quad U^{\dagger}H_{\vec{k}}^{*}U = H_{-\vec{k}},\\
       & H_{3}: \quad \sum_{m,n,l} 
       \left( 
       V_{\vec{k}_3;\vec{k}_1, \vec{k}_2}^{l;m,n}
       \right)
       ^{*} 
       U^{*}_{hn}U^{*}_{sm}U_{lp} = V_{-\vec{k}_3; -\vec{k}_2, -\vec{k}_1}^{p;h,s}, \\
       & H_4: \quad \sum_{m,n,l,s} 
       \left(
       W_{\vec{k}_1, \vec{k}_2; \vec{k}_3, \vec{k}_4}^{l,s;m,n}
       \right)
       ^{*} U^{*}_{hn} U^{*}_{qm}U_{lp}U_{sr} = W_{-\vec{k}_1, -\vec{k}_2; -\vec{k}_3, -\vec{k}_4}^{h,q;p,r}. 
\end{split}
\label{TR conditions}
\end{equation}

The above conditions simplify to those mentioned in the main text for the case of a single-band Hamiltonian. 
We will prove here in detail the relation for $H_3$ as the rest follows similarly. We first notice that the action of $\Theta$ on a specific creation (annihilation) operator results in $\Theta a_{m,\vec{k}}^{\dagger} \Theta^{-1} = \sum_{h} U_{hm}a_{h,-\vec{k}} ^{\dagger}$ $\left(\Theta a_{m,\vec{k}} \Theta^{-1} = \sum_{h} U_{mh}a_{h,-\vec{k}}\right)$. Based on that, we get for $H_3$
\begin{equation}
    \begin{split}
        \Theta H_3 \Theta^{-1} 
        & = \Theta \sum_{\vec{k}_1,\vec{k}_2, \vec{k}_3} \delta_{\vec{k}_1 + \vec{k}_2, \vec{k}_3}\sum_{m,n,l} V_{\vec{k}_3;\vec{k}_1, \vec{k}_2}^{l;m,n}a^{\dagger}_{m, \vec{k}_1}a^{\dagger}_{n, \vec{k}_2}a_{l, \vec{k}_3} \Theta^{-1} \\
        & = 
        \sum_{\vec{k}_1,\vec{k}_2, \vec{k}_3} \delta_{\vec{k}_1 + \vec{k}_2, \vec{k}_3}\sum_{m,n,l} \Theta V_{\vec{k}_3;\vec{k}_1, \vec{k}_2}^{l;m,n} \Theta^{-1} \Theta a^{\dagger}_{m, \vec{k}_1} \Theta^{-1} \Theta a^{\dagger}_{n, \vec{k}_2} \Theta^{-1} \Theta a_{l, \vec{k}_3} \Theta^{-1} \\
        & = 
        \sum_{\vec{k}_1,\vec{k}_2, \vec{k}_3} \delta_{\vec{k}_1 + \vec{k}_2, \vec{k}_3}\sum_{m,n,l} 
        \left(
        V_{\vec{k}_3;\vec{k}_1, \vec{k}_2}^{l;m,n}
        \right)
        ^{*}\sum_{s}U_{sm}^{*}a^{\dagger}_{s, -\vec{k}_1}\sum_{h}U_{hn}^{*}a^{\dagger}_{h, -\vec{k}_2}\sum_{p} U_{lp}a_{p, -\vec{k}_3} \\
        & = 
        \sum_{\vec{k}_1,\vec{k}_2, \vec{k}_3} \delta_{\vec{k}_1 + \vec{k}_2, \vec{k}_3} \sum_{s,h,p} \sum_{m,n,l} 
        \left(
        V_{\vec{k}_3;\vec{k}_1, \vec{k}_2}^{l;m,n}
        \right)
        ^{*} U_{sm}^{*}U_{hn}^{*} U_{lp}a^{\dagger}_{s, -\vec{k}_1} a^{\dagger}_{h, -\vec{k}_2} a_{p, -\vec{k}_3} .
    \end{split}
\end{equation}
For $\Theta H_3 \Theta ^{-1} = H_3$ to hold true, the condition $\sum_{m,n,l} \left( V_{\vec{k}_3;\vec{k}_1, \vec{k}_2}^{l;m,n}\right)^{*} U^{*}_{hn}U^{*}_{sm}U_{lp} = V_{-\vec{k}_3; -\vec{k}_2, -\vec{k}_1}^{p;h,s}$ in Eq.~\eqref{TR conditions} must be met; indeed,
\begin{equation}
\begin{split}
     \Theta H_3 \Theta ^{-1} 
     &=  \sum_{\vec{k}_1,\vec{k}_2, \vec{k}_3} \delta_{\vec{k}_1 + \vec{k}_2, \vec{k}_3} \sum_{s,h,p} \sum_{m,n,l} (V_{\vec{k}_3;\vec{k}_1, \vec{k}_2}^{l;m,n})^{*} U_{sm}^{*}U_{hn}^{*} U_{lp}a^{\dagger}_{s, -\vec{k}_1} a^{\dagger}_{h, -\vec{k}_2} a_{p, -\vec{k}_3}  \\
    & \stackrel{\text{TR}}{=}  
    \sum_{\vec{k}_1,\vec{k}_2, \vec{k}_3} \delta_{\vec{k}_1 + \vec{k}_2, \vec{k}_3} \sum_{s,h,p} V_{-\vec{k}_3; -\vec{k}_2, -\vec{k}_1}^{p;h,s}a^{\dagger}_{s, -\vec{k}_1} a^{\dagger}_{h, -\vec{k}_2} a_{p, -\vec{k}_3} \\
    & = \sum_{\vec{k}_1,\vec{k}_2, \vec{k}_3} \delta_{\vec{k}_1 + \vec{k}_2, \vec{k}_3} \sum_{s,h,p} V_{\vec{k}_3; \vec{k}_2, \vec{k}_1}^{p;h,s}a^{\dagger}_{s, \vec{k}_1} a^{\dagger}_{h, \vec{k}_2} a_{p, \vec{k}_3} \\
    & = H_3.
\end{split}
\end{equation}
In the last step, we derived an expression that has the same form as $H_3$ upon flipping the sign of the momenta in the sum.

\section{Transport theory}
\label{section:Transport theory}
\subsection{Thermal Hall conductivity: Step by step derivation}
We provide a detailed derivation of the longitudinal and transverse thermal conductivities, which can be extracted from the thermal current generated by applying a temperature gradient $\vec{\nabla} T$ to the sample. Our main target is the calculation of the out of equilibrium magnon distribution function $N_{\vec{k}}=N_{\vec{k}}(t, \vec{r}(t))$ that enters the semi-classical microscopic current density 
\begin{equation}
    \vec{j} = \frac{1}{V} \sum_{\vec{k}}\varepsilon_{\vec{k}} \vec{v}_{\vec{k}} N_{\vec{k}},
    \label{microscopic current}
\end{equation}
where $\varepsilon_{\vec{k}}$ is the free magnon dispersion, $\vec{v}_{\vec{k}} = \left( 1 / \hbar\right)\partial \varepsilon_{\vec{k}}/ \partial \vec{k}$ the magnon group velocity, and $V$ the volume of the system.
Before proceeding, we address our choice of current density. (1) First, in general, in a quantum transport theory, special attention must be given to properly accounting for the contribution of energy magnetization to the current, in addition to the conventional Kubo term\cite{EnergyMagnetization_ThermalTransport}. 
The semiclassical current density in Eq.~\eqref{microscopic current} neglects the energy magnetization correction because we only consider a system with a single band. In this case, there are only intraband processes and the energy magnetization correction is zero. It has been argued in Ref.~\onlinecite{ResonantThermalHallEffect} that the magnetization correction is unimportant for extrinsic effects. We therefore expect that the current density in Eq.~\eqref{microscopic current} is sufficient to capture the leading order effects of magnon-magnon skew scattering also in multi-band systems. (2) Second, the treatment of many-body corrections to the magnon spectrum warrants further explanation. As previously mentioned, the magnon energies $\varepsilon_{\vec{k}}$, and thus the magnon group velocities used in Eq.~\eqref{microscopic current}, are taken from the free magnon theory. In principle, diagrammatic perturbation theory could be employed to account for corrections to these free magnon energies arising from magnon-magnon interactions. While including such corrections is necessary for a fully consistent theory, these contributions are proportional to $1/S$, whereas the free magnon energies scale with $S$. Consequently, the resulting impact on the current density is suppressed by a factor of $1/S$, making it a subleading correction compared to the leading order considered here.

From Eq.~\eqref{microscopic current} we can extract the $\mu \nu$ component of the thermal conductivity tensor by comparing to Fourier's law 
\begin{equation}
    j_{\mu} = - \kappa_{\mu \nu} \partial_{\nu}T 
    \label{Fourier's Law}.
\end{equation}
We can compute $N_{\vec{k}}$ by utilizing the semi-classical Boltzmann equation (BE) that reads
\begin{equation}
    \vec{v}_{\vec{k}} \cdot \vec{\nabla}T\frac{\partial N_{\vec{k}}}{\partial T} = I_{\vec{k}}^{\text{coll}}, 
    \label{BE}
\end{equation}
assuming a steady state and a spatially homogenous system.
The left-hand side of Eq.~\eqref{BE} is the diffusion term accounting for the temperature gradient and the right-hand side is the collision integral that quantifies scattering processes between the magnons, which in our case are present due to the magnon-magnon interactions.  To build up the collision integral we need to calculate the full scattering rate $\Gamma$ associated with every possible collision process allowed by the interacting part of the Hamiltonian 
\begin{equation}
    I_{\vec{k}}^{\text{coll}} = \Gamma_{\vec{k}}\left[\left\{N_{\vec{k}'}\right\} \right].
    \label{CI general}
\end{equation}
Equation \eqref{CI general} implies that $\Gamma$ is a function of the distributions functions $N_{\vec{k}'}$ of the magnons with momenta $\vec{k}'$ participating at a scattering process. 
We linearize the BE by expanding
\begin{equation}
    N_{\vec{k}} = \overline{N}_{\vec{k}} + \delta N_{\vec{k}},
    \label{linearize N}
\end{equation}
where $\overline{N}_{\vec{k}}=(\mathrm{e}^{\beta\varepsilon_{\vec{k}}} - 1)^{-1}$ is the Bose-Einstein distribution and $\delta N_{\vec{k}}$ denotes a small part of $N_{\vec{k}}$ that captures the out-of-equilibrium physics, assumed proportional to $\vec{\nabla}T$. Plugging Eq.~\eqref{linearize N} into Eq.~\eqref{BE} and Eq.~\eqref{CI general} and keeping terms that are linear to $\vec{\nabla}T$, we arrive at the following BE
\begin{equation}
     \vec{v}_{\vec{k}} \cdot \vec{\nabla}T \mathrm{e}^{\beta \varepsilon_{\vec{k}}} \varepsilon_{\vec{k}}\frac{\overline{N}^{2}_{\vec{k}}}{k_{\text{B}}T^2} = \sum_{\vec{k}'} C_{\vec{k} \vec{k}'} \delta N_{\vec{k}'},
    \label{linearized BE}
\end{equation}
where $C_{\vec{k} \vec{k}'}$ is the collision kernel and can be regarded as the first order coefficient of the Taylor expansion of $\Gamma_{\vec{k}}[\{N_{\vec{k}_i}\}]$ around the Bose-Einstein equilibrium populations. We proceed as in Refs.~\onlinecite{Hardy1970, Cepellotti2016} and perform Hardy's basis transformation, changing the elements of the collision kernel as 
\begin{equation}
\begin{split}
    &   \mathcal{C}_{\vec{k} \vec{k}'} = \frac{G_{\vec{k}'}}{G_{\vec{k}}} C_{\vec{k} \vec{k}'},  \\
    & G_{\vec{k}} = \sqrt{\overline{N}_{\vec{k}}(\overline{N}_{\vec{k}} + 1)}.
\end{split}
\label{eq: new basis for C}
\end{equation}
Consequently, the right-hand side of Eq.~\eqref{linearized BE} can be written as
\begin{equation}
    \sum_{\vec{k}'} C_{\vec{k} \vec{k}'} \delta N_{\vec{k}'} = \sum_{\vec{k}'} \frac{G_{\vec{k}}}{G_{\vec{k}'}} \mathcal{C}_{\vec{k} \vec{k}'} \delta N_{\vec{k}'} = 
    G_{\vec{k}} \sum_{\vec{k}'}\mathcal{C}_{\vec{k} \vec{k}'} \mathcal{\delta N}_{\vec{k}'}, 
\end{equation}
with $\mathcal{\delta N}_{\vec{k}'} = \delta N_{\vec{k}'} G_{\vec{k}'}^{-1}$ and Eq.~\eqref{linearized BE} takes the form
\begin{equation}
    G_{\vec{k}}^{-1}
    \vec{v}_{\vec{k}} \cdot \vec{\nabla}T \mathrm{e}^{\beta \varepsilon_{\vec{k}}} \varepsilon_{\vec{k}}\frac{\overline{N}^{2}_{\vec{k}}}{k_{\text{B}}T^2} =   \sum_{\vec{k}'}\mathcal{C}_{\vec{k} \vec{k}'} \mathcal{\delta N}_{\vec{k}'}
    \label{linear BE symmetrized}. 
\end{equation}

For a discretized Brillouin zone (BZ) with a total of $N$ momenta, Eq.~\eqref{linear BE symmetrized} can be written in vector form for all $\vec{k}_i$ that belong to the first BZ. This can be done by recasting $\mathcal{\delta N_{\vec{k}}}$ into a vector
\begin{equation}
    \vec{\mathcal{\delta {N}}}  = 
    \begin{pmatrix}
        \mathcal{\delta N}_{\vec{k}_1} \\
        \mathcal{\delta N}_{\vec{k}_2} \\
        \vdots \\
        \mathcal{\delta N}_{\vec{k}} \\
        \vdots \\
        \mathcal{\delta N}_{\vec{k}_N}
    \end{pmatrix},
\end{equation}
and the collision kernel into a matrix
\begin{equation}
    \vec{C} = 
    \begin{pmatrix}
        & D_{\vec{k}_1} & \mathcal{O}_{\vec{k}_1 \vec{k}_2} & \cdots & \mathcal{O}_{\vec{k}_1 \vec{k}} & \cdots & \mathcal{O}_{\vec{k}_1 \vec{k}_N}\\
        & \mathcal{O}_{\vec{k}_2 \vec{k}_1} & D_{\vec{k}_2} & \cdots & \mathcal{O}_{\vec{k}_2 \vec{k}} &\cdots &\mathcal{O}_{\vec{k}_2 \vec{k}_N}\\
        & \vdots & \vdots        & \ddots & \vdots & & \vdots\\
        & \mathcal{O}_{\vec{k} \vec{k}_1} & \mathcal{O}_{\vec{k} \vec{k}_2}  &  \cdots &  D_{\vec{k}} & \cdots & \mathcal{O}_{\vec{k} \vec{k}_N} \\
        & \vdots & \vdots&  & \vdots & \ddots&\vdots\\
        & \mathcal{O}_{\vec{k}_{N}\vec{k}_1}
        & \mathcal{O}_{\vec{k}_{N}\vec{k}_2} & \cdots & \mathcal{O}_{\vec{k}_{N}\vec{k}} &  \cdots & D_{\vec{k}_N}
    \end{pmatrix},
\end{equation}
with $\mathcal{O}_{\vec{k}\vec{k}'}$ denoting the off-diagonal elements and $D_{\vec{k}}$ the diagonal ones. We note here that the diagonal elements of the collision matrix are the same in both the original and the new basis defined by Eq.~\eqref{eq: new basis for C}. The right-hand side of Eq.~\eqref{linearized BE} can now be written as the product of $\mathcal{\vec{\delta N}}$ and $\mathcal{\vec{C}}$ and one can solve the BE in Eq.~\eqref{linearized BE} with respect to $\mathcal{\vec{\delta N}}$ by inverting the collision matrix. Thus, we get for the element $\mathcal{\delta N}_{\vec{k}}$ 
\begin{equation}
    \mathcal{\delta N}_{\vec{k}} = \sum_{k'} G_{\vec{k}'}^{-1}\mathrm{e}^{\beta \varepsilon_{\vec{k}'}} \varepsilon_{\vec{k}'}\frac{\overline{N}^{2}_{\vec{k}'}}{k_{\text{B}}^2T^2}\left[\mathcal{C}^{-1}\right]_{\vec{k}\vec{k}'} \vec{v}_{\vec{k}'} \cdot \vec{\nabla}T, 
    \label{solving BE}
\end{equation}
with $\left[\mathcal{C}^{-1}\right]_{\vec{k}\vec{k}'}$ referring to the $\vec{k}\vec{k}'$ element of the inverse of the collision matrix in Hardy's basis. Combining Eq.~\eqref{solving BE} with the definition of the current density in Eq.~\eqref{microscopic current} we arrive at 
\begin{equation}
    \begin{split}
        j_{\mu} = \frac{1}{Vk_{\text{B}}T^2} \sum_{\vec{k}, \vec{k}'} \frac{G_{\vec{k}}}{G_{\vec{k}'}}\varepsilon_{\vec{k}}\varepsilon_{\vec{k}'} v_{\vec{k}'}^{\nu}v_{\vec{k}}^{\mu}\mathrm{e}^{\beta \varepsilon_{\vec{k}'}} \overline{N}^{2}_{\vec{k}'}\left[\mathcal{C}^{-1}\right]_{\vec{k}\vec{k}'}\partial_{\nu}T.
    \end{split}
\end{equation}
The $\kappa_{\mu\nu}$ component of the conductivity can be extracted by comparison with Fourier's law in Eq.~\eqref{Fourier's Law}:
\begin{align}
    \kappa_{\mu \nu} 
    = 
     -\frac{1}{Vk_{\text{B}}T^2} \sum_{\vec{k}, \vec{k}'} \varepsilon_{\vec{k}}\varepsilon_{\vec{k}'} v_{\vec{k}'}^{\nu} v_{\vec{k}}^{\mu} G_{\vec{k}}G_{\vec{k}'}\left[\mathcal{C}^{-1}\right]_{\vec{k}\vec{k}'}. 
    \label{kappa general}
\end{align}

We now decompose $\vec{\mathcal{C}} = -\vec{D} + \vec{\mathcal{O}}$ into a diagonal part, $\vec{D}$ (not to be confused with the DM vector), and an off-diagonal part, $\vec{\mathcal{O}}$. We assume that the diagonal elements of the collision matrix comprise the dominant contribution to the scattering rates. This is the case when many-body interactions between the magnons are weaker than other physical mechanisms that can lead to additional damping of the magnons (e.g.: boundary scattering, scattering to impurities or to phonons \textit{etc.}). As discussed in Sec.~\ref{subsection: D's in the calculation}, we account for these mechanisms by a phenomenological scattering rate $D_{\text{ph}}$. Within the validity of this approximation we get that 
\begin{equation}
    \vec{\mathcal{C}}^{-1}=(-\vec{D} + \vec{\mathcal{O}})^{-1} =  -\vec{D}^{-1} - \vec{D}^{-1}\vec{\mathcal{O}}\vec{C}^{-1} \approx -\vec{D}^{-1} - \vec{D}^{-1}\vec{\mathcal{O}}\vec{D}^{-1} ,
    \label{dominant D}
\end{equation}
where terms of quadratic or higher order in $\mathcal{\vec{O}}$ have been neglected.
Inserting Eq.~\eqref{dominant D} into  Eq.~\eqref{kappa general} we arrive at
\begin{equation}
    \kappa_{\mu \nu} =  \frac{1}{Vk_{\text{B}}T^2} \sum_{\vec{k}, \vec{k}'} \varepsilon_{\vec{k}}\varepsilon_{\vec{k}'} v_{\vec{k}'}^{\nu} v_{\vec{k}}^{\mu}G_{\vec{k}}G_{\vec{k}'}
    \left(
    \tau_{\vec{k}} \delta_{\vec{k}, \vec{k}'} + \tau_{\vec{k}}\tau_{\vec{k}'}\mathcal{O}_{\vec{k}\vec{k}'}
    \right), 
    \label{kappa component D dominant}
\end{equation}
where we have introduced the magnon relaxation time $\tau_{\vec{k}} =  1/D_{\vec{k}}$. We can now calculate the thermal Hall conductivity by making use of its anti-symmetrized definition, that is,
\begin{align}
    \kappa_{\text{H}} 
         \equiv \frac{\kappa_{\mu \nu} - \kappa_{\nu \mu}}{2} 
\end{align}
to find
\begin{equation}
\begin{split}
    \kappa_{\text{H}} 
         & = 
         \frac{1}{4V k_{\text{B}}T^2 }\sum_{\vec{k}, \vec{k}'}\varepsilon_{\vec{k}}\varepsilon_{\vec{k}'}\tau_{\vec{k}} \tau_{\vec{k}'}
         G_{\vec{k}}G_{\vec{k}'}
         \left(
         v_{\vec{k}}^{\mu} v_{\vec{k}'}^{\nu}- v_{\vec{k}'}^{\mu} v_{\vec{k}}^{\nu}
         \right)
         \left(
         \mathcal{O}_{\vec{k} \vec{k}'} - \mathcal{O}_{\vec{k}' \vec{k}}
         \right) \\
         & = 
          \frac{1}{2V k_{\text{B}} T^2 }\sum_{\vec{k}, \vec{k}'}\varepsilon_{\vec{k}}\varepsilon_{\vec{k}'} \tau_{\vec{k}} \tau_{\vec{k}'} G_{\vec{k}}G_{\vec{k}'}
          \left(
          v_{\vec{k}'}^{\nu} v_{\vec{k}}^{\mu}- v_{\vec{k}}^{\nu} v_{\vec{k}'}^{\mu}
          \right)
          \mathcal{A}_{\vec{k} \vec{k}'},
         \label{kH compoment}
\end{split}
\end{equation}
with 
\begin{equation}
    \mathcal{A}_{\vec{k} \vec{k}'} = \frac{\mathcal{O}_{\vec{k} \vec{k}'} - \mathcal{O}_{\vec{k}' \vec{k}}}{2}
    \label{eq: Antisymmetric part of symmetrized CM}
\end{equation}
being the anti-symmetric part of the collision matrix in Hardy's basis. 
In the derivation of Eq.~\eqref{eq: Antisymmetric part of symmetrized CM} we made the observation that the diagonal contribution proportional to $\delta_{\vec{k}\vec{k}'}\tau_{\vec{k}}$ vanishes identically for the Hall current. The last step consists of writing the result in vector form 
\begin{equation}
    \vec{\kappa}_\text{H} 
     =
     \frac{1}{2Vk_{\text{B}}T^2}\sum_{\vec{k}, \vec{k}'} \vec{v}_{\vec{k}} \times \vec{v}_{\vec{k}'}\varepsilon_{\vec{k}} \varepsilon_{\vec{k}'} \tau_{\vec{k}}\tau_{\vec{k}'} G_{\vec{k}} G_{\vec{k}'} \mathcal{A}_{\vec{k}\vec{k}'},
     \label{eq: Hall vector}
\end{equation}
where in our case of a two-dimensional system only one of the three components of $\vec{\kappa}_\text{H}$ survives. Additionally, the longitudinal conductivity can be computed by taking the case of $\mu = \nu$ in Eq.~\eqref{kappa component D dominant}, yielding 
\begin{equation}
    \kappa_{\mu \mu} =  \frac{1}{Vk_{\text{B}}T^2} \sum_{\vec{k}, \vec{k}'} \varepsilon_{\vec{k}}\varepsilon_{\vec{k}'} v_{\vec{k}'}^{\mu} v_{\vec{k}}^{\mu}G_{\vec{k}} G_{\vec{k}'}
    \left(
    \tau_{\vec{k}}  \delta_{\vec{k}, \vec{k}'} + \tau_{\vec{k}}\tau_{\vec{k}'}\mathcal{O}_{\vec{k}\vec{k}'}
    \right).
    \label{long final}
\end{equation}

We close this section by comparing the thermal Hall conductivity in Eq.~\eqref{kH compoment} to the one obtained in Refs.~\onlinecite{Mangeolle2022PRB, Mangeolle2022PRX}, which is given in the original basis for the collision matrix:
\begin{equation}
\begin{split}
    & \kappa_{\text{H}} =  \frac{1}{2V k_{\text{B}}T^2 }\sum_{\vec{k}, \vec{k}'}\varepsilon_{\vec{k}}\varepsilon_{\vec{k}'}\tau_{\vec{k}} \tau_{\vec{k}'} 
    v_{\vec{k}}^{\mu} v_{\vec{k}'}^{\nu}
    HC_{\vec{k}\vec{k}'}, \\
    & HC_{\vec{k}\vec{k}'} =  \mathrm{e}^{\beta \varepsilon_{\vec{k}'}}\overline{N}^{2}_{\vec{k}'}O_{\vec{k} \vec{k}'} - \mathrm{e}^{\beta \varepsilon_{\vec{k}}}\overline{N}^{2}_{\vec{k}}O_{\vec{k}' \vec{k}}.  
\end{split}
\label{eq: KH from long paper}
\end{equation}
Indeed, one can show that the results are identical by first noting that
\begin{equation}
    \begin{split}
         HC_{\vec{k}\vec{k}'} &= \frac{G_{\vec{k}}}{G_{\vec{k}'}} \mathrm{e}^{\beta \varepsilon_{\vec{k}}'}\overline{N}^{2}_{\vec{k}'}\mathcal{O}_{\vec{k} \vec{k}'} - \frac{G_{\vec{k}'}}{G_{\vec{k}}}\mathrm{e}^{\beta \varepsilon_{\vec{k}}}\overline{N}^{2}_{\vec{k}}\mathcal{O}_{\vec{k}' \vec{k}} \\
         & = \frac{G_{\vec{k}}}{G_{\vec{k}'}} \left(
         \overline{N}_{\vec{k}'}+1
         \right)
         \overline{N}_{\vec{k}'}\mathcal{O}_{\vec{k} \vec{k}'} - \frac{G_{\vec{k}'}}{G_{\vec{k}}}
         \left(
         \overline{N}_{\vec{k}}+1
         \right)
         \overline{N}_{\vec{k}}\mathcal{O}_{\vec{k}' \vec{k}} \\
         & = 
         G_{\vec{k}}G_{\vec{k}'}
         \left(
         \mathcal{O}_{\vec{k} \vec{k}'} - \mathcal{O}_{\vec{k}' \vec{k}}
         \right) \\
         & \equiv 2  G_{\vec{k}}G_{\vec{k}'} \mathcal{A}_{\vec{k} \vec{k}'},
    \end{split}
    \label{HC as A}
\end{equation}
and then rewriting Eq.~\eqref{eq: KH from long paper} in order to get
\begin{align}
\begin{split}
    \kappa_{\text{H}} 
    & =  \frac{1}{2V k_{\text{B}} T^2 }\sum_{\vec{k}, \vec{k}'}\varepsilon_{\vec{k}}\varepsilon_{\vec{k}'}\tau_{\vec{k}} \tau_{\vec{k}'} 
    v_{\vec{k}}^{\mu} v_{\vec{k}'}^{\nu}
    HC_{\vec{k}\vec{k}'} \\
    & 
    =\frac{1}{4V k_{\text{B}} T^2}\sum_{\vec{k}, \vec{k}'}\left(\varepsilon_{\vec{k}}\varepsilon_{\vec{k}'}\tau_{\vec{k}} \tau_{\vec{k}'} 
    v_{\vec{k}}^{\mu} v_{\vec{k}'}^{\nu}
    HC_{\vec{k}\vec{k}'} 
     +  \varepsilon_{\vec{k}}\varepsilon_{\vec{k}'}\tau_{\vec{k}} \tau_{\vec{k}'} 
    v_{\vec{k}}^{\nu} v_{\vec{k}'}^{\mu}
    HC_{\vec{k}'\vec{k}}
    \right) \\
    & 
    = 
    \frac{1}{4Vk_{\text{B}} T^2 }\sum_{\vec{k}, \vec{k}'}\varepsilon_{\vec{k}}\varepsilon_{\vec{k}'}\tau_{\vec{k}} \tau_{\vec{k}'} 
    \left(v_{\vec{k}}^{\mu} v_{\vec{k}'}^{\nu} - 
    v_{\vec{k}'}^{\mu} v_{\vec{k}}^{\nu} \right) 
    HC_{\vec{k}\vec{k}'} \\
    & 
    \stackrel{\eqref{HC as A}}{=}
    \frac{1}{2V k_{\text{B}} T^2 }\sum_{\vec{k}, \vec{k}'}\varepsilon_{\vec{k}}\varepsilon_{\vec{k}'}\tau_{\vec{k}} \tau_{\vec{k}'} 
    G_{\vec{k}}G_{\vec{k}'}
    \left(v_{\vec{k}}^{\mu} v_{\vec{k}'}^{\nu} - 
    v_{\vec{k}'}^{\mu} v_{\vec{k}}^{\nu} \right) 
    \mathcal{A}_{\vec{k} \vec{k}'}.
\end{split}
\end{align}

\subsection{Thermal Hall conductivity and detailed balance}
\label{subsection: DB}
Within the approximation of dominant diagonal scattering, the Hall conductivity vector is given by Eq.~\eqref{eq: Hall vector}, with the anti-symmetric part $\mathcal{A}_{\vec{k}\vec{k}'}$ of the collision matrix in Hardy's basis being the central piece. 
Here, we connect $\mathcal{A}_{\vec{k}\vec{k}'}$ to the detailed balance criterion used in the main text. 

We start by noticing that the elements $\mathcal{O}_{\vec{k}\vec{k}'}$ can be grouped based on wherever the momenta $\vec{k}, \vec{k}'$ are created or destroyed. This is denoted using a superscript notation with pluses and minuses. For example the element $\mathcal{O}^{+-}_{\vec{k}\vec{k}'}$ refers to a process where the momentum $\vec{k}$ is created and the $\vec{k}'$ is destroyed. Processes where $\vec{k}$ is created are referred to as in-processes and their off-diagonal matrix elements read $\mathcal{O}_{\vec{k}\vec{k}'}^{\text{in}}=\mathcal{O}_{\vec{k}\vec{k}'}^{++} + \mathcal{O}_{\vec{k}\vec{k}'}^{+-}$. On the other hand, processes where $\vec{k}$ is destroyed are called out-processes and are associated with elements $\mathcal{O}_{\vec{k}\vec{k}'}^{\text{out}}=\mathcal{O}_{\vec{k}\vec{k}'}^{--} + \mathcal{O}_{\vec{k}\vec{k}'}^{-+}$. In total we can write 
\begin{equation}
    \mathcal{O}_{\vec{k}\vec{k}'} = \mathcal{O}_{\vec{k}\vec{k}'}^{\text{in}} - \mathcal{O}_{\vec{k}\vec{k}'}^{\text{out}} = 
    \left(
    \mathcal{O}_{\vec{k}\vec{k}'}^{++} + \mathcal{O}_{\vec{k}\vec{k}'}^{+-}
    \right) 
    - 
    \left(
    \mathcal{O}_{\vec{k}\vec{k}'}^{--} + \mathcal{O}_{\vec{k}\vec{k}'}^{-+}
    \right).
    \label{eq:O decompose}
\end{equation}
We now calculate $\mathcal{A}_{\vec{k}\vec{k}'}$ based on the notation introduced in Eq.~\eqref{eq:O decompose}. To do that we decompose the $\mathcal{O}_{\vec{k}\vec{k}'}$ further 
\begin{align}
     & 
    \mathcal{O}_{\vec{k}\vec{k}'}^{\text{in}} =  \mathcal{Q}_{\vec{k}\vec{k}'}^{\text{in}}
    \left(
    \overline{N}_{\vec{k}} + 1
    \right), \\
    & 
    \mathcal{O}_{\vec{k}\vec{k}'}^{\text{out}} =  \mathcal{Q}_{\vec{k}\vec{k}'}^{\text{out}}\overline{N}_{\vec{k}},
\end{align}
where $\mathcal{Q}^{\text{in/out}}_{\vec{k}\vec{k}'}$ is a kernel containing microscopic details of the scattering channels, with $\mathcal{Q}^{\text{in/out}}_{\vec{k}\vec{k}'}= {Q}^{\text{in/out}}_{\vec{k}\vec{k}'}G_{\vec{k}'}/G_{\vec{k}}$ to convert between Hardy's and the original basis. To proceed we consider two separate cases. 

\begin{enumerate}
    \item[(i)]
The first case corresponds to scattering channels where the $\vec{k}, \vec{k}'$ are either both created or destroyed with off-diagonal elements $\mathcal{O}_{\vec{k}\vec{k}'}^{\text{in}}= \mathcal{O}_{\vec{k}\vec{k}'}^{++}$ and $\mathcal{O}_{\vec{k}\vec{k}'}^{\text{out}}= \mathcal{O}_{\vec{k}\vec{k}'}^{--}$. Considering these, $\mathcal{A}_{\vec{k} \vec{k}'}$ becomes
\begin{equation}
    \begin{split}
    2\mathcal{A}_{\vec{k}\vec{k}'} 
    &= 
    \left(
    \mathcal{O}_{\vec{k}\vec{k}'}^{++}-\mathcal{O}_{\vec{k}'\vec{k}}^{++}
    \right) 
    - 
    \left(
    \mathcal{O}_{\vec{k}\vec{k}'}^{--}-\mathcal{O}_{\vec{k}'\vec{k}}^{--}
    \right)\\
    & = 
    \left[
    \mathcal{Q}_{\vec{k}\vec{k}'}^{++}
    \left(
    \overline{N}_{\vec{k}} + 1
    \right)
    -\mathcal{Q}_{\vec{k}'\vec{k}}^{++}
    \left(
    \overline{N}_{\vec{k}'} + 1
    \right)
    \right]
    - 
   \left(
    \mathcal{Q}_{\vec{k}\vec{k}'}^{--}\overline{N}_{\vec{k}}-\mathcal{Q}_{\vec{k}'\vec{k}}^{--}\overline{N}_{\vec{k}'}
    \right).
    \end{split}
    \label{eq:A first step}
\end{equation}
We notice that the following relations apply for the $Q_{\vec{k}\vec{k}'}$ that we are considering:
\begin{equation}
    \begin{split}
    & 
        Q_{\vec{k}\vec{k}'}^{++} = Q_{\vec{k}'\vec{k}}^{++}, \\
    & 
       Q_{\vec{k}\vec{k}'}^{--} = Q_{\vec{k}'\vec{k}}^{--}.
    \end{split}
    \label{eq: relations for symmetric Qs}
\end{equation}
Expressed for the $\mathcal{Q}$ we have that
\begin{equation}
    \begin{split}
        \mathcal{Q}_{\vec{k}\vec{k}'}^{\pm\pm} 
        & = \frac{G_{\vec{k}'}}{G_{\vec{k}}} Q_{\vec{k}\vec{k}'}^{\pm\pm} \stackrel{\eqref{eq: relations for symmetric Qs}}{=} \frac{G_{\vec{k}'}}{G_{\vec{k}}} Q_{\vec{k}'\vec{k}}^{\pm\pm} 
        = \left(
        \frac{G_{\vec{k}'}}{G_{\vec{k}}} \right)
        ^2  
        \underbrace{\frac{G_{\vec{k}'}}{G_{\vec{k}}}Q_{\vec{k}'\vec{k}}^{\pm\pm}}_{\mathcal{Q}_{\vec{k}'\vec{k}}^{\pm\pm}} 
        = 
        \left(
        \frac{G_{\vec{k}'}}{G_{\vec{k}}} \right)
        ^2 
        \mathcal{Q}_{\vec{k}'\vec{k}}^{\pm\pm}. 
         \label{eq: relations for symmetric calligraphic Qs}
    \end{split}
\end{equation}
Combining Eq.~\eqref{eq:A first step} with Eq.~\eqref{eq: relations for symmetric calligraphic Qs} and using that $\overline{N}_{\vec{q}}+1 = \mathrm{e}^{\beta \varepsilon_{\vec{q}}}\overline{N}_{\vec{q}}$, we get the following relation for $\mathcal{A}_{\vec{k}\vec{k}'}$:
\begin{equation}
    2 \mathcal{A}_{\vec{k} \vec{k}'} = \frac{\overline{N}_{\vec{k}'}-\overline{N}_{\vec{k}}}{\overline{N}_{\vec{k}'}}
    \left(
    \mathcal{O}_{\vec{k},\vec{k}'}^{++}-\mathrm{e}^{-\beta \varepsilon_{\vec{k}'}}\mathcal{O}_{\vec{k}\vec{k}'}^{--}
    \right). 
    \label{eq: HC symmetric}
\end{equation}
\item[(ii)]
In the second case, we have to consider also the part of the scattering elements where one of the momenta $\vec{k},\vec{k}'$ is destroyed and one is created, namely $\mathcal{O}_{\vec{k}\vec{k}'}^\text{in}=\mathcal{O}_{\vec{k}\vec{k}'}^{+-}$ and $\mathcal{O}_{\vec{k}\vec{k}'}^\text{out} = \mathcal{O}_{\vec{k}\vec{k}'}^{-+}$. For the $Q$ kernels associated with the aforementioned processes we have that
\begin{equation}
    \begin{split}
        &
        Q_{\vec{k}'\vec{k}}^{+-} = Q_{\vec{k}\vec{k}'}^{-+}, \\
        & 
        Q_{\vec{k}'\vec{k}}^{-+} = Q_{\vec{k}\vec{k}'}^{+-}.
    \end{split}
\end{equation}
Following the same procedure as before we get 
\begin{equation}
    2 \mathcal{A}_{\vec{k}\vec{k}'} = \frac{\overline{N}_{\vec{k}} + \overline{N}_{\vec{k}'} + 1}{\overline{N}_{\vec{k}'}}
    \left(
    \mathrm{e}^{-\beta \varepsilon_{\vec{k}'}}\mathcal{O}_{\vec{k}\vec{k}'}^{+-}-\mathcal{O}_{\vec{k}\vec{k}'}^{-+}
    \right).
    \label{eq: HC non-symmetric}
\end{equation}
\end{enumerate}
Combining Eq.~\eqref{eq: HC symmetric} and Eq.~\eqref{eq: HC non-symmetric} we arrive at the final result for the $\mathcal{A}_{\vec{k}\vec{k}'}$ shown in the main text, i.e.,
\begin{equation}
    2 \mathcal{A}_{\vec{k}\vec{k}'} = \frac{\overline{N}_{\vec{k}} + \overline{N}_{\vec{k}'} + 1}{\overline{N}_{\vec{k}'}}
    \left(
    \mathrm{e}^{-\beta \varepsilon_{\vec{k}'}}\mathcal{O}_{\vec{k}\vec{k}'}^{+-}-\mathcal{O}_{\vec{k}\vec{k}'}^{-+}
    \right) 
    + 
    \frac{\overline{N}_{\vec{k}'}-\overline{N}_{\vec{k}}}{\overline{N}_{\vec{k}'}}
    \left(
    \mathcal{O}_{\vec{k}\vec{k}'}^{++}-\mathrm{e}^{-\beta \varepsilon_{\vec{k}'}}\mathcal{O}_{\vec{k}\vec{k}'}^{--}
    \right).
    \label{eq:A-matrix and DB}
\end{equation} 
We read off from Eq.~\eqref{eq:A-matrix and DB} that microscopic detailed balance must be broken to obtain a finite anti-symmetric part of the collision matrix and, hence, a thermal Hall effect. For our model, it can be observed that the detailed balance is broken by a minus sign (referred to as `anti-detailed' balance in Refs. \onlinecite{Mangeolle2022PRB, Mangeolle2022PRX}) because
\begin{equation}
    \begin{split}
        &
        \mathcal{O}_{\vec{k}\vec{k}'}^{++}= -\mathrm{e}^{-\beta \varepsilon_{\vec{k}'}}\mathcal{O}_{\vec{k}\vec{k}'}^{--}
        \\
        & 
        \mathcal{O}_{\vec{k}\vec{k}'}^{+-}= - \mathrm{e}^{\beta \varepsilon_{\vec{k}'}}\mathcal{O}_{\vec{k}\vec{k}'}^{-+}.
    \end{split}
\end{equation}
How this minus sign arises is demonstrated in an example in Sec.~\ref{section: Interference process}. 

\section{Scattering theory}

The aim of this section is to illustrate the calculation of scattering rates corresponding to the first and higher order terms in the interacting part of the Hamiltonian $H_{\text{int}}$. 
To compute higher order terms of the scattering rate $\Gamma_{\text{if}}$ of a specific channel connecting the initial and final state, $\ket{\text{i}}, \ket{\text{f}}$, respectively, we make use of the T-matrix approximation
\begin{equation}
\Gamma_{\text{if}}\left[\left\{N_{\vec{k}'} \right\}\right] = \frac{2\pi}{\hbar} 
\left|
T_{\text{if}}
\right|
^2 \delta
\left(
\varepsilon_{\text{i}}- \varepsilon_{\text{f}}
\right),
    \label{FGR}
\end{equation}
with $N_{\vec{k}'}$ denoting the dependence of the scattering rate on the out of equilibrium magnon populations $N_{\vec{k}'}$.
In the above, $T_{\text{if}}$ is a $T$-matrix element defined through
\begin{equation}
    T_{\text{if}} = \bra{\text{f}} H_{\text{int}} \ket{\text{i}} + \sum_{\nu}\frac{\bra{\text{f}}H_{\text{int}}\ket{\nu}\bra{\nu}H_{\text{int}}\ket{\text{i}}}{\varepsilon_{\text{i}} - \varepsilon_{\nu} + \mathrm{i} \eta} + \hdots = T^{\left(1\right)}_{\text{if}} + T^{\left(2\right)}_{\text{if}} + \ldots, 
    \label{Lippmann}
\end{equation}
with 
\begin{equation}
\begin{split}
    & T^{\left(1\right)}_{\text{if}} = \bra{\text{f}} H_{\text{int}} \ket{\text{i}}, \\
    & T^{\left(2\right)}_{\text{if}} =  \sum_{\nu}\frac{\bra{\text{f}}H_{\text{int}}\ket{\nu}\bra{\nu}H_{\text{int}}\ket{\text{i}}}{\varepsilon_{\text{i}} - \varepsilon_{\nu} + \mathrm{i} \eta}, 
\end{split}
\end{equation}
where $H_{\text{int}}$ is the interacting part of the Hamiltonian, $\eta>0$ a regularization factor, and $\nu$ denotes all possible intermediate states. 
By keeping the terms $T^{\left(1\right)}_{\text{if}}$ and $T^{\left(2\right)}_{\text{if}}$ for the T-matrix element $T_{\text{if}}$, we get for the scattering rate
\begin{equation}
    \Gamma_{\text{i} \text{f}}\left[\left\{N_{\vec{k}'} \right\}\right]
     = 
    \Gamma_{\text{i} \text{f}}^{\left(1\right)} + \Gamma_{\text{i} \text{f}}^{\left(2\right)} + \Gamma_{\text{i} \text{f}}^{\text{interf.}},
\end{equation}
with 
\begin{align}
    \begin{split}
        & \Gamma_{\text{i} \text{f}}^{\left(1\right)} = \frac{2\pi}{\hbar}
        \delta
        \left(
        \varepsilon_{\text{i}}-\varepsilon_{\text{f}}
        \right)
        \left|
        T_{\text{if}}^{(1)}
        \right|
        ^2,  \\
        & \Gamma_{\text{i} \text{f}}^{(2)}=\frac{2\pi}{\hbar}
        \delta
        \left(
        \varepsilon_{\text{i}}-\varepsilon_{\text{f}}
        \right)
        \left|
        T_{\text{if}}^{\left(2\right)}
        \right|
        ^2,  \\
        & \Gamma_{\text{i} \text{f}}^{\text{interf.}} =\frac{4\pi}{\hbar}
    \delta
    \left(
   \varepsilon_{\text{i}}-\varepsilon_{\text{f}}
    \right)
    \left(
    \text{Re}
    T_{\text{if}}^{\left(1\right)}
    \text{Re}
    T_{\text{if}}^{\left(2\right)}
    +\text{Im}
    T_{\text{if}}^{\left(1\right)}
    \text{Im}
    T_{\text{if}}^{\left(2\right)}
    \right).
    \end{split}
    \label{eq: scattering rates general}
\end{align}
Here, $\Gamma_{\text{i}\text{f}}^{\left( 1\right)}$, $\Gamma_{\text{i}\text{f}}^{\left( 2\right)}$, and $\Gamma_{\text{i}\text{f}}^{\text{interf.}}$ are the scattering rates of the lowest order scattering event, the second order scattering event, and the interference between these two events, respectively. Their dependence 
on $N_{\vec{k}'}$ has been omitted to lighted notation. 
From here on, we will be referring to Eq.~\eqref{FGR} for brevity as Fermi's golden rule irrespective of the order of approximation of the $T$-matrix.

\begin{figure}
    \centering
     \begin{tikzpicture}[scale=1.5, decoration={markings, 
        mark= at position 0.65 with {\arrow{stealth}},
        mark= at position 2cm with {\arrow{stealth}}}
        ]   \begin{scope}[xshift=-1.5cm]
            \draw[postaction={decorate}] (1,0.2)--(1.3,0.0);
            \draw[postaction={decorate}] (1,-0.2)--(1.3,0.0);
            \draw[postaction={decorate}] (1.3,0)--(1.7,0.0);
            \draw[fill=black] (1.3,0) circle (0.07);
            \node at (1.95,0) {$+$};
            \node at (1.2, 0.4) {\footnotesize{$\vec{k}$}};
            \node at (1.2, -0.4) {\footnotesize{$\vec{k}_2$}};
            \node at (1.65, 0.2) {\footnotesize{$\vec{k}_1$}};
            \end{scope}

            \begin{scope}[xshift = -0.6cm, yshift=1.2 cm]
                  \draw[postaction={decorate}] (1.4,-1)--(1.7,-1.2);
                  \draw[postaction={decorate}] (1.4,-1.4)--(1.7,-1.2);
                  \draw [postaction={decorate}] (1.7,-1.2) to [out=70,in=110] (2.2,-1.2);
                  \draw [postaction={decorate}] (1.7,-1.2) to [out=-70,in=-110] (2.2,-1.2);
                  \draw[postaction={decorate}] (2.2,-1.2)--(2.7,-1.2);
                  \draw[fill=white] (1.7,-1.2) circle (0.07);
                  \draw[fill=black] (2.2,-1.2) circle (0.07);
                  \node at (2.5, -1) {\footnotesize{$\vec{k}_1$}}; 
                  \node at (2, -0.85) {\footnotesize{$\vec{p}_1$}}; 
                  \node at (2, -1.55){\footnotesize{$\vec{p}_2$}};
                  \node at (1.55, -0.8) {\footnotesize{$\vec{k}$}};
                  \node at (1.55, -1.5) {\footnotesize{$\vec{k}_2$}};
            \end{scope}      
    \end{tikzpicture}
    \caption{Fusion scattering channel studied as an example. The first diagram is the first order fusion term associated with $T^{\left(1\right)}_{\text{if}}$, and the second diagram is one of several possible second order terms, associated with $T^{\left(2\right)}_{\text{if}}$ and containing two scattering events. The black (white) circle denotes the DMI-induced three-magnon (exchange-induced four-magnon) vertex. The momenta $\vec{p}_1$ and $\vec{p}_2$ correspond to intermediate (here: virtual) magnons.}
    \label{fig: bubble example}
\end{figure}
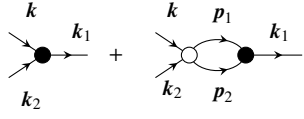

\subsection{First- and higher-order scattering terms}
To elucidate the way that Eq.~\eqref{eq: scattering rates general} should be used to perform calculations we will make use of the example of Fig.~\ref{fig: bubble example}, where a fusion channel containing the first order scattering process and one scattering process of second order in $H_{\text{int}}$ is depicted. We note that since the magnon with momentum $\vec{k}$ is destroyed, the diagrams of Fig.~\ref{fig: bubble example} correspond to an out-process. Since we have no need for its in-counterpart here, we will skip any notation that would otherwise help differentiate between in- and out-processes in this subsection. A notation with ``in'' and ``out'' superscripts will be introduced in the next subsection where the calculation of the Hall conductivity requires the computation of both the in- and out-counterparts of the process. 

In the case of our example, $H_{\text{int}} = H_3 + H_4$ captures three- and four-magnon scattering and the initial and final states read 
\begin{align}
    \ket{\text{i}} = \ket{N_{\vec{k}_{1}}, N_{\vec{k}_{2}}, N_{\vec{k}}}
\end{align}
and
\begin{align}
    \ket{\text{f}} = \ket{N_{\vec{k}}-1, N_{\vec{k}_{2}}-1, N_{\vec{k}_1}+1},
\end{align}
respectively.

To compute the matrix element $T^{\left(1\right)}_{\text{if}}$ in Eq.~\eqref{Lippmann} associated with the first order scattering (three-magnon fusion) process we must connect our initial and final state through the appropriate part of $H_3$ introduced in Eq.~\eqref{3-magnon potential}. We obtain
\begin{equation}
\begin{split}
     T^{\left(1\right)}_{\text{if}} = \bra{\text{f}} H_{3} \ket{\text{i}} & 
    =  \frac{1}{2\sqrt{N}} \bra{\text{f}}  \sum_{\vec{q}_{1}, \vec{q}_{2},\vec{q}_{3}} \delta_{\vec{q}_{1} + \vec{q}_{2}, \vec{q}_{3}}
    V_{\vec{q}_{3}; 
    \vec{q}_{1},\vec{q}_{2}}  a^{\dagger}_{\vec{q}_{1}} a^{\dagger}_{\vec{q_{2}}} a_{\vec{q}_{3}} + \text{h.c.} \ket{\text{i}} \\
    & = 
    \frac{1}{2\sqrt{N}} \sum_{\vec{q}_{1}, \vec{q}_{2},\vec{q}_{3}} \delta_{\vec{q}_{1} + \vec{q}_{2}, \vec{q}_{3}}\left[ V_{\vec{q}_{3}; 
    \vec{q}_{1},\vec{q}_{2}} \underbrace{\bra{\text{f}}   a^{\dagger}_{\vec{q}_{1}} a^{\dagger}_{\vec{q_{2}}} a_{\vec{q}_{3}} \ket{\text{i}}}_{0} + V_{\vec{q}_{3};
    \vec{q}_{1},\vec{q}_{2}}^{*} \bra{\text{f}} a_{\vec{q}_{3}}^{\dagger}a_{\vec{q}_{1}} a_{\vec{q_{2}}} \ket{\text{i}}\right] \\
    & = 
     2\frac{1}{2\sqrt{N}} \sum_{\vec{q}_{1}, \vec{q}_{2},\vec{q}_{3}} \delta_{\vec{q}_{1} + \vec{q}_{2}, \vec{q}_{3}}V_{
     \vec{q}_{3};\vec{q}_{1},\vec{q}_{2}}^{*} \delta_{\vec{q}_1, \vec{k}} \delta_{\vec{q}_2, \vec{k}_2}\delta_{\vec{q}_3, \vec{k}_1}  \bra{\text{f}} a_{\vec{q}_{3}}^{\dagger}a_{\vec{q}_{1}} a_{\vec{q_{2}}} \ket{\text{i}} \\
    & = 
     \frac{1}{\sqrt{N}} V_{\vec{k};\vec{k}_1,\vec{k}_{2}}^{*} 
     \sqrt{(N_{\vec{k}_1}+1)N_{\vec{k}}N_{\vec{k}_2}} \delta_{\vec{k}+\vec{k}_{2}, \vec{k}_1},
\end{split}
\label{3-magnon scattering rate: lowest order}
\end{equation}
where the factor of 2 in the third line accounts for the interchangeability of the momenta $\vec{q}_1$ and $\vec{q}_2$. To obtain the last line in Eq.~\eqref{3-magnon scattering rate: lowest order} we made use of the relations
\begin{equation}
    \begin{split}
        & a_{\vec{k}}^{\dagger}\ket{\hdots,N_{\vec{k}}, \hdots} = \sqrt{N_{\vec{k}}+1}\ket{\hdots,N_{\vec{k}},\hdots}, \\
        & a_{\vec{k}}\ket{\hdots,N_{\vec{k}}, \hdots}=\sqrt{N_{\vec{k}}} \ket{\hdots,N_{\vec{k}}, \hdots}.
    \end{split}
\end{equation}
Next, we compute the second order term in $H_{\text{int}}$ of Eq.~\eqref{Lippmann}, which we evaluate to
\begin{equation}
\begin{split}
    T^{\left(2\right)}_{\text{if}} & 
    = 
    \sum_{\nu}\frac{\bra{f}H_{\text{int}}\ket{\nu}\bra{\nu}H_{\text{int}}\ket{\text{i}}}{\varepsilon_\text{i} - \varepsilon_{\nu} + \mathrm{i}\eta}    
     =
    \sum_{\nu}\frac{\bra{f}H_{3}\ket{\nu}\bra{\nu}H_{4}\ket{\text{i}}}{\varepsilon_\text{i} - \varepsilon_{\nu} + \mathrm{i}\eta}
    \\
    & =
    \frac{1}{8 N^{3/2}}\sum_{\vec{q}_1, \vec{q}_2, \vec{q}_3, \vec{q}_4} \sum_{\vec{k}_1', \vec{k}_2', \vec{k}_3'} \sum_{\nu}\delta_{\vec{k}'_1 + \vec{k}'_2, \vec{k}'_3}\delta_{\vec{q}_1+\vec{q}_2,\vec{q}_3+\vec{q}_4}\frac{V_{\vec{k}'_3;\vec{k}'_1, \vec{k}'_2}^{*} W_{\vec{q}_1,\vec{q}_2;\vec{q}_3, \vec{q}_4}}{\varepsilon_i - \varepsilon_{\nu} + \mathrm{i}\eta} \bra{\text{f}}a_{\vec{k}'_3}^{\dagger}a_{\vec{k}'_2} a_{\vec{k}'_1}\ket{\nu} \bra{\nu}a_{\vec{q}_4}^{\dagger}a_{\vec{q}_3}^{\dagger}a_{\vec{q}_2} a_{\vec{q}_1}\ket{\text{i}}, 
\end{split}
    \label{general lippmann for dH3}
\end{equation}
where terms that contain different combinations of parts of $H_{\text{int}}$ such as 
$\bra{\text{f}}H_{3}\ket{\nu}\bra{\nu}H_{3}\ket{\text{i}}$, $\bra{\text{f}}H_{4}\ket{\nu}\bra{\nu}H_{4}\ket{\text{i}}$, and $\bra{\text{f}}H_{4}\ket{\nu}\bra{\nu}H_{3}\ket{\text{i}}$ vanish since they  cannot connect the chosen final and initial state. The result of Eq.~\eqref{general lippmann for dH3} is general in the sense that it can be used to compute any second order fusion process. To calculate the process that we use as an example (recall Fig.~\ref{fig: bubble example}) we firstly determine the intermediate state, 
\begin{equation}
    \ket{\nu} = \frac{1}{2} \sum_{\vec{p}_1, \vec{p}_2}\ket{N_{\vec{k}_1}, N_{\vec{k}}-1, N_{\vec{k}_2}-1, N_{\vec{p}_1}+1, N_{\vec{p}_2}+1},
    \label{intermediate state}
\end{equation}
where the factor of $1/2$ is required due to the symmetric nature of the diagram in order to avoid double counting associated with the interchange $\vec{p}_1 \leftrightarrow \vec{p}_2$. 
By plugging the intermediate state in Eq.~\eqref{general lippmann for dH3} we get the following result,
\begin{equation}
    T_{\text{if}}^{\left(2\right)} = 
    8\frac{1}{16 N^{3/2}} \sqrt{(N_{\vec{k}_1}+1)N_{\vec{k}}N_{\vec{k}_2}}\delta_{\vec{k}+\vec{k}_2,\vec{k}_1}\sum_{\vec{p}_1, \vec{p}_2}\delta_{\vec{p}_1, \vec{k}_1 - \vec{p}_2}\frac{V_{\vec{k}_1;\vec{p}_1, \vec{p}_2}^{*} W_{\vec{k},\vec{k}_2;\vec{p}_1, \vec{p}_2}}{\varepsilon_{\vec{p}_1} + \varepsilon_{\vec{p}_2} - \varepsilon_{\vec{k}}-\varepsilon_{\vec{k}_2} + \mathrm{i}\eta} 
    \left(
    N_{\vec{p}_2}+1
    \right)
    \left(
    N_{\vec{p}_1}+1
    \right),
    \label{eq: second order T element}
\end{equation}
where the factor of 8 accounts for the equivalence of $\vec{k}'_1$, $\vec{k}'_2$, and of $\vec{q}_1 $, $\vec{q}_2$, and of $\vec{q}_3$, $\vec{q}_4$ in Eq.~\eqref{general lippmann for dH3}. To obtain the scattering rate we use Eq.~\eqref{eq: scattering rates general} and sum over $\vec{k}_1$ and $\vec{k}_2$,
\begin{equation}
    \Gamma_{\vec{k}} 
    =\frac{2 \pi}{\hbar}\sum_{\vec{k}_1, \vec{k}_2}
    \delta
    \left(
    \varepsilon_{\vec{k}}+\varepsilon_{\vec{k}_2}-\varepsilon_{\vec{k}_1}
    \right)
    \left|
    T_{\text{if}}
    \right|
    ^2 
     = 
    \Gamma_{\vec{k}}^{\left(1\right)} + \Gamma_{\vec{k}}^{\left(2\right)} + \Gamma_{\vec{k}}^{\text{interf.}},
\label{eq:full scattering rate: example}
\end{equation}
where 
\begin{equation}
    \begin{split}
        & \Gamma_{\vec{k}}^{\left(1\right)} = \frac{2\pi}{\hbar}\sum_{\vec{k}_1, \vec{k}_2}
        \delta
        \left(
        \varepsilon_{\vec{k}}+\varepsilon_{\vec{k}_2}-\varepsilon_{\vec{k}_1}
        \right)
        \left|
        T_{\text{if}}^{(1)}
        \right|
        ^2, \\
        & \Gamma_{\vec{k}}^{(2)}=\frac{2\pi}{\hbar}\sum_{\vec{k}_1, \vec{k}_2}
        \delta
        \left(
        \varepsilon_{\vec{k}}+\varepsilon_{\vec{k}_2}-\varepsilon_{\vec{k}_1}
        \right)
        \left|
        T_{\text{if}}^{(2)}
        \right|
        ^2, \\
        & \Gamma_{\vec{k}}^{\text{interf.}} =\frac{4\pi}{\hbar}\sum_{\vec{k}_1, \vec{k}_2}
    \delta
    \left(
    \varepsilon_{\vec{k}}+\varepsilon_{\vec{k}_2}-\varepsilon_{\vec{k}_1}
    \right)
    \left(
    \text{Re}
    \left(
    T_{\text{if}}^{\left(1\right)}
    \right)
    \text{Re}
    \left(
    T_{\text{if}}^{\left(2\right)}
    \right)
    +\text{Im}
    \left(
    T_{\text{if}}^{\left(1\right)}
   \right)
    \text{Im}
    \left(
    T_{\text{if}}^{\left(2\right)}
    \right)
    \right).
    \end{split}
    \label{eq: scattering rates example}
\end{equation}

\subsection{Scattering terms and Hall effect}
Here, we will utilize the results of the transport theory presented in Sec.~\ref{section:Transport theory} to identify which of the scattering rates appearing in Eq.~\eqref{eq:full scattering rate: example}
can contribute to a Hall effect and which cannot. 

\subsubsection{First order scattering process}
\label{sec:firstorder}
We begin with the scattering rate corresponding to the first order scattering event,
\begin{equation}
   \Gamma_{\vec{k}}^{\left(1 \right), \text{out}} = 
   \frac{1}{\sqrt{N}}\frac{2\pi}{\hbar}\sum_{\vec{k}_1, \vec{k}_2}
   \delta
   \left(
   \varepsilon_{\vec{k}}+\varepsilon_{\vec{k}_2}-\varepsilon_{\vec{k}_1}
   \right)  
   \left|
   V_{\vec{k}_1;\vec{k},\vec{k}_{2}} ^{*}
   \right|
   ^2
    \left(
    N_{\vec{k}_1}+1
    \right)
    N_{\vec{k}}N_{\vec{k}_2} \delta_{\vec{k}+\vec{k}_{2}, \vec{k}_1},
\end{equation}
where the ``$\text{out}$'' upper-index is a reminder that in order to build the full collision integral one should consider both the in- and out-counterpart of each scattering process. In our case the in-scattering rate corresponds to a split process where the magnon with momentum $\vec{k}_1$ splits into two other magnons with momenta $\vec{k}$ and $\vec{k}_2$. The procedure to obtain the in-scattering rate is identical to the one for its out-counterpart and results in
\begin{equation}
  \Gamma_{\vec{k}}^{\left(1\right),\text{in}} = 
   \frac{1}{\sqrt{N}}\frac{2\pi}{\hbar}
   \sum_{\vec{k}_1, \vec{k}_2}
   \delta
   \left(
   \varepsilon_{\vec{k}}+\varepsilon_{\vec{k}_2}-\varepsilon_{\vec{k}_1}
   \right)
   \left|
   V_{\vec{k}_1;\vec{k},\vec{k}_{2} }
   \right| 
   ^2 
    \left(
    N_{\vec{k}}+1
    \right)
    \left(
    N_{\vec{k}_2}+1
    \right)
    N_{\vec{k}_1}
    \delta_{\vec{k}+\vec{k}_{2}, \vec{k}_1}.  
\end{equation}
In total, the full scattering rate reads
\begin{equation}
    \begin{split}
        \Gamma_{\vec{k}}^{\left(1\right)} & 
        = \Gamma_{\vec{k}}^{\left(1 \right), \text{in}} - \Gamma_{\vec{k}}^{\left(1 \right), \text{out}} \\
        & = \frac{1}{\sqrt{N}}\frac{2\pi}{\hbar}
   \sum_{\vec{k}_1, \vec{k}_2} \delta_{\vec{k}+\vec{k}_{2}, \vec{k}_1}
   \delta
   \left(
   \varepsilon_{\vec{k}}+\varepsilon_{\vec{k}_2}-\varepsilon_{\vec{k}_1}
   \right)
   \left[
   \left|
   V_{\vec{k}_1;\vec{k},\vec{k}_{2}} ^{*}
   \right|
   ^2
    \left(
    N_{\vec{k}_1}+1
    \right)
    N_{\vec{k}}N_{\vec{k}_2} 
    - 
    \left|
   V_{\vec{k}_1;\vec{k},\vec{k}_{2}}
   \right| 
   ^2 
    \left(
    N_{\vec{k}}+1
    \right)
    \left(
    N_{\vec{k}_2}+1
    \right)
    N_{\vec{k}_1} 
   \right] \\
   & = 
   \frac{1}{\sqrt{N}}\frac{2\pi}{\hbar}
   \sum_{\vec{k}_1, \vec{k}_2} \delta_{\vec{k}+\vec{k}_{2}, \vec{k}_1}
   \delta
   \left(
   \varepsilon_{\vec{k}}+\varepsilon_{\vec{k}_2}-\varepsilon_{\vec{k}_1}
   \right)
    \left|
   V_{\vec{k}_1;\vec{k},\vec{k}_{2} }
   \right| 
   ^2 
   \left[
    \left(
    N_{\vec{k}_1}+1
    \right)
    N_{\vec{k}}N_{\vec{k}_2} 
    - 
    \left(
    N_{\vec{k}}+1
    \right)
    \left(
    N_{\vec{k}_2}+1
    \right)
    N_{\vec{k}_1} 
   \right],
    \end{split}
    \label{eq: HC 1BA example}
\end{equation}
where we made use of the relation $\left|
   V_{\vec{k}_1;\vec{k},\vec{k}_{2} }
   \right| 
   ^2 = \left|
   V_{\vec{k}_1;\vec{k},\vec{k}_{2} } ^{*}
   \right|
   ^2$. 
We now proceed in accordance with the transport theory laid out in Sec.~\ref{section:Transport theory}, make the ansatz $N_{\vec{k}_i}= \overline{N}_{{\vec{k}_i}} + \delta N_{{\vec{k}_i}}$ for the out-of-equilibrium populations in Eq.~\eqref{eq: HC 1BA example}, and linearize the scattering rates in 
$\delta N_{{\vec{k}_i}}$. By keeping up to linear terms in the out-of-equilibrium populations $\delta N_{{\vec{k}_i}}$ we arrive at
\begin{equation}
        \Gamma_{\vec{k}}^{\left(1\right)} = 
        C
        \left[
        \{ 
        \overline{N}_{\vec{k}_1}, \overline{N}_{\vec{k}},\overline{N}_{\vec{k}_2} 
        \}
        \right]
        + 
        D_{\vec{k}}\delta N_{\vec{k}} 
        + 
        \sum_{\vec{k}_{1}}O_{\vec{k}\vec{k}_{1}} \delta N_{\vec{k}_{1}}
        + 
        \sum_{\vec{k}_{2}}O_{\vec{k}\vec{k}_{2}} \delta N_{\vec{k}_{2}}, 
        \label{eq:gamma1decomposition}
\end{equation}
with
\begin{equation}
     C
    \left[
    \{ 
    \overline{N}_{\vec{k}}, \overline{N}_{\vec{k}_1},\overline{N}_{\vec{k}_2} 
    \}
    \right]
    = 
   \frac{3}{\sqrt{N}}\frac{2\pi}{\hbar}
   \sum_{\vec{k}_1, \vec{k}_2} \delta_{\vec{k}+\vec{k}_{2}, \vec{k}_1}
   \delta
   \left(
   \varepsilon_{\vec{k}}+\varepsilon_{\vec{k}_2}-\varepsilon_{\vec{k}_1}
   \right)
    \left|
   V_{\vec{k}_1;\vec{k},\vec{k}_{2} }
   \right| 
   ^2 
   \left[
    \left(
    \overline{N}_{\vec{k}_1}+1
    \right)
    \overline{N}_{\vec{k}}\overline{N}_{\vec{k}_2} 
    - 
    \left(
    \overline{N}_{\vec{k}}+1
    \right)
    \left(
    \overline{N}_{\vec{k}_2}+1
    \right)
    \overline{N}_{\vec{k}_1} 
   \right] 
   \label{eq:theconstnat}
\end{equation}
being the constant part of the expansion, and 
\begin{equation}
    \begin{split}
        & D_{\vec{k}} = 
        \frac{1}{\sqrt{N}}\frac{2\pi}{\hbar}
   \sum_{\vec{k}_1, \vec{k}_2} \delta_{\vec{k}+\vec{k}_{2}, \vec{k}_1}
   \delta
   \left(
   \varepsilon_{\vec{k}}+\varepsilon_{\vec{k}_2}-\varepsilon_{\vec{k}_1}
   \right)
    \left|
   V_{\vec{k}_1;\vec{k},\vec{k}_{2} }
   \right| 
   ^2 
   \left[
    (\overline{N}_{\vec{k}_1} + 1)\overline{N}_{\vec{k}_2} 
    - 
    \overline{N}_{\vec{k}_1}
    \left(
    \overline{N}_{\vec{k}_2}+1
    \right) 
   \right], 
   \\
   & 
   O_{\vec{k} \vec{k}_1} = 
    \frac{1}{\sqrt{N}}\frac{2\pi}{\hbar}
   \sum_{\vec{k}_2} \delta_{\vec{k}+\vec{k}_{2}, \vec{k}_1}
   \delta
   \left(
   \varepsilon_{\vec{k}}+\varepsilon_{\vec{k}_2}-\varepsilon_{\vec{k}_1}
   \right)
    \left|
   V_{\vec{k}_1;\vec{k},\vec{k}_{2} }
   \right| 
   ^2 
   \left[
    \overline{N}_{\vec{k}_2}\overline{N}_{\vec{k}}
    - 
    \left(
    \overline{N}_{\vec{k}}+1
    \right) 
    \left(
    \overline{N}_{\vec{k}_2}+1
    \right) 
   \right], 
   \\
   & 
    O_{\vec{k} \vec{k}_2} = 
    \frac{1}{\sqrt{N}}\frac{2\pi}{\hbar}
   \sum_{\vec{k}_1} \delta_{\vec{k}+\vec{k}_{2}, \vec{k}_1}
   \delta
   \left(
   \varepsilon_{\vec{k}}+\varepsilon_{\vec{k}_2}-\varepsilon_{\vec{k}_1}
   \right)
    \left|
   V_{\vec{k}_1;\vec{k},\vec{k}_{2} }
   \right| 
   ^2 
   \left[
   \left(
    \overline{N}_{\vec{k}_1} + 1 
    \right) 
    \overline{N}_{\vec{k}}
    - 
   \left(
    \overline{N}_{\vec{k}} + 1 
    \right) 
    \overline{N}_{\vec{k}_1}
   \right]
    \end{split}
    \label{eq:D and O}
\end{equation}
the diagonal and off-diagonal elements of the scattering matrix corresponding to the studied process, respectively. 
Next, we analyze each of the terms appearing in the expansion in Eq.~\eqref{eq:gamma1decomposition}:
\begin{enumerate}
    \item[(i)] For the constant in Eq.~\eqref{eq:theconstnat}, we find that
\begin{equation}
\begin{split}
     C 
    \left[
    \{ 
    \overline{N}_{\vec{k}}, \overline{N}_{\vec{k}_1},\overline{N}_{\vec{k}_2} 
    \}
    \right] 
    & =
    \frac{3}{\sqrt{N}}\frac{2\pi}{\hbar}
   \sum_{\vec{k}_1, \vec{k}_2} \delta_{\vec{k}+\vec{k}_{2}, \vec{k}_1}
   \delta
   \left(
   \varepsilon_{\vec{k}}+\varepsilon_{\vec{k}_2}-\varepsilon_{\vec{k}_1}
   \right)
    \left|
   V_{\vec{k}_1;\vec{k},\vec{k}_{2} }
   \right| 
   ^2 
   \left[
    \left(
    \overline{N}_{\vec{k}_1}+1
    \right)
    \overline{N}_{\vec{k}}\overline{N}_{\vec{k}_2} 
    - 
    \left(
    \overline{N}_{\vec{k}}+1
    \right)
    \left(
    \overline{N}_{\vec{k}_2}+1
    \right)
    \overline{N}_{\vec{k}_1} 
   \right]   \\
   & = 
   \frac{3}{\sqrt{N}}\frac{2\pi}{\hbar}
   \sum_{\vec{k}_1, \vec{k}_2} \delta_{\vec{k}+\vec{k}_{2}, \vec{k}_1}
   \delta
   \left(
   \varepsilon_{\vec{k}}+\varepsilon_{\vec{k}_2}-\varepsilon_{\vec{k}_1}
   \right)
    \left|
   V_{\vec{k}_1;\vec{k},\vec{k}_{2} }
   \right| 
   ^2 
   \left(
    \mathrm{e}^{\beta \varepsilon_{\vec{k}_1}}\overline{N}_{\vec{k}}
    \overline{N}_{\vec{k}_1}\overline{N}_{\vec{k}_2} 
    - 
     \mathrm{e}^
     {
     \beta 
     (
    \varepsilon_{\vec{k}} + \varepsilon_{\vec{k}_2}
    )
    }
     \overline{N}_{\vec{k}_1}
    \overline{N}_{\vec{k}_2}
    \overline{N}_{\vec{k}}
     \right)
    \\
    & = 
    \frac{3}{\sqrt{N}}\frac{2\pi}{\hbar}
   \sum_{\vec{k}_1, \vec{k}_2} \delta_{\vec{k}+\vec{k}_{2}, \vec{k}}
   \delta
   \left(
   \varepsilon_{\vec{k}}+\varepsilon_{\vec{k}_2}-\varepsilon_{\vec{k}_1}
   \right)
    \left|
   V_{\vec{k}_1;\vec{k},\vec{k}_{2} }
   \right| 
   ^2 
   \mathrm{e}^{\beta \varepsilon_{\vec{k}_1}}
   \left(
   \overline{N}_{\vec{k}}
    \overline{N}_{\vec{k}_1}\overline{N}_{\vec{k}_2} 
    - 
     \overline{N}_{\vec{k}_1}
    \overline{N}_{\vec{k}_2}
    \overline{N}_{\vec{k}}
     \right) \\
     & = 0 ,
\end{split}
\label{eq: example constant of first order element is zero}
\end{equation}
where in the last step energy conservation imposed by the Dirac delta-function $\delta
   \left(
   \varepsilon_{\vec{k}}+\varepsilon_{\vec{k}_2}-\varepsilon_{\vec{k}_1}
   \right)$ was utilized. A zero constant in the collision integral is required for a physically meaningful theory since its existence would imply the presence of net currents in the equilibrium. 
   \item[(ii)] The diagonal element $D_{\vec{k}}$ in Eq.~\eqref{eq:D and O} contributes to a magnon relaxation time $\tau_{\vec{k}}$. However, it is of no further concern for the Hall effect, to which it cannot contribute by definition. [Recall that the thermal Hall vector in Eq.~\eqref{eq: Hall vector} is related to the anti-symmetric part of the scattering matrix.]
   \item[(iii)] 
Finally, we analyze $O_{\vec{k} \vec{k}_1}$ and $O_{\vec{k} \vec{k}_2}$ in Eq.~\eqref{eq:D and O} to determine if they contribute to the Hall effect. We begin with $O_{\vec{k} \vec{k}_2}$. There are two ways to do that: (a) We directly calculate $\mathcal{A}_{\vec{k} \vec{k}_2} = (\mathcal{O}_{\vec{k} \vec{k}_2} - \mathcal{O}_{\vec{k}_2 \vec{k}})/2$, and (b) we check if detailed balance holds. In both cases, we need to switch to Hardy's basis, 
\begin{equation}
    \mathcal{O}_{\vec{k} \vec{k}_2} = \frac{G_{\vec{k}_2}}{G_{\vec{k}}}O_{\vec{k} \vec{k}_2} = \frac{\sqrt{\left( \overline{N}_{\vec{k}_2} + 1 \right)\overline{N}_{\vec{k}_2}}}{\sqrt{\left( \overline{N}_{\vec{k}} + 1 \right)\overline{N}_{\vec{k}}}}O_{\vec{k} \vec{k}_2}. 
    \label{eq: Okk2}
\end{equation}
\begin{enumerate}
    \item[(a)] 
Upon swapping $\vec{k} \leftrightarrow \vec{k}_2$ in Eq.~\eqref{eq: Okk2} one can read off the element $\mathcal{O}_{\vec{k}_2 \vec{k}}$, which is given by 
\begin{equation}
\begin{split}
    \mathcal{O}_{\vec{k}_2 \vec{k}} 
    & = 
    \frac{\sqrt{\left( \overline{N}_{\vec{k}} + 1 \right)\overline{N}_{\vec{k}}}}{\sqrt{\left( \overline{N}_{\vec{k}_2} + 1 \right)\overline{N}_{\vec{k}_2}}}O_{\vec{k}_2 \vec{k}} \\
    & = 
    \frac{1}{\sqrt{N}}\frac{2\pi}{\hbar}
   \sum_{\vec{k}_1} \delta_{\vec{k}+\vec{k}_{2}, \vec{k}_1}
   \delta
   \left(
   \varepsilon_{\vec{k}}+\varepsilon_{\vec{k}_2}-\varepsilon_{\vec{k}_1}
   \right)
    \left|
   V_{\vec{k}_1;\vec{k},\vec{k}_{2} }
   \right| 
   ^2 
   \frac{\sqrt{\left( \overline{N}_{\vec{k}} + 1 \right)\overline{N}_{\vec{k}}}}{\sqrt{\left( \overline{N}_{\vec{k}_2} + 1 \right)\overline{N}_{\vec{k}_2}}}
   \left[
   \left(
    \overline{N}_{\vec{k}_1} + 1 
    \right) 
    \overline{N}_{\vec{k}_2}
    - 
   \left(
    \overline{N}_{\vec{k}_2} + 1 
    \right) 
    \overline{N}_{\vec{k}_1}
   \right]
   \\
   & = 
   \frac{1}{\sqrt{N}}\frac{2\pi}{\hbar}
   \sum_{\vec{k}_1} \delta_{\vec{k}+\vec{k}_{2}, \vec{k}_1}
   \delta
   \left(
   \varepsilon_{\vec{k}}+\varepsilon_{\vec{k}_2}-\varepsilon_{\vec{k}_1}
   \right)
    \left|
   V_{\vec{k}_1;\vec{k},\vec{k}_{2}}
   \right| 
   ^2 
   \frac{\sqrt{\left( \overline{N}_{\vec{k}} + 1 \right)\overline{N}_{\vec{k}}}}{\sqrt{\left( \overline{N}_{\vec{k}_2} + 1 \right)\overline{N}_{\vec{k}_2}}}
   \left(
    \overline{N}_{\vec{k}_1} \mathrm{e}^{\beta \varepsilon_{\vec{k}_1}}
    \overline{N}_{\vec{k}_2}
    - 
    \overline{N}_{\vec{k}_2}\mathrm{e}^{\beta\varepsilon_{\vec{k}_2}}
    \overline{N}_{\vec{k}_1}
   \right) \\
   & = 
   \frac{1}{\sqrt{N}}\frac{2\pi}{\hbar}
   \sum_{\vec{k}_1} \delta_{\vec{k}+\vec{k}_{2}, \vec{k}_1}
   \delta
   \left(
   \varepsilon_{\vec{k}}+\varepsilon_{\vec{k}_2}-\varepsilon_{\vec{k}_1}
   \right)
    \left|
   V_{\vec{k}_1;\vec{k},\vec{k}_{2}}
   \right| 
   ^2 
   \frac{\sqrt{\left( \overline{N}_{\vec{k}} + 1 \right)\overline{N}_{\vec{k}}}}{\sqrt{\left( \overline{N}_{\vec{k}_2} + 1 \right)\overline{N}_{\vec{k}_2}}}
   \overline{N}_{\vec{k}_1} \overline{N}_{\vec{k}_2}\mathrm{e}^{\beta\varepsilon_{\vec{k}_2}}
   \left(
    \mathrm{e}^{\beta (\varepsilon_{\vec{k}_1} - \varepsilon_{\vec{k}_2})}
    - 
   1
   \right) \\
   & = 
   \frac{1}{\sqrt{N}}\frac{2\pi}{\hbar}
   \sum_{\vec{k}_1} \delta_{\vec{k}+\vec{k}_{2}, \vec{k}_1}
   \delta
   \left(
   \varepsilon_{\vec{k}}+\varepsilon_{\vec{k}_2}-\varepsilon_{\vec{k}_1}
   \right)
    \left|
   V_{\vec{k}_1;\vec{k},\vec{k}_{2}}
   \right| 
   ^2 
   \frac{\sqrt{\left( \overline{N}_{\vec{k}} + 1 \right)\overline{N}_{\vec{k}}}}{\sqrt{\left( \overline{N}_{\vec{k}_2} + 1 \right)\overline{N}_{\vec{k}_2}}}
   \overline{N}_{\vec{k}_1} \overline{N}_{\vec{k}_2}\mathrm{e}^{\beta\varepsilon_{\vec{k}_2}} \frac{1}{\overline{N}_{\vec{k}}}
   \\
   & = 
   \frac{1}{\sqrt{N}}\frac{2\pi}{\hbar}
   \sum_{\vec{k}_1} \delta_{\vec{k}+\vec{k}_{2}, \vec{k}_1}
   \delta
   \left(
   \varepsilon_{\vec{k}}+\varepsilon_{\vec{k}_2}-\varepsilon_{\vec{k}_1}
   \right)
    \left|
   V_{\vec{k}_1;\vec{k},\vec{k}_{2}}
   \right| 
   ^2 
    \frac{\sqrt{\overline{N}_{\vec{k}} + 1}}{\sqrt{\overline{N}_{\vec{k}_2}}} \frac{\sqrt{\overline{N}_{\vec{k}_2} + 1}}{\sqrt{\overline{N}_{\vec{k}}}}
   \overline{N}_{\vec{k}_1} 
\end{split}
\end{equation}
and can be seen to be equal to $ \mathcal{O}_{\vec{k} \vec{k}_2} $: 
\begin{equation}
\begin{split}
    \mathcal{O}_{\vec{k} \vec{k}_2} 
    & = 
    \frac{1}{\sqrt{N}}\frac{2\pi}{\hbar}
   \sum_{\vec{k}_1} \delta_{\vec{k}+\vec{k}_{2}, \vec{k}_1}
   \delta
   \left(
   \varepsilon_{\vec{k}}+\varepsilon_{\vec{k}_2}-\varepsilon_{\vec{k}_1}
   \right)
    \left|
   V_{\vec{k}_1;\vec{k},\vec{k}_{2} }
   \right| 
   ^2 
    \frac{\sqrt{\left( \overline{N}_{\vec{k}_2} + 1 \right)\overline{N}_{\vec{k}_2}}}{\sqrt{\left( \overline{N}_{\vec{k}} + 1 \right)\overline{N}_{\vec{k}}}}
   \left[
   \left(
    \overline{N}_{\vec{k}_1} + 1 
    \right) 
    \overline{N}_{\vec{k}}
    - 
   \left(
    \overline{N}_{\vec{k}} + 1 
    \right) 
    \overline{N}_{\vec{k}_1}
   \right] \\
   & = 
    \frac{1}{\sqrt{N}}\frac{2\pi}{\hbar}
   \sum_{\vec{k}_1} \delta_{\vec{k}+\vec{k}_{2}, \vec{k}_1}
   \delta
   \left(
   \varepsilon_{\vec{k}}+\varepsilon_{\vec{k}_2}-\varepsilon_{\vec{k}_1}
   \right)
    \left|
   V_{\vec{k}_1;\vec{k},\vec{k}_{2} }
   \right| 
   ^2 
    \frac{\sqrt{\left( \overline{N}_{\vec{k}_2} + 1 \right)\overline{N}_{\vec{k}_2}}}{\sqrt{\left( \overline{N}_{\vec{k}} + 1 \right)\overline{N}_{\vec{k}}}}
     \overline{N}_{\vec{k}_1}
    \overline{N}_{\vec{k}}
   \left(
    \mathrm{e}^{\beta \varepsilon_{\vec{k}_1}}
    - 
    \mathrm{e}^{\beta \varepsilon_{\vec{k}}}
   \right) \\
   & = 
    \frac{1}{\sqrt{N}}\frac{2\pi}{\hbar}
   \sum_{\vec{k}_1} \delta_{\vec{k}+\vec{k}_{2}, \vec{k}_1}
   \delta
   \left(
   \varepsilon_{\vec{k}}+\varepsilon_{\vec{k}_2}-\varepsilon_{\vec{k}_1}
   \right)
    \left|
   V_{\vec{k}_1;\vec{k},\vec{k}_{2} }
   \right| 
   ^2 
    \frac{\sqrt{\left( \overline{N}_{\vec{k}_2} + 1 \right)\overline{N}_{\vec{k}_2}}}{\sqrt{\left( \overline{N}_{\vec{k}} + 1 \right)\overline{N}_{\vec{k}}}}
     \overline{N}_{\vec{k}_1}
    \overline{N}_{\vec{k}}  \mathrm{e}^{\beta \varepsilon_{\vec{k}}}
   \left(
    \mathrm{e}^{\beta (\varepsilon_{\vec{k}_1} - \varepsilon_{\vec{k}})}
    - 
   1
   \right)\\
   & = 
   \frac{1}{\sqrt{N}}\frac{2\pi}{\hbar}
   \sum_{\vec{k}_1} \delta_{\vec{k}+\vec{k}_{2}, \vec{k}_1}
   \delta
   \left(
   \varepsilon_{\vec{k}}+\varepsilon_{\vec{k}_2}-\varepsilon_{\vec{k}_1}
   \right)
    \left|
   V_{\vec{k}_1;\vec{k},\vec{k}_{2} }
   \right| 
   ^2 
    \frac{\sqrt{\left( \overline{N}_{\vec{k}_2} + 1 \right)\overline{N}_{\vec{k}_2}}}{\sqrt{\left( \overline{N}_{\vec{k}} + 1 \right)\overline{N}_{\vec{k}}}}
     \overline{N}_{\vec{k}_1}
    \overline{N}_{\vec{k}}  \mathrm{e}^{\beta \varepsilon_{\vec{k}}}
    \frac{1}{\overline{N}_{\vec{k}_2}} \\
    & = 
     \frac{1}{\sqrt{N}}\frac{2\pi}{\hbar}
   \sum_{\vec{k}_1} \delta_{\vec{k}+\vec{k}_{2}, \vec{k}_1}
   \delta
   \left(
   \varepsilon_{\vec{k}}+\varepsilon_{\vec{k}_2}-\varepsilon_{\vec{k}_1}
   \right)
    \left|
   V_{\vec{k}_1;\vec{k},\vec{k}_{2} }
   \right| 
   ^2 
    \frac{\sqrt{\overline{N}_{\vec{k}} + 1}}{\sqrt{\overline{N}_{\vec{k}_2}}} \frac{\sqrt{\overline{N}_{\vec{k}_2} + 1}}{\sqrt{\overline{N}_{\vec{k}}}}
   \overline{N}_{\vec{k}_1} \\
   & = \mathcal{O}_{\vec{k}_2 \vec{k} }.  
\end{split}
\end{equation}
Since $\mathcal{O}_{\vec{k} \vec{k}_2} = \mathcal{O}_{\vec{k}_2 \vec{k}}$ we get that $\mathcal{A}_{\vec{k} \vec{k}_2}=0$ and, consequently, this specific term cannot give rise to a finite Hall current. 
    \item[(b)] To see if the in- and out-parts of $\mathcal{O}_{\vec{k} \vec{k}_2}$ meet the detailed balance relation introduced in Sec.~\ref{subsection: DB}, we write
\begin{equation}
    \mathcal{O}_{\vec{k} \vec{k}_2}^{\text{in}} = \mathcal{O}_{\vec{k} \vec{k}_2}^{++} 
    = 
    \frac{1}{\sqrt{N}}\frac{2\pi}{\hbar}
   \sum_{\vec{k}_1} \delta_{\vec{k}+\vec{k}_{2}, \vec{k}_1}
   \delta
   \left(
   \varepsilon_{\vec{k}}+\varepsilon_{\vec{k}_2}-\varepsilon_{\vec{k}_1}
   \right)
    \left|
   V_{\vec{k}_1;\vec{k},\vec{k}_{2} }
   \right| 
   ^2 
    \frac{\sqrt{\left( \overline{N}_{\vec{k}_2} + 1 \right)\overline{N}_{\vec{k}_2}}}{\sqrt{\left( \overline{N}_{\vec{k}} + 1 \right)\overline{N}_{\vec{k}}}}
   \left(
    \overline{N}_{\vec{k}_1} + 1 
    \right) 
    \overline{N}_{\vec{k}}
\end{equation}
and 
\begin{equation}
    \mathcal{O}_{\vec{k} \vec{k}_2}^{\text{out}} = \mathcal{O}_{\vec{k} \vec{k}_2}^{--} 
    = 
    \frac{1}{\sqrt{N}}\frac{2\pi}{\hbar}
   \sum_{\vec{k}_1} \delta_{\vec{k}+\vec{k}_{2}, \vec{k}_1}
   \delta
   \left(
   \varepsilon_{\vec{k}}+\varepsilon_{\vec{k}_2}-\varepsilon_{\vec{k}_1}
   \right)
    \left|
   V_{\vec{k}_1;\vec{k},\vec{k}_{2} }
   \right| 
   ^2 
    \frac{\sqrt{\left( \overline{N}_{\vec{k}_2} + 1 \right)\overline{N}_{\vec{k}_2}}}{\sqrt{\left( \overline{N}_{\vec{k}} + 1 \right)\overline{N}_{\vec{k}}}}
   \left(
    \overline{N}_{\vec{k}} + 1 
    \right) 
    \overline{N}_{\vec{k}_1},
\end{equation}
where the notation of the $\pm \pm$ signs was explained in Sec.~\ref{subsection: DB}. Manipulating the occupation numbers and making use of the energy conservation it can be shown that detailed balance is met, that is to say,
\begin{equation}
\begin{split}
    \mathcal{O}_{\vec{k} \vec{k}_2}^{--} 
   & = 
    \frac{1}{\sqrt{N}}\frac{2\pi}{\hbar}
   \sum_{\vec{k}_1} \delta_{\vec{k}+\vec{k}_{2}, \vec{k}_1}
   \delta
   \left(
   \varepsilon_{\vec{k}}+\varepsilon_{\vec{k}_2}-\varepsilon_{\vec{k}_1}
   \right)
    \left|
   V_{\vec{k}_1;\vec{k},\vec{k}_{2} }
   \right| 
   ^2 
    \frac{\sqrt{\left( \overline{N}_{\vec{k}_2} + 1 \right)\overline{N}_{\vec{k}_2}}}{\sqrt{\left( \overline{N}_{\vec{k}} + 1 \right)\overline{N}_{\vec{k}}}}
   \left(
    \overline{N}_{\vec{k}} + 1 
    \right) 
    \overline{N}_{\vec{k}_1} 
    \\
    & = 
    \frac{1}{\sqrt{N}}\frac{2\pi}{\hbar}
   \sum_{\vec{k}_1} \delta_{\vec{k}+\vec{k}_{2}, \vec{k}_1}
   \delta
   \left(
   \varepsilon_{\vec{k}}+\varepsilon_{\vec{k}_2}-\varepsilon_{\vec{k}_1}
   \right)
    \left|
   V_{\vec{k}_1;\vec{k},\vec{k}_{2} }
   \right| 
   ^2 
    \frac{\sqrt{\left( \overline{N}_{\vec{k}_2} + 1 \right)\overline{N}_{\vec{k}_2}}}{\sqrt{\left( \overline{N}_{\vec{k}} + 1 \right)\overline{N}_{\vec{k}}}}
   \left(
    \overline{N}_{\vec{k}_1} + 1 
    \right) 
    \overline{N}_{\vec{k}} \mathrm{e}^{\beta (\varepsilon_{\vec{k}}-\varepsilon_{\vec{k}_1})} \\
    & = 
    \mathrm{e}^{-\beta\varepsilon_{\vec{k}_2}}\mathcal{O}_{\vec{k} \vec{k}_2}^{++}.
\end{split}
\end{equation}
Since detailed balance is not broken we get again a zero anti-symmetric part and consequently a zero Hall current. 
\end{enumerate}

Finally, we are left with the term $O_{\vec{k} \vec{k}_1}$ in Eq.~\eqref{eq:D and O}. For this term it can also be shown that no Hall current can be generated. The proof of the above statement is more advanced than that of $O_{\vec{k} \vec{k}_2}$. This can be seen by examining its anti-symmetric part $\mathcal{A}_{\vec{k} \vec{k}_1} = \left( \mathcal{O}_{\vec{k} \vec{k}_1} - \mathcal{O}_{\vec{k}_1 \vec{k}}\right) / 2$ and noticing that by interchanging the labels $\vec{k} \leftrightarrow \vec{k}_1$ to get $\mathcal{O}_{\vec{k}_1 \vec{k}}$ the scattering process is changed (to a diagram different from those in Fig.~\ref{fig: bubble example}). In fact, by swapping the labels, we now have an in-process where the two magnons with $\vec{k}_1$ and $\vec{k}_2$ fuse together into the magnon with $\vec{k}$. This is of course a valid scattering process, and its off-diagonal elements can be calculated in the same spirit as described above. Thus, we need to take $O_{\vec{k} \vec{k}_1}$ together with the appropriate off-diagonal element of the in-fusion process to prove the absence of $\mathcal{A}_{\vec{k} \vec{k}_1}$. 
\end{enumerate}

In conclusion, we have shown that the first order scattering event of our example (recall Fig.~\ref{fig: bubble example}) associated with the $T$-matrix element $T_{if}^{\left(1\right)}$ cannot generate a Hall conductivity. This statement holds more generally for the first order scattering rate of every possible channel. A general proof of this statement can be found in Refs.~\cite{Mangeolle2022PRB, Mangeolle2022PRX}. 

\subsubsection{Interference process}
\label{section: Interference process}
We now turn to the scattering rate $\Gamma_{\vec{k}}^{\text{interf.}}$ in Eq.~\eqref{eq: scattering rates example} corresponding to the interference of the first order and second order processes. For convenience, we repeat its general expression:
\begin{equation}
    \Gamma_{\vec{k}}^{\text{interf.}} =\frac{4\pi}{\hbar}\sum_{\vec{k}_1, \vec{k}_2}
    \delta
    \left(
    \varepsilon_{\vec{k}}+\varepsilon_{\vec{k}_2}-\varepsilon_{\vec{k}_1}
    \right)
    \left(
    \text{Re}
    T_{\text{if}}^{(1)}
    \text{Re}
    T_{\text{if}}^{(2)}
    +\text{Im}
    T_{\text{if}}^{(1)}
    \text{Im}
    T_{\text{if}}^{(2)}
    \right).
    \label{eq: Tif^2 repeat}
\end{equation}
From now on, in this subsection, we will omit the ``interf.'' superscript to lighten notation.
For the example in Fig.~\ref{fig: bubble example}, the first and second order $T$-matrix elements have been calculated to be
\begin{equation}
     T_{\text{if}}^{(1),\text{out}} =
     \frac{1}{\sqrt{N}} V_{\vec{k};\vec{k},\vec{k}_{2}}^{*} 
     \sqrt{(N_{\vec{k}_1}+1)N_{\vec{k}}N_{\vec{k}_2}} \delta_{\vec{k}+\vec{k}_{2}, \vec{k}_1},
     \label{eq: Tif^1 repeat}
\end{equation}
and
\begin{equation}
    T_{\text{if}}^{(2),\text{out}} = 
    \frac{1}{2N^{3/2}} \sqrt{(N_{\vec{k}_1}+1)N_{\vec{k}}N_{\vec{k}_2}}\delta_{\vec{k}+\vec{k}_2,\vec{k}_1}\sum_{\vec{p}_1, \vec{p}_2}\delta_{\vec{p}_1, \vec{k}_1 - \vec{p}_2}\frac{V_{\vec{k}_1;\vec{p}_1, \vec{p}_2}^{*} W_{\vec{k},\vec{k}_2;\vec{p}_1, \vec{p}_2}}{\varepsilon_{\vec{p}_1} + \varepsilon_{\vec{p}_2} - \varepsilon_{\vec{k}}-\varepsilon_{\vec{k}_2} + \mathrm{i}\eta} 
    \left(
    N_{\vec{p}_2}+1
    \right)
    \left(
    N_{\vec{p}_1}+1
    \right),
\end{equation}
with ``out'' denoting once more the fact that the scattering process in the example is an out-process. 
To calculate the real and imaginary parts that appear in the scattering rate, we use the Dirac identity
\begin{equation}
    \frac{1}{\Delta E + \mathrm{i} \eta} = \mathcal{P}\left(\frac{1}{\Delta E}\right) - \mathrm{i}\mathrm{\pi}\delta\left(\Delta E\right),
    \label{eq: Dirac's Identity}
\end{equation}
with $\Delta E$ being real, $\eta$ an infinitesimally small positive number, and $\mathcal{P}$ referring to the principle part. Using Eq.~\eqref{eq: Dirac's Identity} we get for the real part of $T_{\text{if}}^{(2),\text{out}}$ that
\begin{equation}
\begin{split}
     \text{Re}\left(T_{\text{if}}^{(2),\text{out}} \right)
     & = 
    \frac{1}{2N^{3/2}} \sqrt{(N_{\vec{k}_1}+1)N_{\vec{k}}N_{\vec{k}_2}}\delta_{\vec{k}+\vec{k}_2,\vec{k}_1}\sum_{\vec{p}_1, \vec{p}_2}\delta_{\vec{p}_1, \vec{k}_1 - \vec{p}_2} W_{\vec{k},\vec{k}_2;\vec{p}_1, \vec{p}_2}\text{Re}\left(\frac{V_{\vec{k}_1; \vec{p}_1, \vec{p}_2}^{*}}{\varepsilon_{\vec{p}_1} + \varepsilon_{\vec{p}_2} - \varepsilon_{\vec{k}}-\varepsilon_{\vec{k}_2} + \mathrm{i}\eta} \right) 
    \left(
    N_{\vec{p}_2}+1
    \right)
    \left(
    N_{\vec{p}_1}+1
    \right) \\
    & = 
    \frac{1}{2N^{3/2}} \sqrt{(N_{\vec{k}_1}+1)N_{\vec{k}}N_{\vec{k}_2}}\delta_{\vec{k}+\vec{k}_2,\vec{k}_1}\sum_{\vec{p}_1, \vec{p}_2}\delta_{\vec{p}_1, \vec{k}_1 - \vec{p}_2} W_{\vec{k},\vec{k}_2;\vec{p}_1, \vec{p}_2} 
      \left(
    N_{\vec{p}_2}+1
    \right)
    \left(
    N_{\vec{p}_1}+1
    \right) \\
    &
    \qquad \times 
    \left[
    \text{Re}\left( V_{\vec{k}_1;\vec{p}_1, \vec{p}_2 }^{*} \right)\mathcal{P}\left( \frac{1}{\varepsilon_{\vec{p}_1} + \varepsilon_{\vec{p}_2} - \varepsilon_{\vec{k}}-\varepsilon_{\vec{k}_2}}\right) + \mathrm{\pi} \text{Im}
    \left( 
    V_{\vec{k}_1;\vec{p}_1, \vec{p}_2}^{*}\right)\delta(\varepsilon_{\vec{p}_1} + \varepsilon_{\vec{p}_2} - \varepsilon_{\vec{k}}-\varepsilon_{\vec{k}_2})
    \right],
    \label{eq: real part of T2}
\end{split}
\end{equation}
and for its imaginary part that
\begin{equation}
\begin{split}
     \text{Im}\left(T_{\text{if}}^{(2),\text{out}} \right) 
     & = 
     \frac{1}{2N^{3/2}} \sqrt{\left(N_{\vec{k}_1}+1\right)N_{\vec{k}}N_{\vec{k}_2}}\delta_{\vec{k}+\vec{k}_2,\vec{k}_1}\sum_{\vec{p}_1, \vec{p}_2}\delta_{\vec{p}_1, \vec{k}_1 - \vec{p}_2} W_{\vec{k},\vec{k}_2;\vec{p}_1, \vec{p}_2} 
      \left(
    N_{\vec{p}_2}+1
    \right)
    \left(
    N_{\vec{p}_1}+1
    \right) \\
    & 
    \qquad \times
    \left[
    \text{Im}\left( V_{\vec{k}_1;\vec{p}_1, \vec{p}_2 }^{*} \right)\mathcal{P}\left( \frac{1}{\varepsilon_{\vec{p}_1} + \varepsilon_{\vec{p}_2} - \varepsilon_{\vec{k}}-\varepsilon_{\vec{k}_2}}\right) - \mathrm{\pi} \text{Re}\left( V_{\vec{k}_1; \vec{p}_1, \vec{p}_2}^{*}\right)\delta(\varepsilon_{\vec{p}_1} + \varepsilon_{\vec{p}_2} - \varepsilon_{\vec{k}}-\varepsilon_{\vec{k}_2})
    \right]. 
\end{split}
    \label{eq: imag part of T2}
\end{equation}
Altogether, we get by combining Eqs.~\eqref{eq: Tif^1 repeat}, \eqref{eq: Tif^2 repeat}, \eqref{eq: real part of T2}, and \eqref{eq: imag part of T2} the scattering-out rate of the interference process, 
\begin{equation}
\begin{split} 
    \Gamma_{\vec{k}}^{\text{out}} 
    & =\frac{4\pi}{\hbar}\frac{1}{2N^2}\sum_{\vec{k}_1, \vec{k}_2}
    (N_{\vec{k}_1}+1)N_{\vec{k}}N_{\vec{k}_2} \delta_{\vec{k}+\vec{k}_2,\vec{k}_1}
    \delta
    \left(
    \varepsilon_{\vec{k}}+\varepsilon_{\vec{k}_2}-\varepsilon_{\vec{k}_1}
    \right) 
     \sum_{\vec{p}_1, \vec{p}_2}\delta_{\vec{p}_1, \vec{k}_1 - \vec{p}_2} W_{\vec{k},\vec{k}_2;\vec{p}_1, \vec{p}_2} 
      \left(
    N_{\vec{p}_2}+1
    \right)
    \left(
    N_{\vec{p}_1}+1
    \right) \\
    & 
    \qquad \times \left[
    \mathcal{R}_{\text{S}}^{\text{out}}\left(\vec{k},\vec{k}_1, \vec{k}_2, \vec{p}_1, \vec{p}_2\right) + \mathcal{R}_{\text{AS}}^{\text{out}}
    \left(
    \vec{k},\vec{k}_1, \vec{k}_2, \vec{p}_1, \vec{p}_2
    \right) 
    \right], 
\end{split}
\label{eq: example interference rate}
\end{equation}
where 
\begin{equation}
\begin{split}
    & 
    \mathcal{R}_{\text{S}}^{\text{out}}\left(
    \vec{k},\vec{k}_1, \vec{k}_2, \vec{p}_1, \vec{p}_2\right) 
     = 
    \mathcal{P}
    \left( 
    \frac{1}{\varepsilon_{\vec{p}_1} + \varepsilon_{\vec{p}_2} - \varepsilon_{\vec{k}}-\varepsilon_{\vec{k}_2}}
    \right)
    \left[
    \text{Re}
    \left( 
    V_{ \vec{k}_1;\vec{k},\vec{k}_{2}}^{*}
    \right) 
    \text{Re}
    \left(
    V_{\vec{k}_1;\vec{p}_1, \vec{p}_2}^{*} 
    \right)
    + 
    \text{Im} \left( 
    V_{\vec{k};\vec{k},\vec{k}_{2}}^{*}
    \right) \
    \text{Im}
    \left( 
    V_{\vec{k}_1;\vec{p}_1, \vec{p}_2}^{*}
    \right) 
    \right], 
    \\
    & 
     \mathcal{R}_{\text{AS}}^{\text{out}}\left(
    \vec{k},\vec{k}_1, \vec{k}_2, \vec{p}_1, \vec{p}_2
    \right) 
     = 
    \mathrm{\pi}\delta
    \left( 
    \varepsilon_{\vec{p}_1} + \varepsilon_{\vec{p}_2} - \varepsilon_{\vec{k}}-\varepsilon_{\vec{k}_2}
    \right)
    \left[
    \text{Re}
    \left( 
    V_{\vec{k}_1;\vec{k},\vec{k}_{2}}^{*}
    \right) 
    \text{Im}
    \left(
    V_{\vec{k}_1;\vec{p}_1, \vec{p}_2}^{*} 
    \right)
    -
    \text{Im} \left( 
    V_{\vec{k}_1;\vec{k},\vec{k}_{2}}^{*}
    \right) \
    \text{Re}
    \left( 
    V_{\vec{k}_1;\vec{p}_1, \vec{p}_2}^{*}
    \right) 
    \right].
\end{split}
\label{eq: scattering rate of example: interferce part both AS and S}
\end{equation}
The subscripts ``AS'' and ``S'' stand for anti-symmetric and symmetric, respectively, already alluding to which part of the scattering rate gives rise to a Hall effect. From here on, any quantity notated with ``AS'' or ``S'' should be understood as being proportional to $\mathcal{R}_{\text{AS}}$ or $\mathcal{R}_{\text{S}}$, respectively. 

The next steps are similar to what we have done in the previous subsection, Sec.~\ref{sec:firstorder}. Namely, we must calculate the in-part of the studied process, linearize both parts with respect to the out-of-equilibrium distribution functions and then switch to Hardy's basis. By doing that we get again four terms in the scattering rates: the constant $\mathcal{C}^{\text{in}/\text{out}}\left[ \{\overline{N}_{\vec{k}_i}\}\right]$, the diagonal part $D_{\vec{k}}^{\text{in}/\text{out}}$, and the off-diagonal elements $\mathcal{O}_{\vec{k} \vec{k}_1}^{\text{in}/\text{out}}$ and $\mathcal{O}_{\vec{k} \vec{k}_2}^{\text{in}/\text{out}}$. As in the case of the first order scattering rate we focus here on the most straightforward calculation involving $\mathcal{O}_{\vec{k} \vec{k}_2}^{\text{in}/\text{out}}$. For the rest of the terms, we only point out that the full constant $\mathcal{C}\left[ \{\overline{N}_{\vec{k}_i}\}\right] = \mathcal{C}^{\text{in}}\left[ \{\overline{N}_{\vec{k}_i}\}\right] - \mathcal{C}^{\text{out}}\left[ \{\overline{N}_{\vec{k}_i}\}\right]$ does \emph{not} vanish. Indeed, the part of $\mathcal{C}\left[ \{\overline{N}_{\vec{k}_i}\}\right]$ proportional to $\mathcal{R}_{\text{AS}} = \mathcal{R}_{\text{AS}}^{\text{in}}- \mathcal{R}_{\text{AS}}^{\text{out}}$ survives while the one proportional to $\mathcal{R}_{\text{S}} = \mathcal{R}_{\text{S}}^{\text{in}}- \mathcal{R}_{\text{S}}^{\text{out}}$
is cancelled due to energy conservation. As explained in Refs.~\cite{Mangeolle2022PRB,Mangeolle2022PRX}, this finite constant indicates that the Bose-Einstein population based on the harmonic spectrum is no longer the exact equilibrium distribution. Indeed, many-body interactions will correct the magnon spectrum and the population in equilibrium. However, since these corrections are perturbatively small (in our case they are $1/S$ corrections, which even come with an additional smallness associated with the small value of $D/J$), they will cause only subleading corrections to the Hall currents. Thus, we drop the constant. 

First, we investigate whether or not $\mathcal{O}_{\vec{k} \vec{k}_2}^{\text{out}}$ and $\mathcal{O}_{\vec{k} \vec{k}_2}^{\text{in}}$ break the detailed balance relation. The two parts are written as
\begin{equation}
\begin{split}
 \mathcal{O}_{\vec{k} \vec{k}_2}^{\text{out}} & 
 = \mathcal{O}_{\vec{k} \vec{k}_2}^{--} \\
 & 
 =\frac{4\pi}{\hbar}\frac{1}{2N^2}\sum_{\vec{k}_1} \frac{G_{\vec{k}_2}}{G_{\vec{k}}}
    \overline{N}_{\vec{k}}
    \left(
    \overline{N}_{\vec{k}_1} + 1 
    \right)
    \delta_{\vec{k}+\vec{k}_2,\vec{k}_1}
    \delta
    \left(
    \varepsilon_{\vec{k}}+\varepsilon_{\vec{k}_2}-\varepsilon_{\vec{k}_1}
    \right) 
     \sum_{\vec{p}_1, \vec{p}_2}\delta_{\vec{p}_1, \vec{k}_1 - \vec{p}_2} W_{\vec{k},\vec{k}_2;\vec{p}_1, \vec{p}_2} 
      \left(
    N_{\vec{p}_2}+1
    \right)
    \left(
    N_{\vec{p}_1}+1
    \right) \\
    & 
    \qquad \times\left[
    \mathcal{R}_{\text{S}}^{\text{out}}\left(\vec{k},\vec{k}_1, \vec{k}_2, \vec{p}_1, \vec{p}_2\right) + \mathcal{R}_{\text{AS}}^{\text{out}}
    \left(
    \vec{k},\vec{k}_1, \vec{k}_2, \vec{p}_1, \vec{p}_2
    \right) 
    \right] \\
    & = 
    \mathcal{O}_{\vec{k} \vec{k}_2}^{\text{out}, \text{AS}} + \mathcal{O}_{\vec{k} \vec{k}_2}^{\text{out}, \text{S}}, 
\end{split}
\label{eq: off-element out interf}
\end{equation}
and 
\begin{equation}
\begin{split}
 \mathcal{O}_{\vec{k} \vec{k}_2}^{\text{in}} & 
 = \mathcal{O}_{\vec{k} \vec{k}_2}^{++} \\
 & 
 =\frac{4\pi}{\hbar}\frac{1}{2N^2}\sum_{\vec{k}_1} \frac{G_{\vec{k}_2}}{G_{\vec{k}}}
    \overline{N}_{\vec{k}_1}
    \left(
    \overline{N}_{\vec{k}} + 1 
    \right)
    \delta_{\vec{k}+\vec{k}_2,\vec{k}_1}
    \delta
    \left(
    \varepsilon_{\vec{k}}+\varepsilon_{\vec{k}_2}-\varepsilon_{\vec{k}_1}
    \right) 
     \sum_{\vec{p}_1, \vec{p}_2}\delta_{\vec{p}_1, \vec{k}_1 - \vec{p}_2} W_{\vec{p}_1, \vec{p}_2; \vec{k},\vec{k}_2} 
      \left(
    N_{\vec{p}_2}+1
    \right)
    \left(
    N_{\vec{p}_1}+1
    \right) \\
    & 
    \qquad \times \left[
    \mathcal{R}_{\text{S}}^{\text{in}}\left(\vec{k},\vec{k}_1, \vec{k}_2, \vec{p}_1, \vec{p}_2\right) + \mathcal{R}_{\text{AS}}^{\text{in}}
    \left(
    \vec{k},\vec{k}_1, \vec{k}_2, \vec{p}_1, \vec{p}_2
    \right) 
    \right] \\
    & = 
    \mathcal{O}_{\vec{k} \vec{k}_2}^{\text{in}, \text{AS}} + \mathcal{O}_{\vec{k} \vec{k}_2}^{\text{in}, \text{S}}, 
\end{split}
\label{eq: off-element in interf}
\end{equation}
with 
\begin{equation}
    \mathcal{R}_{\text{S}}^{\text{in}}\left(
    \vec{k},\vec{k}_1, \vec{k}_2, \vec{p}_1, \vec{p}_2\right) 
     = 
    \mathcal{P}
    \left( 
    \frac{1}{\varepsilon_{\vec{p}_1}+\varepsilon_{\vec{p}_2} - \varepsilon_{\vec{k}_1}}
    \right)
    \left[
    \text{Re}
    \left( 
    V_{\vec{k}_1;\vec{k},\vec{k}_{2}}
    \right) 
    \text{Re}
    \left(
    V_{\vec{k}_1; \vec{p}_1, \vec{p}_2}
    \right)
    + 
     \text{Im}
    \left( 
    V_{\vec{k}_1;\vec{k},\vec{k}_{2}}
    \right) 
    \text{Im}
    \left(
    V_{\vec{k}_1; \vec{p}_1, \vec{p}_2}
    \right)
    \right],
    \end{equation}
and 
    \begin{equation}
    \mathcal{R}_{\text{AS}}^{\text{in}}\left(
    \vec{k},\vec{k}_1, \vec{k}_2, \vec{p}_1, \vec{p}_2
    \right) 
     = 
    \mathrm{\pi}\delta
    \left( 
    \varepsilon_{\vec{p}_1}+\varepsilon_{\vec{p}_2} - \varepsilon_{\vec{k}_1}
    \right)
    \left[
    \text{Re}
    \left( 
    V_{\vec{k}_1;\vec{k},\vec{k}_{2}}
    \right) 
    \text{Im}
    \left(
    V_{\vec{k}_1; \vec{p}_1, \vec{p}_2}
    \right)
    -
     \text{Im}
    \left( 
    V_{\vec{k}_1;\vec{k},\vec{k}_{2}}
    \right) 
    \text{Re}
    \left(
   V_{\vec{k}_1; \vec{p}_1, \vec{p}_2}
    \right)
    \right].
\end{equation}

The symmetric parts of the in- and out-elements respect detailed balance relation, i.e.,
\begin{equation}
\begin{split}
    \mathcal{O}_{\vec{k} \vec{k}_2}^{\text{in}, \text{S}} &
    =\frac{4\pi}{\hbar}\frac{1}{2N^2}\sum_{\vec{k}_1} \frac{G_{\vec{k}_2}}{G_{\vec{k}}}
    \overline{N}_{\vec{k}_1}
    \left(
    \overline{N}_{\vec{k}} + 1 
    \right)
    \delta_{\vec{k}+\vec{k}_2,\vec{k}_1}
    \delta
    \left(
    \varepsilon_{\vec{k}}+\varepsilon_{\vec{k}_2}-\varepsilon_{\vec{k}_1}
    \right) 
     \sum_{\vec{p}_1, \vec{p}_2}\delta_{\vec{p}_1, \vec{k}_1 - \vec{p}_2} W_{\vec{p}_1, \vec{p}_2; \vec{k},\vec{k}_2} 
      \left(
    N_{\vec{p}_2}+1
    \right)
    \left(
    N_{\vec{p}_1}+1
    \right)
    \mathcal{R}_{\text{S}}^{\text{in}} 
    \\
    & = 
    \frac{4\pi}{\hbar}\frac{1}{2N^2}\sum_{\vec{k}_1} \frac{G_{\vec{k}_2}}{G_{\vec{k}}}
    \overline{N}_{\vec{k}_1}
    \left(
    \overline{N}_{\vec{k}} + 1 
    \right)
    \delta_{\vec{k}+\vec{k}_2,\vec{k}_1}
    \delta
    \left(
    \varepsilon_{\vec{k}}+\varepsilon_{\vec{k}_2}-\varepsilon_{\vec{k}_1}
    \right) 
     \sum_{\vec{p}_1, \vec{p}_2}\delta_{\vec{p}_1, \vec{k}_1 - \vec{p}_2} W_{\vec{p}_1, \vec{p}_2; \vec{k},\vec{k}_2} 
      \left(
    N_{\vec{p}_2}+1
    \right)
    \left(
    N_{\vec{p}_1}+1
    \right)
    \\
    & 
    \qquad \times \mathcal{P}
    \left( 
    \frac{1}{\varepsilon_{\vec{p}_1}+\varepsilon_{\vec{p}_2} - \varepsilon_{\vec{k}_1}}
    \right)
    \left[
    \text{Re}
    \left( 
    V_{\vec{k}_1;\vec{k},\vec{k}_{2}}
    \right) 
    \text{Re}
    \left(
   V_{\vec{k}_1; \vec{p}_1, \vec{p}_2}
    \right)
    + 
     \text{Im}
    \left( 
    V_{\vec{k}_1;\vec{k},\vec{k}_{2}}
    \right) 
    \text{Im}
    \left(
   V_{\vec{k}_1; \vec{p}_1, \vec{p}_2}
    \right)
    \right] \\
    & = 
    \frac{4\pi}{\hbar}\frac{1}{2N^2}\sum_{\vec{k}_1} \frac{G_{\vec{k}_2}}{G_{\vec{k}}}
    \mathrm{e}^{\beta (\varepsilon_{\vec{k}} - \varepsilon_{\vec{k}_1})}
    \overline{N}_{\vec{k}}
    \left(
    \overline{N}_{\vec{k}_1} + 1 
    \right)
    \delta_{\vec{k}+\vec{k}_2,\vec{k}_1}
    \delta
    \left(
    \varepsilon_{\vec{k}}+\varepsilon_{\vec{k}_2}-\varepsilon_{\vec{k}_1}
    \right) 
     \sum_{\vec{p}_1, \vec{p}_2}\delta_{\vec{p}_1, \vec{k}_1 - \vec{p}_2} W_{\vec{k},\vec{k}_2 ; \vec{p}_1, \vec{p}_2} 
      \left(
    N_{\vec{p}_2}+1
    \right)
    \left(
    N_{\vec{p}_1}+1
    \right)
    \\
    & 
    \qquad \times \mathcal{P}
    \left( 
    \frac{1}{\varepsilon_{\vec{p}_1}+\varepsilon_{\vec{p}_2} - \varepsilon_{\vec{k}} - \varepsilon_{\vec{k}_2}}
    \right)
    \left[
    \text{Re}
    \left( 
    V_{\vec{k}_1;\vec{k},\vec{k}_{2}}^{*}
    \right) 
    \text{Re}
    \left(
    V_{\vec{k}_1; \vec{p}_1, \vec{p}_2}^{*}
    \right)
    + 
    \left(-
     \text{Im}
    \left( 
     V_{\vec{k}_1;\vec{k},\vec{k}_{2}}^{*}
    \right) 
    \right)
    \left(
    -
    \text{Im}
    \left(
    V_{\vec{k}_1; \vec{p}_1, \vec{p}_2}^{*}
    \right)
    \right)
    \right]
    \\
    & = 
    \mathrm{e}^{-\beta \varepsilon_{\vec{k}_2}}\mathcal{O}_{\vec{k} \vec{k}_2}^{\text{in}, \text{S}}. 
    \end{split}
    \label{eq:symmetricdetailedbalance}
\end{equation}
Thus, they cannot generate a Hall effect.
In the above, we manipulated the Bose-Einstein occupation numbers and used energy conservation to rewrite the argument of the principal part. In addition, we took advantage of the symmetry of the four-magnon vertex in Eq.~\eqref{4-magnon vertex}, namely $W_{\vec{k},\vec{k}_2 ; \vec{p}_1, \vec{p}_2} = W_{\vec{p}_1, \vec{p}_2;\vec{k},\vec{k}_2}$, and the relations $\text{Im}\left( V\right)= - \text{Im}\left( V^{*}\right)$, $\text{Re}\left( V\right)= \text{Re}\left( V^{*}\right)$.

On the other hand, using the same algebraic manipulations, it can be shown that for the anti-symmetric parts, $\mathcal{O}_{\vec{k} \vec{k}_2}^{\text{in}, \text{AS}}$ and $\mathcal{O}_{\vec{k} \vec{k}_2}^{\text{out}, \text{AS}}$, the detailed balance relation is broken:
\begin{equation}
    \mathcal{O}_{\vec{k} \vec{k}_2}^{\text{in}, \text{AS}} = - \mathrm{e}^{-\beta \varepsilon_{\vec{k}_2}}  \mathcal{O}_{\vec{k} \vec{k}_2}^{\text{out},  \text{AS}}. 
    \label{eq: Broken DB AS part}
\end{equation}
The extra minus sign comes from the fact that when relating the in-process to the out-process (and \textit{vice versa}), we have to take the complex conjugate of the three-magnon vertex appearing in the expression. By doing so we get a minus sign from the imaginary part of the vertex, which---unlike the symmetric part in Eq.~\eqref{eq:symmetricdetailedbalance}---does not cancel out, because we have terms of the form $\text{Re}\left( V\right)\text{Im}\left( V\right)$ and not of $\text{Im}\left( V\right)\text{Im}\left( V\right)$ as in Eq.~\eqref{eq:symmetricdetailedbalance}. Finally, by using  Eq.~\eqref{eq: Hall vector} and Eq.~\eqref{eq:A-matrix and DB} the Hall conductivity vector caused by $\mathcal{O}_{\vec{k} \vec{k}_2}$ reads
\begin{equation}
\begin{split}
     \vec{\kappa}_{\text{H}}
     & = 
     \frac{1}{4 k_{\text{B}} T ^2 V}
     \sum_{\vec{k}, \vec{k}_2} G_{\vec{k}} G_{\vec{k}_2} \vec{v}_{\vec{k}} \times \vec{v}_{\vec{k}_2} \varepsilon_{\vec{k}} \varepsilon_{\vec{k}_2} 
    \tau_{\vec{k}}\tau_{\vec{k}_2}
    \frac{\overline{N}_{\vec{k}_2}-\overline{N}_{\vec{k}}}{\overline{N}_{\vec{k}_2}}
    \left( 
    \mathcal{O}_{\vec{k} \vec{k}_2}^{\text{in}, \text{AS}} - \mathrm{e}^{-\beta \varepsilon_{\vec{k}_2}}\mathcal{O}_{\vec{k}, \vec{k}_2}^{\text{out}, \text{AS}}
    \right) \\
    & = 
     \frac{1}{2 k_{\text{B}} T ^2 V}
     \sum_{\vec{k}, \vec{k}_2} G_{\vec{k}} G_{\vec{k}_2} \vec{v}_{\vec{k}} \times \vec{v}_{\vec{k}_2} \varepsilon_{\vec{k}} \varepsilon_{\vec{k}_2} 
    \tau_{\vec{k}}\tau_{\vec{k}_2}
      \frac{\overline{N}_{\vec{k}_2}-\overline{N}_{\vec{k}}}{\overline{N}_{\vec{k}_2}}
    \mathcal{O}_{\vec{k} \vec{k}_2}^{\text{in}, \text{AS}} \\
    & = 
    \frac{1}{4 k_{\text{B}} T ^2 V} \frac{1}{N^2}\frac{\left(2\pi \right)^2}{\hbar}
    \sum_{\vec{k}, \vec{k}_2} G_{\vec{k}_2}^2 \vec{v}_{\vec{k}} \times \vec{v}_{\vec{k}_2} \varepsilon_{\vec{k}} \varepsilon_{\vec{k}_2} 
  \tau_{\vec{k}}\tau_{\vec{k}_2}
  \frac{\overline{N}_{\vec{k}_2}-\overline{N}_{\vec{k}}}{\overline{N}_{\vec{k}_2}}
  \left(
    \overline{N}_{\vec{k}} + 1 
    \right)
    \sum_{\vec{k}_1}
    \delta_{\vec{k}+\vec{k}_2,\vec{k}_1}
    \delta
    \left(
    \varepsilon_{\vec{k}}+\varepsilon_{\vec{k}_2}-\varepsilon_{\vec{k}_1}
    \right)\overline{N}_{\vec{k}_1} \\
    & 
    \qquad \times \sum_{\vec{p}_1, \vec{p}_2}\delta_{\vec{p}_1, \vec{k}_1 - \vec{p}_2} W_{\vec{p}_1, \vec{p}_2; \vec{k},\vec{k}_2} 
      \left(
    N_{\vec{p}_2}+1
    \right)
    \left(
    N_{\vec{p}_1}+1
    \right)
    \delta
    \left( 
    \varepsilon_{\vec{p}_1}+\varepsilon_{\vec{p}_2} - \varepsilon_{\vec{k}_1}
    \right)
    \left[
    \text{Re}
    \left( 
    V_{\vec{k}_1;\vec{k},\vec{k}_{2}}
    \right) 
    \text{Im}
    \left(
    V_{\vec{k}_1; \vec{p}_1, \vec{p}_2}
    \right)
    -
     \text{Im}
    \left( 
    V_{\vec{k}_1;\vec{k},\vec{k}_{2}}
    \right) 
    \text{Re}
    \left(
   V_{\vec{k}_1; \vec{p}_1, \vec{p}_2}
    \right)
    \right].
\end{split}
\end{equation}
A similar result is obtained for the $\mathcal{O}_{\vec{k} \vec{k}_1}$ contribution.

In summary, we have shown that, unlike the first-order scattering process, the  interference between the first- and second-order processes can produce a finite Hall response.
This contribution to the Hall response contains \emph{two} Dirac delta functions, one from the global conservation of energy and one associated with \emph{intermediate resonance scattering}.
This finding is in agreement with the results reported in Ref.~\onlinecite{Mangeolle2022PRB,Mangeolle2022PRX}.

\subsubsection{Second order scattering process}
Finally, we are left with the second order scattering rate in Eq.~\eqref{eq: scattering rates example}, which using Eq.~\eqref{eq: second order T element} can be written as 
\begin{equation}
\begin{split}
     \Gamma_{\vec{k}}^{(2), \text{out}}
     =
     \frac{\pi}{2\hbar} \frac{1}{N^{3}}\sum_{\vec{k}_1, \vec{k}_2}
    \delta_{\vec{k}+\vec{k}_2,\vec{k}_1}
    \delta
    \left(
    \varepsilon_{\vec{k}}+\varepsilon_{\vec{k}_2}-\varepsilon_{\vec{k}_1}
    \right)
    (N_{\vec{k}_1}+1)N_{\vec{k}}N_{\vec{k}_2}
    \left|
    \sum_{\vec{p}_1, \vec{p}_2}
    \delta_{\vec{p}_1, \vec{k}_1 - \vec{p}_2}
    \frac{\Pi\left({\{ \vec{k}_i, \vec{p}_i\}}\right)}{\varepsilon_{\vec{p}_1} + \varepsilon_{\vec{p}_2} - \varepsilon_{\vec{k}}-\varepsilon_{\vec{k}_2} + \mathrm{i}\eta}
    \right|^2 
     , 
    \label{eq: second order rate implicit form}
\end{split}
\end{equation}
with 
\begin{equation}
    \Pi\left({\{ \vec{k}_i, \vec{p}_i\}}\right) = W_{\vec{k},\vec{k}_2;\vec{p}_1, \vec{p}_2} \left(
    N_{\vec{p}_2}+1
    \right)
    \left(
    N_{\vec{p}_1}+1
    \right)
    V_{\vec{k}_1;\vec{p}_1, \vec{p}_2}^{*}. 
    \label{eq: vertex second order Gamma}
\end{equation}
In Eq.~\eqref{eq: vertex second order Gamma}, $\{ \vec{k}_i, \vec{p}_i\}$ is shorthand for $\vec{k}, \vec{k}_1, \vec{k}_2, \vec{p}_1, \vec{p}_2$.
Compared to its first order sibling studied in Sec.~\ref{sec:firstorder}, $\Gamma_{\vec{k}}^{\left( 2\right), \text{out}}$ has a more complicated structure and, due to the complex denominator $\left( \varepsilon_{\vec{p}_1} + \varepsilon_{\vec{p}_2} - \varepsilon_{\vec{k}}-\varepsilon_{\vec{k}_2} + \mathrm{i}\eta\right)^{-1} $, it can contribute to a finite Hall effect. We can see this more clearly by expanding the squared part of Eq.~\eqref{eq: second order rate implicit form} to obtain
\begin{equation}
    \begin{split}
    &\left|
    \sum_{\vec{p}_1, \vec{p}_2}
    \delta_{\vec{p}_1, \vec{k}_1 - \vec{p}_2}
    \frac{\Pi\left({\{ \vec{k}_i, \vec{p}_i\}}\right)}{\varepsilon_{\vec{p}_1} + \varepsilon_{\vec{p}_2} - \varepsilon_{\vec{k}}-\varepsilon_{\vec{k}_2}+ \mathrm{i}\eta}
    \right|^2 
    \\
    &\quad = 
    \left[
    \text{Re}
    \left(
     \sum_{\vec{p}_1, \vec{p}_2}
    \delta_{\vec{p}_1, \vec{k}_1 - \vec{p}_2}
    \frac{\Pi\left({\{ \vec{k}_i, \vec{p}_i\}}\right)}{\varepsilon_{\vec{p}_1} + \varepsilon_{\vec{p}_2} - \varepsilon_{\vec{k}}-\varepsilon_{\vec{k}_2} + \mathrm{i}\eta}
    \right)
    \right]
    ^2 + 
    \left[
    \text{Im}
    \left(
     \sum_{\vec{p}_1, \vec{p}_2}
    \delta_{\vec{p}_1, \vec{k}_1 - \vec{p}_2}
    \frac{\Pi\left({\{ \vec{k}_i, \vec{p}_i\}}\right)}{\varepsilon_{\vec{p}_1} + \varepsilon_{\vec{p}_2} - \varepsilon_{\vec{k}}-\varepsilon_{\vec{k}_2} + \mathrm{i}\eta}
    \right)
    \right]
    ^2 
    \\
    &\quad
     = 
    \sum_{\vec{p}_1, \vec{p}_2}\sum_{\vec{p}'_1, \vec{p}'_2}
     \delta_{\vec{p}_1, \vec{k}_1 - \vec{p}_2}   \delta_{\vec{p}'_1, \vec{k}_1 - \vec{p}'_2}
     \left[
     \text{Re}
     \left(
     \frac{\Pi\left({\{ \vec{k}_i, \vec{p}_i\}}\right)}{\varepsilon_{\vec{p}_1} + \varepsilon_{\vec{p}_2} - \varepsilon_{\vec{k}}-\varepsilon_{\vec{k}_2} + \mathrm{i}\eta}
     \right)
     \text{Re}
     \left(
     \frac{\Pi\left({\{ \vec{k}_i, \vec{p}'_i\}}\right)}{\varepsilon_{\vec{p}'_1} + \varepsilon_{\vec{p}'_2} - \varepsilon_{\vec{k}}-\varepsilon_{\vec{k}_2} + \mathrm{i}\eta}
     \right)
      \right.
     \\
     & \qquad
     \left.
     + 
     \text{Im}
     \left(
     \frac{\Pi\left({\{ \vec{k}_i, \vec{p}_i\}}\right)}{\varepsilon_{\vec{p}_1} + \varepsilon_{\vec{p}_2} - \varepsilon_{\vec{k}}-\varepsilon_{\vec{k}_2} + \mathrm{i}\eta}
     \right)
      \text{Im}
     \left(
     \frac{\Pi\left({\{ \vec{k}_i, \vec{p}'_i\}}\right)}{\varepsilon_{\vec{p}'_1} + \varepsilon_{\vec{p}'_2} - \varepsilon_{\vec{k}}-\varepsilon_{\vec{k}_2} + \mathrm{i}\eta}
     \right)
     \right].
    \end{split}
    \label{eq: expanding square}
\end{equation}
Using Dirac's identity in Eq.~\eqref{eq: Dirac's Identity} we can generate 
the product of two delta-functions corresponding to resonant scattering, and the product of one delta-function from resonant scattering and a principal part. From all these terms we observe that the those proportional to one resonant delta-function,
\begin{equation}
        \sim
        \text{Re}\left[\Pi\left({\{ \vec{k}_i, \vec{p}_i\}}\right) \right] \text{Im} \left[
        \Pi\left({\{ \vec{k}_i, \vec{p}'_i\}}\right)
        \right] 
        \mathcal{P}\left(
        \frac{1}{\varepsilon_{\vec{p}'_1} + \varepsilon_{\vec{p}'_2} - \varepsilon_{\vec{k}}-\varepsilon_{\vec{k}_2}}
        \right)
        \delta\left(
        \varepsilon_{\vec{p}_1} + \varepsilon_{\vec{p}_2} - \varepsilon_{\vec{k}}-\varepsilon_{\vec{k}_2} 
        \right), 
        \label{eq: candidate terms for THE}
\end{equation}
can result in a finite Hall conductivity. 
However, we do not continue with any rigorous derivation of this statement and instead drop $\Gamma_{\vec{k}}^{(2)}$ completely. This neglect is justified because its contribution to the Hall effect is sub-leading compared to the interference terms studied in Sec.~\ref{section: Interference process}.
To see so, recall that the spin-wave expansion carried out in Sec.~\ref{sec:NSWT} is an expansion in $1/\sqrt{S}$. One finds for the free magnon energy $\varepsilon_{\vec{k}} = O(\sqrt{S}^2)$, for the three-magnon vertex $V = O(\sqrt{S})$, and for the four-magnon vertex $W = O(\sqrt{S}^0)$. Counting the powers of $\sqrt{S}$ in the scattering rates, one finds $\Gamma_{\vec{k}}^{(1)} = O(\sqrt{S}^0)$, and $\Gamma_{\vec{k}}^{\text{interf.}} = O(\sqrt{S}^{-2})$, and $\Gamma_{\vec{k}}^{(2)} = O(\sqrt{S}^{-4})$, establishing that $\Gamma_{\vec{k}}^{(2)}$ is sub-leading compared to $\Gamma_{\vec{k}}^{\text{interf.}}$. In other words, $\Gamma_{\vec{k}}^{\text{interf.}}$ provides the leading contribution to the Hall effect. 

\subsection{Table of diagrams contributing to the thermal Hall current}
\label{subsection: diagrams}
Here, we present all \emph{interference diagrams} that we have taken into consideration in order to compute the Hall conductivity. All diagrams are a product of a first-order scattering event and an appropriate second-order contribution. Specifically, we have the following types of diagrams:
\begin{enumerate}
    \item \textbf{Three-magnon in-split diagrams} (Tab.~\ref{tab:in split}): A magnon with momentum $
    \vec{k}_2$ splits into two other magnons with momenta $\vec{k}$ and $\vec{k}_1$, connecting the initial state $\ket{\text{i}} = \ket{N_{\vec{k}_2},N_{\vec{k}}, N_{\vec{k}_1}}$ to the final state $\ket{\text{f}}= \ket{N_{\vec{k}_2} - 1, N_{\vec{k}} + 1, N_{\vec{k}_1} + 1}$. Since two other magnons are involved in the processes apart from the one with momentum $\vec{k}$, we have in total two off-diagonal elements from each diagram, $\mathcal{O}_{\vec{k} \vec{k}_2}$ and $\mathcal{O}_{\vec{k} \vec{k}_1}$. 
    \item \textbf{Three-magnon in-fusion diagrams} (Tab.~\ref{tab:In fusion}): Two magnons with momenta $\vec{k}_1$ and $\vec{k}_2$ fuse into a magnon with momentum $\vec{k}$. These diagrams connect the states $\ket{\text{i}} = \ket{N_{\vec{k}_2},N_{\vec{k}}, N_{\vec{k}_1}}$ and $\ket{\text{f}}= \ket{N_{\vec{k}_2} - 1, N_{\vec{k}} + 1, N_{\vec{k}_1} - 1}$. As in the case of the in-split diagrams, we have here also two off-diagonal elements from each term. 
    \item \textbf{Four-magnon in-diagrams } (Tab.~\ref{tab:4-magnons SM}): Two magnons with momenta $\vec{k}_1$ and $\vec{k}_2$ scatter into two other magnons with momenta $\vec{k}$ and $\vec{k}_3$, respectively. Consequently, the diagrams couple the initial state $\ket{\text{i}} = \ket{N_{\vec{k}_2}, N_{\vec{k}_3}, N_{\vec{k}}, N_{\vec{k}_1}}$ and the final state $\ket{\text{f}}= \ket{N_{\vec{k}_2} - 1, N_{\vec{k}} + 1, N_{\vec{k}_3} + 1, N_{\vec{k}_1} - 1}$. Unlike the three-magnon diagrams described above, here we have in total three off-diagonal elements, $\mathcal{O}_{\vec{k} \vec{k}_1}$, $\mathcal{O}_{\vec{k} \vec{k}_2}$, and $\mathcal{O}_{\vec{k} \vec{k}_3}$. In the four-magnon table, Tab.~\ref{tab:4-magnons SM}, we show only the contributions of $\mathcal{O}_{\vec{k} \vec{k}_1}$ and $\mathcal{O}_{\vec{k} \vec{k}_3}$, since the contribution of $\mathcal{O}_{\vec{k} \vec{k}_2}$ results in duplicate contributions, as explained in more detail in 
    Sec.~\ref{subsubsection: 4-magnons}. 
\end{enumerate}
For the three-magnons diagrams---either in-split or in-fusion---the second order process is generated from the combination of a three-magnon followed by a four-magnon scattering event (or \textit{vice versa}). The three-magnon diagrams contain two intermediate magnons denoted by momenta $\vec{p}_1$ and $\vec{p}_2$, which can be either virtual (forward lines from left to right) or real (backward lines from right to left). On the other hand, the second order diagrams for the four-magnon interference terms stem from two consecutive three-magnon scattering events. Consequently, they contain only one intermediate magnon with momentum $\vec{p}$, which can be either real or virtual. We stress that the notion of real and virtual magnons should be associated only with intermediate magnons.
 
All diagrams selected here scale with the spin and DMI strength as $D^2/ S$. To see this, remember that the bare magnon energies are $O(S)$ and the three-magnon vertices are $O(D\sqrt{S})$. Since each diagram contains the three-magnon vertex twice (contributing $D^2S$), and the bare energies enter in two delta-functions---one from global energy conservation and one from resonant scattering (contributing $1/S^2$)---we find the scaling with $D ^2/S$.

Finally, we note that all diagrams presented correspond to in-processes, because the out- and in-counterparts of a given scattering process can be directly related by the (broken) detailed balance relation introduced in Sec.~\ref{subsection: DB}. 

In the following, we will discuss the diagrams in more detail. We will do so using a diagrammatic language when it facilitates a clearer understanding. From now on, the notation of $\left[ \textit{Diagram}\right]_{\vec{k} \vec{k}'}$ should be understood as the $\mathcal{O}_{\vec{k} \vec{k}'}$ element of the diagram enclosed in the parentheses. We start with the three-magnon diagrams corresponding to the in-split and in-fusion processes, and continue with the four-magnon diagrams. 

\newpage

\subsubsection{Three-magnon diagrams}
We consider the three-magnon diagrams in Tab.~\ref{tab:in split} and Tab.~\ref{tab:In fusion}. The following observations can be made:

\begin{table}
\caption{Three-magnon in-split interference diagrams. A magnon with momentum $\vec{k}_2$ splits into two magnons with momenta $\vec{k}$ and $\vec{k}_1$. This process can occur at first order due to the three-magnon vertex, and at second order due to a combination of three- and four-magnon vertices. The interference of these contributions leads to an off-diagonal part, $\mathcal{O}_{\vec{k}\vec{k}_1}$ and $\mathcal{O}_{\vec{k}\vec{k}_2}$, of the collision matrix, which gives rise to a thermal Hall effect. \label{tab:in split}}
\begin{ruledtabular}
\begin{tabular}
{p{4.3cm} p{4.0cm} p{4.0cm}}
\hspace{1.2 cm}
Diagram 
& \hspace{2.6 cm} $\mathcal{O}_{\vec{k}\vec{k}_1}  $
& \hspace{2.6 cm} $\mathcal{O}_{\vec{k}\vec{k}_2}$ \\ 
\hline 
\begin{align*}
\adjustbox{trim=0 0 -1cm 0}{
\begin{tikzpicture}[scale=1.3, decoration={markings, 
        mark= at position 0.65 with {\arrow{stealth}},
        mark= at position 2cm with {\arrow{stealth}}}
        ]   
        \begin{scope}
          \draw[postaction={decorate}] (1.3,0.0)--(1.7, 0.2);
            \draw[postaction={decorate}] (1.3,0.0)--(1.7, -0.2);
            \draw[postaction={decorate}] (1,0)--(1.3,0);
            \draw[fill=black] (1.3,0) circle (0.07);
            \node at (1.45, 0.3) {\footnotesize{$\vec{k}_1$}};
            \node at (1.45, -0.3) {\footnotesize{$\vec{k}$}};
            \node at (1.85,0) {$\times$};
            \node at (0.8, 0.) {\footnotesize{$\vec{k}_2$}};
        \end{scope}
\end{tikzpicture}
}
\\
\left(
\vcenter{\hbox{
\begin{tikzpicture}[scale=1.3, decoration={markings, 
        mark= at position 0.65 with {\arrow{stealth}},
        mark= at position 2cm with {\arrow{stealth}}}
        ]   
      \begin{scope}
            \draw [postaction={decorate}] 
                (0.9,-1.2) to [out=70,in=110] (0.3,-1.2);
                  \draw [postaction={decorate}] (0.9,-1.2) to [out=-70,in=-110] (0.3,-1.2);
                  \draw[postaction={decorate}] (0.80,-0.9) to [out=0,in=90] (0.9,-1.2);
                  \draw[fill=white] (0.3,-1.2) circle (0.07);
                  \draw[fill=black] (0.9,-1.2) circle (0.07);
                  \draw [postaction={decorate}] 
                  (0.3, -1.27) to [out=-90, in=180] (0.7, -1.8);
                  \draw [postaction={decorate}] 
                  (0.3, -1.13) to [out=90, in=180] (0.7, -0.6); 
                 \node at (0.2, -0.8)   {\footnotesize{$\vec{k}_1$}};
                 \node at (0.2, -1.6)   {\footnotesize{$\vec{k}$}};
                  \node at (1.1, -0.9) {\footnotesize{$\vec{k}_2$}};
                  \node at (0.6, -0.9) {\footnotesize{$\vec{p}_1$}};
                  \node at (0.6, -1.55)
                  {\footnotesize{$\vec{p}_2$}};
                   \node at (1.2, -1.2) {$+$};
        \end{scope}
        \begin{scope}
            \draw[postaction={decorate}] (2.2,-1.2)--(2.5,-1);
                  \draw[postaction={decorate}] (2.2,-1.2)--(2.5,-1.4);
                  \draw [postaction={decorate}] (1.7,-1.2) to [out=70,in=110] (2.2,-1.2);
                  \draw [postaction={decorate}] (1.7,-1.2) to [out=-70,in=-110] (2.2,-1.2);
                  \draw[postaction={decorate}] (1.4,-1.2)--(1.7,-1.2);
                  \draw[fill=black] (1.7,-1.2) circle (0.07);
                  \draw[fill=white] (2.2,-1.2) circle (0.07);
                \node at (1.55, -1) {\footnotesize{$\vec{k}_2$}}; 
                \node at (1.9, -1.5) {\footnotesize{$\vec{p}_1$}}; 
                \node at (1.9, -0.9){\footnotesize{$\vec{p}_2$}};
                \node at (2.6, -0.9) {\footnotesize{$\vec{k}_1$}};
                \node at (2.6, -1.5) {\footnotesize{$\vec{k}$}}; 
        \end{scope}
\end{tikzpicture}
}}
\right)
\end{align*}
& 
\setlength{\abovedisplayskip}{0pt}
\setlength{\belowdisplayskip}{0pt}
{\begin{align*}
\hspace{0.0cm}
\mathcal{O}_{\vec{k}\vec{k}_1}^{++} 
= 
& 16 \frac{(2\pi)^2}{\hbar}\frac{G_{\vec{k}_1}}{G_{\vec{k}}}
\left(
\overline{N}_{\vec{k}}+1
\right) \sum_{\vec{k}_2}\delta_{\vec{k}_2,\vec{k}+\vec{k_1}} \\
& 
\delta
\left(
\varepsilon_{\vec{k}}+\varepsilon_{\vec{k}_1}-\varepsilon_{\vec{k}_2}
\right) \overline{N}_{\vec{k}_2}
\sum_{\vec{p}_1, \vec{p}_2} \delta_{\vec{p}_1 + \vec{p}_2, \vec{k}_2}
\\
&  
\delta 
\left(
\varepsilon_{\vec{k}_1} + \varepsilon_{\vec{k}}-\varepsilon_{\vec{p}_1}-\varepsilon_{\vec{p}_2}
\right) W_{ \vec{p}_1, \vec{p}_2; \vec{k}_1, \vec{k}} \\
& 
\left[
\left(1+ \overline{N}_{\vec{p}_1}\right) \left(1+ \overline{N}_{\vec{p}_2}\right) +  \overline{N}_{\vec{p}_2}\overline{N}_{\vec{p}_1} 
\right]\\ 
& \left[ 
\text{Re}
\left(
V_{\vec{k}_2; \vec{k}, \vec{k}_1}
\right)
\text{Im}
\left(
V_{\vec{k}_2;\vec{p}_1, \vec{p}_2}
\right) 
- 
\right. 
\\
& 
\left. 
\text{Im}
\left(
V_{\vec{k}_2; \vec{k}, \vec{k}_1}
\right)
\text{Re}
\left(
V_{\vec{k}_2;\vec{p}_1, \vec{p}_2}
\right)
\right]
\end{align*}}
 & 
 \setlength{\abovedisplayskip}{0pt}
 \setlength{\belowdisplayskip}{0pt}
{\begin{align*}
\hspace{0.0cm}
\mathcal{O}_{\vec{k}\vec{k}_2}^{+-} 
= 
& 16 \frac{(2\pi)^2}{\hbar}\frac{G_{\vec{k}_2}}{G_{\vec{k}}} \left(
1+\overline{N}_{\vec{k}}
\right) 
\sum_{\vec{k}_1} 
\delta_{\vec{k}_1,\vec{k}_2-\vec{k}} 
\\
& 
 \left(
\overline{N}_{\vec{k}_1} + 1
\right)
\delta
\left(
\varepsilon_{\vec{k}_1}+\varepsilon_{\vec{k}}-\varepsilon_{\vec{k}_2}
\right) 
\sum_{\vec{p}_1, \vec{p}_2} \delta_{\vec{p}_1 + \vec{p}_2, \vec{k}_2}\\
& 
\delta
\left(
\varepsilon_{\vec{k}_1} + \varepsilon_{\vec{k}}-\varepsilon_{\vec{p}_1}-\varepsilon_{\vec{p}_2}
\right)
W_{\vec{p}_1, \vec{p}_2; \vec{k}_1, \vec{k}} 
\\
& 
\left[
\left(
\overline{N}_{\vec{p}_1} +1 
\right) 
\left(
\overline{N}_{\vec{p}_2} + 1
\right)
+ \overline{N}_{\vec{p}_2}\overline{N}_{\vec{p}_1}
\right] \\
& 
\left[ 
\text{Re}\left(V_{\vec{k}_2;\vec{k}_1,\vec{k}} \right)\text{Im}\left(V_{\vec{k}_2;\vec{p}_1,\vec{p}_2}\right) - 
\right. 
\\
& 
\left. 
\text{Im}\left(V_{\vec{k}_2;\vec{k}_1,\vec{k}}\right)\text{Re}\left(V_{\vec{k}_2;\vec{p}_1,\vec{p}_2} 
\right)
\right]
\end{align*}}
\\
\hline 
\begin{align*}
\vcenter{\hbox{
\begin{tikzpicture}[scale=1.3, decoration={markings, 
        mark= at position 0.65 with {\arrow{stealth}},
        mark= at position 2cm with {\arrow{stealth}}}
        ]   
        \begin{scope}
            \draw[postaction={decorate}] (1.3,0.0)--(1.7, 0.2);
            \draw[postaction={decorate}] (1.3,0.0)--(1.7, -0.2);
            \draw[postaction={decorate}] (1,0)--(1.3,0);
            \draw[fill=black] (1.3,0) circle (0.07);
            \node at (1.45, 0.3) {\footnotesize{$\vec{k}_1$}};
            \node at (1.45, -0.3) {\footnotesize{$\vec{k}$}};
            \node at (1.9,0) {$\times$};
            \node at (0.8, 0.) {\footnotesize{$\vec{k}_2$}};
            \begin{scope}
                \draw [postaction={decorate}] 
                  (2.3, 0.07) to [out=90, in=180] (2.9, 0.65);
                  \draw [postaction={decorate}] (2.9,0) to [out=110,in=70] (2.3,0);
                 \draw [postaction={decorate}] (2.3,0) to [out=-70,in=-110] (2.9,0);
                 \draw[postaction={decorate}] (2.9,0) -- (3.4,0);
                 \draw[fill=black] (2.3,0) circle (0.07);
                 \draw[fill=white] (2.9,0) circle (0.07);
                 \draw [postaction={decorate}] 
                  (2.3, -0.65) to [out=0, in=-90] (2.9, -0.07);
                  \node at (2.6, 0.75) {\footnotesize{$\vec{k}$}}; 
                  \node at (3.3, 0.2) {\footnotesize{$\vec{k}_1$}}; 
                  \node at (2.8, -0.65) {\footnotesize{$\vec{k}_2$}}; 
                  \node at (2.5, -0.35) {\footnotesize{$\vec{p}_2$}}; 
                   \node at (2.7, 0.35) {\footnotesize{$\vec{p}_1$}};
            \end{scope}   
        \end{scope}
    \end{tikzpicture}}}
\end{align*}
& 
\setlength{\abovedisplayskip}{0pt}
 \setlength{\belowdisplayskip}{0pt}
{\begin{align*}
\hspace{0.0cm}
\mathcal{O}_{\vec{k}\vec{k}_1}^{++} 
= 
& 16 \frac{(2\pi)^2}{\hbar}\frac{G_{\vec{k}_1}}{G_{\vec{k}}}
\left(
\overline{N}_{\vec{k}}+1
\right)
\sum_{\vec{k}_2}\delta_{\vec{k}_2,\vec{k}+\vec{k_1}} \\
& 
\overline{N}_{\vec{k}_2}
\delta
\left(
\varepsilon_{\vec{k}}+\varepsilon_{\vec{k}_1}-\varepsilon_{\vec{k}_2}
\right) 
\sum_{\vec{p}_1, \vec{p}_2} \delta_{\vec{k}_1 + \vec{p}_2, \vec{p}_1 + \vec{k}_1} \\
& 
\delta 
\left(
 \varepsilon_{\vec{k}}-\varepsilon_{\vec{p}_1} +\varepsilon_{\vec{p}_2}
\right)
 W_{ \vec{p}_1, \vec{k}_1; \vec{k}_2, \vec{p}_2} 
\overline{N}_{\vec{p}_1} 
\\ 
& 
\left(
1+ \overline{N}_{\vec{p}_2}
\right)
\left[ 
\text{Im}\left(V_{\vec{p}_1; \vec{k}, \vec{p}_2}\right)
\text{Re}\left(V_{\vec{k}_2;\vec{k}_1, \vec{k}}\right)-
\right. 
\\ 
& 
\left.
\text{Im}\left(V_{\vec{k}_2; \vec{k}, \vec{k}_1}\right)\text{Re}\left(V_{\vec{p}_1;\vec{k}, \vec{p}_2}\right)
\right]
\end{align*}}
& 
\setlength{\abovedisplayskip}{0pt}
 \setlength{\belowdisplayskip}{0pt}
{\begin{align*}
\hspace{0.0cm}
\mathcal{O}_{\vec{k}\vec{k}_2}^{+-} = 
& 16 \frac{(2\pi)^2}{\hbar}\frac{G_{\vec{k}_2}}{G_{\vec{k}}} \left(
\overline{N}_{\vec{k}}+1
\right)\sum_{\vec{k}_1}\delta_{\vec{k}_2-\vec{k},\vec{k_1}} \\
& 
\delta
\left(
\varepsilon_{\vec{k}}+\varepsilon_{\vec{k}_1}-\varepsilon_{\vec{k}_2}
\right) \left(
\overline{N}_{\vec{k}_1} + 1
\right)
\sum_{\vec{p}_1, \vec{p}_2} \delta_{\vec{k}_1 + \vec{p}_2, \vec{p}_1 + \vec{k}_1} \\
& 
\delta 
\left(
 \varepsilon_{\vec{k}}-\varepsilon_{\vec{p}_1} +\varepsilon_{\vec{p}_2}
\right)
W_{ \vec{p}_1, \vec{k}_1; \vec{k}_2, \vec{p}_2} 
\overline{N}_{\vec{p}_1} 
\\ 
& 
\left(1+ \overline{N}_{\vec{p}_2}
\right)
\left[ 
\text{Im}\left(V_{\vec{p}_1; \vec{k}, \vec{p}_2}\right)\text{Re}\left(V_{\vec{k}_2;\vec{k}_1, \vec{k}}\right)-
\right. \\
& 
\left. 
\text{Im}\left(V_{\vec{k}_2; \vec{k}, \vec{k}_1}\right)\text{Re}\left(V_{\vec{p}_1;\vec{k}, \vec{p}_2}\right)
\right]
\end{align*}}
\\
\hline
\begin{align*}
\vcenter{\hbox{
\begin{tikzpicture}[scale=1.3, decoration={markings, 
        mark= at position 0.65 with {\arrow{stealth}},
        mark= at position 2cm with {\arrow{stealth}}}
        ]   
        \begin{scope}
            \draw[postaction={decorate}] (1.3,0.0)--(1.7, 0.2);
            \draw[postaction={decorate}] (1.3,0.0)--(1.7, -0.2);
            \draw[postaction={decorate}] (1,0)--(1.3,0);
            \draw[fill=black] (1.3,0) circle (0.07);
            \node at (1.45, 0.3) {\footnotesize{$\vec{k}_1$}};
            \node at (1.45, -0.3) {\footnotesize{$\vec{k}$}};
            \node at (1.9,0) {$\times$};
            \node at (0.8, 0.) {\footnotesize{$\vec{k}_2$}};
            \begin{scope}
                \draw [postaction={decorate}] 
                  (2.3, 0.07) to [out=90, in=180] (2.9, 0.65);
                  \draw [postaction={decorate}] (2.9,0) to [out=110,in=70] (2.3,0);
                 \draw [postaction={decorate}] (2.3,0) to [out=-70,in=-110] (2.9,0);
                 \draw[postaction={decorate}] (2.9,0) -- (3.4,0);
                 \draw[fill=black] (2.3,0) circle (0.07);
                 \draw[fill=white] (2.9,0) circle (0.07);
                 \draw [postaction={decorate}] 
                  (2.3, -0.65) to [out=0, in=-90] (2.9, -0.07);
                  \node at (2.6, 0.76) {\footnotesize{$\vec{k}_1$}}; 
                  \node at (3.3, 0.2) {\footnotesize{$\vec{k}$}}; 
                  \node at (2.8, -0.65) {\footnotesize{$\vec{k}_2$}}; 
                  \node at (2.5, -0.35) {\footnotesize{$\vec{p}_2$}}; 
                   \node at (2.7, 0.35) {\footnotesize{$\vec{p}_1$}};
            \end{scope}   
        \end{scope}
    \end{tikzpicture}}}
\end{align*}
\vspace{0.25cm}
& 
\setlength{\abovedisplayskip}{0pt}
 \setlength{\belowdisplayskip}{0pt}
{\begin{align*}
\hspace{0.0cm}
\mathcal{O}_{\vec{k}\vec{k}_1}^{++} = 
& 16 \frac{(2\pi)^2}{\hbar}\frac{G_{\vec{k}_1}}{G_{\vec{k}}} \left(
\overline{N}_{\vec{k}}+1
\right) \sum_{\vec{k}_2}\delta_{\vec{k}_2,\vec{k}+\vec{k_1}} \\
& 
\delta
\left(
\varepsilon_{\vec{k}}+\varepsilon_{\vec{k}_1}-\varepsilon_{\vec{k}_2}
\right) \overline{N}_{\vec{k}_2}
\sum_{\vec{p}_1, \vec{p}_2} \delta_{\vec{k} + \vec{p}_2, \vec{p}_1 + \vec{k}_1} \\
& 
\delta 
\left(
 \varepsilon_{\vec{k}_1}-\varepsilon_{\vec{p}_1} +\varepsilon_{\vec{p}_2}
\right)
W_{ \vec{p}_1, \vec{k}; \vec{k}_2, \vec{p}_2} 
\overline{N}_{\vec{p}_1} 
\\ 
& 
\left(
1+ \overline{N}_{\vec{p}_2}
\right)
\left[ 
\text{Im}\left(V_{\vec{p}_1; \vec{k}_1, \vec{p}_2}\right)
\text{Re}\left(V_{\vec{k}_2;\vec{k}_1, \vec{k}}\right)-
\right. 
\\
& 
\left. 
\text{Im}\left(V_{\vec{k}_2; \vec{k}, \vec{k}_1}\right)\text{Re}\left(V_{\vec{p}_1;\vec{k}_1, \vec{p}_2}\right)
\right]
\end{align*}}
& 
\setlength{\abovedisplayskip}{0pt}
 \setlength{\belowdisplayskip}{0pt}
{\begin{align*}
\hspace{0.0cm}
\mathcal{O}_{\vec{k}\vec{k}_2}^{+-} = 
& 16 \frac{(2\pi)^2}{\hbar}\frac{G_{\vec{k}_2}}{G_{\vec{k}}} \left(
\overline{N}_{\vec{k}}+1
\right)
\sum_{\vec{k}_1}\delta_{\vec{k}_2-\vec{k},\vec{k_1}}  \\
& 
\delta
\left(
\varepsilon_{\vec{k}}+\varepsilon_{\vec{k}_1}-\varepsilon_{\vec{k}_2}
\right) \left(\overline{N}_{\vec{k}_1} + 1 \right)
\sum_{\vec{p}_1, \vec{p}_2} \delta_{\vec{k} + \vec{p}_2, \vec{p}_1 + \vec{k}_1} \\
& 
\delta 
\left(
 \varepsilon_{\vec{k}_1}-\varepsilon_{\vec{p}_1} +\varepsilon_{\vec{p}_2}
\right)
W_{ \vec{p}_1, \vec{k}; \vec{k}_2, \vec{p}_2} 
\overline{N}_{\vec{p}_1} 
\\ 
& 
\left(
1+ \overline{N}_{\vec{p}_2}
\right)
\left[ 
\text{Im}\left(V_{\vec{p}_1; \vec{k}_1, \vec{p}_2}\right)
\text{Re}\left(V_{\vec{k}_2;\vec{k}_1, \vec{k}}\right)-
\right.
\\
& 
\left. 
\text{Im}\left(V_{\vec{k}_2; \vec{k}, \vec{k}_1}\right)\text{Re}\left(V_{\vec{p}_1;\vec{k}_1, \vec{p}_2}\right)
\right]
\end{align*}}
\\
\hline 
\begin{align*}
\vcenter{\hbox{
\begin{tikzpicture}[scale=1.3, decoration={markings, 
        mark= at position 0.65 with {\arrow{stealth}},
        mark= at position 2cm with {\arrow{stealth}}}
        ]   
        \begin{scope}
            \draw[postaction={decorate}] (1.3,0.0)--(1.7, 0.2);
            \draw[postaction={decorate}] (1.3,0.0)--(1.7, -0.2);
            \draw[postaction={decorate}] (1,0)--(1.3,0);
            \draw[fill=black] (1.3,0) circle (0.07);
            \node at (1.45, 0.3) {\footnotesize{$\vec{k}_1$}};
            \node at (1.45, -0.3) {\footnotesize{$\vec{k}$}};
            \node at (1.9,0) {$\times$};
            \node at (0.8, 0.) {\footnotesize{$\vec{k}_2$}};
            \begin{scope}[xshift = 0.3cm]
                \draw [postaction={decorate}] 
                  (2.3, 0.07) to [out=90, in=180] (2.9, 0.65);
                  \draw [postaction={decorate}] (2.9,0) to [out=110,in=70] (2.3,0);
                 \draw [postaction={decorate}] (2.3,0) to [out=-70,in=-110] (2.9,0);
                 \draw[postaction={decorate}] (1.9,0) -- (2.3,0);
                 \draw[fill=white] (2.3,0) circle (0.07);
                 \draw[fill=black] (2.9,0) circle (0.07);
                 \draw[postaction={decorate}] (2.97, 0) -- (3.3, 0); 
                  \node at (2.6, 0.76) {\footnotesize{$\vec{k}$}};   
                  \node at (2.5, -0.35) {\footnotesize{$\vec{p}_1$}}; 
                  \node at (2.7, 0.35) {\footnotesize{$\vec{p}_2$}};
                  \node at (2, 0.2) {\footnotesize{$\vec{k}_2$}}; 
                  \node at (3.15, 0.2) {\footnotesize{$\vec{k}_1$}}; 
            \end{scope}   
        \end{scope}
    \end{tikzpicture}}}
\end{align*}
\vspace{0.25cm}
& 
\setlength{\abovedisplayskip}{0pt}
 \setlength{\belowdisplayskip}{0pt}
{\begin{align*}
\hspace{0.0cm}
\mathcal{O}_{\vec{k}\vec{k}_1}^{++} = 
& 16 \frac{(2\pi)^2}{\hbar}\frac{G_{\vec{k}_1}}{G_{\vec{k}}}\left(
\overline{N}_{\vec{k}}+1
\right) \sum_{\vec{k}_2}\delta_{\vec{k}_2,\vec{k}+\vec{k_1}} \\
& 
\delta
\left(
\varepsilon_{\vec{k}}+\varepsilon_{\vec{k}_1}-\varepsilon_{\vec{k}_2}
\right) \overline{N}_{\vec{k}_2}
\sum_{\vec{p}_1, \vec{p}_2} \delta_{\vec{k} + \vec{p}_2, \vec{p}_1 + \vec{k}_1} \\
& 
\delta 
\left(
 \varepsilon_{\vec{k}_2} + \varepsilon_{\vec{p}_2}-\varepsilon_{\vec{p}_1}  - \varepsilon_{\vec{k}}
\right)
W_{ \vec{p}_1, \vec{k}; \vec{k}_2, \vec{p}_2} 
\overline{N}_{\vec{p}_2} 
\\ 
& 
\left(1+ \overline{N}_{\vec{p}_1}\right)
\left[ 
\text{Im}\left(V_{\vec{p}_1; \vec{k}_1, \vec{p}_2}\right)
\text{Re}\left(V_{\vec{k}_2;\vec{k}_1, \vec{k}}\right)- 
\right. 
\\
& 
\left. 
\text{Im}\left(V_{\vec{k}_2; \vec{k}, \vec{k}_1}\right)\text{Re}\left(V_{\vec{p}_1;\vec{k}_1, \vec{p}_2}\right)
\right]
\end{align*}}
& 
\setlength{\abovedisplayskip}{0pt}
 \setlength{\belowdisplayskip}{0pt}
{\begin{align*}
\hspace{0.0cm}
\mathcal{O}_{\vec{k}\vec{k}_2}^{+-} = 
& 16 \frac{(2\pi)^2}{\hbar}\frac{G_{\vec{k}_2}}{G_{\vec{k}}} \left(
\overline{N}_{\vec{k}}+1
\right)\sum_{\vec{k}_1}\delta_{\vec{k}_2-\vec{k},\vec{k_1}}  \\
& 
\delta
\left(
\varepsilon_{\vec{k}}+\varepsilon_{\vec{k}_1}-\varepsilon_{\vec{k}_2}
\right) \left(\overline{N}_{\vec{k}_1} + 1 \right)
\sum_{\vec{p}_1, \vec{p}_2} \delta_{\vec{k} + \vec{p}_2, \vec{p}_1 + \vec{k}_1} \\
& 
\delta 
\left(
 \varepsilon_{\vec{k}_2} + \varepsilon_{\vec{p}_2}-\varepsilon_{\vec{p}_1}  - \varepsilon_{\vec{k}}
\right)
W_{ \vec{p}_1, \vec{k}; \vec{k}_2, \vec{p}_2} 
\overline{N}_{\vec{p}_2} 
\\ 
& 
\left(
1+ \overline{N}_{\vec{p}_1}
\right)
\left[ 
\text{Im}\left(V_{\vec{p}_1; \vec{k}_1, \vec{p}_2}\right)
\text{Re}\left(V_{\vec{k}_2;\vec{k}_1, \vec{k}}\right)-
\right. 
\\
& 
\left. 
\text{Im}\left(V_{\vec{k}_2; \vec{k}, \vec{k}_1}\right)\text{Re}\left(V_{\vec{p}_1;\vec{k}_1, \vec{p}_2}\right)
\right]
\end{align*}}
\\
\hline 
\begin{align*}
\vcenter{\hbox{
\begin{tikzpicture}[scale=1.3, decoration={markings, 
        mark= at position 0.65 with {\arrow{stealth}},
        mark= at position 2cm with {\arrow{stealth}}}
        ]   
        \begin{scope}
            \draw[postaction={decorate}] (1.3,0.0)--(1.7, 0.2);
            \draw[postaction={decorate}] (1.3,0.0)--(1.7, -0.2);
            \draw[postaction={decorate}] (1,0)--(1.3,0);
            \draw[fill=black] (1.3,0) circle (0.07);
            \node at (1.45, 0.3) {\footnotesize{$\vec{k}_1$}};
            \node at (1.45, -0.3) {\footnotesize{$\vec{k}$}};
            \node at (1.9,0) {$\times$};
            \node at (0.8, 0.) {\footnotesize{$\vec{k}_2$}};
            \begin{scope}[xshift = 0.3cm]
                \draw [postaction={decorate}] 
                  (2.3, 0.07) to [out=90, in=180] (2.9, 0.65);
                  \draw [postaction={decorate}] (2.9,0) to [out=110,in=70] (2.3,0);
                 \draw [postaction={decorate}] (2.3,0) to [out=-70,in=-110] (2.9,0);
                 \draw[postaction={decorate}] (1.9,0) -- (2.3,0);
                 \draw[fill=white] (2.3,0) circle (0.07);
                 \draw[fill=black] (2.9,0) circle (0.07);
                 \draw[postaction={decorate}] (2.97, 0) -- (3.3, 0); 
                  \node at (2.6, 0.76) {\footnotesize{$\vec{k}_1$}};   
                  \node at (2.5, -0.35) {\footnotesize{$\vec{p}_1$}}; 
                  \node at (2.7, 0.35) {\footnotesize{$\vec{p}_2$}};
                  \node at (2, 0.2) {\footnotesize{$\vec{k}_2$}}; 
                  \node at (3.15, 0.2) {\footnotesize{$\vec{k}$}}; 
            \end{scope}   
        \end{scope}
    \end{tikzpicture}}}
\end{align*}
\vspace{0.25cm}
& 
\setlength{\abovedisplayskip}{0pt}
 \setlength{\belowdisplayskip}{0pt}
{\begin{align*}
\hspace{0.0cm}
\mathcal{O}_{\vec{k}\vec{k}_1}^{++} =
& 16 \frac{(2\pi)^2}{\hbar}\frac{G_{\vec{k}_1}}{G_{\vec{k}}} 
\left(
\overline{N}_{\vec{k}}+1
\right) \left(
\overline{N}_{\vec{k}}+1
\right) \sum_{\vec{k}_2}\delta_{\vec{k}_2,\vec{k}+\vec{k_1}}\\
& 
\delta
\left(
\varepsilon_{\vec{k}}+\varepsilon_{\vec{k}_1}-\varepsilon_{\vec{k}_2}
\right) \overline{N}_{\vec{k}_2}
\sum_{\vec{p}_1, \vec{p}_2} \delta_{\vec{k}_1 + \vec{p}_2, \vec{p}_1 + \vec{k}_1} \\
& 
\delta 
\left(
 \varepsilon_{\vec{k}_1} +\varepsilon_{\vec{p}_1} - \varepsilon_{\vec{p}_2} - \varepsilon_{\vec{k}_2}
\right)
W_{ \vec{p}_1, \vec{k}_1; \vec{k}_2, \vec{p}_2} 
\overline{N}_{\vec{p}_2} 
\\ 
& 
\left(1+ \overline{N}_{\vec{p}_1}
\right)
\left[ 
\text{Im}\left(V_{\vec{p}_1; \vec{k}, \vec{p}_2}\right)
\text{Re}\left(V_{\vec{k}_2;\vec{k}_1, \vec{k}}\right)-
\right. 
\\
& 
\left. 
\text{Im}\left(V_{\vec{k}_2; \vec{k}, \vec{k}_1}\right)\text{Re}\left(V_{\vec{p}_1;\vec{k}, \vec{p}_2}\right)
\right]
\end{align*}}
& 
\setlength{\abovedisplayskip}{0pt}
 \setlength{\belowdisplayskip}{0pt}
{\begin{align*}
\hspace{0.0cm}
\mathcal{O}_{\vec{k}\vec{k}_2}^{+-} = 
& 16 \frac{(2\pi)^2}{\hbar}\frac{G_{\vec{k}_2}}{G_{\vec{k}}} 
\left(
\overline{N}_{\vec{k}_1} + 1
\right) 
\sum_{\vec{k}_1}\delta_{\vec{k}_2-\vec{k},\vec{k_1}}  \\
& 
\delta
\left(
\varepsilon_{\vec{k}}+\varepsilon_{\vec{k}_1}-\varepsilon_{\vec{k}_2}
\right)
\sum_{\vec{p}_1, \vec{p}_2} \delta_{\vec{k}_1 + \vec{p}_2, \vec{p}_1 + \vec{k}_1}
\\
& 
\delta 
\left(
 \varepsilon_{\vec{k}_1} +\varepsilon_{\vec{p}_1} - \varepsilon_{\vec{p}_2} - \varepsilon_{\vec{k}_2}
\right)
W_{ \vec{p}_1, \vec{k}_1; \vec{k}_2, \vec{p}_2} 
\overline{N}_{\vec{p}_2} 
\\ 
& 
\left(
1+ \overline{N}_{\vec{p}_1}
\right)
\left[ 
\text{Im}\left(V_{\vec{p}_1; \vec{k}, \vec{p}_2}\right)\text{Re}\left(V_{\vec{k}_2;\vec{k}_1, \vec{k}}\right)-
\right. 
\\
& 
\left. 
\text{Im}\left(V_{\vec{k}_2; \vec{k}, \vec{k}_1}\right)\text{Re}\left(V_{\vec{p}_1;\vec{k}, \vec{p}_2}\right)
\right]
\end{align*}}
\end{tabular}
\end{ruledtabular}
\end{table}

\begin{table}
\caption{Three-magnon in-fusion interference diagrams. Two magnons with momenta $\vec{k}_1$ and $\vec{k}_2$ fuse into a magnons with momentum $\vec{k}$. This process can occur at first order due to the three-magnon vertex, and at second order due to a combination of three- and four-magnon vertices. The interference of these contributions leads to an off-diagonal part, $\mathcal{O}_{\vec{k}\vec{k}_1}$ and $\mathcal{O}_{\vec{k}\vec{k}_2}$, of the collision matrix, which gives rise to a thermal Hall effect. \label{tab:In fusion}}
\begin{ruledtabular}
\begin{tabular}
{p{4.3cm} p{6.0cm} p{6.0cm}}
\hspace{1.2 cm} Diagram 
& \hspace{2.6 cm} $\mathcal{O}_{\vec{k}\vec{k}_1}  $
& \hspace{2.6 cm} $\mathcal{O}_{\vec{k}\vec{k}_2}$ \\ 
\hline

\begin{align*}
\adjustbox{trim=0 0 -1cm 0}{
\begin{tikzpicture}[scale=1.3, decoration={markings, 
        mark= at position 0.65 with {\arrow{stealth}},
        mark= at position 2cm with {\arrow{stealth}}}
        ]   
        \begin{scope}
           \draw[postaction={decorate}] (1,0.2)--(1.3,0.0);
            \draw[postaction={decorate}] (1,-0.2)--(1.3,0.0);
            \draw[postaction={decorate}] (1.3,0)--(1.7,0.0);
            \draw[fill=black] (1.3,0) circle (0.07);
            \node at (1.85,0) {$\times$};
            \node at (1.2, 0.4) {\footnotesize{$\vec{k}_1$}};
            \node at (1.2, -0.4) {\footnotesize{$\vec{k}_2$}};
            \node at (1.65, 0.2) {\footnotesize{$\vec{k}$}};
            \end{scope}
\end{tikzpicture}}
\\
\left(
\vcenter{\hbox{
\begin{tikzpicture}[scale=1.3, decoration={markings, 
        mark= at position 0.65 with {\arrow{stealth}},
        mark= at position 2cm with {\arrow{stealth}}}
        ]   
        \begin{scope}
                 \draw [postaction={decorate}] (0.9,-1.2) to [out=70,in=110] (0.3,-1.2);
                 \draw [postaction={decorate}] (0.9,-1.2) to [out=-70,in=-110] (0.3,-1.2);
                 \draw[postaction={decorate}] (0.3,-1.2) -- (0.6,-1.2);
                 \draw[fill=black] (0.3,-1.2) circle (0.07);
                 \draw[fill=white] (0.9,-1.2) circle (0.07);
                 \draw [postaction={decorate}] 
                  (0.3, -0.7) to [out=0, in=90] (0.9, -1.13);
                  \draw [postaction={decorate}] 
                  (0.3, -1.7) to [out=0, in=-90] (0.9, -1.27);
                  \node at (1, -0.8) {\footnotesize{$\vec{k}_1$}};
                  \node at (1, -1.6)  {\footnotesize{$\vec{k}_2$}};
                   \node at (0.7, -1.2) {\footnotesize{$\vec{k}$}};
                   \node at (0.35, -0.9){\footnotesize{$\vec{p}_1$}};
                   \node at (0.35, -1.5)
                 {\footnotesize{$\vec{p}_2$}};
                   \node at (1.2, -1.2) {$+$};
                 \end{scope}
        \begin{scope}
             \draw[postaction={decorate}] (1.4,-1)--(1.7,-1.2);
                  \draw[postaction={decorate}] (1.4,-1.4)--(1.7,-1.2);
                  \draw [postaction={decorate}] (1.7,-1.2) to [out=70,in=110] (2.2,-1.2);
                  \draw [postaction={decorate}] (1.7,-1.2) to [out=-70,in=-110] (2.2,-1.2);
                  \draw[postaction={decorate}] (2.2,-1.2)--(2.7,-1.2);
                  \draw[fill=white] (1.7,-1.2) circle (0.07);
                  \draw[fill=black] (2.2,-1.2) circle (0.07);
                  \node at (2.5, -1) {\footnotesize{$\vec{k}$}}; 
                  \node at (2, -0.85) {\footnotesize{$\vec{p}_1$}}; 
                  \node at (2, -1.55){\footnotesize{$\vec{p}_2$}};
                  \node at (1.55, -0.8) {\footnotesize{$\vec{k}_1$}};
                  \node at (1.55, -1.5) {\footnotesize{$\vec{k}_2$}};
        \end{scope}
\end{tikzpicture}
}}
\right)
\end{align*}
 & 
 \setlength{\abovedisplayskip}{0pt}
 \setlength{\belowdisplayskip}{0pt}
 {\begin{align*}
\hspace{0.0cm}
\mathcal{O}_{\vec{k}\vec{k}_1}^{+-} = 
& 16 \frac{\left(2\pi\right)^2}{\hbar}\frac{G_{\vec{k}_1}}{G_{\vec{k}}}
\left(
\overline{N}_{\vec{k}}+1
\right)
\sum_{\vec{k}_2}\delta_{\vec{k}_1,\vec{k}-\vec{k_2}} 
\\
& 
\overline{N}_{\vec{k}_2}
\delta
\left(
\varepsilon_{\vec{k}}-\varepsilon_{\vec{k}_1}-\varepsilon_{\vec{k}_2}
\right)
\sum_{\vec{p}_1, \vec{p}_2} \delta_{\vec{p}_1 + \vec{p}_2, \vec{k}_1 + \vec{k}_2} \\
& 
\delta
\left(\varepsilon_{\vec{k}_1} + \varepsilon_{\vec{k}_2}-\varepsilon_{\vec{p}_1}-\varepsilon_{\vec{p}_2}
\right)
W_{\vec{k}_1, \vec{k}_2; \vec{p}_1, \vec{p}_2}  \\
& 
\left[
\left(
\overline{N}_{\vec{p}_1} + 1\right)
\left(
\overline{N}_{\vec{p}_2} + 1
\right)
+ \overline{N}_{\vec{p}_1} \overline{N}_{\vec{p}_2}
\right] \\
& 
\left[ \text{Re}\left(V^*_{\vec{k};\vec{k}_1, \vec{k}_2 }\right)\text{Im}\left(V^*_{\vec{k};\vec{p}_1, \vec{p}_2}\right)-
\right.
\\
& 
\left. 
\text{Im}\left(V^*_{\vec{k};\vec{k}_1, \vec{k}_2}\right)\text{Re}\left(V^*_{\vec{k};\vec{p}_1, \vec{p}_2 }\right)\right]
\end{align*}}
 & 
 \setlength{\abovedisplayskip}{0pt}
 \setlength{\belowdisplayskip}{0pt}
 {\begin{align*}
\hspace{0.0cm}
\mathcal{O}_{\vec{k}\vec{k}_2}^{+-} = 
& 16 \frac{\left(2\pi\right)^2}{\hbar}\frac{G_{\vec{k}_2}}{G_{\vec{k}}}
\left(
\overline{N}_{\vec{k}}+1
\right)
\sum_{\vec{k}_1}\delta_{\vec{k}_1,\vec{k}-\vec{k_2}} 
\\
& 
\overline{N}_{\vec{k}_1}
\delta
\left(
\varepsilon_{\vec{k}}-\varepsilon_{\vec{k}_1}-\varepsilon_{\vec{k}_2}
\right)
\sum_{\vec{p}_1, \vec{p}_2} \delta_{\vec{p}_1 + \vec{p}_2, \vec{k}_1 + \vec{k}_2} \\
& 
\delta
\left(\varepsilon_{\vec{k}_1} + \varepsilon_{\vec{k}_2}-\varepsilon_{\vec{p}_1}-\varepsilon_{\vec{p}_2}
\right)
W_{\vec{k}_1, \vec{k}_2; \vec{p}_1, \vec{p}_2}  \\
& 
\left[
\left(
\overline{N}_{\vec{p}_1} + 1\right)
\left(
\overline{N}_{\vec{p}_2} + 1
\right)
+ \overline{N}_{\vec{p}_1} \overline{N}_{\vec{p}_2}
\right] \\
& 
\left[ \text{Re}\left(V^*_{\vec{k};\vec{k}_1, \vec{k}_2}\right)\text{Im}\left(V^*_{\vec{k};\vec{p}_1, \vec{p}_2}\right)-
\right.
\\
& 
\left. 
\text{Im}\left(V^*_{\vec{k};\vec{k}_1, \vec{k}_2}\right)\text{Re}\left(V^*_{\vec{k};\vec{p}_1, \vec{p}_2}\right)\right]
\end{align*}}
\\
\hline 
\vspace{0.25 cm}
\begin{align*}
\vcenter{\hbox{
    \begin{tikzpicture}[scale=1.3, decoration={markings, 
        mark= at position 0.65 with {\arrow{stealth}},
        mark= at position 2cm with {\arrow{stealth}}}
        ]   
            \begin{scope}
            \draw[postaction={decorate}] (1,0.2)--(1.3,0.0);
            \draw[postaction={decorate}] (1,-0.2)--(1.3,0.0);
            \draw[postaction={decorate}] (1.3,0)--(1.7,0.0);
            \draw[fill=black] (1.3,0) circle (0.07);
            \node at (1.95,0) {$\times$};
            \node at (1.2, 0.4) {\footnotesize{$\vec{k}_1$}};
            \node at (1.2, -0.4) {\footnotesize{$\vec{k}_2$}};
            \node at (1.65, 0.2) {\footnotesize{$\vec{k}$}};
            \end{scope}
            \begin{scope}
                \draw[postaction={decorate}] (2.25, 0)--(2.6, 0);
                 \draw[fill=black] (2.67,0) circle (0.07);
                \draw[postaction={decorate}] (3.27, 0) to [in=70, out =110](2.67, 0); 
                 \draw[postaction={decorate}] (2.67, 0) to [in=-110, out =-70](3.27, 0);
                 \draw[fill=white] (3.27,0) circle (0.07);
                \draw[postaction={decorate}] (3.34, 0)--(3.69, 0);
                \draw[postaction = {decorate}] (2.67, 0.65) to [ out = 0, in=90] (3.27, 0.07); 
                \node at (2.4, 0.2) {\footnotesize{$\vec{k}_1$}};
                \node at (3.27, 0.65) {\footnotesize{$\vec{k}_2$}};
                \node at (3.5, 0.2) {\footnotesize{$\vec{k}$}};
                \node at (3.2, -0.3) {\footnotesize{$\vec{p}_2$}};
                \node at (2.8, 0.35)
                {\footnotesize{$\vec{p}_1$}};
            \end{scope}
    \end{tikzpicture}}}
\end{align*}    
& 
\setlength{\abovedisplayskip}{0pt}
 \setlength{\belowdisplayskip}{0pt}
{\begin{align*}
\hspace{0.0cm}
\mathcal{O}_{\vec{k}\vec{k}_1}^{+-} = 
& 16 \frac{\left(2\pi\right)^2}{\hbar}\frac{G_{\vec{k}_1}}{G_{\vec{k}}}\left(
\overline{N}_{\vec{k}}+1
\right)\sum_{\vec{k}_2}\delta_{\vec{k}_2,\vec{k}-\vec{k_1}}\\
&
\overline{N}_{\vec{k}_2}
\delta
\left(
\varepsilon_{\vec{k}}-\varepsilon_{\vec{k}_1}-\varepsilon_{\vec{k}_2}
\right) 
\sum_{\vec{p}_1, \vec{p}_2} \delta_{\vec{p}_1 + \vec{k}_1, \vec{p}_2}\\
& 
\delta
\left(\varepsilon_{\vec{k}_1} + \varepsilon_{\vec{p}_1}-\varepsilon_{\vec{p}_2}
\right)
W_{\vec{k}_1, \vec{k}_2; \vec{p}_1, \vec{p}_2} 
\overline{N}_{\vec{p}_1}
\\
& 
\left(
\overline{N}_{\vec{p}_2} + 1
\right)
\left[ \text{Re}\left(V^*_{\vec{k};\vec{k}_1, \vec{k}_2}\right)\text{Im}\left(V^*_{\vec{p}_2;\vec{p}_1, \vec{k}_1}\right)-
\right. 
\\
& 
\left. 
\text{Im}
\left(
V^*_{\vec{k};\vec{k}_1, \vec{k}_2}
\right)
\text{Re}
\left(
V^*_{\vec{p}_2;\vec{p}_1, \vec{k}_1}
\right)
\right]
\end{align*}}
& 
\setlength{\abovedisplayskip}{0pt}
 \setlength{\belowdisplayskip}{0pt}
{\begin{align*}
\hspace{0.0cm}
\mathcal{O}_{\vec{k}\vec{k}_2}^{+-} = 
& 16 \frac{\left(2\pi\right)^2}{\hbar}\frac{G_{\vec{k}_2}}{G_{\vec{k}}}\left(
\overline{N}_{\vec{k}}+1
\right)\sum_{\vec{k}_1}\delta_{\vec{k}_1,\vec{k}-\vec{k_2}} 
\\
&
\delta
\left(
\varepsilon_{\vec{k}}-\varepsilon_{\vec{k}_1}-\varepsilon_{\vec{k}_2}
\right) 
\overline{N}_{\vec{k}_1}
\sum_{\vec{p}_1, \vec{p}_2} \delta_{\vec{p}_1 + \vec{k}_1, \vec{p}_2}
\\
& 
\delta
\left(\varepsilon_{\vec{k}_1} + \varepsilon_{\vec{p}_1}-\varepsilon_{\vec{p}_2}
\right)
W_{\vec{k}_1, \vec{k}_2; \vec{p}_1, \vec{p}_2} 
\overline{N}_{\vec{p}_1}
\\
& 
\left(
\overline{N}_{\vec{p}_2} + 1
\right)
\left[ \text{Re}\left(V^*_{\vec{k}; \vec{k}_1, \vec{k}_2}\right)\text{Im}\left(V^*_{ \vec{p}_2;\vec{p}_1, \vec{k}_1}\right)-
\right. 
\\
& 
\left. 
\text{Im}
\left(
V^*_{\vec{k};\vec{k}_1, \vec{k}_2}
\right)
\text{Re}
\left(
V^*_{\vec{p}_2;\vec{p}_1, \vec{k}_1}
\right)
\right]
\end{align*}}
\\
\hline 
\vspace{0.25 cm}
\begin{align*}
\vcenter{\hbox{
\begin{tikzpicture}[scale=1.3, decoration={markings, 
        mark= at position 0.65 with {\arrow{stealth}},
        mark= at position 2cm with {\arrow{stealth}}}
        ]   
            \begin{scope}
            \draw[postaction={decorate}] (1,0.2)--(1.3,0.0);
            \draw[postaction={decorate}] (1,-0.2)--(1.3,0.0);
            \draw[postaction={decorate}] (1.3,0)--(1.7,0.0);
            \draw[fill=black] (1.3,0) circle (0.07);
            \node at (1.95,0) {$\times$};
            \node at (1.2, 0.4) {\footnotesize{$\vec{k}_1$}};
            \node at (1.2, -0.4) {\footnotesize{$\vec{k}_2$}};
            \node at (1.65, 0.2) {\footnotesize{$\vec{k}$}};
            \end{scope}
            \begin{scope}[xshift = 0.3cm]
                  \draw [postaction={decorate}] (2.9,0) to [out=110,in=70] (2.3,0);
                 \draw [postaction={decorate}] (2.3,0) to [out=-70,in=-110] (2.9,0);
                 \draw[postaction={decorate}] (1.9,0) -- (2.3,0);
                 \draw[fill=white] (2.3,0) circle (0.07);
                 \draw[fill=black] (2.9,0) circle (0.07);
                 \draw[postaction = {decorate}] (2.4, 0.65) to [ out = 0, in=90] (2.9, 0.07); 
                \draw[postaction = {decorate}]  (2.3, -0.07) to [out =-90, in = -180] (2.7, -0.6);
                  \node at (2.9, 0.6) {\footnotesize{$\vec{k}_1$}};   
                  \node at (2.7, -0.35) {\footnotesize{$\vec{p}_1$}}; 
                  \node at (2.6, 0.35) {\footnotesize{$\vec{p}_2$}};
                  \node at (2.1, 0.2) {\footnotesize{$\vec{k}_2$}}; 
                  \node at (2.2, -0.5) {\footnotesize{$\vec{k}$}}; 
            \end{scope}
    \end{tikzpicture}}}
\end{align*}
& 
\setlength{\abovedisplayskip}{0pt}
 \setlength{\belowdisplayskip}{0pt}
{\begin{align*}
\hspace{0.0cm}
\mathcal{O}_{\vec{k}\vec{k}_1}^{+-} = 
& 16 \frac{\left(2\pi\right)^2}{\hbar}\frac{G_{\vec{k}_1}}{G_{\vec{k}}}\left(
\overline{N}_{\vec{k}}+1
\right) \sum_{\vec{k}_2}\delta_{\vec{k}_2,\vec{k}-\vec{k_1}} \\
& 
\delta
\left(
\varepsilon_{\vec{k}}-\varepsilon_{\vec{k}_1}-\varepsilon_{\vec{k}_2}
\right) 
\overline{N}_{\vec{k}_2}
\sum_{\vec{p}_1, \vec{p}_2} \delta_{\vec{p}_1 + \vec{k}_1, \vec{p}_2} 
\\
& 
\delta
\left(\varepsilon_{\vec{k}_1} + \varepsilon_{\vec{p}_1}-\varepsilon_{\vec{p}_2} - \varepsilon_{\vec{k}_2}
\right)
W_{\vec{k}_1, \vec{k}_2; \vec{p}_1, \vec{p}_2} 
\overline{N}_{\vec{p}_2}
\\
& 
\left(
\overline{N}_{\vec{p}_1} + 1
\right)
\left[ \text{Re}\left(V^*_{\vec{k};\vec{k}_1, \vec{k}_2 }\right)\text{Im}\left(V^*_{\vec{p}_2;\vec{p}_1, \vec{k}_1}\right)-
\right. 
\\
& 
\left. 
\text{Im}
\left(
V^*_{\vec{k};\vec{k}_1, \vec{k}_2}
\right)
\text{Re}
\left(
V^*_{\vec{p}_2; \vec{p}_1, \vec{k}}
\right)
\right]
\end{align*}}
& 
\setlength{\abovedisplayskip}{0pt}
 \setlength{\belowdisplayskip}{0pt}
{\begin{align*}
\hspace{0.0cm}
\mathcal{O}_{\vec{k}\vec{k}_2}^{+-} = 
& 16 \frac{\left(2\pi\right)^2}{\hbar}\frac{G_{\vec{k}_2}}{G_{\vec{k}}} \left(
\overline{N}_{\vec{k}}+1 
\right)\sum_{\vec{k}_1}\delta_{\vec{k}_1,\vec{k}-\vec{k_2}}\\
& 
\delta
\left(
\varepsilon_{\vec{k}}-\varepsilon_{\vec{k}_1}-\varepsilon_{\vec{k}_2}
\right) \overline{N}_{\vec{k}_1}
\sum_{\vec{p}_1, \vec{p}_2} \delta_{\vec{p}_1 + \vec{k}_1, \vec{p}_2} \\
& 
\delta
\left(\varepsilon_{\vec{k}_1} + \varepsilon_{\vec{p}_1}-\varepsilon_{\vec{p}_2} - \varepsilon_{\vec{k}_2}
\right)
W_{\vec{k}_1, \vec{k}_2; \vec{p}_1, \vec{p}_2} 
\overline{N}_{\vec{p}_2}
\\
& 
\left(
\overline{N}_{\vec{p}_1} + 1
\right)
\left[ \text{Re}\left(V^*_{\vec{k}; \vec{k}_1, \vec{k}_2 }\right)\text{Im}\left(V^*_{ \vec{p}_2;\vec{p}_1, \vec{k}_1}\right)-
\right.
\\
& 
\left. 
\text{Im}
\left(
V^*_{\vec{k};\vec{k}_1, \vec{k}_2}
\right)
\text{Re}
\left(
V^*_{\vec{p}_2;\vec{p}_1, \vec{k}_1}
\right)
\right]
\end{align*}}
\end{tabular}
\end{ruledtabular}
\end{table}

\begin{enumerate}
    \item 
The two contributions, $\mathcal{O}_{\vec{k} \vec{k}_1}$ and $\mathcal{O}_{\vec{k} \vec{k}_2}$, stemming from the diagram 
\begin{equation}
\vcenter{\hbox{
    \begin{tikzpicture}[scale=1.3, decoration={markings, 
        mark= at position 0.65 with {\arrow{stealth}},
        mark= at position 2cm with {\arrow{stealth}}}
        ]   
        \begin{scope}
        \draw[postaction={decorate}] (1,0.2)--(1.3,0.0);
        \draw[postaction={decorate}] (1,-0.2)--(1.3,0.0);
        \draw[postaction={decorate}] (1.3,0)--(1.7,0.0);
        \draw[fill=black] (1.3,0) circle (0.07);
        \node at (1.2, 0.4) {\footnotesize{$\vec{k}_1$}};
        \node at (1.2, -0.4) {\footnotesize{$\vec{k}_2$}};
        \node at (1.65, 0.2) {\footnotesize{$\vec{k}$}};
        \node at (1.85, 0) {$\times$};
        \end{scope}
        \end{tikzpicture}}}
\left( 
    \vcenter{\hbox{
    \begin{tikzpicture}[scale=1.3, decoration={markings, 
        mark= at position 0.65 with {\arrow{stealth}},
        mark= at position 2cm with {\arrow{stealth}}}
        ]   
        \begin{scope}
        \draw [postaction={decorate}] (0.9,-1.2) to [out=70,in=110] (0.3,-1.2);
        \draw [postaction={decorate}] (0.9,-1.2) to [out=-70,in=-110] (0.3,-1.2);
        \draw[postaction={decorate}] (0.3,-1.2) -- (0.6,-1.2);
        \draw[fill=white] (0.3,-1.2) circle (0.07);
        \draw[fill=black] (0.9,-1.2) circle (0.07);
        \draw [postaction={decorate}] 
        (0.3, -0.7) to [out=0, in=90] (0.9, -1.2);
        \draw [postaction={decorate}] 
        (0.3, -1.7) to [out=0, in=-90] (0.9, -1.2);
        \node at (1, -0.8) {\footnotesize{$\vec{k}_1$}};
        \node at (1, -1.6)  {\footnotesize{$\vec{k}_2$}};
        \node at (0.7, -1.2) {\footnotesize{$\vec{k}$}};
        \node at (0.35, -0.9){\footnotesize{$\vec{p}_1$}};
        \node at (0.35, -1.5)
        {\footnotesize{$\vec{p}_2$}};
        \node at (1.2, -1.2) {$+$};
        \end{scope}
        \begin{scope}
        \draw[postaction={decorate}] (1.4,-1)--(1.7,-1.2);
        \draw[postaction={decorate}] (1.4,-1.4)--(1.7,-1.2);
        \draw [postaction={decorate}] (1.7,-1.2) to [out=70,in=110] (2.2,-1.2);
        \draw [postaction={decorate}] (1.7,-1.2) to [out=-70,in=-110] (2.2,-1.2);
        \draw[postaction={decorate}] (2.2,-1.2)--(2.7,-1.2);
        \draw[fill=white] (1.7,-1.2) circle (0.07);
        \draw[fill=black] (2.2,-1.2) circle (0.07);
        \node at (2.5, -1) {\footnotesize{$\vec{k}$}}; 
        \node at (2, -0.85) {\footnotesize{$\vec{p}_1$}}; 
        \node at (2, -1.55){\footnotesize{$\vec{p}_2$}};
        \node at (1.55, -0.8) {\footnotesize{$\vec{k}_1$}};
        \node at (1.55, -1.5) {\footnotesize{$\vec{k}_2$}};
        \end{scope}
    \label{diagram: symmetric in-fusion}
    \end{tikzpicture}}}
\right)  
\end{equation} 
are duplicates of each other, owing to the symmetric nature of the diagram. Indeed, flipping the role of $\vec{k}_1$ and $\vec{k}_2$ in Eq.~\eqref{diagram: symmetric in-fusion} leaves the diagram unchanged.

\item 
The symmetric in-split diagram with element  
\begin{equation}
    \mathcal{O}_{\vec{k} \vec{k}_2}^{\text{in-split}} = 
    \left[
    \vcenter{\hbox{
       \begin{tikzpicture}[scale=1.3, decoration={markings, 
        mark= at position 0.65 with {\arrow{stealth}},
        mark= at position 2cm with {\arrow{stealth}}}
        ]   
            \begin{scope}
            \draw[postaction={decorate}] (1.3,0.0)--(1.7, 0.2);
            \draw[postaction={decorate}] (1.3,0.0)--(1.7, -0.2);
            \draw[postaction={decorate}] (1,0)--(1.3,0);
            \draw[fill=black] (1.3,0) circle (0.07);
            \node at (1.45, 0.3) {\footnotesize{$\vec{k}_1$}};
            \node at (1.45, -0.3) {\footnotesize{$\vec{k}$}};
            \node at (0.8, 0.) {\footnotesize{$\vec{k}_2$}};
            \node at (1.8, 0) {$\times$};
            \end{scope}
        \end{tikzpicture}
            }}
            \left(
            \vcenter{\hbox{
            \begin{tikzpicture}[scale=1.3, decoration={markings, 
        mark= at position 0.65 with {\arrow{stealth}},
        mark= at position 2cm with {\arrow{stealth}}}
        ]   
            \begin{scope}[xshift=2.1cm, yshift=1.2cm]
                \begin{scope}
                  \draw [postaction={decorate}] (0.9,-1.2) to [out=70,in=110] (0.3,-1.2);
                  \draw [postaction={decorate}] (0.9,-1.2) to [out=-70,in=-110] (0.3,-1.2);
                  \draw[postaction={decorate}] (0.80,-0.9) to [out=0,in=90] (0.9,-1.2);
                  \draw[fill=white] (0.3,-1.2) circle (0.07);
                  \draw[fill=black] (0.9,-1.2) circle (0.07);
                  \draw [postaction={decorate}] 
                  (0.3, -1.27) to [out=-90, in=180] (0.7, -1.8);
                  \draw [postaction={decorate}] 
                  (0.3, -1.13) to [out=90, in=180] (0.7, -0.6);

                 \node at (0.2, -0.8)   {\footnotesize{$\vec{k}_1$}};
                 \node at (0.2, -1.6)   {\footnotesize{$\vec{k}$}};
                  \node at (1.1, -0.9) {\footnotesize{$\vec{k}_2$}};
                  \node at (0.6, -0.9) {\footnotesize{$\vec{p}_1$}};
                  \node at (0.6, -1.55)
                  {\footnotesize{$\vec{p}_2$}};
                   \node at (1.2, -1.2) {$+$};
                 \end{scope}
                 \begin{scope}
                   \draw[postaction={decorate}] (2.2,-1.2)--(2.5,-1);
                  \draw[postaction={decorate}] (2.2,-1.2)--(2.5,-1.4);
                  \draw [postaction={decorate}] (1.7,-1.2) to [out=70,in=110] (2.2,-1.2);
                  \draw [postaction={decorate}] (1.7,-1.2) to [out=-70,in=-110] (2.2,-1.2);
                  \draw[postaction={decorate}] (1.4,-1.2)--(1.7,-1.2);
                  \draw[fill=black] (1.7,-1.2) circle (0.07);
                  \draw[fill=white] (2.2,-1.2) circle (0.07);
                \node at (1.55, -1) {\footnotesize{$\vec{k}_2$}}; 
                \node at (1.9, -1.5) {\footnotesize{$\vec{p}_1$}}; 
                \node at (1.9, -0.9){\footnotesize{$\vec{p}_2$}};
                \node at (2.6, -0.9) {\footnotesize{$\vec{k}_1$}};
                \node at (2.6, -1.5) {\footnotesize{$\vec{k}$}};
                 \end{scope}
            \end{scope}      
    \end{tikzpicture}}}
    \right)
    \hspace{0.2 cm}
    \right]_{\vec{k} \vec{k}_2}
\label{diagram: bubble split in}
\end{equation}
yields a contribution to the Hall conductivity equal to the one of the symmetric in-fusion diagram. To see that, we recall the formula for the Hall conductivity for this particular contribution, 
\begin{equation}
\begin{split}
    \vec{\kappa}_{\text{H}}^{\text{in-split}}& = \frac{1}{2 k_{\text{B}}T^2} \sum_{\vec{k}, \vec{k}_2} \vec{v}_{\vec{k}} \times \vec{v}_{\vec{k}_2} \varepsilon_{\vec{k}} \varepsilon_{\vec{k}_2} \tau_{\vec{k}} \tau_{\vec{k}_2}G_{\vec{k}} G_{\vec{k}_2} \mathcal{A}_{\vec{k} \vec{k}_2} \\
     & = 
     \frac{1}{2} \frac{1}{2 k_{\text{B}}T^2} \sum_{\vec{k}, \vec{k}_2} \vec{v}_{\vec{k}} \times \vec{v}_{\vec{k}_2} \varepsilon_{\vec{k}} \varepsilon_{\vec{k}_2} \tau_{\vec{k}} \tau_{\vec{k}_2}G_{\vec{k}} G_{\vec{k}_2} \frac{\overline{N}_{\vec{k}_2} + \overline{N}_{\vec{k}} + 1}{\overline{N}_{\vec{k}_2}} \left( \mathrm{e}^{-\beta \varepsilon_{\vec{k}_2}}\mathcal{O}_{\vec{k} \vec{k}_2}^{\text{in-split}} - \mathcal{O}_{\vec{k} \vec{k}_2}^{\text{out-fusion}} \right) \\
     & = 
     \frac{1}{2 k_{\text{B}}T^2} \sum_{\vec{k}, \vec{k}_2} \vec{v}_{\vec{k}} \times \vec{v}_{\vec{k}_2} \varepsilon_{\vec{k}} \varepsilon_{\vec{k}_2} \tau_{\vec{k}} \tau_{\vec{k}_2} G_{\vec{k}} G_{\vec{k}_2} \frac{\overline{N}_{\vec{k}_2} + 1 +  \overline{N}_{\vec{k}}}{\overline{N}_{\vec{k}_2}} \mathrm{e}^{-\beta \varepsilon_{\vec{k}_2}}\mathcal{O}_{\vec{k} \vec{k}_2}^{\text{in-split}}, \\
\end{split}
\label{eq: KH split-in bubble}
\end{equation}
where in the last step we utilized the result of Sec.~\ref{subsection: DB}
stating that for the Hall effect only the part of the process breaking the detailed balance relation by a minus sign, $\mathcal{O}_{\vec{k} \vec{k}_2}^{\text{in-split}} = -\mathrm{e}^{\beta \varepsilon_{\vec{k}_2}} \mathcal{O}_{\vec{k} \vec{k}_2}^{\text{out-fusion}}$, results in a finite contribution. 
Flipping the role of the two momenta, we see that Eq.~\eqref{diagram: bubble split in} transforms into the following out-split term
\begin{equation}
\begin{split}
\mathcal{O}_{\vec{k} \vec{k}_2}^{\text{in-split}}
& \stackrel{\vec{k} \leftrightarrow \vec{k}_2}{=}
\mathcal{O}_{\vec{k}_2 \vec{k}}^{\text{in-split}} = 
\frac{G_{\vec{k}}}{G_{\vec{k}_2}} O_{\vec{k}_2 \vec{k}} ^{\text{In-split}} =
\frac{G_{\vec{k}}}{G_{\vec{k}_2}} \left( \overline{N}_{\vec{k}_2} + 1\right)Q_{\vec{k}_2 \vec{k}} ^{\text{in-split}} = 
\frac{G_{\vec{k}}}{G_{\vec{k}_2}} \frac{\left( \overline{N}_{\vec{k}_2} + 1\right)}{\overline{N}_{\vec{k}}}\overline{N}_{\vec{k}}Q_{\vec{k} \vec{k}_2} ^{\text{out-split}}
\\
& = 
\frac{\left( \overline{N}_{\vec{k}_2} + 1\right)}{\overline{N}_{\vec{k}}}
\left(\frac{G_{\vec{k}}}{G_{\vec{k}_2}}\right)^2 
\mathcal{O}_{\vec{k} \vec{k}_2}^{\text{out-split}} = 
\left(\frac{G_{\vec{k}}}{G_{\vec{k}_2}}\right)^2 \frac{\left( \overline{N}_{\vec{k}_2} + 1\right)}{\overline{N}_{\vec{k}}}
\left[
\hspace{-0.1cm}
\vcenter{\hbox{
    \begin{tikzpicture}[scale=1.3, decoration={markings, 
        mark= at position 0.65 with {\arrow{stealth}},
        mark= at position 2cm with {\arrow{stealth}}}
        ]   
        \begin{scope}
           \draw[postaction={decorate}] (1.3,0.0)--(1.7, 0.2);
           \draw[postaction={decorate}] (1.3,0.0)--(1.7, -0.2);
            \draw[postaction={decorate}] (1,0)--(1.3,0);
            \draw[fill=black] (1.3,0) circle (0.07);
            \node at (1.45, 0.3) {\footnotesize{$\vec{k}_1$}};
            \node at (1.45, -0.3) {\footnotesize{$\vec{k}_2$}};
            \node at (0.8, 0.) {\footnotesize{$\vec{k}$}};
            \node at (1.8, 0) {$+$}; 
            \end{scope} 
    \end{tikzpicture}
            }}
    \left(
            \vcenter{\hbox{
    \begin{tikzpicture}[scale=1.3, decoration={markings, 
        mark= at position 0.65 with {\arrow{stealth}},
        mark= at position 2cm with {\arrow{stealth}}}
        ]   
            \begin{scope}[xshift=2.1cm, yshift=1.2cm]
                \begin{scope}
                  \draw [postaction={decorate}] (0.9,-1.2) to [out=70,in=110] (0.3,-1.2);
                  \draw [postaction={decorate}] (0.9,-1.2) to [out=-70,in=-110] (0.3,-1.2);
                  \draw[postaction={decorate}] (0.80,-0.9) to [out=0,in=90] (0.9,-1.2);
                  \draw[fill=white] (0.3,-1.2) circle (0.07);
                  \draw[fill=black] (0.9,-1.2) circle (0.07);
                  \draw [postaction={decorate}] 
                  (0.3, -1.27) to [out=-90, in=180] (0.7, -1.8);
                  \draw [postaction={decorate}] 
                  (0.3, -1.13) to [out=90, in=180] (0.7, -0.6);
                 \node at (0.2, -0.8){\footnotesize{$\vec{k}_1$}};
                 \node at (0.2, -1.6){\footnotesize{$\vec{k}_2$}};
                  \node at (1.1, -0.9){\footnotesize{$\vec{k}$}};
                  \node at (0.6, -0.9){\footnotesize{$\vec{p}_1$}};
                  \node at (0.6, -1.55)
                  {\footnotesize{$\vec{p}_2$}};
                   \node at (1.2, -1.2) {$+$};
                 \end{scope}
                 \begin{scope}
                   \draw[postaction={decorate}] (2.2,-1.2)--(2.5,-1);
                  \draw[postaction={decorate}] (2.2,-1.2)--(2.5,-1.4);
                  \draw [postaction={decorate}] (1.7,-1.2) to [out=70,in=110] (2.2,-1.2);
                  \draw [postaction={decorate}] (1.7,-1.2) to [out=-70,in=-110] (2.2,-1.2);
                  \draw[postaction={decorate}] (1.4,-1.2)--(1.7,-1.2);
                  \draw[fill=black] (1.7,-1.2) circle (0.07);
                  \draw[fill=white] (2.2,-1.2) circle (0.07);
                \node at (1.55, -1) {\footnotesize{$\vec{k}$}}; 
                \node at (1.9, -1.5) {\footnotesize{$\vec{p}_1$}}; 
                \node at (1.9, -0.9){\footnotesize{$\vec{p}_2$}};
                \node at (2.6, -0.9) {\footnotesize{$\vec{k}_1$}};
                \node at (2.6, -1.5) {\footnotesize{$\vec{k}_2$}};
                 \end{scope}
            \end{scope}     
    \end{tikzpicture}}}
    \right)
    \hspace{0.2 cm}
    \right]_{\vec{k} \vec{k}_2}.
\end{split}
    \label{eq: swapping momenta for split-in}
\end{equation}
Making again use of the broken detailed balance relation we can connect the out-split diagram in Eq.~\eqref{eq: swapping momenta for split-in} to its in-counterpart,
\begin{equation}
    \mathcal{O}_{\vec{k} \vec{k}_2}^{\text{out-split}} = - \mathrm{e} ^ {-\beta \varepsilon_{\vec{k}_2}} \mathcal{O}_{\vec{k} \vec{k}_2}^{\text{in-fusion}}, 
\end{equation}
or, in a diagrammatic language,
\begin{equation}
    \left[
\hspace{-0.1cm}
\vcenter{\hbox{
    \begin{tikzpicture}[scale=1.3, decoration={markings, 
        mark= at position 0.65 with {\arrow{stealth}},
        mark= at position 2cm with {\arrow{stealth}}}
        ]   
        \begin{scope}
           \draw[postaction={decorate}] (1.3,0.0)--(1.7, 0.2);
           \draw[postaction={decorate}] (1.3,0.0)--(1.7, -0.2);
            \draw[postaction={decorate}] (1,0)--(1.3,0);
            \draw[fill=black] (1.3,0) circle (0.07);
            \node at (1.45, 0.3) {\footnotesize{$\vec{k}_1$}};
            \node at (1.45, -0.3) {\footnotesize{$\vec{k}$}};
            \node at (0.8, 0.) {\footnotesize{$\vec{k}_2$}};
            \node at (1.8, 0) {$\times$}; 
            \end{scope} 
    \end{tikzpicture}
            }}
    \left(
            \vcenter{\hbox{
    \begin{tikzpicture}[scale=1.3, decoration={markings, 
        mark= at position 0.65 with {\arrow{stealth}},
        mark= at position 2cm with {\arrow{stealth}}}
        ]   
            \begin{scope}[xshift=2.1cm, yshift=1.2cm]
                \begin{scope}
                  \draw [postaction={decorate}] (0.9,-1.2) to [out=70,in=110] (0.3,-1.2);
                  \draw [postaction={decorate}] (0.9,-1.2) to [out=-70,in=-110] (0.3,-1.2);
                  \draw[postaction={decorate}] (0.80,-0.9) to [out=0,in=90] (0.9,-1.2);
                  \draw[fill=white] (0.3,-1.2) circle (0.07);
                  \draw[fill=black] (0.9,-1.2) circle (0.07);
                  \draw [postaction={decorate}] 
                  (0.3, -1.27) to [out=-90, in=180] (0.7, -1.8);
                  \draw [postaction={decorate}] 
                  (0.3, -1.13) to [out=90, in=180] (0.7, -0.6);
                 \node at (0.2, -0.8){\footnotesize{$\vec{k}_1$}};
                 \node at (0.2, -1.6){\footnotesize{$\vec{k}$}};
                  \node at (1.1, -0.9){\footnotesize{$\vec{k}_2$}};
                  \node at (0.6, -0.9){\footnotesize{$\vec{p}_1$}};
                  \node at (0.6, -1.55)
                  {\footnotesize{$\vec{p}_2$}};
                   \node at (1.2, -1.2) {$+$};
                 \end{scope}
                 \begin{scope}
                   \draw[postaction={decorate}] (2.2,-1.2)--(2.5,-1);
                  \draw[postaction={decorate}] (2.2,-1.2)--(2.5,-1.4);
                  \draw [postaction={decorate}] (1.7,-1.2) to [out=70,in=110] (2.2,-1.2);
                  \draw [postaction={decorate}] (1.7,-1.2) to [out=-70,in=-110] (2.2,-1.2);
                  \draw[postaction={decorate}] (1.4,-1.2)--(1.7,-1.2);
                  \draw[fill=black] (1.7,-1.2) circle (0.07);
                  \draw[fill=white] (2.2,-1.2) circle (0.07);
                \node at (1.55, -1) {\footnotesize{$\vec{k}_2$}}; 
                \node at (1.9, -1.5) {\footnotesize{$\vec{p}_1$}}; 
                \node at (1.9, -0.9){\footnotesize{$\vec{p}_2$}};
                \node at (2.6, -0.9) {\footnotesize{$\vec{k}_1$}};
                \node at (2.6, -1.5) {\footnotesize{$\vec{k}$}};
                 \end{scope}
            \end{scope}     
    \end{tikzpicture}}}
    \right)
    \hspace{0.2 cm}
    \right]_{\vec{k} \vec{k}_2}
    = 
    - \mathrm{e} ^ {-\beta \varepsilon_{\vec{k}_2}} 
    \left[
    \vcenter{\hbox{
    \begin{tikzpicture}[scale=1.3, decoration={markings, 
        mark= at position 0.65 with {\arrow{stealth}},
        mark= at position 2cm with {\arrow{stealth}}}
        ]   
        \begin{scope}
        \draw[postaction={decorate}] (1,0.2)--(1.3,0.0);
        \draw[postaction={decorate}] (1,-0.2)--(1.3,0.0);
        \draw[postaction={decorate}] (1.3,0)--(1.7,0.0);
        \draw[fill=black] (1.3,0) circle (0.07);
        \node at (1.2, 0.4) {\footnotesize{$\vec{k}_1$}};
        \node at (1.2, -0.4) {\footnotesize{$\vec{k}_2$}};
        \node at (1.65, 0.2) {\footnotesize{$\vec{k}$}};
        \node at (1.85, 0) {$\times$};
        \end{scope}
        \end{tikzpicture}}}
\left( 
    \vcenter{\hbox{
    \begin{tikzpicture}[scale=1.3, decoration={markings, 
        mark= at position 0.65 with {\arrow{stealth}},
        mark= at position 2cm with {\arrow{stealth}}}
        ]   
        \begin{scope}
        \draw [postaction={decorate}] (0.9,-1.2) to [out=70,in=110] (0.3,-1.2);
        \draw [postaction={decorate}] (0.9,-1.2) to [out=-70,in=-110] (0.3,-1.2);
        \draw[postaction={decorate}] (0.3,-1.2) -- (0.6,-1.2);
        \draw[fill=white] (0.3,-1.2) circle (0.07);
        \draw[fill=black] (0.9,-1.2) circle (0.07);
        \draw [postaction={decorate}] 
        (0.3, -0.7) to [out=0, in=90] (0.9, -1.2);
        \draw [postaction={decorate}] 
        (0.3, -1.7) to [out=0, in=-90] (0.9, -1.2);
        \node at (1, -0.8) {\footnotesize{$\vec{k}_1$}};
        \node at (1, -1.6)  {\footnotesize{$\vec{k}_2$}};
        \node at (0.7, -1.2) {\footnotesize{$\vec{k}$}};
        \node at (0.35, -0.9){\footnotesize{$\vec{p}_1$}};
        \node at (0.35, -1.5)
        {\footnotesize{$\vec{p}_2$}};
        \node at (1.2, -1.2) {$+$};
        \end{scope}
        \begin{scope}
        \draw[postaction={decorate}] (1.4,-1)--(1.7,-1.2);
        \draw[postaction={decorate}] (1.4,-1.4)--(1.7,-1.2);
        \draw [postaction={decorate}] (1.7,-1.2) to [out=70,in=110] (2.2,-1.2);
        \draw [postaction={decorate}] (1.7,-1.2) to [out=-70,in=-110] (2.2,-1.2);
        \draw[postaction={decorate}] (2.2,-1.2)--(2.7,-1.2);
        \draw[fill=white] (1.7,-1.2) circle (0.07);
        \draw[fill=black] (2.2,-1.2) circle (0.07);
        \node at (2.5, -1) {\footnotesize{$\vec{k}$}}; 
        \node at (2, -0.85) {\footnotesize{$\vec{p}_1$}}; 
        \node at (2, -1.55){\footnotesize{$\vec{p}_2$}};
        \node at (1.55, -0.8) {\footnotesize{$\vec{k}_1$}};
        \node at (1.55, -1.5) {\footnotesize{$\vec{k}_2$}};
        \end{scope}
    \end{tikzpicture}}}
\right) 
\hspace{0.2 cm}
\right]_{\vec{k} \vec{k}_2}. 
\end{equation}
Consequently, upon swapping $\vec{k} \leftrightarrow \vec{k}_2$, the Hall conductivity in Eq.~\eqref{eq: KH split-in bubble} takes the form  
\begin{equation}
    \begin{split}
        \vec{\kappa}_{\text{H}}^{\text{in-split}} & = 
        \frac{1}{2 k_{\text{B}}T^2} \sum_{\vec{k}, \vec{k}_2} \vec{v}_{\vec{k}} \times \vec{v}_{\vec{k}_2} \varepsilon_{\vec{k}} \varepsilon_{\vec{k}_2} \tau_{\vec{k}} \tau_{\vec{k}_2}G_{\vec{k}} G_{\vec{k}_2} \frac{\overline{N}_{\vec{k}_2} + \overline{N}_{\vec{k}} + 1}{\overline{N}_{\vec{k}_2}} \mathrm{e}^{-\beta \varepsilon_{\vec{k}_2}}\mathcal{O}_{\vec{k} \vec{k}_2}^{\text{in-split}}\\
        & \stackrel{\vec{k} \leftrightarrow \vec{k}_2}{=}
        \frac{1}{2 k_{\text{B}}T^2} \sum_{\vec{k}, \vec{k}_2} \vec{v}_{\vec{k}_2} \times \vec{v}_{\vec{k}} \varepsilon_{\vec{k}} \varepsilon_{\vec{k}_2} \tau_{\vec{k}} \tau_{\vec{k}_2}G_{\vec{k}} G_{\vec{k}_2} \frac{\overline{N}_{\vec{k}_2} + \overline{N}_{\vec{k}} + 1}{\overline{N}_{\vec{k}}} \mathrm{e}^{-\beta \varepsilon_{\vec{k}}}\mathcal{O}_{\vec{k}_2 \vec{k}}^{\text{in-split}} \\
        & = -\frac{1}{2 k_{\text{B}}T^2} \sum_{\vec{k}, \vec{k}_2} \vec{v}_{\vec{k}} \times \vec{v}_{\vec{k}_2} \varepsilon_{\vec{k}} \varepsilon_{\vec{k}_2} \tau_{\vec{k}} \tau_{\vec{k}_2}G_{\vec{k}} G_{\vec{k}_2} \left(\frac{G_{\vec{k}}}{G_{\vec{k}_2}} \right)^2 \frac{1 + \overline{N}_{\vec{k}_2}}{\overline{N}_{\vec{k}}} \frac{\overline{N}_{\vec{k}_2} + \overline{N}_{\vec{k}} + 1}{\overline{N}_{\vec{k}}} \mathrm{e}^{-\beta \varepsilon_{\vec{k}}}\mathcal{O}_{\vec{k} \vec{k}_2}^{\text{out-split}} \\
        & = \frac{1}{2 k_{\text{B}}T^2} \sum_{\vec{k}, \vec{k}_2} \vec{v}_{\vec{k}} \times \vec{v}_{\vec{k}_2} \varepsilon_{\vec{k}} \varepsilon_{\vec{k}_2} \tau_{\vec{k}} \tau_{\vec{k}_2}G_{\vec{k}} G_{\vec{k}_2} \left(\frac{G_{\vec{k}}}{G_{\vec{k}_2}} \right)^2 \frac{1 + \overline{N}_{\vec{k}_2}}{\overline{N}_{\vec{k}}} \mathrm{e}^{-\beta \varepsilon_{\vec{k}_2}}\mathrm{e}^{-\beta \varepsilon_{\vec{k}}}\frac{\overline{N}_{\vec{k}} + \overline{N}_{\vec{k}_2} + 1}{\overline{N}_{\vec{k}}} \mathcal{O}_{\vec{k} \vec{k}_2}^{\text{in-fusion}} \\
        & = 
        \frac{1}{2 k_{\text{B}}T^2} \sum_{\vec{k}, \vec{k}_2} \vec{v}_{\vec{k}} \times \vec{v}_{\vec{k}_2} \varepsilon_{\vec{k}} \varepsilon_{\vec{k}_2} \tau_{\vec{k}} \tau_{\vec{k}_2}G_{\vec{k}} G_{\vec{k}_2} \mathrm{e}^{-\beta \varepsilon_{\vec{k}_2}}\frac{\overline{N}_{\vec{k}} + \overline{N}_{\vec{k}_2} + 1}{\overline{N}_{\vec{k}}} \mathcal{O}_{\vec{k} \vec{k}_2}^{\text{in-fusion}} \\
        & = 
        \vec{\kappa}^{\text{in-fusion}}_{\text{H}}.
    \end{split}
\end{equation}
We have thus proven that $\mathcal{O}_{\vec{k} \vec{k}_2}^{\text{in-split}}$ and $\mathcal{O}_{\vec{k} \vec{k}_2}^{\text{in-fusion}}$ provide the same contribution to the Hall conductivity.

\item There are some additional three-magnon diagrams that have not been explicitly presented in Tab.~\ref{tab:in split} and Tab.~\ref{tab:In fusion}. These diagrams are ``tadpole-like'' and ``Hartree-like'' diagrams and, as we will briefly discuss, they are identically zero. Since there are a large number of both, we will discuss only one example of each case. 

We begin with the following ``Hartree-like'' interference diagram:
\begin{equation}
\vcenter{\hbox{
    \begin{tikzpicture}
        [scale=1.3, decoration={markings, 
        mark= at position 0.65 with {\arrow{stealth}},
        mark= at position 2cm with {\arrow{stealth}}}
        ]   
            \begin{scope}
           \draw[postaction={decorate}] (1,0.2)--(1.3,0.0);
            \draw[postaction={decorate}] (1,-0.2)--(1.3,0.0);
            \draw[postaction={decorate}] (1.3,0)--(1.7,0.0);
            \draw[fill=black] (1.3,0) circle (0.07);
            \node at (1.2, 0.4) {\footnotesize{$\vec{k}_1$}};
            \node at (1.2, -0.4) {\footnotesize{$\vec{k}_2$}};
            \node at (1.65, 0.2) {\footnotesize{$\vec{k}$}};
            \node at (1.9, 0) {$\times$}; 
            \draw[postaction={decorate}] (2.2, 0.2)--(2.5, 0); 
            \draw[postaction={decorate}] (2.2, -0.2)--(2.5, 0); 
            \draw[postaction={decorate}]  (2.5, 0) -- (2.9,0); 
            \draw [postaction={decorate}] (2.9, 0) -- (3.3, 0); 
            \draw[postaction={decorate}] (2.9,0.47) to[out=180, in=130] (2.9, 0.07);
            \draw[postaction={decorate}] (2.9, 0.07) to[out=50,in=0] (2.9,0.47);
            \draw[fill=white] (2.9, 0) circle (0.07); 
            \draw[fill=black] (2.5, 0) circle (0.07); 
            \node at (3.2, 0.2) {\footnotesize{$\vec{k}$}} ; 
            \node at (2.7, -0.2) {\footnotesize{$\vec{p}_1$}};
            \node at (3.1, 0.55) {\footnotesize{$\vec{p}_2$}};
            \node at (2.2, 0.4) {\footnotesize{$\vec{k}_2$}};
             \node at (2.2, -0.4) {\footnotesize{$\vec{k}_1$}};
            \end{scope}
    \end{tikzpicture}}}
.
\label{diagr: Hartree}
\end{equation}
Using the dictionary developed in Sec.~\ref{subsection: lexicon}, we get the off-diagonal term contributing to the Hall conductivity to be 
\begin{equation}
\begin{split}
      \mathcal{O}_{\vec{k}\vec{k}_1} &=  
      16 \frac{\left( 2 \pi\right)^2}{\hbar}
      \frac{G_{\vec{k}_1}}{G_{\vec{k}}} \left( \overline{N}_{\vec{k}} + 1 \right) \sum_{\vec{k}_2} \delta_{\vec{k}, \vec{k}_1 + \vec{k}_2}\overline{N}_{\vec{k}_1} \delta\left( \varepsilon_{\vec{k}_1} + \varepsilon_{\vec{k}_2} - \varepsilon_{\vec{k}}\right) 
      \sum_{\vec{p}_1, \vec{p}_2} 
      \delta_{\vec{k}, \vec{p}_1} W_{\vec{k}, \vec{p}_2; \vec{p}_2, \vec{p}_1} \delta\left( \varepsilon_{\vec{p}_1} - \varepsilon_{\vec{k}}\right) \\
      &\quad \times \left[
    \text{Re} \left( V_{\vec{k};\vec{k}_1,\vec{k}_2}^{*}\right)  \text{Im} \left( V_{\vec{p}_1;\vec{k}_1,\vec{k}_2}^{*}\right) - 
      \text{Re} \left( V_{\vec{p}_1;\vec{k}_1,\vec{k}_2}^{*}\right)  \text{Im} \left( V_{\vec{k}; \vec{k}_1,\vec{k}_2}^{*}\right)
    \right] \left( \overline{N}_{\vec{p}_1} + 1\right)\overline{N}_{\vec{p}_2}
    \\
    &= 0,
\end{split}
\end{equation}
where we made use of the fact that $V_{\vec{k};\vec{k}_1,\vec{k}_2 }^{*} = V_{\vec{p}_1; \vec{k}_1,\vec{k}_2}^{*}$. The other possible contribution, $\mathcal{O}_{\vec{k} \vec{k}_2}$, is equal to $\mathcal{O}_{\vec{k} \vec{k}_1}$ due to the symmetric nature of the diagram. It can be shown that all the ``Hartree-like'' diagrams contain a vertex function of the form $\text{Re}\left( V\right)\text{Im}\left(V \right) - \text{Re}\left( V \right)\text{Im}\left( V\right)$, which vanishes identically. Consequently, they are all zero. 

Finally, we consider the ``tadpole-like'' diagrams, a representative example of which is the following in-fusion diagram:
\begin{equation}
    \vcenter{\hbox{
    \begin{tikzpicture} [scale=1.3, decoration={markings, 
        mark= at position 0.65 with {\arrow{stealth}},
        mark= at position 2cm with {\arrow{stealth}}}
        ] 
        \begin{scope}[xshift = -1.75cm]
         \draw[postaction={decorate}] (1,0.2)--(1.3,0.0);
            \draw[postaction={decorate}] (1,-0.2)--(1.3,0.0);
            \draw[postaction={decorate}] (1.3,0)--(1.7,0.0);
            \draw[fill=black] (1.3,0) circle (0.07);
            \node at (1.2, 0.4) {\footnotesize{$\vec{k}_1$}};
            \node at (1.2, -0.4) {\footnotesize{$\vec{k}_2$}};
            \node at (1.65, 0.2) {\footnotesize{$\vec{k}$}};
            \node at (1.9, 0) {$\times$}; 
        \end{scope}

        \begin{scope}
        \draw[postaction={decorate}] (1.3,0.47) to[out=180, in=130] (1.3, 0.07);
        \draw[postaction={decorate}] (1.3, 0.07) to[out=50,in=0] (1.3,0.47);
        \draw[fill=black] (1.3, 0) circle (0.07); 
        \draw[postaction={decorate}] (0.4,0.2)--(0.7,0.0);
        \draw[postaction={decorate}] (0.4,-0.2)--(0.7,0.0);
        \draw[postaction={decorate}] (0.7, 0) -- (1.3, 0);
        \draw[postaction={decorate}] (0.7, 0) to [out = 90, in=180] (1, 0.3) ;
        \draw [fill=white] (0.7, 0) circle (0.07); 
        \node at (0.5, 0.35) {\footnotesize{$\vec{k}_2$}}; 
        \node at (0.5, -0.35) {\footnotesize{$\vec{k}_1$}};
        \node at (0.95, 0.45) {\footnotesize{$\vec{k}$}};
        \node at (1.1, -0.2) {\footnotesize{$\vec{p}_1$}};
         \node at (1.6, 0.4) {\footnotesize{$\vec{p}_2$}};
        \end{scope}
    \end{tikzpicture}}}
    .
\end{equation}
Translating the diagram into mathematical expressions, we get for the $\mathcal{O}_{\vec{k} \vec{k}_1}$ contribution
\begin{equation}
    \begin{split}
        \mathcal{O}_{\vec{k} \vec{k}_1} 
        &= 16 \frac{\left( 2 \pi \right)^2}{\hbar} \frac{G_{\vec{k}_1}}{G_{\vec{k}}} \left( \overline{N}_{\vec{k}} + 1\right) 
        \sum_{\vec{k}_2}\delta_{\vec{k}_2, \vec{k}_1 + \vec{k}} \delta\left( \varepsilon_{\vec{k}_1} + \varepsilon_{\vec{k}_2} - \varepsilon_{\vec{k}}\right) \sum_{\vec{p}_1 \vec{p}_2} \delta \left(
        \varepsilon_{\vec{p}_1} + \varepsilon_{\vec{k}}- \varepsilon_{\vec{k}_1}-\varepsilon_{\vec{k}_2}\right)\delta_{\vec{p}_1, 0} W_{\vec{k}_1, \vec{k}_2; \vec{k}, \vec{p}_1}\left( \overline{N}_{\vec{p}_1} + 1\right)\overline{N}_{\vec{p}_2} \\
        & 
        \quad \times \left[ 
        \text{Re}\left( V_{\vec{k};\vec{k}_1, \vec{k}_2}^{*}\right) \text{Im}\left( V_{\vec{p}_2;\vec{p}_1, \vec{p}_2}^{*} \right) -  \text{Re}\left(V_{\vec{p}_2;\vec{p}_1, \vec{p}_2}^{*}\right) \text{Im}\left( V_{\vec{k};\vec{k}_1, \vec{k}_2}^{*}\right)
        \right] \\
        & = 
        A\frac{\left( 2 \pi \right)^2}{\hbar} \frac{G_{\vec{k}_1}}{G_{\vec{k}}} \left( \overline{N}_{\vec{k}} + 1\right) 
        \sum_{\vec{k}_2}\delta_{\vec{k}_2, \vec{k}_1 + \vec{k}} \delta\left( \varepsilon_{\vec{k}_1} + \varepsilon_{\vec{k}_2} - \varepsilon_{\vec{k}}\right) 
        \left( \overline{N}_{\vec{p}_1=0} + 1\right) \delta \left(
        \varepsilon_{\vec{p}_1=0} + \varepsilon_{\vec{k}}- \varepsilon_{\vec{k}_1}-\varepsilon_{\vec{k}_2}\right)W_{\vec{k}_1, \vec{k}_2; \vec{k}, \vec{p}_1 =0}\\
        & 
        \quad \times \sum_{\vec{p}_2} 
        \overline{N}_{\vec{p}_2}
        \left[
        \sin p_{2}^{x} \left( \sin k_2^y + \sin k_1^y\right) - \sin p_{2}^{y} \left( \sin k_2^x + \sin k_1^x\right)
        \right] 
        \\
        &= 0 ,
    \end{split}
\end{equation}
where we noticed that the sum over $\vec{p}_2$ must vanish, because $\overline{N}_{\vec{p}_2} \left[\sin p_{2}^{x} \left( \sin k_2^y + \sin k_1^y\right) - \sin p_{2}^{y} \left( \sin k_2^x + \sin k_1^x\right)\right]$ is odd in $\vec{p}_2$. Alternatively, one can make use of the phase-space restrictions imposed by $\delta \left(
\varepsilon_{\vec{p}_1=0} + \varepsilon_{\vec{k}}- \varepsilon_{\vec{k}_1}-\varepsilon_{\vec{k}_2}\right)$ and $\delta\left( \varepsilon_{\vec{k}_1} + \varepsilon_{\vec{k}_2} - \varepsilon_{\vec{k}}\right)$ to show that the contribution of the diagram is zero. Indeed, the combination of the two delta-functions leads to $\varepsilon_{\vec{p}_1=0} = 0 \Rightarrow \Delta = 0$, which is not feasible, since our theory holds only for the field polarized state of the chiral magnet where $\Delta > 0.84 S D^2 / J$. Finally, the other contribution, $\mathcal{O}_{\vec{k} \vec{k}_2}$, as well as any off-diagonal element coming from any other tadpole diagram, can be shown to be zero for the same reason. 

\item We point out that every diagram shown in the in-split (Tab.~\ref{tab:in split}) and in-fusion tables (Tab.~\ref{tab:In fusion}) could  also be considered with the magnon momenta $\vec{k}_1$ and $\vec{k}_2$ swapped. This would result in duplicate processes for each contribution, since we are calculating both $\mathcal{O}_{\vec{k} \vec{k}_2}$ and $\mathcal{O}_{\vec{k} \vec{k}_1}$ for each diagram. Thus, such diagrams should not be taken into account. 
\end{enumerate}

\subsubsection{Four-magnon diagrams}
We consider the four-magnon diagrams in Tab.~\ref{tab:4-magnons SM}. The following observations can be made:

\begin{table}
\caption{Four-magnon scattering-in interference terms. Two magnons with momenta $\vec{k}_1$ and $\vec{k}_2$ scatter into two other magnons with momenta $\vec{k}$ and $\vec{k}_3$. This process can occur at first order due to the four-magnon vertex, and at second order due to two subsequent three-magnon vertices. The interference of these contributions leads to an off-diagonal part, $\mathcal{O}_{\vec{k}\vec{k}_1}$, $\mathcal{O}_{\vec{k}\vec{k}_2}$, and $\mathcal{O}_{\vec{k}\vec{k}_3}$, of the collision matrix, which gives rise to a thermal Hall effect. The contribution $\mathcal{O}_{\vec{k}\vec{k}_2}$ is not explicitly given because it is related to the other contributions (see text for details). \label{tab:4-magnons SM}}
\begin{ruledtabular}
\begin{tabular}
{p{4.3cm} p{6.0cm} p{6.0cm}}
\hspace{1.2 cm} Diagram 
& \hspace{2.6 cm} $\mathcal{O}_{\vec{k}\vec{k}_1}$
& \hspace{2.6 cm} $\mathcal{O}_{\vec{k}\vec{k}_3}$ \\
\hline 

\begin{align*}
\adjustbox{trim=0 0 -1cm 0}{
\begin{tikzpicture}[scale=1.3, decoration={markings, 
        mark= at position 0.65 with {\arrow{stealth}},
        mark= at position 2cm with {\arrow{stealth}}}
        ]   
        \begin{scope}
            \draw[postaction={decorate}] (1, 0.3) -- (1.3, 0); 
            \draw[postaction={decorate}] (1, -0.3) -- (1.3, 0);  
            \draw[postaction={decorate}] (1.3, 0) -- (1.6, 0.3); 
             \draw[postaction={decorate}] (1.3, 0) -- (1.6, -0.3);
             \draw[fill=white] (1.3,0) circle (0.07);
             \node at (1, 0.45) {\footnotesize{$\vec{k}_1$}}; 
              \node at (1, -0.45) {\footnotesize{$\vec{k}_2$}}; 
              \node at (1.6, 0.45) {\footnotesize{$\vec{k}_3$}}; 
               \node at (1.6, -0.45) {\footnotesize{$\vec{k}$}}; 
            \node at (1.8, 0) {$\times$}; 
        \end{scope}
\end{tikzpicture}
}
\\
\left(
\vcenter{\hbox{
\begin{tikzpicture}[scale=1.3, decoration={markings, 
        mark= at position 0.65 with {\arrow{stealth}},
        mark= at position 2cm with {\arrow{stealth}}}
        ]   
      \begin{scope}
            \draw[postaction={decorate}] (0.5, -0.8) -- (0.8, -1.1); 
            \draw[postaction={decorate}] (0.5, -1.4) -- (0.8, -1.1);
            \draw[fill=black] (0.8, -1.1) circle (0.07); 
            \draw[postaction={decorate}] (0.8, -1.1)-- (1.3, -1.1); 
             \draw [postaction={decorate}] (1.3, -1.1) -- (1.6, -0.8) ;
              \draw [postaction={decorate}] (1.3, -1.1) -- (1.6, -1.4) ;
            \draw[fill=black] (1.3, -1.1) circle (0.07); 
            \node at (1.8, -1.1) {$+$}; 
            \node at (0.7, -0.7 ) {$\vec{k}_1$}; 
            \node at (0.7, -1.5 ) {$\vec{k}_2$}; 
            \node at (1.05, -0.9) {$\vec{p}$}; 
             \node at (1.4, -0.7 ) {$\vec{k}$}; 
            \node at (1.4, -1.5 ) {$\vec{k}_3$}; 
        \end{scope}
        \begin{scope}
            \draw[postaction={decorate}]  (2.3, -1.1) -- (2, -0.8); 
            \draw[postaction={decorate}] (2.3, -1.1) -- (2, -1.4) ;
            \draw[fill=black] (2.3, -1.1) circle (0.07); 
            \draw[postaction={decorate}] (2.8, -1.1) -- (2.3, -1.1); 
             \draw [postaction={decorate}]  (3.1, -0.8) -- (2.8, -1.1) ;
              \draw [postaction={decorate}]  (3.1, -1.4) -- (2.8, -1.1) ;
            \draw[fill=black] (2.8, -1.1) circle (0.07); 
              \node at (2.2, -0.7 ) {$\vec{k}$}; 
            \node at (2.2, -1.5 ) {$\vec{k}_3$}; 
            \node at (2.55, -0.9) {$\vec{p}$}; 
             \node at (2.9, -0.7 ) {$\vec{k}_1$}; 
            \node at (2.9, -1.5 ) {$\vec{k}_2$}; 
        \end{scope}
\end{tikzpicture}
}}
\right)
\end{align*}  
& 
\setlength{\abovedisplayskip}{0pt}
 \setlength{\belowdisplayskip}{0pt}
{\begin{align*}
\hspace{0.0cm}
\mathcal{O}_{\vec{k} \vec{k}_1}^{+-}  
= 
& 
16 \frac{\left( 2 \pi \right)^2}{\hbar}\frac{G_{\vec{k}_1}}{G_{\vec{k}}} \left( 
\overline{N}_{\vec{k}} + 1
\right)\sum_{\vec{k}_2, \vec{k}_3}  \delta_{\vec{k}_2, \vec{k}_3 + \vec{k} - \vec{k}_1} \\
& 
\delta 
\left(
\varepsilon_{\vec{k}} + \varepsilon_{\vec{k}_3} - \varepsilon_{\vec{k}_1} - \varepsilon_{\vec{k}_2}
\right) 
\left(
\overline{N}_{\vec{k}_3} + 1
\right) 
\overline{N}_{\vec{k}_2}
\\
& 
W_{\vec{k}_1, \vec{k}_2 ; \vec{k}_3, \vec{k}} 
\sum_{\vec{p}} \delta_{\vec{k}_1 + \vec{k}_2, \vec{p}} 
\\
& 
\text{Im}
\left( 
V_{\vec{p};\vec{k}_1, \vec{k}_2}^{*} V_{\vec{p}; \vec{k}, \vec{k}_3} 
\right) 
\left[
\overline{N}_{\vec{p}} \delta 
\left(
\varepsilon_{\vec{k}} + \varepsilon_{\vec{k}_3} - \varepsilon_{\vec{p}}
\right)
+
\right. 
\\
& 
\left. 
\left(
\overline{N}_{\vec{p}} + 1
\right) 
\delta 
\left(
\varepsilon_{\vec{k}_1} + \varepsilon_{\vec{k}_2} - \varepsilon_{\vec{p}}
\right) 
\right] 
\end{align*}}
& 
\setlength{\abovedisplayskip}{0pt}
 \setlength{\belowdisplayskip}{0pt}
{\begin{align*}
\hspace{0.0cm}
\mathcal{O}_{\vec{k} \vec{k}_3}^{++} 
= 
& 
16 \frac{\left( 2 \pi \right)^2}{\hbar}\frac{G_{\vec{k}_3}}{G_{\vec{k}}}  \left( 
\overline{N}_{\vec{k}} + 1
\right) \sum_{\vec{k}_2, \vec{k}_1} \delta_{\vec{k}_2, \vec{k}_3 + \vec{k} - \vec{k}_1}\\
& 
\delta 
\left(
\varepsilon_{\vec{k}} + \varepsilon_{\vec{k}_3} - \varepsilon_{\vec{k}_1} - \varepsilon_{\vec{k}_2}
\right) 
\overline{N}_{\vec{k}_1} \overline{N}_{\vec{k}_2}\\
& 
W_{\vec{k}_1, \vec{k}_2 ; \vec{k}_3, \vec{k}} 
\sum_{\vec{p}} \delta_{\vec{k}_1 + \vec{k}_2, \vec{p}}  \\
& 
\text{Im}
\left( 
V_{\vec{p};\vec{k}_1, \vec{k}_2}^{*} V_{\vec{p}; \vec{k}, \vec{k}_3} 
\right)
\left[
\overline{N}_{\vec{p}} \delta 
\left(
\varepsilon_{\vec{k}} + \varepsilon_{\vec{k}_3} - \varepsilon_{\vec{p}}
\right) 
\right. 
\\
& 
\left. 
+ 
\left(
\overline{N}_{\vec{p}} + 1
\right) 
\delta 
\left(
\varepsilon_{\vec{k}_1} + \varepsilon_{\vec{k}_2} - \varepsilon_{\vec{p}}
\right) 
\right] 
\end{align*}}
\\ 
\hline 
\begin{align*}
\adjustbox{trim=0 0 -1cm 0}{
\begin{tikzpicture}[scale=1.3, decoration={markings, 
        mark= at position 0.65 with {\arrow{stealth}},
        mark= at position 2cm with {\arrow{stealth}}}
        ]   
        \begin{scope}[xshift = 0.3cm, yshift = 0.25cm]
            \draw[postaction={decorate}] (1, 0.3) -- (1.3, 0); 
            \draw[postaction={decorate}] (1, -0.3) -- (1.3, 0);  
            \draw[postaction={decorate}] (1.3, 0) -- (1.6, 0.3); 
             \draw[postaction={decorate}] (1.3, 0) -- (1.6, -0.3);
             \draw[fill=white] (1.3,0) circle (0.07);
             \node at (1, 0.45) {\footnotesize{$\vec{k}_1$}}; 
              \node at (1, -0.45) {\footnotesize{$\vec{k}_2$}}; 
              \node at (1.6, 0.45) {\footnotesize{$\vec{k}_3$}}; 
               \node at (1.6, -0.45) {\footnotesize{$\vec{k}$}}; 
            \node at (1.8, 0) {$\times$}; 
        \end{scope}
\end{tikzpicture}}
\\
\left(
\vcenter{\hbox{
\begin{tikzpicture}[scale=1.3, decoration={markings, 
        mark= at position 0.65 with {\arrow{stealth}},
        mark= at position 2cm with {\arrow{stealth}}}
        ]   
        \begin{scope}
            \draw[postaction={decorate}] (0.5, -0.8) -- (0.8, -1.1); 
            \draw[postaction={decorate}]  (0.8, -1.1) -- (0.5, -1.4);
            \draw[fill=black] (0.8, -1.1) circle (0.07); 
            \draw[postaction={decorate}] (0.8, -1.1)-- (1.3, -1.1); 
             \draw [postaction={decorate}] (1.3, -1.1) -- (1.6, -0.8) ;
              \draw [postaction={decorate}](1.6, -1.4) -- (1.3, -1.1) ;
            \draw[fill=black] (1.3, -1.1) circle (0.07); 
            \node at (1.8, -1.1) {$+$}; 
            \node at (0.7, -0.7 ) {$\vec{k}_1$}; 
            \node at (0.7, -1.5 ) {$\vec{k}_3$}; 
            \node at (1.05, -0.9) {$\vec{p}$}; 
             \node at (1.4, -0.7 ) {$\vec{k}$}; 
            \node at (1.4, -1.5 ) {$\vec{k}_2$}; 
        \end{scope}
        \begin{scope}
            \draw[postaction={decorate}]  (2.3, -1.1) -- (2, -0.8); 
            \draw[postaction={decorate}] (2, -1.4) --(2.3, -1.1);
            \draw[fill=black] (2.3, -1.1) circle (0.07); 
            \draw[postaction={decorate}] (2.8, -1.1) -- (2.3, -1.1); 
             \draw [postaction={decorate}]  (3.1, -0.8) -- (2.8, -1.1) ;
              \draw [postaction={decorate}]  (2.8, -1.1) -- (3.1, -1.4) ;
            \draw[fill=black] (2.8, -1.1) circle (0.07); 
              \node at (2.2, -0.7 ) {$\vec{k}$}; 
            \node at (2.2, -1.5 ) {$\vec{k}_2$}; 
            \node at (2.55, -0.9) {$\vec{p}$}; 
             \node at (2.9, -0.7 ) {$\vec{k}_1$}; 
            \node at (2.9, -1.5 ) {$\vec{k}_3$}; 
        \end{scope}
\end{tikzpicture}
}}
\right)
\end{align*}
& 
\setlength{\abovedisplayskip}{0pt}
 \setlength{\belowdisplayskip}{0pt}
{\begin{align*}
\hspace{0.0cm}
\mathcal{O}_{\vec{k} \vec{k}_1}^{+-} 
= 
& 
16 \frac{\left( 2 \pi \right)^2}{\hbar}\frac{G_{\vec{k}_1}}{G_{\vec{k}}} \left( 
\overline{N}_{\vec{k}} + 1
\right) 
\sum_{\vec{k}_2, \vec{k}_3} \delta_{\vec{k}_2, \vec{k}_3 + \vec{k} - \vec{k}_1} 
\\
& 
\delta 
\left(
\varepsilon_{\vec{k}} + \varepsilon_{\vec{k}_3} - \varepsilon_{\vec{k}_1} - \varepsilon_{\vec{k}_2}
\right) 
\left(
\overline{N}_{\vec{k}_3} + 1
\right)  \\
& 
W_{\vec{k}_1, \vec{k}_2 ; \vec{k}_3, \vec{k}} \overline{N}_{\vec{k}_2}
\sum_{\vec{p}} \delta_{\vec{k}_1 - \vec{k}_3, \vec{p}} 
\\
& 
\text{Im}
\left( 
V_{\vec{k};\vec{p}, \vec{k}_2}^{*} V_{\vec{k}_1; \vec{p}, \vec{k}_3} 
\right)
\left[
\overline{N}_{\vec{p}} \delta 
\left(
\varepsilon_{\vec{k}} - \varepsilon_{\vec{k}_2} - \varepsilon_{\vec{p}}
\right) 
+ 
\right. 
\\
& 
\left. 
\left(
\overline{N}_{\vec{p}} + 1
\right) 
\delta 
\left(
\varepsilon_{\vec{k}_1} - \varepsilon_{\vec{k}_3} - \varepsilon_{\vec{p}}
\right) 
\right] 
\end{align*}}
& 
\setlength{\abovedisplayskip}{0pt}
 \setlength{\belowdisplayskip}{0pt}
{\begin{align*}
\hspace{0.0cm}
\mathcal{O}_{\vec{k} \vec{k}_3}^{++} 
= &
16 \frac{\left( 2 \pi \right)^2}{\hbar}\frac{G_{\vec{k}_3}}{G_{\vec{k}}}  \left( 
\overline{N}_{\vec{k}} + 1
\right) \sum_{\vec{k}_2, \vec{k}_1} \delta_{\vec{k}_2, \vec{k}_3 + \vec{k} - \vec{k}_1}) 
\\
& 
\delta 
\left(
\varepsilon_{\vec{k}} + \varepsilon_{\vec{k}_3} - \varepsilon_{\vec{k}_1} - \varepsilon_{\vec{k}_2}
\right) 
\overline{N}_{\vec{k}_2}\overline{N}_{\vec{k}_1}
 \\
& 
W_{\vec{k}_1, \vec{k}_2 ; \vec{k}_3, \vec{k}} 
\sum_{\vec{p}} \delta_{\vec{k}_1 - \vec{k}_3, \vec{p}}
\\
& 
\text{Im}
\left( 
V_{\vec{k};\vec{p}, \vec{k}_2}^{*} V_{\vec{k}_1; \vec{p}, \vec{k}_3} 
\right) 
\left[
\overline{N}_{\vec{p}} \delta 
\left(
\varepsilon_{\vec{k}} - \varepsilon_{\vec{k}_2} - \varepsilon_{\vec{p}}
\right) 
+ 
\right. 
\\
& 
\left. 
\left(
\overline{N}_{\vec{p}} + 1
\right) 
\delta 
\left(
\varepsilon_{\vec{k}_1} - \varepsilon_{\vec{k}_3} - \varepsilon_{\vec{p}}
\right) 
\right] 
\end{align*}}
\\
\hline 
\begin{align*}
\adjustbox{trim=0 0 -1cm 0}{
\begin{tikzpicture}[scale=1.3, decoration={markings, 
        mark= at position 0.65 with {\arrow{stealth}},
        mark= at position 2cm with {\arrow{stealth}}}
        ]   
        \begin{scope}[xshift = 0.3cm, yshift = 0.25cm]
            \draw[postaction={decorate}] (1, 0.3) -- (1.3, 0); 
            \draw[postaction={decorate}] (1, -0.3) -- (1.3, 0);  
            \draw[postaction={decorate}] (1.3, 0) -- (1.6, 0.3); 
             \draw[postaction={decorate}] (1.3, 0) -- (1.6, -0.3);
             \draw[fill=white] (1.3,0) circle (0.07);
             \node at (1, 0.45) {\footnotesize{$\vec{k}_1$}}; 
              \node at (1, -0.45) {\footnotesize{$\vec{k}_2$}}; 
              \node at (1.6, 0.45) {\footnotesize{$\vec{k}_3$}}; 
               \node at (1.6, -0.45) {\footnotesize{$\vec{k}$}}; 
            \node at (1.8, 0) {$\times$}; 
        \end{scope}
\end{tikzpicture}}
\\
\left(
\vcenter{\hbox{
\begin{tikzpicture}[scale=1.3, decoration={markings, 
        mark= at position 0.65 with {\arrow{stealth}},
        mark= at position 2cm with {\arrow{stealth}}}
        ]   
        \begin{scope}
            \draw[postaction={decorate}] (0.5, -0.8) -- (0.8, -1.1); 
            \draw[postaction={decorate}]  (0.8, -1.1) -- (0.5, -1.4);
            \draw[fill=black] (0.8, -1.1) circle (0.07); 
            \draw[postaction={decorate}] (0.8, -1.1)-- (1.3, -1.1); 
             \draw [postaction={decorate}] (1.3, -1.1) -- (1.6, -0.8) ;
              \draw [postaction={decorate}](1.6, -1.4) -- (1.3, -1.1) ;
            \draw[fill=black] (1.3, -1.1) circle (0.07); 
            \node at (1.8, -1.1) {$+$}; 
            \node at (0.7, -0.7 ) {$\vec{k}_2$}; 
            \node at (0.7, -1.5 ) {$\vec{k}_3$}; 
            \node at (1.05, -0.9) {$\vec{p}$}; 
             \node at (1.4, -0.7 ) {$\vec{k}$}; 
            \node at (1.4, -1.5 ) {$\vec{k}_1$}; 
        \end{scope}
        \begin{scope}
            \draw[postaction={decorate}]  (2.3, -1.1) -- (2, -0.8); 
            \draw[postaction={decorate}] (2, -1.4) --(2.3, -1.1);
            \draw[fill=black] (2.3, -1.1) circle (0.07); 
            \draw[postaction={decorate}] (2.8, -1.1) -- (2.3, -1.1); 
             \draw [postaction={decorate}]  (3.1, -0.8) -- (2.8, -1.1) ;
              \draw [postaction={decorate}]  (2.8, -1.1) -- (3.1, -1.4) ;
            \draw[fill=black] (2.8, -1.1) circle (0.07); 
              \node at (2.2, -0.7 ) {$\vec{k}$}; 
            \node at (2.2, -1.5 ) {$\vec{k}_1$}; 
            \node at (2.55, -0.9) {$\vec{p}$}; 
             \node at (2.9, -0.7 ) {$\vec{k}_2$}; 
            \node at (2.9, -1.5 ) {$\vec{k}_3$}; 
        \end{scope}
\end{tikzpicture}
}}
\right)
\end{align*}
& 
\setlength{\abovedisplayskip}{0pt}
 \setlength{\belowdisplayskip}{0pt}
{\begin{align*}
\hspace{0.0cm}
\mathcal{O}_{\vec{k} \vec{k}_1}^{+-} 
= 
& 16 \frac{\left( 2 \pi \right)^2}{\hbar}\frac{G_{\vec{k}_1}}{G_{\vec{k}}} \left( 
\overline{N}_{\vec{k}} + 1
\right) \sum_{\vec{k}_2, \vec{k}_3} \delta_{\vec{k}_2, \vec{k}_3 + \vec{k} - \vec{k}_1} \\
& 
\delta 
\left(
\varepsilon_{\vec{k}} + \varepsilon_{\vec{k}_3} - \varepsilon_{\vec{k}_1} - \varepsilon_{\vec{k}_2}
\right) 
\left(
\overline{N}_{\vec{k}_3} + 1
\right)  \\
& 
W_{\vec{k}_1, \vec{k}_2 ; \vec{k}_3, \vec{k}} \overline{N}_{\vec{k}_2}
\sum_{\vec{p}} \delta_{\vec{k}_2 - \vec{k}_3, \vec{p}} \\
& 
\text{Im}
\left( 
V_{\vec{k};\vec{p}, \vec{k}_1}^{*} V_{\vec{k}_2; \vec{p}, \vec{k}_3} 
\right) 
\left[
\overline{N}_{\vec{p}} \delta 
\left(
\varepsilon_{\vec{k}} - \varepsilon_{\vec{k}_1} - \varepsilon_{\vec{p}}
\right) 
+ 
\right. 
\\
& 
\left.
\left(
\overline{N}_{\vec{p}} + 1
\right) 
\delta 
\left(
\varepsilon_{\vec{k}_2} - \varepsilon_{\vec{k}_3} - \varepsilon_{\vec{p}}
\right) 
\right] 
\end{align*}}
& 
\setlength{\abovedisplayskip}{0pt}
 \setlength{\belowdisplayskip}{0pt}
{\begin{align*}
\hspace{0.0cm}
\mathcal{O}_{\vec{k} \vec{k}_3}^{++} 
= 
& 16 \frac{\left( 2 \pi \right)^2}{\hbar}\frac{G_{\vec{k}_3}}{G_{\vec{k}}} \left( 
\overline{N}_{\vec{k}} + 1
\right) \sum_{\vec{k}_2, \vec{k}_1} \delta_{\vec{k}_2, \vec{k}_3 + \vec{k} - \vec{k}_1} \\
& 
\delta 
\left(
\varepsilon_{\vec{k}} + \varepsilon_{\vec{k}_3} - \varepsilon_{\vec{k}_1} - \varepsilon_{\vec{k}_2}
\right) 
\overline{N}_{\vec{k}_1} \overline{N}_{\vec{k}_2}\\
& 
W_{\vec{k}_1, \vec{k}_2 ; \vec{k}_3, \vec{k}} 
\sum_{\vec{p}} \delta_{\vec{k}_2 - \vec{k}_3, \vec{p}} \\
& 
\text{Im}
\left( 
V_{\vec{k};\vec{p}, \vec{k}_1}^{*} V_{\vec{k}_2; \vec{p}, \vec{k}_3} 
\right) 
\left[
\overline{N}_{\vec{p}} \delta 
\left(
\varepsilon_{\vec{k}} - \varepsilon_{\vec{k}_1} - \varepsilon_{\vec{p}}
\right) 
+ 
\right. 
\\
& 
\left. 
\left(
\overline{N}_{\vec{p}} + 1
\right) 
\delta 
\left(
\varepsilon_{\vec{k}_2} - \varepsilon_{\vec{k}_3} - \varepsilon_{\vec{p}}
\right) 
\right] 
\end{align*}}
\\
\hline
\begin{align*}
\adjustbox{trim=0 0 -1cm 0}{
\begin{tikzpicture}[scale=1.3, decoration={markings, 
        mark= at position 0.65 with {\arrow{stealth}},
        mark= at position 2cm with {\arrow{stealth}}}
        ]   
        \begin{scope}[xshift = 0.3cm, yshift = 0.25cm]
            \draw[postaction={decorate}] (1, 0.3) -- (1.3, 0); 
            \draw[postaction={decorate}] (1, -0.3) -- (1.3, 0);  
            \draw[postaction={decorate}] (1.3, 0) -- (1.6, 0.3); 
             \draw[postaction={decorate}] (1.3, 0) -- (1.6, -0.3);
             \draw[fill=white] (1.3,0) circle (0.07);
             \node at (1, 0.45) {\footnotesize{$\vec{k}_1$}}; 
              \node at (1, -0.45) {\footnotesize{$\vec{k}_2$}}; 
              \node at (1.6, 0.45) {\footnotesize{$\vec{k}_3$}}; 
               \node at (1.6, -0.45) {\footnotesize{$\vec{k}$}}; 
            \node at (1.8, 0) {$\times$}; 
        \end{scope}
\end{tikzpicture}}
\\
\left(
\vcenter{\hbox{
\begin{tikzpicture}[scale=1.3, decoration={markings, 
        mark= at position 0.65 with {\arrow{stealth}},
        mark= at position 2cm with {\arrow{stealth}}}
        ]   
        \begin{scope}
            \draw[postaction={decorate}] (0.5, -0.8) -- (0.8, -1.1); 
            \draw[postaction={decorate}]  (0.8, -1.1) -- (0.5, -1.4);
            \draw[fill=black] (0.8, -1.1) circle (0.07); 
            \draw[postaction={decorate}] (0.8, -1.1)-- (1.3, -1.1); 
             \draw [postaction={decorate}] (1.3, -1.1) -- (1.6, -0.8) ;
              \draw [postaction={decorate}](1.6, -1.4) -- (1.3, -1.1) ;
            \draw[fill=black] (1.3, -1.1) circle (0.07); 
            \node at (1.8, -1.1) {$+$}; 
            \node at (0.7, -0.7 ) {$\vec{k}_1$}; 
            \node at (0.7, -1.5 ) {$\vec{k}$}; 
            \node at (1.05, -0.9) {$\vec{p}$}; 
             \node at (1.4, -0.7 ) {$\vec{k}_3$}; 
            \node at (1.4, -1.5 ) {$\vec{k}_2$}; 
        \end{scope}
        \begin{scope}
            \draw[postaction={decorate}]  (2.3, -1.1) -- (2, -0.8); 
            \draw[postaction={decorate}] (2, -1.4) --(2.3, -1.1);
            \draw[fill=black] (2.3, -1.1) circle (0.07); 
            \draw[postaction={decorate}] (2.8, -1.1) -- (2.3, -1.1); 
             \draw [postaction={decorate}]  (3.1, -0.8) -- (2.8, -1.1) ;
              \draw [postaction={decorate}]  (2.8, -1.1) -- (3.1, -1.4) ;
            \draw[fill=black] (2.8, -1.1) circle (0.07); 
              \node at (2.2, -0.7 ) {$\vec{k}_3$}; 
            \node at (2.2, -1.5 ) {$\vec{k}_2$}; 
            \node at (2.55, -0.9) {$\vec{p}$}; 
             \node at (2.9, -0.7 ) {$\vec{k}_1$}; 
            \node at (2.9, -1.5 ) {$\vec{k}$}; 
        \end{scope}
\end{tikzpicture}
}}
\right)
\end{align*}
&
\setlength{\abovedisplayskip}{0pt}
 \setlength{\belowdisplayskip}{0pt}
{\begin{align*}
\hspace{0.0cm}
\mathcal{O}_{\vec{k} \vec{k}_1}^{+-} 
=
& 16 \frac{\left( 2 \pi \right)^2}{\hbar}\frac{G_{\vec{k}_1}}{G_{\vec{k}}} \left( 
\overline{N}_{\vec{k}} + 1
\right) \sum_{\vec{k}_2, \vec{k}_3} \delta_{\vec{k}_2, \vec{k}_3 + \vec{k} - \vec{k}_1} \\
& 
\delta 
\left(
\varepsilon_{\vec{k}} + \varepsilon_{\vec{k}_3} - \varepsilon_{\vec{k}_1} - \varepsilon_{\vec{k}_2}
\right) 
\left(
\overline{N}_{\vec{k}_3} + 1
\right)  \\
& 
W_{\vec{k}_1, \vec{k}_2 ; \vec{k}_3, \vec{k}} \overline{N}_{\vec{k}_2}
\sum_{\vec{p}} \delta_{\vec{k}_1 - \vec{k}, \vec{p}} \\
& 
\text{Im}
\left( 
V_{\vec{k}_3;\vec{p}, \vec{k}_2}^{*} V_{\vec{k}_1; \vec{p}, \vec{k}} 
\right) 
\left[
\overline{N}_{\vec{p}} \delta 
\left(
\varepsilon_{\vec{k}_3} - \varepsilon_{\vec{k}_2} - \varepsilon_{\vec{p}}
\right) 
+ 
\right.
\\
& 
\left. 
\left(
\overline{N}_{\vec{p}} + 1
\right) 
\delta 
\left(
\varepsilon_{\vec{k}_1} - \varepsilon_{\vec{k}} - \varepsilon_{\vec{p}}
\right) 
\right] 
\end{align*}}
& 
\setlength{\abovedisplayskip}{0pt}
 \setlength{\belowdisplayskip}{0pt}
{\begin{align*}
\hspace{0.0cm}
\mathcal{O}_{\vec{k} \vec{k}_3}^{++} 
= 
& 16 \frac{\left( 2 \pi \right)^2}{\hbar}\frac{G_{\vec{k}_3}}{G_{\vec{k}}} \left( 
\overline{N}_{\vec{k}} + 1
\right)
\sum_{\vec{k}_2, \vec{k}_1} \delta_{\vec{k}_2, \vec{k}_3 + \vec{k} - \vec{k}_1} \\
& 
\delta 
\left(
\varepsilon_{\vec{k}} + \varepsilon_{\vec{k}_3} - \varepsilon_{\vec{k}_1} - \varepsilon_{\vec{k}_2}
\right) 
\overline{N}_{\vec{k}_1}  \overline{N}_{\vec{k}_2} \\
& 
W_{\vec{k}_1, \vec{k}_2 ; \vec{k}_3, \vec{k}}
\sum_{\vec{p}} \delta_{\vec{k}_1 - \vec{k}_3, \vec{p}} \\
& 
\text{Im}
\left( 
V_{\vec{k}_3;\vec{p},\vec{k}_2}^{*} V_{\vec{k}_1; \vec{p}, \vec{k}} 
\right) 
\left[
\overline{N}_{\vec{p}} \delta 
\left(
\varepsilon_{\vec{k}_3} - \varepsilon_{\vec{k}_2} - \varepsilon_{\vec{p}}
\right) 
+
\right.
\\
& 
\left. 
\left(
\overline{N}_{\vec{p}} + 1
\right) 
\delta 
\left(
\varepsilon_{\vec{k}_1} - \varepsilon_{\vec{k}} - \varepsilon_{\vec{p}}
\right) 
\right] 
\end{align*}}
\\
\hline
\begin{align*}
\adjustbox{trim=0 0 -1cm 0}{
\begin{tikzpicture}[scale=1.3, decoration={markings, 
        mark= at position 0.65 with {\arrow{stealth}},
        mark= at position 2cm with {\arrow{stealth}}}
        ]   
        \begin{scope}[xshift = 0.3cm, yshift = 0.25cm]
            \draw[postaction={decorate}] (1, 0.3) -- (1.3, 0); 
            \draw[postaction={decorate}] (1, -0.3) -- (1.3, 0);  
            \draw[postaction={decorate}] (1.3, 0) -- (1.6, 0.3); 
             \draw[postaction={decorate}] (1.3, 0) -- (1.6, -0.3);
             \draw[fill=white] (1.3,0) circle (0.07);
             \node at (1, 0.45) {\footnotesize{$\vec{k}_1$}}; 
              \node at (1, -0.45) {\footnotesize{$\vec{k}_2$}}; 
              \node at (1.6, 0.45) {\footnotesize{$\vec{k}_3$}}; 
               \node at (1.6, -0.45) {\footnotesize{$\vec{k}$}}; 
            \node at (1.8, 0) {$\times$}; 
        \end{scope}
\end{tikzpicture}}
\\
\left(
\vcenter{\hbox{
\begin{tikzpicture}[scale=1.3, decoration={markings, 
        mark= at position 0.65 with {\arrow{stealth}},
        mark= at position 2cm with {\arrow{stealth}}}
        ]   
        \begin{scope}
            \draw[postaction={decorate}] (0.5, -0.8) -- (0.8, -1.1); 
            \draw[postaction={decorate}]  (0.8, -1.1) -- (0.5, -1.4);
            \draw[fill=black] (0.8, -1.1) circle (0.07); 
            \draw[postaction={decorate}] (0.8, -1.1)-- (1.3, -1.1); 
             \draw [postaction={decorate}] (1.3, -1.1) -- (1.6, -0.8) ;
              \draw [postaction={decorate}](1.6, -1.4) -- (1.3, -1.1) ;
            \draw[fill=black] (1.3, -1.1) circle (0.07); 
            \node at (1.8, -1.1) {$+$}; 
            \node at (0.7, -0.7 ) {$\vec{k}_2$}; 
            \node at (0.7, -1.5 ) {$\vec{k}$}; 
            \node at (1.05, -0.9) {$\vec{p}$}; 
             \node at (1.4, -0.7 ) {$\vec{k}_3$}; 
            \node at (1.4, -1.5 ) {$\vec{k}_1$}; 
        \end{scope}
        \begin{scope}
            \draw[postaction={decorate}]  (2.3, -1.1) -- (2, -0.8); 
            \draw[postaction={decorate}] (2, -1.4) --(2.3, -1.1);
            \draw[fill=black] (2.3, -1.1) circle (0.07); 
            \draw[postaction={decorate}] (2.8, -1.1) -- (2.3, -1.1); 
             \draw [postaction={decorate}]  (3.1, -0.8) -- (2.8, -1.1) ;
              \draw [postaction={decorate}]  (2.8, -1.1) -- (3.1, -1.4) ;
            \draw[fill=black] (2.8, -1.1) circle (0.07); 
              \node at (2.2, -0.7 ) {$\vec{k}_3$}; 
            \node at (2.2, -1.5 ) {$\vec{k}_1$}; 
            \node at (2.55, -0.9) {$\vec{p}$}; 
             \node at (2.9, -0.7 ) {$\vec{k}_2$}; 
            \node at (2.9, -1.5 ) {$\vec{k}$}; 
        \end{scope}
\end{tikzpicture}
}}
\right)
\end{align*}
&
\setlength{\abovedisplayskip}{0pt}
 \setlength{\belowdisplayskip}{0pt}
{\begin{align*}
\hspace{0.0cm}
\mathcal{O}_{\vec{k} \vec{k}_1}^{+-} 
= 
&
16 \frac{\left( 2 \pi \right)^2}{\hbar}\frac{G_{\vec{k}_1}}{G_{\vec{k}}} \left( 
\overline{N}_{\vec{k}} + 1
\right)\sum_{\vec{k}_2, \vec{k}_3} \delta_{\vec{k}_2, \vec{k}_3 + \vec{k} - \vec{k}_1} \\
& 
\delta 
\left(
\varepsilon_{\vec{k}} + \varepsilon_{\vec{k}_3} - \varepsilon_{\vec{k}_1} - \varepsilon_{\vec{k}_2}
\right) 
\left(
\overline{N}_{\vec{k}_3} + 1
\right)  \\
& 
W_{\vec{k}_1, \vec{k}_2 ; \vec{k}_3, \vec{k}} \overline{N}_{\vec{k}_2}
\sum_{\vec{p}} \delta_{\vec{k}_2 - \vec{k}, \vec{p}} \\
& 
\text{Im}
\left( 
V_{\vec{k}_3;\vec{p}, \vec{k}_1}^{*} V_{\vec{k}_2; \vec{p}, \vec{k}} 
\right) 
\left[
\overline{N}_{\vec{p}} \delta 
\left(
\varepsilon_{\vec{k}_3} - \varepsilon_{\vec{k}_1} - \varepsilon_{\vec{p}}
\right) 
+ 
\right. 
\\
&
\left.
\left(
\overline{N}_{\vec{p}} + 1
\right) 
\delta 
\left(
\varepsilon_{\vec{k}_2} - \varepsilon_{\vec{k}} - \varepsilon_{\vec{p}}
\right) 
\right] 
\end{align*}}
& 
\setlength{\abovedisplayskip}{0pt}
 \setlength{\belowdisplayskip}{0pt}
{\begin{align*}
\hspace{0.0cm}
\mathcal{O}_{\vec{k} \vec{k}_3}^{++} 
= 
& 16 \frac{\left( 2 \pi \right)^2}{\hbar}\frac{G_{\vec{k}_3}}{G_{\vec{k}}} \left( 
\overline{N}_{\vec{k}} + 1
\right) \sum_{\vec{k}_2, \vec{k}_1} 
\delta_{\vec{k}_2, \vec{k}_3 + \vec{k} - \vec{k}_1}\\
& 
\delta 
\left(
\varepsilon_{\vec{k}} + \varepsilon_{\vec{k}_3} - \varepsilon_{\vec{k}_1} - \varepsilon_{\vec{k}_2}
\right) 
\overline{N}_{\vec{k}_1}  \overline{N}_{\vec{k}_2}\\
& 
W_{\vec{k}_1, \vec{k}_2 ; \vec{k}_3, \vec{k}}
\sum_{\vec{p}} \delta_{\vec{k}_2 - \vec{k}_3, \vec{p}} \\
& 
\text{Im}
\left( 
V_{\vec{k}_3;\vec{p}, \vec{k}_1}^{*} V_{\vec{k}_2; \vec{p}, \vec{k}} 
\right) 
\left[
\overline{N}_{\vec{p}} \delta 
\left(
\varepsilon_{\vec{k}_3} - \varepsilon_{\vec{k}_1} - \varepsilon_{\vec{p}}
\right) 
+
\right. 
\\
& 
\left.
\left(
\overline{N}_{\vec{p}} + 1
\right) 
\delta 
\left(
\varepsilon_{\vec{k}_2} - \varepsilon_{\vec{k}} - \varepsilon_{\vec{p}}
\right) 
\right] 
\end{align*}}
\end{tabular}
\end{ruledtabular}
\end{table}

\label{subsubsection: 4-magnons}
\begin{enumerate}
    \item As already discussed, the four-magnon table (Tab.~\ref{tab:4-magnons SM}) does not contain the $\mathcal{O}_{\vec{k} \vec{k}_2}$ contribution. The reason for this omission is that including $\mathcal{O}_{\vec{k} \vec{k}_2}$ results in duplicate processes. To see this, consider the first four-magnon diagram of the table:
    \begin{equation}
\text{Symmetric Diagram} = 
\vcenter{\hbox{
\begin{tikzpicture}[scale=1.3, decoration={markings, 
        mark= at position 0.65 with {\arrow{stealth}},
        mark= at position 2cm with {\arrow{stealth}}}
        ]   
        \begin{scope}
            \draw[postaction={decorate}] (1, 0.3) -- (1.3, 0); 
            \draw[postaction={decorate}] (1, -0.3) -- (1.3, 0);  
            \draw[postaction={decorate}] (1.3, 0) -- (1.6, 0.3); 
             \draw[postaction={decorate}] (1.3, 0) -- (1.6, -0.3);
             \draw[fill=white] (1.3,0) circle (0.07);
             \node at (1, 0.45) {\footnotesize{$\vec{k}_1$}}; 
              \node at (1, -0.45) {\footnotesize{$\vec{k}_2$}}; 
              \node at (1.6, 0.45) {\footnotesize{$\vec{k}_3$}}; 
               \node at (1.6, -0.45) {\footnotesize{$\vec{k}$}}; 
            \node at (1.8, 0) {$\times$}; 
        \end{scope}
\end{tikzpicture}}}
\left(
\vcenter{\hbox{
\begin{tikzpicture}[scale=1.3, decoration={markings, 
        mark= at position 0.65 with {\arrow{stealth}},
        mark= at position 2cm with {\arrow{stealth}}}
        ]   
      \begin{scope}
            \draw[postaction={decorate}] (0.5, -0.8) -- (0.8, -1.1); 
            \draw[postaction={decorate}] (0.5, -1.4) -- (0.8, -1.1);
            \draw[fill=black] (0.8, -1.1) circle (0.07); 
            \draw[postaction={decorate}] (0.8, -1.1)-- (1.3, -1.1); 
             \draw [postaction={decorate}] (1.3, -1.1) -- (1.6, -0.8) ;
              \draw [postaction={decorate}] (1.3, -1.1) -- (1.6, -1.4) ;
            \draw[fill=black] (1.3, -1.1) circle (0.07); 
            \node at (1.8, -1.1) {$+$}; 
            \node at (0.7, -0.7 ) {\footnotesize{$\vec{k}_1$}}; 
            \node at (0.7, -1.5 ) {\footnotesize{$\vec{k}_2$}}; 
            \node at (1.05, -0.9) {\footnotesize{$\vec{p}$}}; 
             \node at (1.4, -0.7 ) {\footnotesize{$\vec{k}$}}; 
            \node at (1.4, -1.5 ) {\footnotesize{$\vec{k}_3$}}; 
        \end{scope}
        \begin{scope}
            \draw[postaction={decorate}]  (2.3, -1.1) -- (2, -0.8); 
            \draw[postaction={decorate}] (2.3, -1.1) -- (2, -1.4) ;
            \draw[fill=black] (2.3, -1.1) circle (0.07); 
            \draw[postaction={decorate}] (2.8, -1.1) -- (2.3, -1.1); 
             \draw [postaction={decorate}]  (3.1, -0.8) -- (2.8, -1.1) ;
              \draw [postaction={decorate}]  (3.1, -1.4) -- (2.8, -1.1) ;
            \draw[fill=black] (2.8, -1.1) circle (0.07); 
              \node at (2.2, -0.7 ) {\footnotesize{$\vec{k}$}}; 
            \node at (2.2, -1.5 ) {\footnotesize{$\vec{k}_3$}}; 
            \node at (2.55, -0.9) {\footnotesize{$\vec{p}$}}; 
             \node at (2.9, -0.7 ) {\footnotesize{$\vec{k}_1$}}; 
            \node at (2.9, -1.5 ) {\footnotesize{$\vec{k}_2$}}; 
        \end{scope}
\end{tikzpicture}
}}
\right),
\label{diagr: symmetric DH4}
    \end{equation}
where we immediately see that due to the symmetric nature of the diagram swapping $\vec{k}_1$ and $\vec{k}_2$ leaves the diagram unaffected. We have thus $\mathcal{O}_{\vec{k} \vec{k}_1}\equiv  \mathcal{O}_{\vec{k} \vec{k}_2}$. 
The rest of the four-magnon diagrams---four in total---lack the symmetric nature of Eq.~\eqref{diagr: symmetric DH4}. However, we notice that two of the diagrams can be generated by the other two by swapping $\vec{k}_1$ with $\vec{k}_2$. For example, we have this connection between the diagrams
\begin{equation}
\text{Diagram \# 1} = 
\vcenter{\hbox{
\begin{tikzpicture}[scale=1.3, decoration={markings, 
        mark= at position 0.65 with {\arrow{stealth}},
        mark= at position 2cm with {\arrow{stealth}}}
        ]   
        \begin{scope}[xshift = 0.3cm, yshift = 0.25cm]
            \draw[postaction={decorate}] (1, 0.3) -- (1.3, 0); 
            \draw[postaction={decorate}] (1, -0.3) -- (1.3, 0);  
            \draw[postaction={decorate}] (1.3, 0) -- (1.6, 0.3); 
             \draw[postaction={decorate}] (1.3, 0) -- (1.6, -0.3);
             \draw[fill=white] (1.3,0) circle (0.07);
             \node at (1, 0.45) {\footnotesize{$\vec{k}_1$}}; 
              \node at (1, -0.45) {\footnotesize{$\vec{k}_2$}}; 
              \node at (1.6, 0.45) {\footnotesize{$\vec{k}_3$}}; 
               \node at (1.6, -0.45) {\footnotesize{$\vec{k}$}}; 
            \node at (1.8, 0) {$\times$}; 
        \end{scope}
\end{tikzpicture}}}
\left(
\vcenter{\hbox{
\begin{tikzpicture}[scale=1.3, decoration={markings, 
        mark= at position 0.65 with {\arrow{stealth}},
        mark= at position 2cm with {\arrow{stealth}}}
        ]   
        \begin{scope}
            \draw[postaction={decorate}] (0.5, -0.8) -- (0.8, -1.1); 
            \draw[postaction={decorate}]  (0.8, -1.1) -- (0.5, -1.4);
            \draw[fill=black] (0.8, -1.1) circle (0.07); 
            \draw[postaction={decorate}] (0.8, -1.1)-- (1.3, -1.1); 
             \draw [postaction={decorate}] (1.3, -1.1) -- (1.6, -0.8) ;
              \draw [postaction={decorate}](1.6, -1.4) -- (1.3, -1.1) ;
            \draw[fill=black] (1.3, -1.1) circle (0.07); 
            \node at (1.8, -1.1) {$+$}; 
            \node at (0.7, -0.7 ) {\footnotesize{$\vec{k}_1$}}; 
            \node at (0.7, -1.5 ) {\footnotesize{$\vec{k}_3$}}; 
            \node at (1.05, -0.9) {\footnotesize{$\vec{p}$}}; 
             \node at (1.4, -0.7 ) {\footnotesize{$\vec{k}$}}; 
            \node at (1.4, -1.5 ) {\footnotesize{$\vec{k}_2$}}; 
        \end{scope}
        \begin{scope}
            \draw[postaction={decorate}]  (2.3, -1.1) -- (2, -0.8); 
            \draw[postaction={decorate}] (2, -1.4) --(2.3, -1.1);
            \draw[fill=black] (2.3, -1.1) circle (0.07); 
            \draw[postaction={decorate}] (2.8, -1.1) -- (2.3, -1.1); 
             \draw [postaction={decorate}]  (3.1, -0.8) -- (2.8, -1.1) ;
              \draw [postaction={decorate}]  (2.8, -1.1) -- (3.1, -1.4) ;
            \draw[fill=black] (2.8, -1.1) circle (0.07); 
              \node at (2.2, -0.7 ) {\footnotesize{$\vec{k}$}}; 
            \node at (2.2, -1.5 ) {\footnotesize{$\vec{k}_2$}}; 
            \node at (2.55, -0.9) {\footnotesize{$\vec{p}$}}; 
             \node at (2.9, -0.7 ) {\footnotesize{$\vec{k}_1$}}; 
            \node at (2.9, -1.5 ) {\footnotesize{$\vec{k}_3$}}; 
        \end{scope}
\end{tikzpicture}
}}
\right)
\end{equation}
and 
\begin{equation}
\text{Diagram \# 2} = 
\vcenter{\hbox{
\begin{tikzpicture}[scale=1.3, decoration={markings, 
        mark= at position 0.65 with {\arrow{stealth}},
        mark= at position 2cm with {\arrow{stealth}}}
        ]   
        \begin{scope}[xshift = 0.3cm, yshift = 0.25cm]
            \draw[postaction={decorate}] (1, 0.3) -- (1.3, 0); 
            \draw[postaction={decorate}] (1, -0.3) -- (1.3, 0);  
            \draw[postaction={decorate}] (1.3, 0) -- (1.6, 0.3); 
             \draw[postaction={decorate}] (1.3, 0) -- (1.6, -0.3);
             \draw[fill=white] (1.3,0) circle (0.07);
             \node at (1, 0.45) {\footnotesize{$\vec{k}_1$}}; 
              \node at (1, -0.45) {\footnotesize{$\vec{k}_2$}}; 
              \node at (1.6, 0.45) {\footnotesize{$\vec{k}_3$}}; 
               \node at (1.6, -0.45) {\footnotesize{$\vec{k}$}}; 
            \node at (1.8, 0) {$\times$}; 
        \end{scope}
\end{tikzpicture}}}
\left(
\vcenter{\hbox{
\begin{tikzpicture}[scale=1.3, decoration={markings, 
        mark= at position 0.65 with {\arrow{stealth}},
        mark= at position 2cm with {\arrow{stealth}}}
        ]   
        \begin{scope}
            \draw[postaction={decorate}] (0.5, -0.8) -- (0.8, -1.1); 
            \draw[postaction={decorate}]  (0.8, -1.1) -- (0.5, -1.4);
            \draw[fill=black] (0.8, -1.1) circle (0.07); 
            \draw[postaction={decorate}] (0.8, -1.1)-- (1.3, -1.1); 
             \draw [postaction={decorate}] (1.3, -1.1) -- (1.6, -0.8) ;
              \draw [postaction={decorate}](1.6, -1.4) -- (1.3, -1.1) ;
            \draw[fill=black] (1.3, -1.1) circle (0.07); 
            \node at (1.8, -1.1) {$+$}; 
            \node at (0.7, -0.7 ) {\footnotesize{$\vec{k}_2$}}; 
            \node at (0.7, -1.5 ) {\footnotesize{$\vec{k}_3$}}; 
            \node at (1.05, -0.9) {\footnotesize{$\vec{p}$}}; 
             \node at (1.4, -0.7 ) {\footnotesize{$\vec{k}$}}; 
            \node at (1.4, -1.5 ) {\footnotesize{$\vec{k}_1$}}; 
        \end{scope}
        \begin{scope}
            \draw[postaction={decorate}]  (2.3, -1.1) -- (2, -0.8); 
            \draw[postaction={decorate}] (2, -1.4) --(2.3, -1.1);
            \draw[fill=black] (2.3, -1.1) circle (0.07); 
            \draw[postaction={decorate}] (2.8, -1.1) -- (2.3, -1.1); 
             \draw [postaction={decorate}]  (3.1, -0.8) -- (2.8, -1.1) ;
              \draw [postaction={decorate}]  (2.8, -1.1) -- (3.1, -1.4) ;
            \draw[fill=black] (2.8, -1.1) circle (0.07); 
              \node at (2.2, -0.7 ) {\footnotesize{$\vec{k}$}}; 
            \node at (2.2, -1.5 ) {\footnotesize{$\vec{k}_1$}}; 
            \node at (2.55, -0.9) {\footnotesize{$\vec{p}$}}; 
             \node at (2.9, -0.7 ) {\footnotesize{$\vec{k}_2$}}; 
            \node at (2.9, -1.5 ) {\footnotesize{$\vec{k}_3$}}; 
        \end{scope}
\end{tikzpicture}
}}
\right)\stackrel{\vec{k}_1 \leftrightarrow \vec{k}_2}{=} \text{Diagram \# 1}. 
\end{equation}
One can already see that the contribution $\mathcal{O}_{\vec{k} \vec{k}_1}^{\#1}$ is identical to $\mathcal{O}_{\vec{k} \vec{k}_2}^{\#2}$, since the magnon with $\vec{k}_1$ of diagram $\#$1 and the magnon with $\vec{k}_2$ of diagram $\#$2 are completely equivalent (they enter the same vertex). Consequently, to avoid over-counting processes, one should not calculate both $\mathcal{O}_{\vec{k} \vec{k}_1}$ and $\mathcal{O}_{\vec{k} \vec{k}_2}$ for each diagram, but only one of the two (either  $\mathcal{O}_{\vec{k} \vec{k}_1}$ or $\mathcal{O}_{\vec{k} \vec{k}_2}$).

\item We now proceed discussing the $\mathcal{O}_{\vec{k} \vec{k}_3}^{++}$ contribution from the four-magnon diagrams, apart from the symmetric one shown in Eq.~\eqref{diagr: symmetric DH4}. We firstly note that some of the $\mathcal{O}_{\vec{k} \vec{k}_3}^{++}$ correspond to identical processes that should not be accounted for when calculating the conductivities. These identical $\mathcal{O}_{\vec{k} \vec{k}_3}^{++}$ come from a pair of diagrams that can be  mapped into each other by swapping $\vec{k}_1$ and $\vec{k}_2$. We point here that such swapping is allowed since the expression for $\mathcal{O}_{\vec{k} \vec{k}_3}$ contains a summation over both $\vec{k}_1$ and $\vec{k}_2$. The first pair is the one of Diagram $\#1$ and Diagram $\#$2 where one can see that  $\mathcal{O}_{\vec{k} \vec{k}_3}^{\# 1}$ and $\mathcal{O}_{\vec{k} \vec{k}_3}^{\# 2}$ are identical by simply swapping the dummy indices $\vec{k}_1$ and $\vec{k}_2$. The same logic applies to the two other diagrams, 
\begin{equation}
\text{Diagram \# 3} = 
\vcenter{\hbox{
\begin{tikzpicture}[scale=1.3, decoration={markings, 
        mark= at position 0.65 with {\arrow{stealth}},
        mark= at position 2cm with {\arrow{stealth}}}
        ]   
        \begin{scope}[xshift = 0.3cm, yshift = 0.25cm]
            \draw[postaction={decorate}] (1, 0.3) -- (1.3, 0); 
            \draw[postaction={decorate}] (1, -0.3) -- (1.3, 0);  
            \draw[postaction={decorate}] (1.3, 0) -- (1.6, 0.3); 
             \draw[postaction={decorate}] (1.3, 0) -- (1.6, -0.3);
             \draw[fill=white] (1.3,0) circle (0.07);
             \node at (1, 0.45) {\footnotesize{$\vec{k}_1$}}; 
              \node at (1, -0.45) {\footnotesize{$\vec{k}_2$}}; 
              \node at (1.6, 0.45) {\footnotesize{$\vec{k}_3$}}; 
               \node at (1.6, -0.45) {\footnotesize{$\vec{k}$}}; 
            \node at (1.8, 0) {$\times$}; 
        \end{scope}
\end{tikzpicture}}}
\left(
\vcenter{\hbox{
\begin{tikzpicture}[scale=1.3, decoration={markings, 
        mark= at position 0.65 with {\arrow{stealth}},
        mark= at position 2cm with {\arrow{stealth}}}
        ]   
        \begin{scope}
            \draw[postaction={decorate}] (0.5, -0.8) -- (0.8, -1.1); 
            \draw[postaction={decorate}]  (0.8, -1.1) -- (0.5, -1.4);
            \draw[fill=black] (0.8, -1.1) circle (0.07); 
            \draw[postaction={decorate}] (0.8, -1.1)-- (1.3, -1.1); 
             \draw [postaction={decorate}] (1.3, -1.1) -- (1.6, -0.8) ;
              \draw [postaction={decorate}](1.6, -1.4) -- (1.3, -1.1) ;
            \draw[fill=black] (1.3, -1.1) circle (0.07); 
            \node at (1.8, -1.1) {$+$}; 
            \node at (0.7, -0.7 ) {\footnotesize{$\vec{k}_1$}}; 
            \node at (0.7, -1.5 ) {\footnotesize{$\vec{k}$}}; 
            \node at (1.05, -0.9) {\footnotesize{$\vec{p}$}}; 
             \node at (1.4, -0.7 ) {\footnotesize{$\vec{k}_3$}}; 
            \node at (1.4, -1.5 ) {\footnotesize{$\vec{k}_2$}}; 
        \end{scope}
        \begin{scope}
            \draw[postaction={decorate}]  (2.3, -1.1) -- (2, -0.8); 
            \draw[postaction={decorate}] (2, -1.4) --(2.3, -1.1);
            \draw[fill=black] (2.3, -1.1) circle (0.07); 
            \draw[postaction={decorate}] (2.8, -1.1) -- (2.3, -1.1); 
             \draw [postaction={decorate}]  (3.1, -0.8) -- (2.8, -1.1) ;
              \draw [postaction={decorate}]  (2.8, -1.1) -- (3.1, -1.4) ;
            \draw[fill=black] (2.8, -1.1) circle (0.07); 
              \node at (2.2, -0.7 ) {\footnotesize{$\vec{k}_3$}}; 
            \node at (2.2, -1.5 ) {\footnotesize{$\vec{k}_2$}}; 
            \node at (2.55, -0.9) {\footnotesize{$\vec{p}$}}; 
             \node at (2.9, -0.7 ) {\footnotesize{$\vec{k}_1$}}; 
            \node at (2.9, -1.5 ) {\footnotesize{$\vec{k}$}}; 
        \end{scope}
\end{tikzpicture}
}}
\right)
\label{diagr: DH4 3}
\end{equation}
and 
\begin{equation}
\text{Diagram \# 4}=
\vcenter{\hbox{
\begin{tikzpicture}[scale=1.3, decoration={markings, 
        mark= at position 0.65 with {\arrow{stealth}},
        mark= at position 2cm with {\arrow{stealth}}}
        ]   
        \begin{scope}[xshift = 0.3cm, yshift = 0.25cm]
            \draw[postaction={decorate}] (1, 0.3) -- (1.3, 0); 
            \draw[postaction={decorate}] (1, -0.3) -- (1.3, 0);  
            \draw[postaction={decorate}] (1.3, 0) -- (1.6, 0.3); 
             \draw[postaction={decorate}] (1.3, 0) -- (1.6, -0.3);
             \draw[fill=white] (1.3,0) circle (0.07);
             \node at (1, 0.45) {\footnotesize{$\vec{k}_1$}}; 
              \node at (1, -0.45) {\footnotesize{$\vec{k}_2$}}; 
              \node at (1.6, 0.45) {\footnotesize{$\vec{k}_3$}}; 
               \node at (1.6, -0.45) {\footnotesize{$\vec{k}$}}; 
            \node at (1.8, 0) {$\times$}; 
        \end{scope}
\end{tikzpicture}}}
\left(
\vcenter{\hbox{
\begin{tikzpicture}[scale=1.3, decoration={markings, 
        mark= at position 0.65 with {\arrow{stealth}},
        mark= at position 2cm with {\arrow{stealth}}}
        ]   
        \begin{scope}
            \draw[postaction={decorate}] (0.5, -0.8) -- (0.8, -1.1); 
            \draw[postaction={decorate}]  (0.8, -1.1) -- (0.5, -1.4);
            \draw[fill=black] (0.8, -1.1) circle (0.07); 
            \draw[postaction={decorate}] (0.8, -1.1)-- (1.3, -1.1); 
             \draw [postaction={decorate}] (1.3, -1.1) -- (1.6, -0.8) ;
              \draw [postaction={decorate}](1.6, -1.4) -- (1.3, -1.1) ;
            \draw[fill=black] (1.3, -1.1) circle (0.07); 
            \node at (1.8, -1.1) {$+$}; 
            \node at (0.7, -0.7 ) {\footnotesize{$\vec{k}_2$}}; 
            \node at (0.7, -1.5 ) {\footnotesize{$\vec{k}$}}; 
            \node at (1.05, -0.9) {\footnotesize{$\vec{p}$}}; 
             \node at (1.4, -0.7 ) {\footnotesize{$\vec{k}_3$}}; 
            \node at (1.4, -1.5 ) {\footnotesize{$\vec{k}_1$}}; 
        \end{scope}
        \begin{scope}
            \draw[postaction={decorate}]  (2.3, -1.1) -- (2, -0.8); 
            \draw[postaction={decorate}] (2, -1.4) --(2.3, -1.1);
            \draw[fill=black] (2.3, -1.1) circle (0.07); 
            \draw[postaction={decorate}] (2.8, -1.1) -- (2.3, -1.1); 
             \draw [postaction={decorate}]  (3.1, -0.8) -- (2.8, -1.1) ;
              \draw [postaction={decorate}]  (2.8, -1.1) -- (3.1, -1.4) ;
            \draw[fill=black] (2.8, -1.1) circle (0.07); 
              \node at (2.2, -0.7 ) {\footnotesize{$\vec{k}_3$}}; 
            \node at (2.2, -1.5 ) {\footnotesize{$\vec{k}_1$}}; 
            \node at (2.55, -0.9) {\footnotesize{$\vec{p}$}}; 
             \node at (2.9, -0.7 ) {\footnotesize{$\vec{k}_2$}}; 
            \node at (2.9, -1.5 ) {\footnotesize{$\vec{k}$}}; 
        \end{scope}
\end{tikzpicture}
}}
\right)
\stackrel{\vec{k}_1 \leftrightarrow \vec{k}_2}{=} \text{Diagram \# 3}
. 
\end{equation}
The rest of the $\mathcal{O}_{\vec{k} \vec{k}_3}^{++}$ come from four-magnon diagrams that are related to each other by swapping the momenta $\vec{k}$ and $\vec{k}_3$. These $\mathcal{O}_{\vec{k} \vec{k}_3}^{++}$ are unique contributions but they can be shown to result in the same contribution for the $\kappa_{\text{H}}$. 
To prove our statement we begin with the formula for the $\mathcal{O}_{\vec{k} \vec{k}_3}$ contribution to the vector Hall conductivity of the Diagram $\#3$,
\begin{equation}
    \vec{\kappa}_{\text{H}}^{\# 3} = 16\frac{\left(2 \pi\right)^2}{\hbar} \frac{1}{2 k_{\text{B}} T ^2}\sum_{\vec{k}, \vec{k}_3} \vec{v_{\vec{k}}} \times \vec{v}_{\vec{\vec{k}_3}}  \varepsilon_{\vec{k}} \varepsilon_{\vec{k}_3} \tau_{\vec{k}} \tau_{\vec{k}_3} G_{\vec{k}} G_{\vec{k}_3}\frac{\overline{N}_{\vec{k}_3} - \overline{N}_{\vec{k}}}{\overline{N}_{\vec{k}_3}} \mathcal{O}_{\vec{k} \vec{k}_3}^{++, \# 3},
    \label{eq: KH Okk3 DH4 3}
\end{equation}
and observe that diagram $\#1$ and $\#3$ are related to each other by the interchange of the momenta $\vec{k}$ and $\vec{k}_3$. Indeed, we find
\begin{equation}
\begin{split}
\mathcal{O}_{\vec{k}\vec{k}_3}^{++, \# 3}
& = 
\left[
\vcenter{\hbox{
\begin{tikzpicture}[scale=1.3, decoration={markings, 
        mark= at position 0.65 with {\arrow{stealth}},
        mark= at position 2cm with {\arrow{stealth}}}
        ]   
        \begin{scope}[xshift = 0.3cm, yshift = 0.25cm]
            \draw[postaction={decorate}] (1, 0.3) -- (1.3, 0); 
            \draw[postaction={decorate}] (1, -0.3) -- (1.3, 0);  
            \draw[postaction={decorate}] (1.3, 0) -- (1.6, 0.3); 
             \draw[postaction={decorate}] (1.3, 0) -- (1.6, -0.3);
             \draw[fill=white] (1.3,0) circle (0.07);
             \node at (1, 0.45) {\footnotesize{$\vec{k}_1$}}; 
              \node at (1, -0.45) {\footnotesize{$\vec{k}_2$}}; 
              \node at (1.6, 0.45) {\footnotesize{$\vec{k}_3$}}; 
               \node at (1.6, -0.45) {\footnotesize{$\vec{k}$}}; 
            \node at (1.8, 0) {$\times$}; 
        \end{scope}
\end{tikzpicture}}}
\left(
\vcenter{\hbox{
\begin{tikzpicture}[scale=1.3, decoration={markings, 
        mark= at position 0.65 with {\arrow{stealth}},
        mark= at position 2cm with {\arrow{stealth}}}
        ]   
        \begin{scope}
            \draw[postaction={decorate}] (0.5, -0.8) -- (0.8, -1.1); 
            \draw[postaction={decorate}]  (0.8, -1.1) -- (0.5, -1.4);
            \draw[fill=black] (0.8, -1.1) circle (0.07); 
            \draw[postaction={decorate}] (0.8, -1.1)-- (1.3, -1.1); 
             \draw [postaction={decorate}] (1.3, -1.1) -- (1.6, -0.8) ;
              \draw [postaction={decorate}](1.6, -1.4) -- (1.3, -1.1) ;
            \draw[fill=black] (1.3, -1.1) circle (0.07); 
            \node at (1.8, -1.1) {$+$}; 
            \node at (0.7, -0.7 ) {\footnotesize{$\vec{k}_1$}}; 
            \node at (0.7, -1.5 ) {\footnotesize{$\vec{k}$}}; 
            \node at (1.05, -0.9) {\footnotesize{$\vec{p}$}}; 
             \node at (1.4, -0.7 ) {\footnotesize{$\vec{k}_3$}}; 
            \node at (1.4, -1.5 ) {\footnotesize{$\vec{k}_2$}}; 
        \end{scope}
        \begin{scope}
            \draw[postaction={decorate}]  (2.3, -1.1) -- (2, -0.8); 
            \draw[postaction={decorate}] (2, -1.4) --(2.3, -1.1);
            \draw[fill=black] (2.3, -1.1) circle (0.07); 
            \draw[postaction={decorate}] (2.8, -1.1) -- (2.3, -1.1); 
             \draw [postaction={decorate}]  (3.1, -0.8) -- (2.8, -1.1) ;
              \draw [postaction={decorate}]  (2.8, -1.1) -- (3.1, -1.4) ;
            \draw[fill=black] (2.8, -1.1) circle (0.07); 
              \node at (2.2, -0.7 ) {\footnotesize{$\vec{k}_3$}}; 
            \node at (2.2, -1.5 ) {\footnotesize{$\vec{k}_2$}}; 
            \node at (2.55, -0.9) {\footnotesize{$\vec{p}$}}; 
             \node at (2.9, -0.7 ) {\footnotesize{$\vec{k}_1$}}; 
            \node at (2.9, -1.5 ) {\footnotesize{$\vec{k}$}}; 
        \end{scope}
\end{tikzpicture}
}}
\right)
\hspace{0.2 cm}
\right]_{\vec{k} \vec{k}_3} 
= \frac{G_{\vec{k}_3}}{G_{\vec{k}}} \left( \overline{N}_{\vec{k}} + 1\right) Q_{\vec{k} \vec{k}_3}^{++, \# 3} \\
& 
\stackrel{\vec{k} \leftrightarrow \vec{k}_3}{=} 
\frac{G_{\vec{k}}}{G_{\vec{k}_3}} \left( \overline{N}_{\vec{k}_3} + 1\right) Q_{\vec{k}_3 \vec{k}}^{++, \# 3} = 
\left(\frac{G_{\vec{k}}}{G_{\vec{k}_3}}\right)^2 \frac{\overline{N}_{\vec{k}_3} + 1}{\overline{N}_{\vec{k}} + 1}\left(\overline{N}_{\vec{k}} + 1\right) \mathcal{Q}_{\vec{k} \vec{k}_3}^{++, \# 1} \\
& 
= 
\left(\frac{G_{\vec{k}}}{G_{\vec{k}_3}}\right)^2 \frac{\overline{N}_{\vec{k}_3} + 1}{\overline{N}_{\vec{k}} + 1} \mathcal{O}_{\vec{k} \vec{k}_3}^{++, \# 1}
\\
& 
= 
\left(\frac{G_{\vec{k}}}{G_{\vec{k}_3}}\right)^2 \frac{\overline{N}_{\vec{k}_3} + 1}{\overline{N}_{\vec{k}} + 1}
\left[
\vcenter{\hbox{
\begin{tikzpicture}[scale=1.3, decoration={markings, 
        mark= at position 0.65 with {\arrow{stealth}},
        mark= at position 2cm with {\arrow{stealth}}}
        ]   
        \begin{scope}[xshift = 0.3cm, yshift = 0.25cm]
            \draw[postaction={decorate}] (1, 0.3) -- (1.3, 0); 
            \draw[postaction={decorate}] (1, -0.3) -- (1.3, 0);  
            \draw[postaction={decorate}] (1.3, 0) -- (1.6, 0.3); 
             \draw[postaction={decorate}] (1.3, 0) -- (1.6, -0.3);
             \draw[fill=white] (1.3,0) circle (0.07);
             \node at (1, 0.45) {\footnotesize{$\vec{k}_1$}}; 
              \node at (1, -0.45) {\footnotesize{$\vec{k}_2$}}; 
              \node at (1.6, 0.45) {\footnotesize{$\vec{k}_3$}}; 
               \node at (1.6, -0.45) {\footnotesize{$\vec{k}$}}; 
            \node at (1.8, 0) {$\times$}; 
        \end{scope}
\end{tikzpicture}}}
\left(
\vcenter{\hbox{
\begin{tikzpicture}[scale=1.3, decoration={markings, 
        mark= at position 0.65 with {\arrow{stealth}},
        mark= at position 2cm with {\arrow{stealth}}}
        ]   
        \begin{scope}
            \draw[postaction={decorate}] (0.5, -0.8) -- (0.8, -1.1); 
            \draw[postaction={decorate}]  (0.8, -1.1) -- (0.5, -1.4);
            \draw[fill=black] (0.8, -1.1) circle (0.07); 
            \draw[postaction={decorate}] (0.8, -1.1)-- (1.3, -1.1); 
             \draw [postaction={decorate}] (1.3, -1.1) -- (1.6, -0.8) ;
              \draw [postaction={decorate}](1.6, -1.4) -- (1.3, -1.1) ;
            \draw[fill=black] (1.3, -1.1) circle (0.07); 
            \node at (1.8, -1.1) {$+$}; 
            \node at (0.7, -0.7 ) {\footnotesize{$\vec{k}_1$}}; 
            \node at (0.7, -1.5 ) {\footnotesize{$\vec{k}_3$}}; 
            \node at (1.05, -0.9) {\footnotesize{$\vec{p}$}}; 
             \node at (1.4, -0.7 ) {\footnotesize{$\vec{k}$}}; 
            \node at (1.4, -1.5 ) {\footnotesize{$\vec{k}_2$}}; 
        \end{scope}
        \begin{scope}
            \draw[postaction={decorate}]  (2.3, -1.1) -- (2, -0.8); 
            \draw[postaction={decorate}] (2, -1.4) --(2.3, -1.1);
            \draw[fill=black] (2.3, -1.1) circle (0.07); 
            \draw[postaction={decorate}] (2.8, -1.1) -- (2.3, -1.1); 
             \draw [postaction={decorate}]  (3.1, -0.8) -- (2.8, -1.1) ;
              \draw [postaction={decorate}]  (2.8, -1.1) -- (3.1, -1.4) ;
            \draw[fill=black] (2.8, -1.1) circle (0.07); 
              \node at (2.2, -0.7 ) {\footnotesize{$\vec{k}$}}; 
            \node at (2.2, -1.5 ) {\footnotesize{$\vec{k}_2$}}; 
            \node at (2.55, -0.9) {\footnotesize{$\vec{p}$}}; 
             \node at (2.9, -0.7 ) {\footnotesize{$\vec{k}_1$}}; 
            \node at (2.9, -1.5 ) {\footnotesize{$\vec{k}_3$}}; 
        \end{scope}
\end{tikzpicture}
}}
\right)
\hspace{0.2 cm}
\right]_{\vec{k} \vec{k}_3}. 
\end{split}
\label{eq: relation DH4 1 and 3}
\end{equation}
By flipping the two momenta in Eq.~\eqref{eq: KH Okk3 DH4 3}, we get
\begin{equation}
\begin{split}
\vec{\kappa}_{\text{H}}^{\# 3}
& = 
16\frac{\left(2 \pi\right)^2}{\hbar} \frac{1}{2 k_{\text{B}} T ^2}\sum_{\vec{k}, \vec{k}_3} \vec{v_{\vec{k}_3}} \times \vec{v}_{\vec{\vec{k}}} \varepsilon_{\vec{k}} \varepsilon_{\vec{k}_3} \tau_{\vec{k}} \tau_{\vec{k}_3} G_{\vec{k}} G_{\vec{k}_3}\frac{\overline{N}_{\vec{k}_3} - \overline{N}_{\vec{k}}}{\overline{N}_{\vec{k}_3}} \mathcal{O}_{\vec{k} \vec{k}_3}^{++, \# 3} \\
& \stackrel{\vec{k}_3 \leftrightarrow \vec{k}}{=}
16\frac{\left(2 \pi\right)^2}{\hbar} \frac{1}{2 k_{\text{B}} T ^2}\sum_{\vec{k}, \vec{k}_3} \vec{v_{\vec{k}_3}} \times \vec{v}_{\vec{\vec{k}}}\varepsilon_{\vec{k}} \varepsilon_{\vec{k}_3} \tau_{\vec{k}} \tau_{\vec{k}_3} G_{\vec{k}} G_{\vec{k}_3}\frac{\overline{N}_{\vec{k}} - \overline{N}_{\vec{k}_3}}{\overline{N}_{\vec{k}}} \mathcal{O}_{\vec{k}_3 \vec{k}}^{++, \# 3} \\
& 
\stackrel{\eqref{eq: relation DH4 1 and 3}}{=}
16\frac{\left(2 \pi\right)^2}{\hbar} \frac{1}{2 k_{\text{B}} T ^2}\sum_{\vec{k}, \vec{k}_3} \vec{v_{\vec{k}}} \times \vec{v}_{\vec{\vec{k}_3}} \varepsilon_{\vec{k}} \varepsilon_{\vec{k}_3} \tau_{\vec{k}} \tau_{\vec{k}_3} G_{\vec{k}} G_{\vec{k}_3}\frac{\overline{N}_{\vec{k}_3} - \overline{N}_{\vec{k}}}{\overline{N}_{\vec{k}}} \left(\frac{G_{\vec{k}}}{G_{\vec{k}_3}}\right)^2 \frac{\overline{N}_{\vec{k}_3} + 1}{\overline{N}_{\vec{k}} + 1}\mathcal{O}_{\vec{k} \vec{k}_3}^{++, \# 1} \\
& = 
16\frac{\left(2 \pi\right)^2}{\hbar} \frac{1}{2 k_{\text{B}} T ^2}\sum_{\vec{k}, \vec{k}_3} \vec{v_{\vec{k}}} \times \vec{v}_{\vec{\vec{k}_3}} \varepsilon_{\vec{k}} \varepsilon_{\vec{k}_3} \tau_{\vec{k}} \tau_{\vec{k}_3} G_{\vec{k}} G_{\vec{k}_3}\frac{\overline{N}_{\vec{k}_3} - \overline{N}_{\vec{k}}}{\overline{N}_{\vec{k}_3}} \mathcal{O}_{\vec{k} \vec{k}_3}^{++, \# 1} \\
& 
= \vec{\kappa}_{\text{H}}^{\# 1}. 
\end{split}
\end{equation}
The same result can be shown to hold between the diagrams $\#4$ and $\#1$. 
We have thus proven that the non-symmetric four-magnon diagrams that are related to each other by $\vec{k} \leftrightarrow \vec{k}_3$ have a $\mathcal{O}_{\vec{k} \vec{k}_3}$ part that contributes equally to the Hall conductivity. 
\end{enumerate}

\subsubsection{Note on neglected diagrams}
Below, we comment on some other diagrams to explain why they were neglected.
\begin{enumerate}
    \item We emphasize that we have only considered diagrams up to second order in $H_\text{int}$. At third order in $H_\text{int}$ one would encounter, for example, the triangle diagram and had to evaluate the interference term
    \begin{equation}
    \vcenter{\hbox{
    \begin{tikzpicture}
        [scale=1.3, decoration={markings, 
        mark= at position 0.65 with {\arrow{stealth}},
        mark= at position 2cm with {\arrow{stealth}}}
        ]   
            \begin{scope}
           \draw[postaction={decorate}] (1,0.2)--(1.3,0.0);
            \draw[postaction={decorate}] (1,-0.2)--(1.3,0.0);
            \draw[postaction={decorate}] (1.3,0)--(1.7,0.0);
            \draw[fill=black] (1.3,0) circle (0.07);
            \node at (1.2, 0.4) {\footnotesize{$\vec{k}_1$}};
            \node at (1.2, -0.4) {\footnotesize{$\vec{k}_2$}};
            \node at (1.65, 0.2) {\footnotesize{$\vec{k}$}};
            \node at (1.9, 0) {$\times$}; 
            \draw[postaction={decorate}] (2.2, 0.2)--(2.5, 0.2); 
            \draw[postaction={decorate}] (2.2, -0.2)--(2.5, -0.2); 
            \draw[postaction={decorate}]  (2.5, 0.2) -- (2.9,0); 
            \draw[postaction={decorate}]  (2.5, -0.2) -- (2.9,0);
            \draw[postaction={decorate}]  (2.5, -0.2) -- (2.5, 0.2);
            \draw [postaction={decorate}] (2.9, 0) -- (3.3, 0); 
            \draw[fill=black] (2.9, 0) circle (0.07); 
            \draw[fill=black] (2.5, 0.2) circle (0.07); 
            \draw[fill=black] (2.5, -0.2) circle (0.07); 
            \node at (3.2, 0.2) {\footnotesize{$\vec{k}$}} ; 
            \node at (2.7, -0.3) {\footnotesize{$\vec{p}_1$}};
            \node at (2.7, 0.3) {\footnotesize{$\vec{p}_2$}};
            \node at (2.3, 0) {\footnotesize{$\vec{p}_3$}};
            \node at (2.2, 0.4) {\footnotesize{$\vec{k}_1$}};
             \node at (2.2, -0.4) {\footnotesize{$\vec{k}_2$}};
            \end{scope}
    \end{tikzpicture}}}
    .
    \label{diagr:triangle}
    \end{equation}
    Its contribution to the off-diagonal scattering rate is of the same order in $1/\sqrt{S}$ as that of the interference diagrams containing one four-magnon and two three-magnon vertices. However, since there are four three-magnon vertices in Eq.~\eqref{diagr:triangle}, the triangle diagram interference is $O(D^4)$, which is subleading compared to the $O(D^2)$ contributions.
    
    \item We have only considered scattering processes that keep the number of magnons constant or change it by $\pm 1$. In principle, one could consider interference diagrams of the form
    \begin{equation}
    \vcenter{\hbox{
    \begin{tikzpicture}
        [scale=1.3, decoration={markings, 
        mark= at position 0.65 with {\arrow{stealth}},
        mark= at position 2cm with {\arrow{stealth}}}
        ]   
            \begin{scope}
            \draw[postaction={decorate}] (1,0.2)--(1.3,0.0);
            \draw[postaction={decorate}] (1,0.0)--(1.3,0.0);
            \draw[postaction={decorate}] (1,-0.2)--(1.3,0.0);
            \draw[postaction={decorate}] (1.3,0)--(1.7,0.0);
            \draw[fill=white] (1.3,0) circle (0.07);
            \node at (1.2, 0.4) {\footnotesize{$\vec{k}_1$}};
            \node at (1.2, -0.4) {\footnotesize{$\vec{k}_2$}};
            \node at (0.8, 0.0) {\footnotesize{$\vec{k}_3$}};
            \node at (1.65, 0.2) {\footnotesize{$\vec{k}$}};
            \node at (1.9, 0) {$\times$}; 
            \begin{scope}[xshift=0.3cm]
            \draw[postaction={decorate}] (2.2, 0.2)--(2.5, 0.1); 
            \draw[postaction={decorate}] (2.2, 0)--(2.5, 0.1);
            \draw[postaction={decorate}] (2.5, 0.1)--(2.9, 0);
            \draw[postaction={decorate}] (2.2, -0.2)--(2.9, 0); 
            \draw[postaction={decorate}]  (2.9, 0) -- (3.3,0); 
            \draw[fill=black] (2.5, 0.1) circle (0.07); 
            \draw[fill=black] (2.9, 0) circle (0.07); 
            \node at (3.2, 0.2) {\footnotesize{$\vec{k}$}} ; 
            \node at (2.7, 0.25) {\footnotesize{$\vec{p}$}};
            \node at (2.2, 0.4) {\footnotesize{$\vec{k}_1$}};
             \node at (2.2, -0.4) {\footnotesize{$\vec{k}_2$}};
             \node at (2.0, 0) {\footnotesize{$\vec{k}_3$}};
            \end{scope}
             \end{scope}
    \end{tikzpicture}}}
    ,
    \label{diagr:DeltaN2}
    \end{equation}
    which change the number by $-2$. For such diagrams to occur, the four-magnon vertex must have a number non-conserving component, as it occurs when a Bogoliubov transformation is necessary to diagonalize the harmonic theory. This is not the case in our model.

    \item Interference diagrams with only four-magnon vertices, e.g.,
    \begin{equation}
    \vcenter{\hbox{
    \begin{tikzpicture}[scale=1.3, decoration={markings, 
        mark= at position 0.65 with {\arrow{stealth}},
        mark= at position 2cm with {\arrow{stealth}}}
        ]   
        \begin{scope}[xshift = 0.3cm, yshift = 0.25cm]
            \draw[postaction={decorate}] (1, 0.3) -- (1.3, 0); 
            \draw[postaction={decorate}] (1, -0.3) -- (1.3, 0);  
            \draw[postaction={decorate}] (1.3, 0) -- (1.6, 0.3); 
             \draw[postaction={decorate}] (1.3, 0) -- (1.6, -0.3);
             \draw[fill=white] (1.3,0) circle (0.07);
             \node at (1, 0.45) {\footnotesize{$\vec{k}_1$}}; 
              \node at (1, -0.45) {\footnotesize{$\vec{k}_2$}}; 
              \node at (1.6, 0.45) {\footnotesize{$\vec{k}_3$}}; 
               \node at (1.6, -0.45) {\footnotesize{$\vec{k}$}}; 
            \node at (1.8, 0) {$\times$}; 
        \end{scope}
    \end{tikzpicture}}}
    \vcenter{\hbox{
    \begin{tikzpicture}[scale=1.3, decoration={markings, 
        mark= at position 0.65 with {\arrow{stealth}},
        mark= at position 2cm with {\arrow{stealth}}}
        ]   
        \begin{scope}
            \draw[postaction={decorate}] (0.5, -0.8) -- (0.8, -1.1); 
            \draw[postaction={decorate}] (0.5, -1.4) -- (0.8, -1.1);
            \draw[postaction={decorate}] (0.8, -1.1) to[out=40, in=140] (1.3, -1.1); 
            \draw[postaction={decorate}] (0.8, -1.1) to[out=-40, in=220] (1.3, -1.1);
             \draw [postaction={decorate}] (1.3, -1.1) -- (1.6, -0.8) ;
              \draw [postaction={decorate}] (1.3, -1.1) -- (1.6, -1.4) ;
            \draw[fill=white] (0.8, -1.1) circle (0.07);
            \draw[fill=white] (1.3, -1.1) circle (0.07); 
            \node at (0.7, -0.7 ) {\footnotesize $\vec{k}_1$}; 
            \node at (0.7, -1.5 ) {\footnotesize $\vec{k}_2$}; 
            \node at (1.05, -0.8) {\footnotesize $\vec{p}_1$}; 
            \node at (1.05, -1.4) {\footnotesize $\vec{p}_2$}; 
             \node at (1.4, -0.7 ) {\footnotesize $\vec{k}_3$}; 
            \node at (1.4, -1.5 ) {\footnotesize $\vec{k}$}; 
        \end{scope}
    \end{tikzpicture}
    }},
    \label{diagr: only 4s}
    \end{equation}
    could also contribute to the thermal Hall effect, in principle. However, since in our model the four-magnon vertex does not break the TR symmetry, the interference process in Eq.~\eqref{diagr: only 4s} does not contribute. Furthermore, even if it contributed, it would be of higher order in $1/\sqrt{S}$ than the considered interference diagrams.

    \item
    We have neglected any diagrams that do not change the particle number and only consist of a single in-coming and a single out-going magnon propagator, e.g.,
    \begin{equation}
    \vcenter{\hbox{
        \begin{tikzpicture}[scale=1.3, decoration={markings, 
        mark= at position 0.65 with {\arrow{stealth}},
        mark= at position 2cm with {\arrow{stealth}}}
        ]   
        \begin{scope}
        \draw[postaction={decorate}] (1.2,-1.2)--(1.7,-1.2);
        \draw [postaction={decorate}] (1.7,-1.2) to [out=70,in=110] (2.2,-1.2);
        \draw [postaction={decorate}] (1.7,-1.2) to [out=-70,in=-110] (2.2,-1.2);
        \draw[postaction={decorate}] (2.2,-1.2)--(2.7,-1.2);
        \draw[fill=black] (1.7,-1.2) circle (0.07);
        \draw[fill=black] (2.2,-1.2) circle (0.07);
        \node at (2.5, -1) {\footnotesize{$\vec{k}$}}; 
        \node at (2, -0.85) {\footnotesize{$\vec{p}_1$}}; 
        \node at (2, -1.55){\footnotesize{$\vec{p}_2$}};
        \node at (1.4, -1.0) {\footnotesize{$\vec{k}_1$}};
        \end{scope}
    \end{tikzpicture}}}.
    \end{equation}
    Taking the absolute square, this diagram can only contribute at order $O(D^4)$ to the \emph{diagonal} part of the collision kernel because momentum does not change. When interfering with another diagram, say
    \begin{equation}
    \vcenter{\hbox{
        \begin{tikzpicture}[scale=1.3, decoration={markings, 
        mark= at position 0.65 with {\arrow{stealth}},
        mark= at position 2cm with {\arrow{stealth}}}
        ]   
        \begin{scope}
        \draw[postaction={decorate}] (1.2,-1.2)--(1.7,-1.2);
        \draw [postaction={decorate}] (1.7,-1.2) to [out=80,in=120] (2.2,-1.2);
        \draw [postaction={decorate}] (1.7,-1.2) to [out=-80,in=-120] (2.2,-1.2);
        \draw [postaction={decorate}] (2.2,-1.2) -- (1.7,-1.2);
        \draw[postaction={decorate}] (2.2,-1.2)--(2.7,-1.2);
        \draw[fill=white] (1.7,-1.2) circle (0.07);
        \draw[fill=white] (2.2,-1.2) circle (0.07);
        \node at (2.5, -1) {\footnotesize{$\vec{k}$}}; 
        \node at (2, -0.85) {\footnotesize{$\vec{p}_1$}}; 
        \node at (2, -1.55){\footnotesize{$\vec{p}_2$}};
        \node at (1.4, -1.0) {\footnotesize{$\vec{k}_1$}};
        \end{scope}
        ,
    \end{tikzpicture}
    }},
    \end{equation}
    there is still no contribution to the Hall effect and their contribution to the diagonal part of the collision kernel has a higher order in $1/\sqrt{S}$  than that of the leading diagrams.
\end{enumerate}

\subsection{Time reversal symmetry and broken detailed balance}
\label{sub: TR and DB}

We connect here the TR  symmetry to the off-diagonal elements $\mathcal{O}_{\vec{k} \vec{k}'}$ of the scattering matrix that break the detailed balance relation and contribute to the Hall conductivity. More specifically, we show that by assuming that the system respects TR symmetry the vertices of $\mathcal{O}_{\vec{k} \vec{k}'}$ vanish identically. One can see from the tables of Sec.~\ref{subsection: diagrams} that we have off-diagonal elements containing two different forms of vertices:
        \begin{equation}
            \begin{split}
                & \text{Im}\left( V_{\tilde{\vec{q}}}^{*}V_{\tilde{\vec{q}}'}\right),  \\
                & \text{Re}\left( V_{\tilde{\vec{q}}}\right)\text{Im}\left(V_{\tilde{\vec{q}}'} \right) - \text{Re}\left( V_{\tilde{\vec{q}}'}\right)\text{Im}\left( V_{\tilde{\vec{q}}}\right),
            \end{split}
        \end{equation}
with $\tilde{\vec{q}} \neq \tilde{\vec{q}'}$ abbreviating the magnon momenta participating in the three-magnon process quantified by the vertex $V$. We will perform the calculation for the second case by using the example in Eq.~\eqref{eq: example interference rate}, where the vertex of the calculated element reads
\begin{equation}
    \left[
    \text{Re}
    \left( 
    V_{ \vec{k}_1;\vec{k},\vec{k}_{2}}^{*}
    \right) 
    \text{Im}
    \left(
    V_{\vec{k}_1;\vec{p}_1, \vec{p}_2}^{*} 
    \right)
    -
    \text{Im} \left( 
    V_{\vec{k}_1;\vec{k},\vec{k}_{2}}^{*}
    \right) \
    \text{Re}
    \left( 
    V_{\vec{k}_1;\vec{p}_1, \vec{p}_2}^{*}
    \right) 
    \right]. 
    \label{eq: vertex of example}
\end{equation}
The TR-condition for the three-magnon interaction vertex in Eq.~\eqref{TR conditions} simplifies for the one-band case to 
\begin{equation}
    V_{\tilde{\vec{q}}}^{*} = \mathrm{e}^{\mathrm{i}\phi}V_{-\tilde{\vec{q}}},
    \label{eq: H3 TR condition for 1 band}
\end{equation}
with $\phi$ being a \textit{global} phase. Consequently, by assuming that the system respects TR symmetry we can use Eq.~\eqref{eq: H3 TR condition for 1 band} to rewrite the first part of the vertex as 
\begin{equation}
    \begin{split}
        \text{Re}
    \left( 
    V_{\vec{k}_1;\vec{k},\vec{k}_{2}}^{*}
    \right) 
    \text{Im}
    \left(
    V_{\vec{k}_1;\vec{p}_1, \vec{p}_2}^{*} 
    \right) 
    & = 
    \text{Re}
    \left( 
    \mathrm{e}^{\mathrm{i}\phi}
     V_{-\vec{k}_1;-\vec{k},-\vec{k}_{2}}
    \right) 
    \text{Im}
    \left( 
    \mathrm{e}^{\mathrm{i}\phi}
     V_{-\vec{k}_1;-\vec{p}_1, -\vec{p}_2}
    \right) \\
    & = 
    \left[
    \cos \phi 
    \text{Re}
    \left(
    V_{-\vec{k}_1;-\vec{k},-\vec{k}_{2}}
    \right)
    - 
    \sin \phi \text{Im}
    \left( 
     V_{-\vec{k}_1;-\vec{p}_1, -\vec{p}_2}
    \right) 
    \right] 
    \left[
    \cos \phi \text{Im}
    \left( 
    V_{-\vec{k}_1;-\vec{p}_1, -\vec{p}_2}
    \right)
    + 
     \sin \phi \text{Re}
    \left( 
     V_{-\vec{k}_1;-\vec{p}_1, -\vec{p}_2}
    \right) 
    \right] \\
    & = 
    \cos^2\phi \text{Re}
    \left(
    V_{-\vec{k}_1;-\vec{k},-\vec{k}_{2}}
    \right)
    \text{Im}
    \left( 
    V_{ -\vec{k}_1;-\vec{p}_1, -\vec{p}_2}
    \right)
    - \sin ^2 \phi 
    \text{Im}
    \left(
    V_{-\vec{k}_1;-\vec{k},-\vec{k}_{2}}
    \right)
    \text{Re}
    \left( 
    V_{-\vec{k}_1;-\vec{p}_1, -\vec{p}_2}
    \right)
    \\
    & 
    \quad - \sin \phi \cos \phi 
    \left[
    \text{Im}
    \left(
    V_{-\vec{k}_1;-\vec{k},-\vec{k}_{2}}
    \right)\text{Im}
    \left( 
    V_{-\vec{k}_1;-\vec{p}_1, -\vec{p}_2}
    \right)
    + 
    \text{Re}
    \left(
    V_{-\vec{k}_1;-\vec{k},-\vec{k}_{2}}
    \right)
    \text{Re}
    \left( 
    V_{-\vec{k}_1;-\vec{p}_1, -\vec{p}_2}
    \right)
    \right].
    \end{split}
\end{equation}
The second part of the vertex,  $ \text{Im} \left( 
    V_{\vec{k}_1;\vec{k},\vec{k}_{2}}^{*}
    \right) \
    \text{Re}
    \left( 
    V_{\vec{k}_1;\vec{p}_1, \vec{p}_2}^{*}
    \right)$, can be generated from the first one by swapping the real and imaginary parts of the $V$-vertices. Doing so, subtracting the two parts from each other, and noticing that the terms proportional to $\sin \phi \cos\phi$ cancel each other out, we get for the full vertex
\begin{equation}
\begin{split}
    \text{Re}
    \left( 
    V_{ \vec{k}_1;\vec{k},\vec{k}_{2}}^{*}
    \right) 
    \text{Im}
    \left(
    V_{\vec{k}_1;\vec{p}_1, \vec{p}_2}^{*} 
    \right)
    -
    \text{Im} \left( 
    V_{\vec{k}_1;\vec{k},\vec{k}_{2}}^{*}
    \right) \
    \text{Re}
    \left( 
    V_{\vec{k}_1;\vec{p}_1, \vec{p}_2}^{*}
    \right) 
    & = 
    \left( 
    \cos^2\phi + \sin^2\phi 
    \right) 
    \left[ 
    \text{Re}
    \left( 
    V_{-\vec{k}_1;-\vec{k},-\vec{k}_{2}}
    \right) 
    \text{Im}
    \left(
    V_{ -\vec{k}_1;-\vec{p}_1, -\vec{p}_2}
    \right) 
    \right.
    \\
    & 
    \quad -
    \left.
    \text{Im} \left( 
    V_{-\vec{k}_1;-\vec{k},-\vec{k}_{2}}
    \right) \
    \text{Re}
    \left( 
    V_{-\vec{k}_1;-\vec{p}_1, -\vec{p}_2}
    \right)
    \right] \\
    & = 
    - \left[ 
    \text{Re}
    \left( 
    V_{-\vec{k}_1;-\vec{k},-\vec{k}_{2}}^{*}
    \right) 
    \text{Im}
    \left(
    V_{-\vec{k}_1;-\vec{p}_1, -\vec{p}_2}^{*}
    \right) 
    -
    \text{Im} \left( 
    V_{-\vec{k}_1;-\vec{k},-\vec{k}_{2}}^{*}
    \right) \
    \text{Re}
    \left( 
    V_{ -\vec{k}_1;-\vec{p}_1, -\vec{p}_2}^{*}
    \right)
    \right]
    \\
    & = 
    -
    \left[
    \text{Re}
    \left( 
    V_{\vec{k}_1;\vec{k},\vec{k}_{2}}^{*}
    \right) 
    \text{Im}
    \left(
    V_{ \vec{k}_1;\vec{p}_1, \vec{p}_2}^{*} 
    \right)
    -
    \text{Im} \left( 
    V_{\vec{k}_1;\vec{k},\vec{k}_{2}}^{*}
    \right) \
    \text{Re}
    \left( 
    V_{\vec{k}_1;\vec{p}_1, \vec{p}_2}^{*}
    \right) 
    \right],
\end{split}
\end{equation}
where we made use of the relations $\text{Im}\left( V_{\tilde{\vec{q}}}\right) = -\text{Im}\left( V_{\tilde{\vec{q}}}^{*}\right)$ and $\text{Re}\left( V_{\tilde{\vec{q}}}\right) = \text{Re}\left( V_{\tilde{\vec{q}}}^{*}\right)$ as well as of the oddness of the $V$-vertices as a function of the momenta. We have proven that the vertex of Eq.~\eqref{eq: vertex of example} must be identical to zero for TR to be respected. The same conclusion can be derived for this particular type of vertex regardless of the process. Furthermore, the vertex of the first case, $\text{Im}\left( V_{\tilde{\vec{q}}}^{*}V_{\tilde{\vec{q}}'}\right)$, can also be seen to vanish identically following the exact same logic. In this case, the derivation is even simpler since the global phases cancel out from the very beginning of the calculation. With all these we have shown that for the case of our model TR symmetry must be violated in order to produce off-diagonal elements that break the detail balance condition. 

Closing this part, we stress that our conclusion that $\mathcal{A}_{\vec{k} \vec{k}'} = 0$ in the presence of TR symmetry is specific to our model. Since in general there is no direct connection between TR symmetry and (broken) detailed balance\cite{Mangeolle2022PRB,Mangeolle2022PRX}, one could generate elements that break the detailed balance---and thus generate a finite anti-symmetric part of the collision matrix $\mathcal{A}_{\vec{k} \vec{k}'}$---but respect TR symmetry: $\mathcal{A}_{\vec{k} \vec{k}'} = - \mathcal{A}_{-\vec{k},- \vec{k}'}$. The Hall conductivity would then vanish upon summation over $\vec{k}$ and $\vec{k}'$. 

\begin{subsection}{Diagrammatic lexicon}
\label{subsection: lexicon}
Here, we provide a small diagrammatic lexicon to translate diagrams into mathematical expressions. We will go step by step and first describe how one can extract the scattering rate $\Gamma_{\vec{k}}\left[\left\{ N_{\vec{k}'} \right\}\right]$ of a particular diagram. 

Given a diagrammatic representation of a scattering process, the scattering rate can be extracted according to the following rules:
\begin{enumerate}
    \item Each magnon mode with momentum $\vec{k}'$ entering a vertex (circle) is destroyed and gives rise to a factor of $\sqrt{N_{\vec{k}'}}$.  Each magnon mode $\vec{k}'$ that leaves a vertex is created and associated with a factor of $\sqrt{N_{\vec{k}'}+1}$. We stress here that the $N_{\vec{k}'}$'s are the out-of-equilibrium occupation numbers. 
    \item Intermediate magnon modes with momenta $\vec{p}_i$ give rise to a factor of $\overline{N}_{\vec{p}_i}+1$ if they are virtual and a factor of $\overline{N}_{\vec{p}_i}$ if they are real. The real intermediate magnons are drawn with backward moving lines from right to left, while the virtual intermediate magnons are drawn with forward moving lines from left to right. We stress that the notion of virtual and real magnons depicted with forward and backward lines refers only to the intermediate $\vec{p}_i$ magnons. 
    \item Each vertex is associated either with the three-magnon potential in Eq.~\eqref{3-magnon potential} (black circle) or with the four-magnon potential in Eq.~\eqref{4-magnon vertex} (white circle). At each vertex, momentum conservation must be met, leading to a Kronecker delta connecting the magnon momenta coming in and out of a vertex. 
    \item All intermediate momenta $\vec{p}_i$ must be summed over. 
    \item  Higher order scattering diagrams must be weighted by the complex kernel $\left( \varepsilon_{i} - \varepsilon_{\nu} + \mathrm{i}\eta \right)^{-1}$, where $\varepsilon_{i}$ denotes the total energy of the magnons that enter the first vertex of the diagram, and $\varepsilon_{\nu}$ denotes the total energy of the magnons associated with the intermediate state $\ket{\nu}$. 
    \item For expressions corresponding to the sum of first order and higher order scattering events, the squared magnitude $|...|^2$ of the expression must be computed.  For interference terms, the formula of Eq.~\eqref{eq: scattering rates general} must be used to compute the scattering rate. 
    \item All topologically equivalent diagrams must be taken into account, giving rise to additional multiplicity factors. A guide on how to derive these factors is provided in Sec.~\ref{subsection: multiplicity factors}. A factor of $\frac{2\pi}{\hbar}$ must be associated with each term, stemming from 
    Fermi's golden rule. 
    \item A delta-function enforcing global energy conservation $\delta ( \varepsilon_{i} - \varepsilon_{f} )$ must be added, stemming from Fermi's golden rule. 
    \item Finally, all momenta $\vec{k}' \neq \vec{k}$ must be summed over. 
\end{enumerate}
Going through the above instructions, we obtain $\Gamma_{\vec{k}}\left[\left\{ N_{\vec{k}'}\right\}\right] $ for a particular diagram. With the such obtained expression for $\Gamma_{\vec{k}}\left[\left\{ N_{\vec{k}_i} \right\} \right]$, one must do the following to get its off-diagonal element $\mathcal{O}_{\vec{k} \vec{k}'}$:
\begin{enumerate}
    \item  $\Gamma_{\vec{k}}\left[\left\{ N_{\vec{k}'}\right\}\right] $ must be linearized with respect to $N_{\vec{k}'} = \overline{N}_{\vec{k}'} + \delta N_{\vec{k}'}$. By Taylor-expanding around the Bose Einstein populations and keeping only terms linear  in the deviations $\delta N_{\vec{k}'}$, we get that $\Gamma_{\vec{k}}\left[\left\{ N_{\vec{k}'} \right\} \right] = \sum_{\vec{k}'}C_{\vec{k} \vec{k}'}\delta N_{\vec{k}'}$. Recall that $O_{\vec{k} \vec{k}'}$ corresponds to the off diagonal part of $C_{\vec{k} \vec{k}'}$.
    \item Since the calculations are done in the symmetrized basis of the collision matrix a factor of $\frac{G_{\vec{k}'}}{G_{\vec{k}}}$ must multiply the expression for $O_{\vec{k} \vec{k}'}$ to obtain $\mathcal{O}_{\vec{k} \vec{k}'}$. 
    \item Finally, the part of $\mathcal{O}_{\vec{k} \vec{k}'}$ that contributes to the Hall conductivity is the one that breaks the detail balanced relation as explained in Sec.~\ref{subsection: DB}. For the interference terms, this part always comes with resonant intermediate scattering, containing one additional delta-function coming from the imaginary part of  $\left( \varepsilon_{i} - \varepsilon_{\nu} + \mathrm{i}\eta \right)^{-1}$. The first order scattering terms are unable to generate such parts. 
\end{enumerate}
\end{subsection}

\begin{subsection}{Numerical Factors}
\label{subsection: multiplicity factors}
We explain here how
to compute the numerical factors of the diagrams introduced in Sec.~\ref{subsection: diagrams}. In principle, there are four different sources leading to accumulation of numerical factors:
\begin{enumerate}
    \item Factors from the ``topology'' of the diagrams stemming from the calculation of the $T$-matrix elements.
    \item Factors of $1/2$ needed to account for duplicate processes upon summing the scattering rates over the momenta $\vec{k}'$ with $\vec{k}' \neq \vec{k}$. 
    \item A factor of $2 \pi / \hbar$ coming from Fermi's golden rule. 
    \item Finally, an additional factor of $2\pi$ associated with all interference diagrams that contribute to the Hall conductivity is needed. In this case, the factor of $\pi$ comes from the imaginary part of $\text{Im}\left[\left( \Delta E + \mathrm{i}\eta \right)^{-1}\right] = - \pi \delta\left( \Delta E\right)$ and the factor of 2 stems from the scattering rate of the interference process and it is generated when the squared magnitude $|...|^2$ of the total scattering channel is expanded. 
\end{enumerate}
To demonstrate the first three potential sources of numerical factors, we will provide simple examples. 
The first example regards the calculation of the first order four-magnon scattering rate using Fermi's golden rule:
\begin{equation}
    \Gamma_{\vec{k}}^4 = \frac{2 \pi}{\hbar}\sum_{\vec{k}_1, \vec{k}_2, \vec{k}_3} \delta \left( \varepsilon_{\vec{k}_1} + \varepsilon_{\vec{k}_2} - \varepsilon_{\vec{k}} - \varepsilon_{\vec{k}_3}\right)  
    \left|
    \vcenter{\hbox{
    \begin{tikzpicture}[scale=1.3, decoration={markings, 
        mark= at position 0.65 with {\arrow{stealth}},
        mark= at position 2cm with {\arrow{stealth}}}
        ]   
        \begin{scope}[xshift = 0.3cm, yshift = 0.25cm]
            \draw[postaction={decorate}] (1, 0.3) -- (1.3, 0); 
            \draw[postaction={decorate}] (1, -0.3) -- (1.3, 0);  
            \draw[postaction={decorate}] (1.3, 0) -- (1.6, 0.3); 
             \draw[postaction={decorate}] (1.3, 0) -- (1.6, -0.3);
             \draw[fill=white] (1.3,0) circle (0.07);
             \node at (1, 0.45) {\footnotesize{$\vec{k}_1$}}; 
              \node at (1, -0.45) {\footnotesize{$\vec{k}_2$}}; 
              \node at (1.6, 0.45) {\footnotesize{$\vec{k}_3$}}; 
               \node at (1.6, -0.45) {\footnotesize{$\vec{k}$}}; 
        \end{scope}
\end{tikzpicture}}}
    \right|^2.
\end{equation}
We calculate the squared amplitude of the four-magnon matrix element by connecting the appropriate initial state, $ \ket{\text{i}} =\ket{N_{\vec{k}}, N_{\vec{k}_1}, N_{\vec{k}_2}, N_{\vec{k}_3}}$, and final state, $ \ket{\text{f}} =\ket{N_{\vec{k}}+1, N_{\vec{k}_1}-1, N_{\vec{k}_2}-1, N_{\vec{k}_3}+1}$, via the four-magnon vertex of Eq.~\eqref{4-magnon vertex}:
\begin{equation}
    \begin{split}
    \left|
    \vcenter{\hbox{
    \begin{tikzpicture}[scale=1.3, decoration={markings, 
        mark= at position 0.65 with {\arrow{stealth}},
        mark= at position 2cm with {\arrow{stealth}}}
        ]   
        \begin{scope}[xshift = 0.3cm, yshift = 0.25cm]
            \draw[postaction={decorate}] (1, 0.3) -- (1.3, 0); 
            \draw[postaction={decorate}] (1, -0.3) -- (1.3, 0);  
            \draw[postaction={decorate}] (1.3, 0) -- (1.6, 0.3); 
             \draw[postaction={decorate}] (1.3, 0) -- (1.6, -0.3);
             \draw[fill=white] (1.3,0) circle (0.07);
             \node at (1, 0.45) {\footnotesize{$\vec{k}_1$}}; 
              \node at (1, -0.45) {\footnotesize{$\vec{k}_2$}}; 
              \node at (1.6, 0.45) {\footnotesize{$\vec{k}_3$}}; 
               \node at (1.6, -0.45) {\footnotesize{$\vec{k}$}}; 
        \end{scope}
\end{tikzpicture}}}
    \right|^2
    &= 
    \left|\bra{\text{f}} H_4  \ket{\text{i}} \right| ^2 
    = \frac{1}{16 N ^2} \sum_{\vec{q}_1, \vec{q}_2, \vec{q}_3, \vec{q}_4} \delta_{\vec{q}_1 +\vec{q}_2, \vec{q}_3 + \vec{q}_4}W_{\vec{q}_1, \vec{q}_2; \vec{q}_3, \vec{q}_4}\left|\bra{\text{f}} a^{\dagger}_{\vec{q}_1} a^{\dagger}_{\vec{q}_2}a_{\vec{q}_3}a_{\vec{q}_4}\ket{\text{i}}\right|^2 \\
    & = 
    \frac{1}{16 N ^2} \sum_{\vec{q}_1, \vec{q}_2, \vec{q}_3, \vec{q}_4} \delta_{\vec{q}_1 +\vec{q}_2, \vec{q}_3 + \vec{q}_4}\left| W_{\vec{q}_1, \vec{q}_2; \vec{q}_3, \vec{q}_4}\right|^2
    \left( N_{\vec{q}_3} + 1 \right)\left( N_{\vec{q}_4} + 1 \right) N_{\vec{q}_1} N_{\vec{q}_2}
    \\
    & 
    \quad\times \left[
    \delta_{\vec{q}_1, \vec{k}_1} \delta_{\vec{q}_2, \vec{k}_2} \delta_{\vec{q}_3, \vec{k}_3} \delta_{\vec{q}_4, \vec{k}} + \delta_{\vec{q}_1, \vec{k}_2} \delta_{\vec{q}_2, \vec{k}_1} \delta_{\vec{q}_3, \vec{k}_3} \delta_{\vec{q}_4, \vec{k}} + \left( \vec{q} \leftrightarrow \vec{q}_4\right)   
    \right]  \\
    & = 
    \frac{4}{16 N ^2} \delta_{\vec{k}_1 +\vec{k}_2, \vec{k}_3 + \vec{k}} \left|W_{\vec{k}_1, \vec{k}_2; \vec{k}_3, \vec{k}_4}\right|^2
    \left( N_{\vec{k}_3} + 1 \right)\left( N_{\vec{k}} + 1 \right) N_{\vec{k}_1} N_{\vec{k}_2}. 
    \end{split}
\end{equation}
We got a numerical factor of 4 from the interchangeability of the $\vec{q}_1, \vec{q}_2$ and $\vec{q}_3, \vec{q}_4$ momenta. 
Finally, to get the full scattering rate, we have to sum over the momenta $\vec{k}_1, \vec{k}_2, \vec{k}_3$. We need an additional factor of $1/2$ since the role of $\vec{k}_1$ and $\vec{k}_2$ is identical. We have in total
\begin{equation}
    \Gamma_{\vec{k}}^4 = \frac{2 \pi}{\hbar}\frac{1}{8N^2}\sum_{\vec{k}_1, \vec{k}_2, \vec{k}_3} \delta_{\vec{k}_1 +\vec{k}_2, \vec{k}_3 + \vec{k}} \left|W_{\vec{k}_1, \vec{k}_2; \vec{k}_3, \vec{k}}\right|^2
    \left( N_{\vec{k}_3} + 1 \right)\left( N_{\vec{k}} + 1 \right) N_{\vec{k}_1} N_{\vec{k}_2} \delta \left( \varepsilon_{\vec{k}_1} + \varepsilon_{\vec{k}_2} - \varepsilon_{\vec{k}} - \varepsilon_{\vec{k}_3}\right). 
\end{equation}
\end{subsection}

As a second example, we will contrast the above calculation of the four-magnon scattering rate with that of the three-magnon in-split process,
\begin{equation}
    \vcenter{\hbox{
    \begin{tikzpicture}[scale=1.3, decoration={markings, 
        mark= at position 0.65 with {\arrow{stealth}},
        mark= at position 2cm with {\arrow{stealth}}}
        ]   
            \draw[postaction={decorate}] (1.3,0.0)--(1.7, 0.2);
            \draw[postaction={decorate}] (1.3,0.0)--(1.7, -0.2);
            \draw[postaction={decorate}] (1,0)--(1.3,0);
            \draw[fill=black] (1.3,0) circle (0.07);
            \node at (1.45, 0.3) {\footnotesize{$\vec{k}_1$}};
            \node at (1.45, -0.3) {\footnotesize{$\vec{k}$}};
            \node at (0.8, 0.) {\footnotesize{$\vec{k}_2$}};
    \end{tikzpicture}}}.
\end{equation}
Here, by calculating
the squared amplitude, we get only a factor of 2, because we have two magnon operators that commute with each other (two creation operators). Additionally, since the role of $\vec{k}_1$ and $\vec{k}_2$ is not interchangeable, no factor of $1/2$ is needed when we perform the sum over the two momenta.

\section{Technical details of the numerical implementation}
\label{subsection: D's in the calculation}
We elaborate on specific aspects regarding the numerical implementation of the calculation. We begin by discussing which terms were taken into account for the calculation of the relaxation time $\tau_{\vec{k}}$ that enters the longitudinal and Hall conductivities. Then we proceed with some numerical aspects of the calculation of the two conductivities. 
     \begin{enumerate}
         \item   
         To start, we recall that the Hall and longitudinal conductivities defined in the main text are given by the expressions (cf.~also Sec.~\ref{section:Transport theory})
         \begin{equation}
             \begin{split}
                 & \kappa_{\text{H}} \equiv \frac{\kappa_{\mu \nu} - \kappa_{\nu \mu}}{2} = 
                 \frac{1}{2V T^2 k_{\text{B}}}\sum_{\vec{k}, \vec{k}'}\varepsilon_{\vec{k}}\varepsilon_{\vec{k}'}\tau_{\vec{k}} \tau_{\vec{k}'} 
                 G_{\vec{k}}G_{\vec{k}'}
                 \left(v_{\vec{k}}^{\mu} v_{\vec{k}'}^{\nu} - 
                  v_{\vec{k}'}^{\mu} v_{\vec{k}}^{\nu} \right) 
                  \mathcal{A}_{\vec{k} \vec{k}'}, \\
                & 
                \kappa_{\text{L}} \equiv \kappa_{\mu \mu} = \frac{1}{Vk_{\text{B}}T^2} \sum_{\vec{k}, \vec{k}'} \varepsilon_{\vec{k}}\varepsilon_{\vec{k}'} v_{\vec{k}'}^{\mu} v_{\vec{k}}^{\mu}G_{\vec{k}} G_{\vec{k}'}
    \left(
    \tau_{\vec{k}}  \delta_{\vec{k}, \vec{k}'} + \tau_{\vec{k}}\tau_{\vec{k}'}\mathcal{O}_{\vec{k}\vec{k}'}
    \right),
         \end{split}
         \label{eq: conductitivites restated}
         \end{equation}
    respectively, where $\varepsilon_{\vec{k}}$ is the free magnon energy, $v_{\vec{k}}^{\mu} = \left( 1 / \hbar\right)\partial \varepsilon_{\vec{k}} / \partial k_{\mu}$ the $\mu$-component of the magnon velocity, $\mathcal{O}_{\vec{k} \vec{k}'}$ an off-diagonal and $\mathcal{A}_{\vec{k} \vec{k}'} = \left(\mathcal{O}_{\vec{k} \vec{k}'}- \mathcal{O}_{\vec{k}' \vec{k}}\right)/2$ an anti-symmetric element of the collision matrix in Hardy's basis ( $G_{\vec{k}} = \sqrt{\left( \overline{N}_{\vec{k}} + 1\right)\overline{N}_{\vec{k}}}$), and $\tau_{\vec{k}} = 1 / D_{\vec{k}}$ the relaxation time defined as the inverse of the diagonal collision matrix element $D_{\vec{k}}$.  
   According to Eq.~\eqref{eq: conductitivites restated} both $\kappa_{\text{L}}$ and $ \kappa_{\text{H}}$ require knowledge of the relaxation time $\tau_{\vec{k}} = 1/ D_{\vec{k}}$.
   In our numerical implementation, we have set
   \begin{align}
       D_{\vec{k}} \approx D_{\vec{k}}^{3,1} + D_{\vec{k}}^{3,2} + D_{\vec{k}}^{4} + D_{\vec{k}}^\text{ph},
       \label{eq:definition of D}
   \end{align}
   where $D_{\vec{k}}^{3,1}$ and $D_{\vec{k}}^{3,2}$ derive from three-magnon interactions $\left( H_3\right)$ and $D_{\vec{k}}^{4}$ from four-magnon interactions $\left( H_4 \right)$. To construct these one must consider both the in- and out-counterpart of each process and then subtract them to derive the total diagonal element. Since for the three-magnon interactions we have two different possibilities (fusion and split processes), we have $D_{\vec{k}}^{3, 1} = D_{\vec{k}}^{\text{in-fusion}} - D_{\vec{k}}^{\text{out-split}}$, and $D_{\vec{k}}^{3, 2} = D_{\vec{k}}^{\text{in-split}} - D_{\vec{k}}^{\text{out-fusion}}$. On the other hand, there is only one four-magnon first order contribution $D_{4} = D_{4}^{\text{in}}-D_{4}^{\text{out}}$. We also include by hand a phenomenological Gilbert damping $D_{\vec{k}}^\text{ph}$, as obtained by calculating spin-wave energies from the linearized Landau-Lifshitz-Gilbert equation, to account for crystal imperfections and other sources of damping. The explicit expressions read as
         \begin{equation}
             \begin{split}
                 & D_{\vec{k}}^{3, 1} = -\frac{1}{2N}\frac{2 \pi}{\hbar}\sum_{\vec{k}_1, \vec{k}_2} \delta_{\vec{k}_1 + \vec{k}_2, \vec{k}} 
                 \left| 
                 V_{\vec{k}; \vec{k}_1, \vec{k}_2}
                 \right|^2 
                 \delta 
                 \left( 
                 \varepsilon_{\vec{k}} - \varepsilon_{\vec{k}_1} - \varepsilon_{\vec{k}_2}
                 \right) 
                 \left[
                 \overline{N}_{\vec{k}_1} \overline{N}_{\vec{k}_2} - \left(\overline{N}_{\vec{k}_1} + 1
                 \right) 
                 \left(
                 \overline{N}_{\vec{k}_2}  + 1
                 \right) 
                 \right],  \\
                 & D_{\vec{k}}^{3, 2} = -\frac{2}{N}\frac{2 \pi}{\hbar}\sum_{\vec{k}_1, \vec{k}_2} \delta_{\vec{k}_1 + \vec{k}, \vec{k}_2}
                 \left| 
                 V_{\vec{k}_2; \vec{k}_1, \vec{k}}
                 \right|^2
                 \delta 
                 \left( 
                 \varepsilon_{\vec{k}} + \varepsilon_{\vec{k}_1} - \varepsilon_{\vec{k}_2}
                 \right) 
                 \left[
                 \left(\overline{N}_{\vec{k}_1} + 1 \right)\overline{N}_{\vec{k}_2} - \left(\overline{N}_{\vec{k}_2} + 1 \right)\overline{N}_{\vec{k}_1} 
                 \right], 
                 \\
                 & 
                 D_{\vec{k}}^{4} = -\frac{1}{2N^2}\frac{2 \pi}{\hbar}\sum_{\vec{k}_1, \vec{k}_2, \vec{k}_3} 
                 \delta_{\vec{k}_3 + \vec{k}, \vec{k}_2 + \vec{k}_1} 
                 \delta 
                 \left( 
                 \varepsilon_{\vec{k}_3} +  \varepsilon_{\vec{k}} -  \varepsilon_{\vec{k}_2} -  \varepsilon_{\vec{k}_1}
                 \right) 
                 \left| 
                 W_{\vec{k}_1, \vec{k}_2 ; \vec{k}_3, \vec{k}}
                 \right|^2 
                 \left[
                 \overline{N}_{\vec{k}_1} \overline{N}_{\vec{k}_2} 
                 \left( 
                 \overline{N}_{\vec{k}_3} + 1
                 \right)
                 - 
                 \overline{N}_{\vec{k}_3} 
                 \left( 
                 \overline{N}_{\vec{k}_1} + 1
                 \right)
                 \left( 
                 \overline{N}_{\vec{k}_2} + 1
                 \right)
                 \right], \\
            & 
            D_{\vec{k}}^\text{ph} = \alpha_\text{G} \frac{\varepsilon_{\vec{k}}}{\hbar},
             \end{split}
         \label{eq: Ds for calculations}
         \end{equation}
         where $\alpha_\text{G}$ is the Gilbert damping constant. We set $\alpha_\text{G} = 10^{-3}$ in the calculations. We emphasize that according to Eq.~\eqref{eq:definition of D} the numerically implemented relaxation time does only include up to first-order scattering processes. 
         
         Regarding the off-diagonal elements $\mathcal{O}_{\vec{k} \vec{k}'}$ we have taken into account the terms shown in the three- and four-magnon tables of Sec.~\ref{subsection: diagrams} that are necessary to generate a finite $\mathcal{A}_{\vec{k} \vec{k}'}$. Moreover, the off-diagonal part of $\kappa_{\text{L}}$ has been dropped, simplifying the formula for the longitudinal conductivity to
         \begin{equation}
             \kappa_{\text{L}} 
             =  
             \frac{1}{Vk_{\text{B}}T^2} \sum_{\vec{k}, \vec{k}'} \varepsilon_{\vec{k}}\varepsilon_{\vec{k}'} v_{\vec{k}'}^{\mu} v_{\vec{k}}^{\mu}G_{\vec{k}}G_{\vec{k}'}
            \tau_{\vec{k}} \delta_{\vec{k}, \vec{k}'}
            =
            \frac{1}{V k_{\text{B}}T^2} \sum_{\vec{k}} 
            \tau_{\vec{k}}
            \left( \varepsilon_{\vec{k}}  v_{\vec{k}}^{\mu}G_{\vec{k}} \right)^2
            . 
         \end{equation} 
\item 
We continue explaining how we have numerically implemented the momentum sums appearing in the expressions for the conductivities [see Eq.~\eqref{eq: conductitivites restated}], in the inverse relaxation times [see Eq.~\eqref{eq: Ds for calculations}], and the off-diagonal elements $\mathcal{O}_{\vec{k} \vec{k}'}$ of the scattering kernel (see tables in Sec.~\ref{subsection: diagrams}). First, we discretize the first Brillouin zone using an $N \times N$ $\vec{k}$-mesh. The delta-functions appearing in the expressions of the diagonal and off-diagonal elements of the scattering matrix were replaced by Lorentzian functions, that is,
\begin{equation}
    \delta\left(\Delta E \right) \rightarrow 
    \frac{1}{\pi}\frac{\gamma}{\gamma ^2 + \Delta E ^2 },
\end{equation}
with $\gamma$ being the smearing (width) of the Lorentzian. The smearing $\gamma$ enters as an additional parameter and requires some optimization:
Too large a smearing would over-smooth the Lorentzian and capture processes that do not satisfy the kinematic constraints, while too small a smearing may miss an important amount of scattering channels. 
To estimate $\gamma$ we compute the steepest possible change to the energy of the free magnon spectrum, $\Delta E_\text{max}$, that results from changing the magnon momentum by $\Delta \vec{k}$. In our code, $\Delta \vec{k}$ is the distance between two neighboring $\vec{k}$-points and for a two-dimensional square Brillouin zone containing a total of $N\times N$ points, it is given by $\Delta \vec{k} = \frac{2 \pi}{N}\left(\vec{\hat{x}} + \vec{\hat{y}} \right)$. We can compute $\Delta E_\text{max}$ as
\begin{equation}
\begin{split}
     \Delta E_\text{max} &= 
     \vec{\Delta \vec{k}} \cdot \vec{v}_{\vec{k}}^{\text{max}} \\
     & 
     = 2JS \frac{2 \pi}{N}\left.\left( \sin k_x + \sin k_y \right)\right|_{\text{max}} \\
     & 
     = 
     8\pi\frac{JS}{N}.
\end{split}
\end{equation}
Since at low temperatures the dominant contributions to the scattering come from the parabolic bottom of the band, where the change in the energy $\Delta E$ is relatively small, we can take for $\gamma$ a value smaller than the estimated $\Delta E_{\text{max}}$. More specifically, we observe that for a number of points in the range of $N \sim 50 - 100$ and $J=1\,\text{meV}$, $S=1$, a value of $\gamma \sim 0.1 \Delta E_\text{max}$ produces well-converged results. 
Finally, since the calculation of the off-diagonal elements scales as $N^6$, we introduced an energy cut-off $E_{\text{cut-off}}$ for the Lorentzian and discarded processes with $\Delta E>E_{\text{cut-off}}$ to lower the computational cost of the calculation. 

The results in the main text were obtained for $\gamma/J = 0.038$ and $E_{\text{cut-off}}/J = 0.075$.
\end{enumerate}

\begin{section}{Additional Results}

\begin{subsection}{Temperature dependence and energy-scale relations of $\kappa_{\text{H}}$ and $\kappa_{\text{L}}$}
\label{sec:Temperaturedependence}
Here we discuss two relevant scaling behaviors of the longitudinal ($\kappa_{\text{L}}$) and Hall conductivity ($\kappa_{\text{H}}$). The mathematical expressions of the conductivities are rewritten below, to facilitate the smoother reading of this section:
\begin{equation}
    \begin{split}
        & 
        \kappa_{\text{L}} \equiv 
        \kappa_{\mu \mu}
        =
        \frac{1}{V k_{\text{B}}T^2} \sum_{\vec{k}} \varepsilon^2_{\vec{k}} \left( v_{\vec{k}}^{\mu}\right)^2 G^2_{\vec{k}}
        \tau_{\vec{k}}, \\
        & 
        \kappa_{\text{H}} 
        \equiv 
        \kappa_{\mu \nu} = 
                 \frac{1}{2V T^2 k_{\text{B}}}\sum_{\vec{k}, \vec{k}'}\varepsilon_{\vec{k}}\varepsilon_{\vec{k}'}\tau_{\vec{k}} \tau_{\vec{k}'} 
                 G_{\vec{k}}G_{\vec{k}'}
                 \left(v_{\vec{k}}^{\mu} v_{\vec{k}'}^{\nu} - 
                  v_{\vec{k}'}^{\mu} v_{\vec{k}}^{\nu} \right) 
                  \mathcal{A}_{\vec{k} \vec{k}'}.
    \end{split}
\label{eq: conductivities restated}
\end{equation}
Note that we have dropped the off-diagonal contributions to $\kappa_{\text{L}}$, as explained in Sec.~\ref{subsection: D's in the calculation}.
We first analyze the temperature dependence of $\kappa_{\text{H}}$ and $\kappa_{\text{L}}$ for $k_{\text{B}}T \gg \varepsilon_{\vec{k}}$ and $k_{\text{B}}T \ll \varepsilon_{\vec{k}}$, where $\varepsilon_{\vec{k}}$ 
is the free-magnon dispersion. Second, we analyze how the conductivities scale upon rescaling of the magnetic exchange parameters.

\subsubsection{Temperature dependence}
Apart from the factor of $1/T^2$ multiplying each conductivity, temperature enters the Bose-Einstein occupation numbers
\begin{equation}
    \overline{N}_{\vec{k}} = \frac{1}{\mathrm{e}^{\beta \varepsilon_{\vec{k}}} - 1}
\end{equation}
($\beta = 1/ k_\text{B} T$), which are contained in (i) the $G_{\vec{k}} = \sqrt{\left(\overline{N}_{\vec{k}} + 1 \right) \overline{N}_{\vec{k}}}$ factors coming from expressing the collision matrix in Hardy's basis, and (ii) the elements of the collision matrix. The bare magnon energies $\varepsilon_{\vec{k}}$ do not carry any temperature dependence, because many-body renormalizations of the spectrum have been neglected. 

There are two types of elements of the collision matrix: the diagonal ones $D_{\vec{k}} = 1 / \tau_{\vec{k}}$, where $\tau_{\vec{k}}$ is the relaxation time, and the off-diagonal ones $\mathcal{O}_{\vec{k} \vec{k}'}$. As it can be seen in Eq.~\eqref{eq: conductivities restated}, $D_{\vec{k}}$ appears in both $\kappa_{\text{L}}$ and $\kappa_{\text{H}}$ while $\mathcal{O}_{\vec{k} \vec{k}'}$ only in the Hall conductivity, as it is contained in the anti-symmetric part of the collision matrix $\mathcal{A}_{\vec{k} \vec{k}'} = \left(\mathcal{O}_{\vec{k} \vec{k}'} - \mathcal{O}_{\vec{k}' \vec{k}}\right) / 2$. 
Regarding the diagonal elements, we recall our approximation to only include first order scattering events, setting $D_{\vec{k}} \approx D_{\vec{k}}^{\text{ph}} + D_{\vec{k}}^{3, 1} + D_{\vec{k}}^{3, 2} + D_{\vec{k}}^{4}$ (see Eq.~\eqref{eq: Ds for calculations} in Sec.~\ref{subsection: D's in the calculation}). Only $D_{\vec{k}}^{\text{ph}}$ is not temperature dependent. Using the global energy conservation for each of the terms in the $D_{\vec{k}}$, as well as the relation $\overline{N}_{\vec{k}} \mathrm{e}^{\beta \varepsilon_{\vec{k}}} =  \overline{N}_{\vec{k}} + 1$, the diagonal elements can be written in the more compact form 
\begin{equation}
    \begin{split}
        & D_{\vec{k}}^{3, 1} = \frac{1}{2N}\frac{2 \pi}{\hbar}\sum_{\vec{k}_1, \vec{k}_2} \delta_{\vec{k}_1 + \vec{k}_2, \vec{k}} 
                 \left| 
                 V_{\vec{k}; \vec{k}_1, \vec{k}_2}
                 \right|^2 
                 \delta 
                 \left( 
                 \varepsilon_{\vec{k}} - \varepsilon_{\vec{k}_1} - \varepsilon_{\vec{k}_2}
                 \right) 
                  \overline{N}_{\vec{k}_1} \overline{N}_{\vec{k}_2}
                  \frac{1}{\overline{N}_{\vec{k}}}, 
                 \\
                 & D_{\vec{k}}^{3, 2} = \frac{2}{N}\frac{2 \pi}{\hbar}\sum_{\vec{k}_1, \vec{k}_2} \delta_{\vec{k}_1 + \vec{k}, \vec{k}_2}
                 \left| 
                 V_{\vec{k}_2; \vec{k}_1, \vec{k}}
                 \right|^2
                 \delta 
                 \left( 
                 \varepsilon_{\vec{k}} + \varepsilon_{\vec{k}_1} - \varepsilon_{\vec{k}_2}
                 \right) 
                 \overline{N}_{\vec{k}_1}\overline{N}_{\vec{k}_2} \frac{1}{\overline{N}_{\vec{k}}} \mathrm{e}^{\beta \varepsilon_{\vec{k}_1}}
                 , 
                 \\
                 & 
                 D_{\vec{k}}^{4} = \frac{1}{2N^2}\frac{2 \pi}{\hbar}\sum_{\vec{k}_1, \vec{k}_2, \vec{k}_3} 
                 \delta_{\vec{k}_3 + \vec{k}, \vec{k}_2 + \vec{k}_1} 
                 \delta 
                 \left( 
                 \varepsilon_{\vec{k}_3} +  \varepsilon_{\vec{k}} -  \varepsilon_{\vec{k}_2} -  \varepsilon_{\vec{k}_1}
                 \right) 
                  \overline{N}_{\vec{k}_1} \overline{N}_{\vec{k}_2} 
                 \overline{N}_{\vec{k}_3}
                 \left| 
                 W_{\vec{k}_1, \vec{k}_2 ; \vec{k}_3, \vec{k}}
                 \right|^2 
                \frac{1}{\overline{N}_{\vec{k}}} \mathrm{e}^{\beta \varepsilon_{\vec{k}}}. 
    \label{eq: D's compact}
    \end{split}
\end{equation}
For $\mathcal{O}_{\vec{k} \vec{k}'}$ (and equivalently for $\mathcal{A}_{\vec{k} \vec{k}'}$) we have many different processes that contribute, all of them coming from the interference of first and second order scattering channels. To understand the temperature dependence of the different interference processes we must consider how many occupation number each different kind of process can contain. The ones with the largest number will provide the leading contribution in the high-$T$ limit since the occupation numbers acquire large values for large temperatures. On the other hand,  the ones with the smallest amount of occupation numbers will account for the leading contribution in the low-$T$ limit, where $\overline{N}_{\vec{k}}$ are small. By taking a look at the diagram tables of Sec.~\ref{subsection: diagrams}, we observe that the leading term for both the high and low temperature regime can be generated from in-processes of three-magnon diagrams that contain two virtual magnons. In the temperature analysis that follows we will use as a representative contribution to the off-diagonal elements the $\mathcal{O}_{\vec{k} \vec{k}_1}$ element that comes from the following in-fusion diagram:
\begin{equation}
\begin{split}
\mathcal{O}_{\vec{k} \vec{k}_1}^{+-}
& = 
\left(
\vcenter{\hbox{
    \begin{tikzpicture}[scale=1.3, decoration={markings, 
        mark= at position 0.65 with {\arrow{stealth}},
        mark= at position 2cm with {\arrow{stealth}}}
        ]   
            \begin{scope}
           \draw[postaction={decorate}] (1,0.2)--(1.3,0.0);
            \draw[postaction={decorate}] (1,-0.2)--(1.3,0.0);
            \draw[postaction={decorate}] (1.3,0)--(1.7,0.0);
            \draw[fill=black] (1.3,0) circle (0.07);
            \node at (1.2, 0.4) {\footnotesize{$\vec{k}_1$}};
            \node at (1.2, -0.4) {\footnotesize{$\vec{k}_2$}};
            \node at (1.65, 0.2) {\footnotesize{$\vec{k}$}};
            \end{scope}
            \begin{scope}[xshift = 0.9cm, yshift=1.2cm]
                 \begin{scope}
                  \draw[postaction={decorate}] (1.4,-1)--(1.7,-1.2);
                  \draw[postaction={decorate}] (1.4,-1.4)--(1.7,-1.2);
                  \draw [postaction={decorate}] (1.7,-1.2) to [out=70,in=110] (2.2,-1.2);
                  \draw [postaction={decorate}] (1.7,-1.2) to [out=-70,in=-110] (2.2,-1.2);
                  \draw[postaction={decorate}] (2.2,-1.2)--(2.7,-1.2);
                  \draw[fill=white] (1.7,-1.2) circle (0.07);
                  \draw[fill=black] (2.2,-1.2) circle (0.07);
                  \node at (2.5, -1) {\footnotesize{$\vec{k}$}}; 
                  \node at (2, -0.85) {\footnotesize{$\vec{p}_1$}}; 
                  \node at (2, -1.55){\footnotesize{$\vec{p}_2$}};
                  \node at (1.55, -0.8) {\footnotesize{$\vec{k}_1$}};
                  \node at (1.55, -1.5) {\footnotesize{$\vec{k}_2$}};
                 \end{scope}
            \end{scope}  
            \node at (2, 0) {$\times$};
    \end{tikzpicture}}} 
\right)_{\vec{k} \vec{k}_1} \\
& = 
16 \frac{\left(2\pi\right)^2}{\hbar}\frac{G_{\vec{k}_1}}{G_{\vec{k}}}
\left(
\overline{N}_{\vec{k}}+1
\right)
\sum_{\vec{k}_2}\delta_{\vec{k}_1,\vec{k}-\vec{k_2}} 
V_{\vec{k}; \vec{k}_1, \vec{k}_2}^{*}
\overline{N}_{\vec{k}_2}
\delta
\left(
\varepsilon_{\vec{k}}-\varepsilon_{\vec{k}_1}-\varepsilon_{\vec{k}_2}
\right) \\
& \quad \times 
\sum_{\vec{p}_1, \vec{p}_2} \delta_{\vec{p}_1 + \vec{p}_2, \vec{k}_1 + \vec{k}_2} 
\delta
\left(\varepsilon_{\vec{k}_1} + \varepsilon_{\vec{k}_2}-\varepsilon_{\vec{p}_1}-\varepsilon_{\vec{p}_2}
\right)
W_{\vec{k}_1, \vec{k}_2; \vec{p}_1, \vec{p}_2}  
\left(
\overline{N}_{\vec{p}_1} + 1\right)
\left(
\overline{N}_{\vec{p}_2} + 1
\right).
\end{split}
\end{equation}

\begin{enumerate}
    \item{\textbf{High temperature regime}} \\
    For the high-$T$ limit corresponding to $k_{\text{B}} T \gg \varepsilon_{\vec{k}}$, the Bose-Einstein population can be approximated by 
    $
        \overline{N}_{\vec{k}} \approx k_\text{B}T / \varepsilon_{\vec{k}}.
    $ 
    and the exponential $\mathrm{e}^{\beta \varepsilon_{\vec{k}}}$ by  $\mathrm{e}^{\beta \varepsilon_{\vec{k}}} \approx 1$. With these relations in mind the components of $D_{\vec{k}}$ and the off-diagonal element $\mathcal{O}_{\vec{k} \vec{k}_1}^{+-}$ yield the following high-temperature dependence:
\begin{equation}
    \begin{split}
       & 
       D_{\vec{k}}^{3,1} \sim T, \\
       & 
       D_{\vec{k}}^{3,2} \sim T, \\
       & 
       D_{\vec{k}}^{4} \sim T^2, \\
       & 
       \mathcal{O}_{\vec{k} \vec{k}_1}^{+-} \sim T^4.
    \end{split}
\end{equation}
By taking the $D_{\vec{k}}^{4} \sim T^2$ as the leading component of $D_{\vec{k}}$ and considering that $G_{\vec{k}} \sim T$ for high temperatures, we get from Eq.~\eqref{eq: conductitivites restated} the following leading temperature dependence of the conductivities:
\begin{equation}
\begin{split}
   & \kappa_{\text{L}} \sim \frac{1}{T^2}, \\
   & \kappa_{\text{H}} \sim \text{constant}.
\end{split}
\end{equation}
The $1/T^2$ decay of the longitudinal conductivity is a reasonable result and can be interpreted to arise from Umklapp scattering. On the other hand, the constant for the Hall conductivity is \textit{not} physically meaningful and it arises as a byproduct of the approximations that we have implemented. Below, we explain the origin of, and what to do to lift the incorrect high-temperature limit of $\kappa_{\text{H}}$.

By including only first-order scattering processes in $D_{\vec{k}}$ (recall Sec.~\ref{subsection: D's in the calculation}) it lacks contributions from exactly those higher-order scattering processes that give rise to a finite $\mathcal{A}_{\vec{k} \vec{k}'}$. Since higher-order scattering comes with a higher power in temperature at large temperatures, the factor $\tau_{\vec{k}} \tau_{\vec{k}'} \mathcal{A}_{\vec{k} \vec{k}'} = D^{-1}_{\vec{k}} D^{-1}_{\vec{k}'} \mathcal{A}_{\vec{k} \vec{k}'}$ in the Hall conductivity [e.g., see Eq.~\eqref{eq: conductivities restated}] gets spuriously dominated by $\mathcal{A}_{\vec{k} \vec{k}'}$ as $T \to \infty$, leading to an overestimation of the Hall effect. Thus, if we included higher-order scattering in $D_{\vec{k}}$, such that $D_{\vec{k}} \sim T^\gamma$ with $\gamma > 2$, we would obtain $\kappa_\text{H} \sim 1/T^{2\gamma-4}$ as $T \to \infty$. Specifically, by including interference terms between first and second order scattering, we would find $\gamma = 4$ and 
\begin{equation}
\begin{split}
   & \kappa_{\text{L}} \sim \frac{1}{T^4}, \\
   & \kappa_{\text{H}} \sim \frac{1}{T^4},
\end{split}
\end{equation}
as $T \to \infty$, causing both conductivities to vanish with the same exponent.

(At temperatures of the order of the exchange interaction, one would expect thermal fluctuations to melt the magnetic order and the expansion in magnons around a magnetically ordered state to break down. Within our transport theory, the many-body renormalization effects of the spectrum could be taken into account partially by rescaling the bare magnon energies. One would have to replace the spin by the average magnetization, $S \rightarrow \langle M\left(T \right) \rangle$, and solve self-consistently. We have not attempted to implement any self consistent corrections, because the aforementioned strategy already provides a cure of the high-temperature limit within the assumption of an unrenormalized magnon spectrum, which is possible because the external field always partially polarizes the magnet, rendering the magnon picture useful.)
 
\item {\textbf{Low temperature regime}} \\
We now analyze the regime of low temperatures corresponding to $k_\text{B}T \ll J, \Delta$. In this regime the Bose-Einstein occupation numbers can be approximated by $\overline{N}_{\vec{k}} \sim \mathrm{e}^{-\beta \varepsilon_{\vec{k}}} \approx \mathrm{e}^{-\beta \Delta} \ll 1$, where we have retained only the very bottom of the magnon spectrum (the magnetic field $\Delta$) in the exponent. We continue by observing that the leading contribution to the $D_{\vec{k}}$ is in this case the temperature independent Gilbert-like damping $D_{\vec{k}} \sim D_{\vec{k}}^{\text{ph}} = a_{\text{G}} \varepsilon_{\vec{k}} / \hbar$. Regarding the exemplary off-diagonal element $\mathcal{O}_{\vec{k} \vec{k}_1}$, its leading component is the one with the lowest amount of occupation numbers and bears the dependence of $\mathcal{O}_{\vec{k} \vec{k}_1} \sim \mathrm{e}^{-\beta \Delta}$. By considering that in this regime $G_{\vec{k}}^2 \sim \mathrm{e}^{-\beta \Delta}$ we have for the two conductivities the following scaling behavior with the temperature
\begin{equation}
\begin{split}
    & 
    \kappa_{\text{L}} = \frac{\mathrm{e}^{-\beta \Delta}}{T^2}, \\
    & 
    \kappa_{\text{H}} = \frac{\mathrm{e}^{-2 \beta \Delta}}{T^2}.
\end{split}
\end{equation}

Importantly, the above result for the Hall conductivity $\kappa_{\text{H}} \sim \mathrm{e}^{-2 \beta \Delta}/T^2$ has been deducted for a specific off-diagonal element based on the fact that it contains the least amount of occupation factors. Although this can provide an intuitive understanding of the temperature scaling, it ignores the fact that each off-diagonal element is a complicated function of the magnon momenta and energies, containing three-and four-magnon vertex functions as well as kinematic restrictions from two delta functions. The total contribution of each diagram cannot be estimated only based on their temperature scaling; off-diagonal elements with seemingly sub-leading $T$-scaling can be equally large as the leading ones due to the interplay of the scattering vertices as well as the kinematic restrictions imposed on them. In principle on could expect that the realistic temperature scaling of the Hall conductivity is one of $\left(a_1\mathrm{e}^{-r_1 \beta \Delta} + a_2\mathrm{e}^{-r_2 \beta \Delta} \right)/T^2$ with $r_1, r_2 >1$ and $a_1, a_2$ two temperature independent constants. 

Thus, the main message for the low-temperature regime is that there is a thermal activation window for both conductivities, but with different activation behavior. Since the exponent of $\kappa_\text{H}$ is larger than that of $\kappa_\text{L}$, higher temperatures are required to induce a sizable Hall conductivity in comparison to the longitudinal one. 

(Regarding our choice to keep only the magnetic field from the magnon energies $\varepsilon_{\vec{k}} = 2JS\left(2 - \cos k_x - \cos k_y \right) + \Delta$. We expect that in the parabolic approximation limit of $\varepsilon_{\vec{k}} = 2JS\left(2 - \cos k_x - \cos k_y \right) + \Delta \approx JS\vec{k}^2 + \Delta$---valid for low temperatures---the $\vec{k}$-dependent part of the spectrum will give rise to an additional temperature dependence of $T^{-r'}$ ($r'>1$) upon integration over the $\vec{k}$-space. Such additional dependence will be overshadowed by the strong exponential decay arising from the Boltzmann activation factors that are connected to the bottom of the dispersion, $\mathrm{e}^{-\beta \Delta}$.)
\end{enumerate}

\subsubsection{Behavior of the conductivities upon scaling of energy parameters}
\label{subsub: scaling of conductivities}
We investigate here how $\kappa_\text{H}$ and $\kappa_{\text{L}}$ behave when the energy parameters of the system are scaled by a dimensionless positive number $\lambda$:
\begin{equation}
    \begin{split}
        & J \rightarrow \lambda J, \\
        & D \rightarrow \lambda D, \\
        & \Delta \rightarrow \lambda \Delta, \\
        & T \rightarrow \lambda T.
    \end{split}
\label{eq: scale relations}
\end{equation}
Under this scaling, the magnon energies get also rescaled, $\varepsilon_{\vec{k}} \rightarrow \lambda \varepsilon_{\vec{k}}$, and, consequently, the Bose-Einstein occupation numbers, $\overline{N}_{\vec{k}} = ( \mathrm{e}^{\beta \varepsilon_{\vec{k}}} -1)^{-1}$ remain unchanged.
As a result, the diagonal elements of the scattering kernel, $D_{\vec{k}}$, appearing in Eq.~\eqref{eq: D's compact}, experience the scaling only via the four- and three-magnon vertices and the delta-functions imposing global energy conservation. Regarding the vertices, we have for the four-magnon vertex $W \to \lambda W$ because of $J \to \lambda J$, and for the three-magnon vertex $V \to \lambda V$ because of $D \to \lambda D$. 
Furthermore, the delta-functions have units of inverse energy and thus scale as $\delta \left(\Delta E \right) \rightarrow \frac{1}{\lambda} \delta \left(\Delta E \right)$. In total we obtain
\begin{equation}
    \begin{split}
        & 
        D_{\vec{k}}^{4} \rightarrow \lambda  D_{\vec{k}}^{4}, \\
        & 
        D_{\vec{k}}^{3, 1} \rightarrow \lambda  D_{\vec{k}}^{3, 1}, \\
        & 
        D_{\vec{k}}^{3, 2} \rightarrow \lambda  D_{\vec{k}}^{3, 2}, \\
        & 
        D_{\vec{k}}^{\text{ph}} \rightarrow \lambda  D_{\vec{k}}^{\text{ph}}.
    \end{split}
\end{equation}
Note that the phenomenological Gilbert-like damping term scales also in the same way as the rest since it is proportional to the magnon energy, $ D_{\vec{k}}^{\text{ph}} = a_G \varepsilon_{\vec{k}} / \hbar$. Thus, the total diagonal element can be seen to scale as
\begin{equation}
    D_{\vec{k}} \rightarrow \lambda D_{\vec{k}}.
    \label{eq: scaling of total D}
\end{equation}
The off-diagonal elements of the collision kernel, $\mathcal{O}_{\vec{k} \vec{k}'}$, resulting from the interference processes shown in the tables of Sec.~\ref{subsection: diagrams} scale as $\mathcal{O}_{\vec{k} \vec{k}'} \sim J D^2$ and contain two delta-functions (one from the global energy conservation and one from resonant scattering). Consequently we get that the anti-symmetric part of the collision matrix $\mathcal{A}_{\vec{k} \vec{k}'} = \left(\mathcal{O}_{\vec{k} \vec{k}'}- \mathcal{O}_{\vec{k}' \vec{k}} \right) / 2$ scales as 
\begin{equation}
    \mathcal{A}_{\vec{k} \vec{k}'} \rightarrow \frac{\lambda^3}{\lambda^2} \mathcal{A}_{\vec{k} \vec{k}'} = \lambda \mathcal{A}_{\vec{k} \vec{k}'}.
    \label{eq: scaling of A}
\end{equation}
Combining Eqs.~\eqref{eq: scale relations}, \eqref{eq: scaling of total D}, and \eqref{eq: scaling of A}, and considering that $\tau_{\vec{k}} = 1 / D_{\vec{k}}$, we get for the Hall and longitudinal conductivities shown in Eq.~\eqref{eq: conductitivites restated} that
\begin{equation}
    \begin{split}
        & \kappa_\text{L} \rightarrow \lambda \kappa_\text{L}, \\
        & \kappa_\text{H} \rightarrow \lambda \kappa_\text{H}.
    \end{split}
\end{equation}
\end{subsection}

\begin{subsection}{Thermal conductivities as a function of Gilbert damping}
\label{subsec: values of GD}
 \begin{figure}
        \centering
        \includegraphics{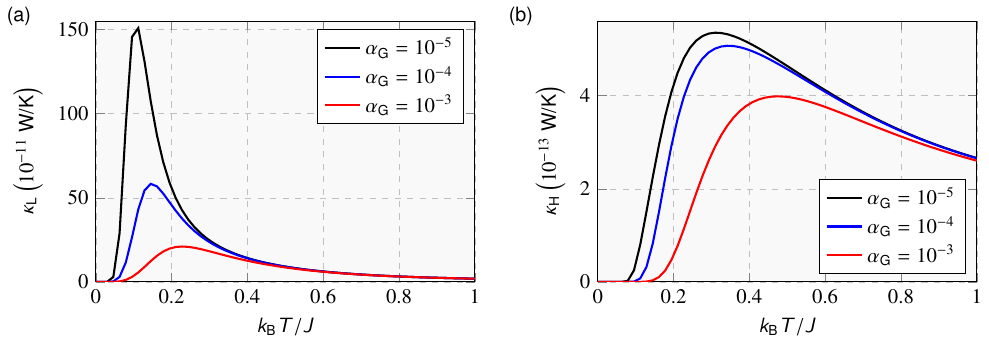}
        \caption{Temperature dependence of (a) the longitudinal thermal conductivity $\kappa_\text{L}$, and (b) the thermal Hall $\kappa_\text{H}$ for selected parameters of Gilbert damping $\alpha_\text{G}$. 
        The magnetic field is kept constant, $\Delta = 0.4 J$.} 
        \label{fig:a_G dependence}
        \end{figure}
As explained in Sec.~\ref{subsection: D's in the calculation}, our inverse relaxation times include a phenomenological Gilbert damping-type contribution $D_{\vec{k}}^{\text{ph}}=\alpha_\text{G} \varepsilon_{\vec{k}} /\hbar$. In the rest of the manuscript, we have set $\alpha_\text{G} = 10^{-3}$. In Fig.~\ref{fig:a_G dependence}, we explore the conductivities as a function of $\alpha_\text{G}$. The longitudinal conductivity $\kappa_{\text{L}}$ in Fig.~\ref{fig:a_G dependence}(a) exhibits a sharp increase at low temperatures as $\alpha_\text{G}$ is reduced by an order of magnitude. At low temperatures, many-body scattering is thermally suppressed, the Gilbert damping remains as the only source of resistivity, and one finds $\kappa_\text{L} \propto \alpha_\text{G}^{-1} \exp(-\beta \Delta)$. In contrast, the Hall conductivity $\kappa_{\text{H}}$ does not exhibit a similarly drastic increase with decreasing $\alpha_\text{G}$. Analytically, we find $\kappa_\text{H} \propto \alpha_\text{G}^{-2} \exp(-2\beta \Delta)$. Although there is the $1/\alpha^2_\text{G}$ scaling at low temperatures, the increased activation gap (recall discussion in Sec.~\ref{sec:Temperaturedependence}) completely suppresses the low-temperature signal. Once the activation gap is overcome, temperature is already so large that many-body scattering takes over as the leading resistive channel. As a result, $\kappa_\text{H}$ stays relatively unaffected by a change of $\alpha_\text{G}$.
\end{subsection}

\begin{subsection}{Freezing of scattering phase space by increasing magnetic field}
\label{subsection: frozen phase space}
In the main text, we have shown that the thermal Hall conductivity  $\kappa_\text{H}$ decreases with increasing magnetic field $\tilde{\Delta} = \Delta / J$. Here, we provide a detailed explanation of this observation.

$\kappa_\text{H}$ is generated by the anti-symmetric part of the collision matrix, $\mathcal{A}_{\vec{k} \vec{k}'}$, which, to lowest order in $1/S$, arises from the interference of first and second order scattering events. According to the arguments in Sec.~\ref{subsection: diagrams}, there are two types of interference processes:
\begin{enumerate}
    \item A first-order three-magnon process interferes with a second-order three-magnon process that is a combination of a three-magnon and a four-magnon scattering event (recall Tabs.~\ref{tab:in split} and \ref{tab:In fusion}).
    \item A first-order four-magnon process interferes with a second-order four-magnon process built from two three-magnon scattering events (recall Tab.~\ref{tab:4-magnons SM}). 
\end{enumerate}
Thus, all interference terms are necessarily associated with some three-magnon scattering events. Since the interference processes can only cause a finite $\mathcal{A}_{\vec{k} \vec{k}'}$ if the intermediate scattering is \emph{resonant} (recall Sec.~\ref{section: Interference process}), all interference terms come with a three-magnon energy conservation.

There are two options for three-magnon energy conservation. Either a magnon with energy $\varepsilon_{\vec{k}}$ \textit{decays} into two other magnons with energies $\varepsilon_{\vec{q}}$ and $\varepsilon_{\vec{k}-\vec{q}}$ or it \textit{collides} with a magnon with energy $\varepsilon_{\vec{q}}$ to fuse into a magnon with $\varepsilon_{\vec{k}+\vec{q}}$. Whether these processes are kinematically possible, that is, whether there are final states available to scatter into, is dictated by the respective density of states (DOS). The first decay process is possible for a magnon with energy $\varepsilon_{\vec{k}}$ if the two-magnon DOS
\begin{align}
    \mathcal{D}_{\vec{k}} (\omega)
    = 
    \frac{1}{N} \sum_{\vec{q}} \delta\left(\omega -\varepsilon_{\vec{q}} - \varepsilon_{\vec{k} - \vec{q}} \right)
\end{align}
is finite at $\omega = \varepsilon_{\vec{k}}$; this condition is well known in the context of spontaneous magnon decays \cite{zhitomirskychernyshev2013} and the manifold of the two-magnon states are also referred to as the decay continuum. The second \textit{collision} process is governed by the collision DOS (also: collision continuum)
\begin{align}
    \mathcal{D}'_{\vec{k}}(\omega) = \frac{1}{N}\sum_{\vec{q}}\delta \left(\omega + \varepsilon_{\vec{q}} -\varepsilon_{\vec{q} + \vec{k}}\right),
\end{align}
which also has to be finite at $\omega = \varepsilon_{\vec{k}}$.

Importantly, since the magnon energies $\varepsilon_{\vec{k}} = 2JS\left(2- \cos k_x - \cos k_y \right) + \Delta$ are functions of the magnetic field $\Delta$ and both $\mathcal{D}_{\vec{k}}(\varepsilon_{\vec{k}})$ and $\mathcal{D}'_{\vec{k}}(\varepsilon_{\vec{k}})$ contain delta functions with \emph{three} magnons, the two DOS bear a dependence on the magnetic field. Specifically, upon changing the field by $\Delta \rightarrow \Delta + a$, the decay continuum moves relatively to the single-magnon energies by $a$, and single-magnon energies move relatively to the collision continuum by $a$ as well. Consequently, for large enough magnetic fields, specifically $\Delta > \Delta' = 2JS$, none of the continua overlap anymore with the single magnon band, and the DOS available to the two kinds of three-magnon scattering events vanishes: $\mathcal{D}_{\vec{k}}(\varepsilon_{\vec{k}}), \mathcal{D}'_{\vec{k}}(\varepsilon_{\vec{k}}) \rightarrow 0$.

\begin{figure}
     \centering
   \includegraphics[width=\textwidth]{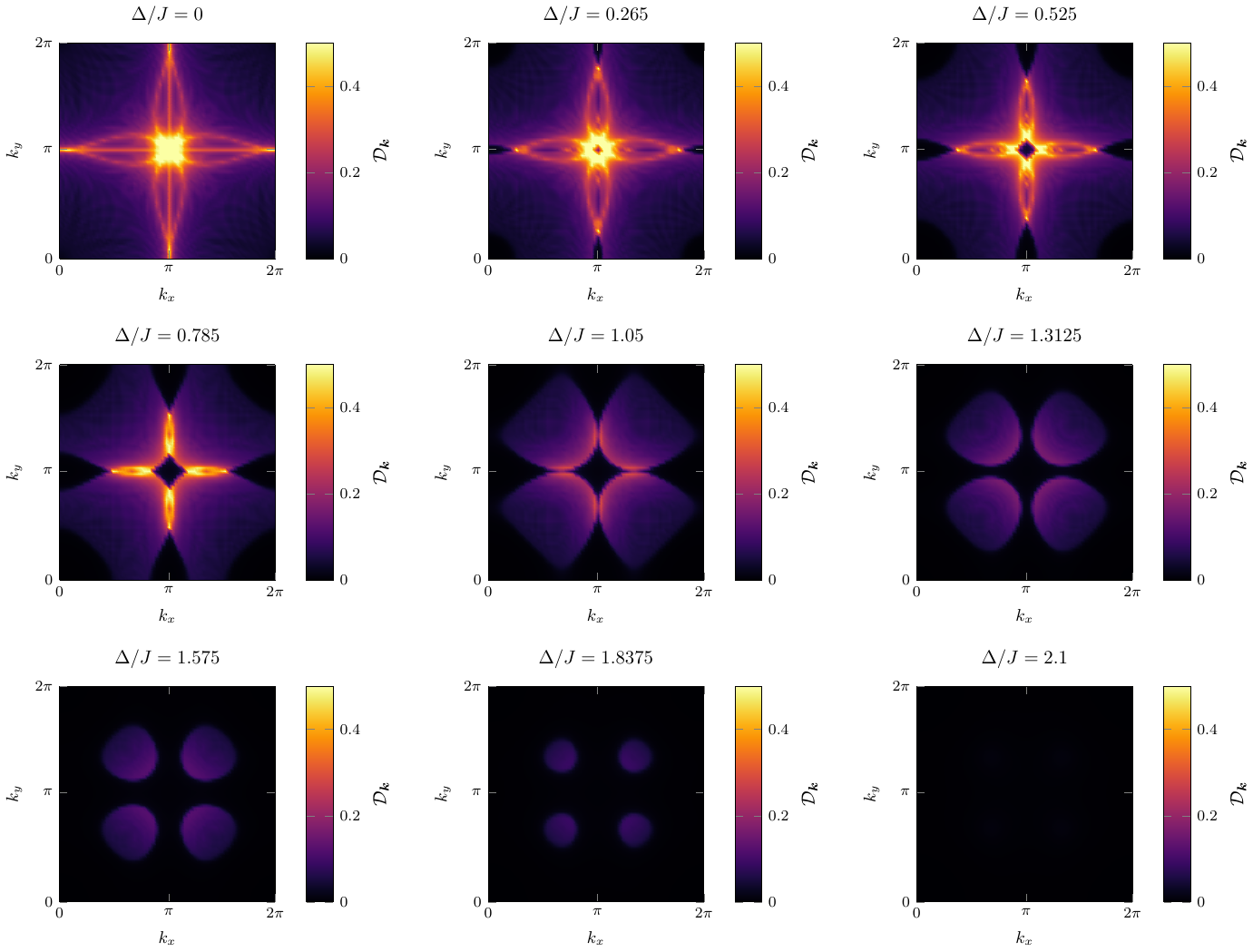}
     \caption{The two-magnon DOS $\mathcal{D}_{\vec{k}}(\varepsilon_{\vec{k}})$, quantifying the number of available final states for a two-magnon decay, as a function of magnetic field $\Delta$ ($S=1$). The DOS is shown as a color-map in the first Brillouin zone for a growing magnetic field $\Delta / J$, with a minimum value of $\Delta / J = 0$ and a maximum of $\Delta / J = 2.1$.}
    \label{fig:DOS decay}
\end{figure} 

\begin{figure}
    \centering
     \includegraphics[width=\textwidth]{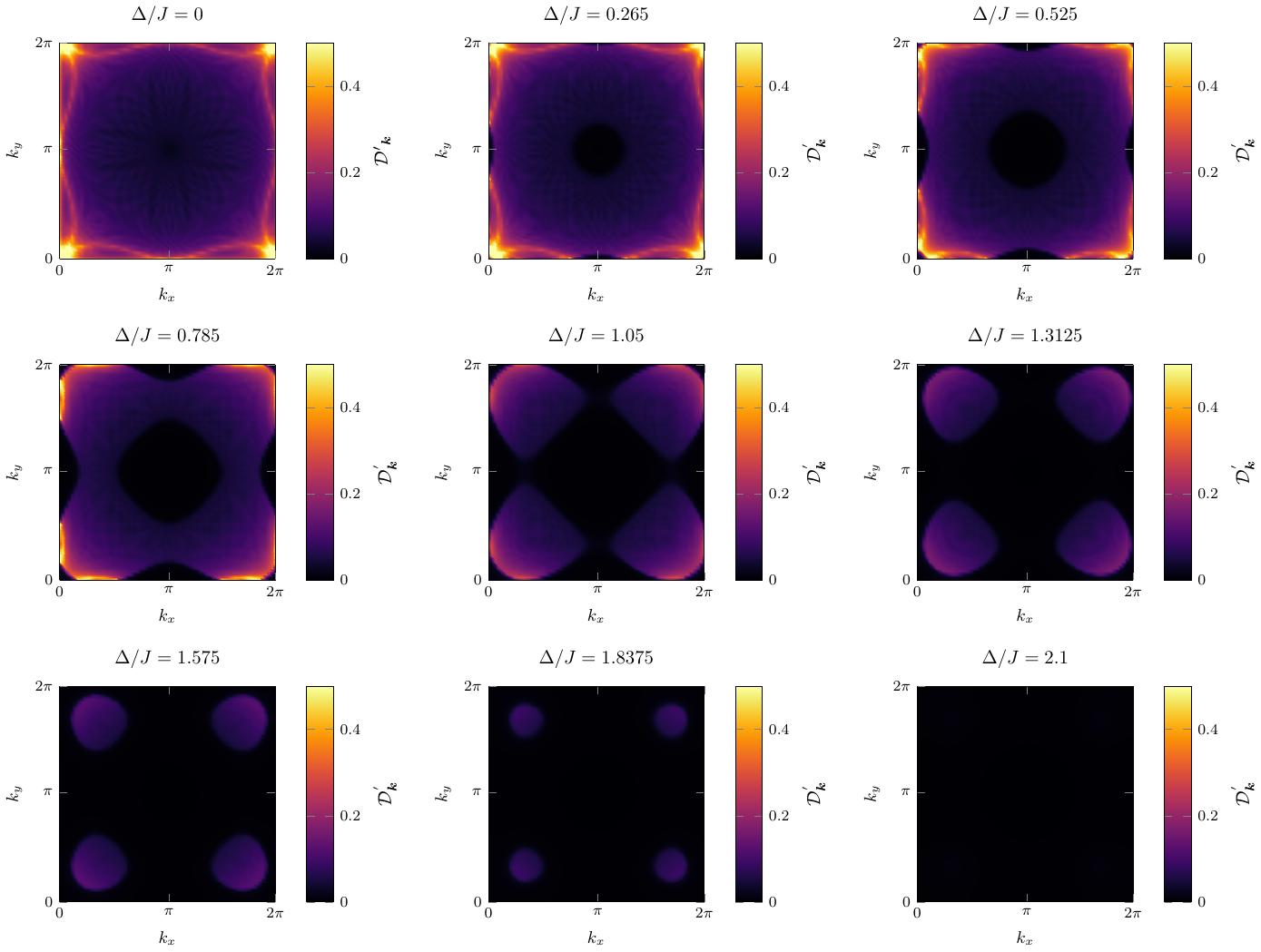}
     \caption{The collision DOS $\mathcal{D}'_{\vec{k}}(\varepsilon_{\vec{k}})$, quantifying the number of available final states for a magnon to collide with another magnon, as a function of magnetic field $\Delta$ ($S=1$). The DOS is shown as a color-map in the first Brillouin zone for a growing magnetic field $\Delta / J$, with a minimum value of $\Delta / J = 0$ and a maximum of $\Delta / J = 2.1$.}
    \label{fig: DOS collision}
\end{figure} 

To illustrate this magnetic field dependence of the two scattering DOS, we plot $\mathcal{D}_{\vec{k}}(\varepsilon_{\vec{k}})$ and $\mathcal{D}'_{\vec{k}}(\varepsilon_{\vec{k}})$ in Figs.~\ref{fig:DOS decay} and \ref{fig: DOS collision}, respectively. The decrease and eventual vanishing of the DOS with increasing $\Delta$ is clearly visible. Besides the critical field $\Delta' = 2JS$, we identify $\Delta^*=JS$ as a special value because there is a Lifshitz-type transition from a single area of finite DOS to four disconnected areas. To analyze the number of available final scattering states further, we calculate the integrated DOS
\begin{align}
    \begin{split}
        L(\Delta) &= \frac{1}{N} \sum_{\vec{k}} \mathcal{D}_{\vec{k}}(\varepsilon_{\vec{k}}),
        \\
        L'(\Delta) &= \frac{1}{N} \sum_{\vec{k}} \mathcal{D}'_{\vec{k}}(\varepsilon_{\vec{k}}),
    \end{split}
\end{align}
which we plot in Fig.~\ref{fig:KH and DOS}(a) as a function of $\Delta$, where $\Delta^* = J$ (for $S=1$) can be identified as an inflection point.

\begin{figure}
     \centering
    \includegraphics[scale=1]{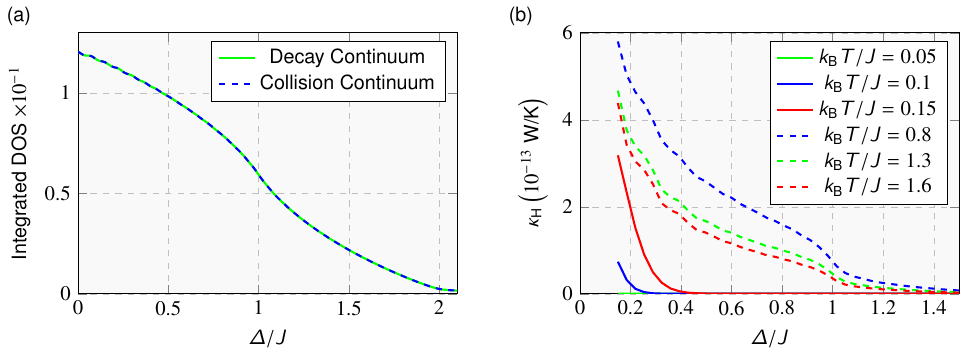}
    \caption{(a) Integrated DOS $L^{(')}$ as a function of the magnetic field $\Delta / J$, exhibiting an inflection point at $\Delta = \Delta^* = J$ and vanishing at $\Delta = \Delta' = 2J$. (b) Thermal Hall conductivity $\kappa_\text{H}$ as a function of the magnetic field $\Delta / J$ for six different temperatures. While for temperatures smaller than $\Delta$, $\kappa_\text{H}$ is too small to make out the inflection point (solid lines), there is a pronounced drop of $\kappa_\text{H}$ at $\Delta^*$ at temperatures as large as the magnetic field (dashed lines). We set $S=1$.}
    \label{fig:KH and DOS}
\end{figure}

This inflection point causes a direct signature in the thermal Hall conductivity $\kappa_\text{H}$, which is depicted in Fig.~\ref{fig:KH and DOS}(b). First, note that at temperatures much smaller than $\Delta^*$ (solid lines), $\kappa_\text{H}$ is suppressed by the spin-wave gap and, hence, too small to exhibit the inflection point. However, for temperatures as large as $\Delta^*$, the inflection point leads to a pronounced drop of $\kappa_\text{H}$ (dashed lines). How exactly the inflection point translates into the behavior of $\kappa_\text{H}$ cannot be answered by kinematic phase space arguments alone because they neglect the scattering matrix elements and the thermal populations. However, we can conclude from these results that the magnetic field is the central external parameter to control thermal Hall transport, as it can completely freeze out any processes giving rise to magnon-magnon skew scattering.
\end{subsection}

\newpage

\begin{subsection}{Temperature-magnetic field maps of the thermal conductivities}
\label{sub: T-B maps}
\begin{figure}
    \centering
    \includegraphics{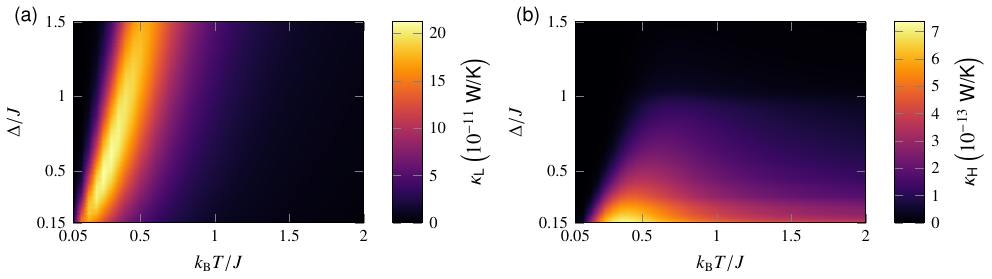}
    \caption{Temperature-field maps of (a) the longitudinal thermal conductivity $\kappa_\text{L}$, and (b) the thermal Hall conductivity $\kappa_\text{H}$. 
    }
    \label{fig:color plots}
\end{figure}

In the main text, we have shown selected constant-field cuts of the thermal conductivities. Here, we provide the full temperature and field dependence of $\kappa_\text{L}$ and $\kappa_\text{H}$ in color-map form, see Fig.~\ref{fig:color plots}. The calculation is based on Eq.~\eqref{eq: conductitivites restated} and adopts the approximations described in Sec.~\ref{subsection: D's in the calculation}. A $\vec{k}$-mesh of $80 \times 80$ points was used, for a total of $N_T=40$ equidistant $\tilde{T}$-values and a total of $N_\Delta=40$ equidistant magnetic field values. See Sec.~\ref{subsection: D's in the calculation} for further technical details.

For $\kappa_\text{L}$ in Fig.~\ref{fig:color plots}(a), we can clearly see how it peaks as a function of temperature, with the peak being a result of the low-temperature activation behavior and the high-temperature Umklapp scattering. As $\Delta$ increases, the peak gets shifted to higher temperatures because the spin-wave gap increases. For a further discussion of the trends of $\kappa_\text{L}$, see Sec.~\ref{sec:Temperaturedependence}.

The thermal Hall conductivity $\kappa_\text{H}$ in Fig.~\ref{fig:color plots}(b) exhibits a different behavior. First, the activation-like low-temperature trend is more pronounced, which is the result of $\kappa_\text{H}$'s being caused by higher-order scattering (recall Sec.~\ref{sec:Temperaturedependence}). Second, the high-temperature tail does not vanish, which, as explained in Sec.~\ref{sec:Temperaturedependence}, is an artifact of the approximations made. Finally, we identify the pronounced decrease of $\kappa_\text{H}$ at the magnetic field $\Delta = J$, which is a result of scattering phase-space freezing, as discussed in Sec.~\ref{subsection: frozen phase space}.

As mentioned in the main text, Refs.~\onlinecite{Hou2017thermally, Carnahan2021} have considered the classical version of a chiral two-dimensional ferromagnet based on spin Hamiltonian \eqref{eq:fullhamil} (with spin operators replaced by three-dimensional unit vectors). It was found that thermal fluctuations give rise to a finite thermodynamic average of the scalar spin chirality $\langle \chi \rangle = \langle \vec{S}_i \cdot (\vec{S}_j \times \vec{S}_k)\rangle$ even when the averaged magnetic moments do not form a skyrmion crystal \cite{Hou2017thermally}. At a magnetic field of $\Delta / J_\text{cl} = 0.2$, a peak of $\langle \chi \rangle$ was identified just below $T/J_\text{cl} \approx 1$ \cite{Hou2017thermally}. We use $J_\text{cl}$ to denote the exchange constant assumed in the classical simulations. A finite scalar spin chirality is a typical source of Hall effects and, indeed, classical spin dynamics simulations based on the stochastic Landau-Lifshitz-Gilbert equation have revealed a finite $\kappa_\text{H}$ in the fluctuating phase, peaking around $T/J_\text{cl} \approx 0.75$ at $\Delta / J_\text{cl} = 0.2$ \cite{Carnahan2021}. 
In our calculations, we find that $\kappa_\text{H}$ peaks around $T/J \approx 0.4$ at $\Delta / J = 0.2$ for $S=1$. We attribute the difference in the peak position to the difference in the classical and quantum spin Hamiltonians, with the appropriate rescaling between the two requiring further analysis. We therefore suggest that the results of Refs.~\onlinecite{Carnahan2021} are the classical limit of our quantum theory of magnon-magnon skew scattering: The time-reversal symmetry breaking of the microscopic three-magnon interaction vertices gives rise to a finite average of the scalar spin chirality as well as the thermal Hall effect.

\end{subsection}
\end{section}

\begin{section}{Effective Berry curvature theory at large magnetic fields}
\label{section: anomalous berry curvature}
The many-body magnon-magnon skew scattering was found to vanish when the magnetic field $\Delta$ exceeds the critical value $\Delta' = 2JS$. This occurs because no three-magnon scattering processes can be resonant, which was identified as a necessary condition for generating the anti-symmetric component of the scattering kernel. Consequently, for $\Delta > \Delta'$, particle number non-conserving many-body interactions contribute exclusively through off-resonant effects.

In the harmonic theory, the system is characterized by a single magnon band. While a single band would preclude Berry curvature for electrons, the situation is more nuanced for magnons. Unlike electrons, magnons can exhibit anomalous coupling or ``pairing'' effects due to their non-conserved number. Such anomalous terms, of the form $a^\dagger a^\dagger$, couple the particle and hole sectors. If these couplings break time-reversal symmetry, Berry curvature can emerge even in the presence of a single physical band.

In the current model, the harmonic theory lacks anomalous couplings because the classical field-polarized state is the exact quantum mechanical ground state. As a result, the magnon number is spuriously conserved within the harmonic approximation. However, this spurious symmetry does not extend to the interacting theory. Below, we show that three-magnon scattering, apart from inducing asymmetric scattering rates that lead to the mechanism of MMSS, can also generate temperature-dependent anomalous self-energies.
The resulting coupling of particle and hole sectors induces a Berry curvature in an effective harmonic theory because the three-magnon interactions break the time-reversal symmetry. As a result, there is a small but finite band-geometric contribution to the thermal Hall conductivity, denoted by $\kappa_{\text{H}}^{\text{geo}}$. 

The main idea is to rewrite the harmonic theory explicitly in Bogoliubov-de-Gennes form, that is,
\begin{align}
    H_2 = \frac{1}{2} \sum_{\vec{k}}
    (a^\dagger_{\vec{k}}, a_{-\vec{k}})
    \mathcal{E}_{\vec{k}}
    \begin{pmatrix}
        a_{\vec{k}}  \\ a^\dagger_{-\vec{k}} 
    \end{pmatrix} 
    - 
    \frac{1}{2} \sum_{\vec{k}} \varepsilon_{\vec{k}}, 
    \label{eq:harmonic-in-BdG}
\end{align}
where
\begin{align}
    \mathcal{E}_{\vec{k}} =
    \begin{pmatrix}
        \varepsilon_{\vec{k}} & 0 \\ 0 & \varepsilon_{-\vec{k}}
    \end{pmatrix},
\end{align}
and then carry out a many-body perturbation theory. Thus, one defines a temperature Green's function matrix
\begin{align}
    \mathcal{G}(\tau,k) = 
    - \left\langle \mathcal{T}_\tau \begin{pmatrix}
        a_{\vec{k}}(\tau) a^\dagger_{\vec{k}} & a_{\vec{k}}(\tau) a_{-\vec{k}}  \\ 
        a^\dagger_{-\vec{k}} (\tau) a^\dagger_{\vec{k}} &  a^\dagger_{-\vec{k}} (\tau)  a_{-\vec{k}}
    \end{pmatrix}
    \right\rangle,
\end{align}
with $\mathcal{T}_\tau$ denoting the ordering of imaginary time $\tau$. The angular brackets indicate the thermal average. To transform between imaginary time and Matsubara frequency space, one uses
\begin{align}
    \mathcal{G}(\mathrm{i}\omega_n, {\vec{k}}) = \int_0^\beta \mathrm{e}^{\mathrm{i}\omega_n \tau} \mathcal{G}(\tau,{\vec{k}}) \, \mathrm{d}\tau ,
\end{align}
and its inverse transformation 
\begin{align}
    \mathcal{G}(\tau,\vec{k})
    =
    \frac{1}{\beta} \sum_n \mathrm{e}^{-\mathrm{i}\omega_n\tau} \mathcal{G}(\mathrm{i}\omega_n, \vec{k}).
\end{align}
The non-interacting Green's function is given by
\begin{align}
    \mathcal{G}^{(0)}(\mathrm{i}\omega_n, \vec{k}) 
    &= 
    - \int_0^\beta \mathrm{e}^{\mathrm{i}\omega_n \tau} 
    \begin{pmatrix}
        (1+\overline{N}_{\vec{k}}) \mathrm{e}^{-\varepsilon_{\vec{k}} \tau} & 0 \\
        0 & \overline{N}_{-\vec{k}} \mathrm{e}^{\varepsilon_{-\vec{k}} \tau}
    \end{pmatrix} \, \mathrm{d}\tau 
    \nonumber \\
    &=
    \left( \sigma_3 \mathrm{i}\omega_n -  \mathcal{E}_{\vec{k}} \right)^{-1},
\end{align}
where $\sigma_3 = \text{diag}(1,-1)$ is the third Pauli matrix, and $\overline{N}_{\vec{k}} = (\mathrm{e}^{\beta \varepsilon_{\vec{k}}} - 1)^{-1}$ the Bose-Einstein function. Only the diagonal entries of $\mathcal{G}^{(0)}(\mathrm{i}\omega_n, \vec{k})$, that is, the \textit{normal} Green's functions, are nonzero in the non-interacting limit. The off-diagonal entries, called the anomalous Green's functions, are zero.
The interacting Green's function is given by
\begin{align}
    \mathcal{G}^{-1}(\mathrm{i} \omega_n,\vec{k}) 
    = 
    \sigma_3\mathrm{i}\omega_n - \mathcal{E}_{\vec{k}} - \Sigma(\mathrm{i} \omega_n,\vec{k}),
\end{align}
with the self-energy matrix
\begin{align}
    \Sigma(\mathrm{i} \omega_n,\vec{k})
    =
    \begin{pmatrix}
        \Sigma_{11}(\mathrm{i} \omega_n,\vec{k}) & \Sigma_{12}(\mathrm{i} \omega_n,\vec{k}) \\ \Sigma_{21}(\mathrm{i} \omega_n,\vec{k}) & \Sigma_{22}(\mathrm{i} \omega_n,\vec{k})
    \end{pmatrix}.
\end{align}
Its diagonal (off-diagonal) entries are normal (anomalous) self-energies.

To leading order in the $1/S$ expansion there are three-magnon and four-magnon contributions. The interacting Green's function can be approximated by $\mathcal{G}(\tau,\vec{k}) \approx \mathcal{G}^{(0)}(\tau,\vec{k}) + \mathcal{G}^{(1)}(\tau,\vec{k}) + \mathcal{G}^{(2)}(\tau,\vec{k})$, where the first-order term $\mathcal{G}^{(1)}(\tau,\vec{k})$ contains a four-magnon vertex, and the second-order term contains two three-magnon vertices,
\begin{align}
    \mathcal{G}^{(2)}(\tau,\vec{k}) \nonumber 
    =
    - \frac{1}{2}
    \int_0^\beta \mathrm{d}\tau_1  \int_0^\beta \mathrm{d}\tau_2
    \left\langle \mathcal{T}_\tau 
    H_3(\tau_1) H_3(\tau_2)
    \begin{pmatrix}
        a_{\vec{k}}(\tau) a^\dagger_{\vec{k}} & a_{\vec{k}}(\tau) a_{-\vec{k}}  \\ 
        a^\dagger_{-\vec{k}} (\tau) a^\dagger_{\vec{k}} &  a^\dagger_{-\vec{k}} (\tau)  a_{-\vec{k}}
    \end{pmatrix}
    \right\rangle_0^\text{c}.
\end{align}
Here, $\langle \cdot \rangle_0^\text{c}$ indicates the thermal average over the non-interacting theory of connected diagrams.
Since we are after the effect of time-reversal symmetry breaking, we drop the four-magnon interactions in $\mathcal{G}^{(1)}(\tau,\vec{k})$, and only consider the three-magnon interactions in $\mathcal{G}^{(2)}(\tau,\vec{k})$. The relevant self-energies arise from the bubble diagrams. The corresponding normal self-energies are given by
\begin{align}
    \Sigma_{11}(\mathrm{i} \omega,\vec{k}) = &\frac{1}{2N}
    \sum_{\vec{q}}
    \left|V_{\vec{k}; \vec{q}, \vec{k}-\vec{q}} \right|^2 
    \frac{\overline{N}_{\vec{q}}+\overline{N}_{\vec{k}-\vec{q}} + 1}{\mathrm{i}\omega-\varepsilon_{\vec{q}} - \varepsilon_{\vec{k}-\vec{q}}} \nonumber
    +
    \frac{1}{N}
    \sum_{\vec{q}} 
    \left| V_{\vec{k} + \vec{q};\vec{k}, \vec{q} } \right|^2
    \frac{\overline{N}_{\vec{q}} - \overline{N}_{\vec{k}+\vec{q}}}{\mathrm{i}\omega + \varepsilon_{\vec{q}}-\varepsilon_{\vec{k}+\vec{q}}},
    \label{eq:normal_self_energy}
\end{align}
and
\begin{align}
    \Sigma_{22}(\mathrm{i}\omega,\vec{k})
    =
    \frac{1}{2N}
    \sum_{\vec{q}} 
    \left| V_{-\vec{k};\vec{q}, -\vec{k}-\vec{q}} \right|^2
    \frac{-\overline{N}_{\vec{q}} - \overline{N}_{-\vec{k}-\vec{q}} - 1 }{\mathrm{i}\omega + \varepsilon_{\vec{q}} + \varepsilon_{-\vec{k}-\vec{q}}}
    \nonumber 
    +\frac{1}{N} \sum_{\vec{q}} \left| V_{-\vec{k} + \vec{q};-\vec{k}, \vec{q}} \right|^2 
    \frac{\overline{N}_{-\vec{k}+\vec{q}}-\overline{N}_{\vec{q}}}{
    \mathrm{i}\omega+\varepsilon_{-\vec{k}+\vec{q}}-\varepsilon_{\vec{q}}
    }.
\end{align}
They are finite even at zero temperature.
In contrast, anomalous self-energies only appear at finite temperatures because the classical ferromagnetic state is the exact ground state;
\begin{equation}
    \Sigma_{12}(\mathrm{i}\omega,\vec{k})
    =
    \frac{1}{N}
    \sum_{ \vec{q}} 
    V_{\vec{q};-\vec{k}, \vec{k}+\vec{q}} 
    V_{\vec{k} + \vec{q};\vec{k}, \vec{q}}
    \frac{\overline{N}_{\vec{q}}-\overline{N}_{\vec{k}+\vec{q}}}{\mathrm{i}\omega +\varepsilon_{\vec{q}}
    -\varepsilon_{\vec{k}+\vec{q}}},
    \label{eq:anomalous_self_energy}
\end{equation}
and
\begin{equation}
    \Sigma_{21}(\mathrm{i}\omega,\vec{k})
    =
    \frac{1}{N} \sum_{\vec{q}} 
    V^\ast_{\vec{q};\vec{k}, \vec{q}-\vec{k}} 
    V^\ast_{ -\vec{k} + \vec{q};-\vec{k}, \vec{q}} \frac{\overline{N}_{\vec{q}-\vec{k}}-\overline{N}_{\vec{q}}}{\mathrm{i}\omega + \varepsilon_{\vec{q}-\vec{k}} - \varepsilon_{\vec{q}}}.
\end{equation}
Beyond the dependence on temperature, the main difference between normal and anomalous self-energies is that the complex three-magnon interaction vertices form an absolute square in the normal self-energies but not so in the anomalous self-energies.

In the spirit of Refs.~\cite{McClarty2019,mook2021}, we build an effective harmonic Hamiltonian $H_{\vec{k}}^{\text{eff}}$ from $\mathcal{E}_{\vec{k}} +  \Sigma(\mathrm{i} \omega,\vec{k})$. We evaluate the self-energies on-shell and perform the analytical continuation $\mathrm{i} \omega \to \omega + \mathrm{i}\eta^+$. 
We consider two cases, (i) $\Delta > \Delta' = 2 J$ and (ii) $\Delta < \Delta'$.
In the first case, the damping effects of the three-magnon interactions are kinematically frozen and the regularization parameter $\eta$ of the analytical continuation can be ignored. In the second case, the energy denominator of the self energies exhibits a finite imaginary part and $\mathcal{E}_{\vec{k}} + \Sigma(\omega+\mathrm{i}\eta,\vec{k})$ becomes non-Hermitian, which can be interpreted as a magnon lifetime. In principle, the resulting spectral broadening can be incorporated into the band-geometric thermal Hall conductivity \cite{Koyama2024}. However, since we expect the broadening to cause an overall suppression of transport, we decide to neglect it, and only consider the Hermitian part of the effective Hamiltonian. Therefore, we define
\begin{equation}
\begin{split}
   & 
    \Sigma_{\vec{k}}
    \equiv
    \Sigma_{11}(\varepsilon_{\vec{k}},\vec{k}) 
    = \frac{1}{2N}
    \sum_{\vec{q}}
    \left|V_{\vec{k};\vec{q}, \vec{k}-\vec{q}} \right|^2 
    \left(\overline{N}_{\vec{q}}+\overline{N}_{\vec{k}-\vec{q}} + 1 \right)\mathcal{P}\left[
    \frac{1}{\varepsilon_{\vec{k}} -\varepsilon_{\vec{q}} - \varepsilon_{\vec{k}-\vec{q}}} \right] \\
    &\qquad \qquad \qquad \qquad+
    \frac{1}{N}
    \sum_{\vec{q}} 
    \left| V_{\vec{k} + \vec{q};\vec{k}, \vec{q}} \right|^2 \left( \overline{N}_{\vec{q}} - \overline{N}_{\vec{k}+\vec{q}}\right) 
    \mathcal{P}\left[\frac{1}{\varepsilon_{\vec{k}} + \varepsilon_{\vec{q}}-\varepsilon_{\vec{k}+\vec{q}}}\right], \\
    & 
    F_{\vec{k}}
    \equiv
    \Sigma_{12}(\varepsilon_{\vec{k}},\vec{k})
    =
    \frac{1}{N}
    \sum_{ \vec{q}} 
    V_{\vec{q};-\vec{k}, \vec{k}+\vec{q}} 
    V_{\vec{k} + \vec{q};\vec{k}, \vec{q}}
    \left( \overline{N}_{\vec{q}}-\overline{N}_{\vec{k} 
    +\vec{q}}\right)\mathcal{P}\left[\frac{1}{\varepsilon_{\vec{k}} + \varepsilon_{\vec{q}}
    -\varepsilon_{\vec{k}+\vec{q}}}\right],
\end{split}
\label{eq: on shell self energies}
\end{equation}
where $\mathcal{P}$ denotes the principal part, and build the effective Hamilton matrix
\begin{equation}
    H^\text{eff}_{\vec{k}} 
    = 
    \begin{pmatrix}
        \varepsilon_{\vec{k}} + \Sigma_{\vec{k}} &  F_{\vec{k}} \\  F^\ast_{\vec{k}} & \varepsilon_{\vec{k}} + \Sigma_{\vec{k}}
    \end{pmatrix},
    \label{eq:effective-ham}
\end{equation}
which replaces $\mathcal{E}_{\vec{k}}$ in Eq.~\eqref{eq:harmonic-in-BdG}.
In the above we made use of $\Sigma_{-\vec{k}} = \Sigma_{\vec{k}}$, $\varepsilon_{\vec{k}} = \varepsilon_{-\vec{k}}$, and $V_{\vec{k}_3;\vec{k}_1, \vec{k}_2} = -V_{-\vec{k}_3;-\vec{k}_1, -\vec{k}_2}$.
The effective Hamilton matrix in Eq.~\eqref{eq:effective-ham} breaks time-reversal symmetry because $F^\ast_{\vec{k}} \ne F_{-\vec{k}}$, and depends on temperature. In particular, the effective pairing term $F_{\vec{k}}$ requires finite temperatures. 

Next, we carry out a Bogoliubov transformation to diagonalize $H^\text{eff}_{\vec{k}}$, which is done by means of the paraunitary matrix $T_{\vec{k}}$. To keep the Bose-Einstein statistics of the magnons intact we require that the relation 
\begin{equation}
    T^\dagger_{\vec{k}} \sigma_3 T_{\vec{k}} = \sigma_3,
    \label{eq: BE statistics}
\end{equation}
holds. In our case it is easy to prove that the following matrix 
\begin{align}
    T_{\vec{k}} 
    = 
    \begin{pmatrix}
        \mathrm{e}^{\mathrm{i} \gamma_{\vec{k}}} 
        \cosh\frac{\zeta_{\vec{k}}}{2} 
        &
        \sinh\frac{\zeta_{\vec{k}}}{2} 
        \\
        \sinh\frac{\zeta_{\vec{k}}}{2}
        &
        \mathrm{e}^{-\mathrm{i} \gamma_{\vec{k}}} \cosh\frac{\zeta_{\vec{k}}}{2} 
    \end{pmatrix},
\end{align}
with $\cosh\zeta_{\vec{k}} = \left(\varepsilon_{\vec{k}} + \Sigma_{\vec{k}} \right) / \varepsilon^\text{eff}_{\vec{k}}$,  $\;\sinh\zeta_{\vec{k}} = - |F_{\vec{k}}| / \varepsilon^\text{eff}_{\vec{k}} $ and $ F_{\vec{k}} = | F_{\vec{k}} | \mathrm{e}^{\mathrm{i} \gamma_{\vec{k}}} $, 
obeys the relation of Eq.~\eqref{eq: BE statistics}.
Using $T_{\vec{k}}$ and $T_{\vec{K}}^{\dagger}$ we can also diagonalize the effective Hamilton kernel 
\begin{align}
    \mathcal{E}^\text{eff}_{\vec{k}}
    \equiv
    T^\dagger_{\vec{k}} H^\text{eff}_{\vec{k}}  T_{\vec{k}}
    =
    \begin{pmatrix}
        \varepsilon^\text{eff}_{\vec{k}}
        &
        0 
        \\
        0 
        &
        \varepsilon^\text{eff}_{-\vec{k}}
    \end{pmatrix},
\end{align}
where 
\begin{align}
    \varepsilon^\text{eff}_{\vec{k}} = \sqrt{ (\varepsilon_{\vec{k}} + \Sigma_{\vec{k}})^2 - |F_{\vec{k}} |^2 },
    \label{eq:corrected_spectrum}
\end{align}
is the interaction-renormalized magnon energy. 
It holds $\varepsilon^\text{eff}_{\vec{k}} = \varepsilon^\text{eff}_{-\vec{k}}$. Note that $\varepsilon^\text{eff}_{\vec{k}}$ is not strictly consistent in $1/S$ because $\Sigma_{\vec{k}}$ provides a $1/S$ correction to $\varepsilon_{\vec{k}}$ and $F_{\vec{k}}$ a $1/S^2$ correction, as can be seen by expanding the square root in Eq.~\eqref{eq:corrected_spectrum}. Thus, consistency in $1/S$ would require adding $1/S^2$ corrections to the normal self-energy. However, as far as generating a finite Berry curvature is concerned, the anomalous $F_{\vec{k}}$ is the leading contribution. 

The Berry curvature of the magnon bands can be computed numerically using the formula \cite{Shindou13, THE_of_magnons_collinear_antiferromagnets}
\begin{equation}
    \Omega_{\vec{k}, n}^{xy} = - 2 \sum_{\substack{m = 1,2 \\ m\neq n}}\text{Im} \left[\frac{\left(\sigma_3 T_{\vec{k}}^{\dagger} \partial_x H_{\vec{k}}^{\text{eff}} T_{\vec{k}} \right)_{nm} \left(\sigma_3 T_{\vec{k}}^{\dagger} \partial_y H_{\vec{k}}^{\text{eff}} T_{\vec{k}} \right)_{mn}}{\left[\left(\sigma_3  \mathcal{E}_{\vec{k}}^{\text{eff}} \right)_{nn} - \left(\sigma_3  \mathcal{E}_{\vec{k}}^{\text{eff}} \right)_{mm} \right]^2} \right]. 
    \label{eq: Berry curvature numerical 1}
\end{equation}
Since only  the Berry curvature of the particle sector is physically meaningful, we get from the above 
\begin{equation}
    \Omega_{\vec{k}} \equiv \Omega_{\vec{k}, 1}^{xy}   = 
    - \frac{1}{2\left(\varepsilon^\text{eff}_{\vec{k}}\right)^2}  \text{Im}\left[\left(\sigma_3 T_{\vec{k}}^{\dagger} \partial_x H_{\vec{k}}^{\text{eff}} T_{\vec{k}} \right)_{12} \left(\sigma_3 T_{\vec{k}}^{\dagger} \partial_y H_{\vec{k}}^{\text{eff}} T_{\vec{k}} \right)_{21}\right]. \
    \label{eq: Berry curvature numerical 2}
\end{equation}

\begin{figure}[h]
    \centering
    \includegraphics[width=1\linewidth]{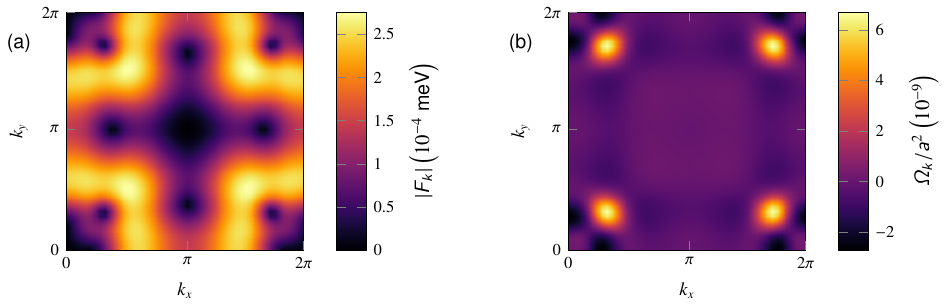}
    \caption{(a) The magnitude of the anomalous self energy $F_{\vec{k}}$ in units of $\text{meV}$ and (b) the Berry curvature divided by $a^2$, where $a$ is the lattice constant, plotted within the first Brillouin zone. Parameters read $\Delta = 2.3 \; J$, $D=0.1J$, $k_\text{B}T/J = 2$, and $S=1$, with $J = 1 \; \text{meV}$.}
    \label{fig:Berry and F}
\end{figure}

For the numerical evaluation of the self-energies, we use a finite regularization parameter $\eta = 0.1 J$ for the case $\Delta < \Delta'$, and set $\eta =0$ at $\Delta > \Delta'$. 
In Fig.~\ref{fig:Berry and F}(a,b), we respectively plot the magnitude of the anomalous self energy $\left|F_{\vec{k}}\right|$ and the Berry curvature $\Omega_{\vec{k}}$ at $\Delta = 2.3 \; J$, $k_\text{B}T/J = 2$, $D=0.1 J$, and $S=1$, with $J = 1 \; \text{meV}$. Both $\left| F_{\vec{k}}\right|$ and $\Omega_{\vec{k}}$ respect the four-fold rotation symmetry of the square lattice, but are very small: $\left|F_{\vec{k}}\right|$ is four to five orders of magnitude smaller than the bare magnon spectrum, while $\Omega_{\vec{k}}$ is of the order of magnitude of $10 ^{-9}$ in units of $a^2$, where $a$ is the lattice constant of the square lattice. For comparison, a magnon band with finite Chern number would carry an average Berry curvature of $1/(2\pi) \approx 1.6 \times 10 ^{-1}$ in units of $a^2$.
Lowering the magnetic field, increases the thermal population of magnons, and, in turn, $\left|F_{\vec{k}}\right|$ and $\Omega_{\vec{k}}$.
In the limit of $\Delta < \Delta'$, the Berry curvature grows up to $\Omega_{\vec{k}} / a^2 \sim 10^{-4}$.
\begin{figure}
    \centering
    \includegraphics[width=1\linewidth]{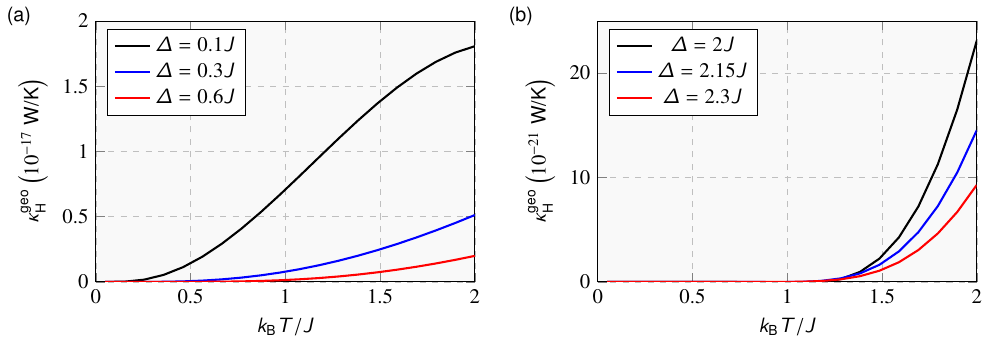}
    \caption{The Hall band geometric conductivity induced by the many-body corrections to the bare magnon spectrum, plotted as a function of the rescaled temperature and for values of the magnetic field correspoding to the two different physical regimes of $\Delta < \Delta' = 2\;J$ and of $\Delta > \Delta' = 2\;J$ . Since the corrections are thermally activated and stem from the DMI, the resulting Hall conductivity is quite small for both regimes.}
    \label{fig: anomalous geometric Hall}
\end{figure}

We evaluate the band geometric thermal Hall conductivity $\kappa_{\text{H}}^{\text{geo}}$ by means of its expression within the noninteracting theory, that is, \cite{Matsumoto2011PRL}
\begin{equation}
    \kappa_{\text{H}}^{\text{geo}} = -\frac{k_{\text{B}}^2 T}{\hbar A} \sum_{ \vec{k}}c_2\left( \overline{N}_{\vec{k}}\right) \Omega_{\vec{k}},
    \label{eq: geo kH}
\end{equation}
where $c_2\left(x\right) = \left( 1+ x\right)\left(\log\frac{1 + x}{x} \right)^2 - \left(\log x \right)^2 - 2\text{Li}_2\left(-x\right)$, and $\text{Li}_2\left(x\right)$ is the dilogarithm function. $A$ is the area of the sample and $\overline{N}_{\vec{k}} = \overline{N}\left(\varepsilon^{\text{eff}}_{\vec{k}} \right)$ is the Bose-Einstein distribution which takes the effective energy given in Eq.~\eqref{eq:corrected_spectrum} as input.
In Fig.~\ref{fig: anomalous geometric Hall}(a,b), $\kappa_{\text{H}}^{\text{geo}}$ is shown as a function of temperature for selected values of $\Delta / J$ in the limits $\Delta < \Delta'$ and $\Delta > \Delta'$, respectively. In the latter limit, $\kappa_{\text{H}}^{\text{geo}}$ is completely negligible, as expected from the tiny Berry curvature. 
Even for $\Delta < \Delta'$, we find $\kappa_{\text{H}}^{\text{geo}} \sim 10^{-17} \;W/K$, around four orders of magnitude smaller than the thermal Hall conductivity obtained from MMSS derived in the main part of our work. We conclude that the contribution to the thermal Hall effect from magnon band geometry is negligible in our model.
\end{section}

\end{widetext}

\bibliography{bib}

\end{document}